\documentclass[9pt]{amsart}

\usepackage{latexsym,pifont}
\usepackage{epsfig}
\usepackage{rotating}

\usepackage{graphicx}

\usepackage{color}

\usepackage{ifpdf}
\ifpdf
\usepackage{epstopdf}
\usepackage{hyperref}
\else
\usepackage[hypertex]{hyperref}
\fi

\usepackage{amsfonts,amsmath,amssymb,mathrsfs,extarrows,MnSymbol}
\usepackage{tikzsymbols}
\usepackage[all]{xy}

\xyoption{knot}

\usepackage{tikz}
\usetikzlibrary{matrix,arrows}

\usepackage[makeroom]{cancel}

\newtheorem{theorem}{Theorem}[section]
\newtheorem*{theorem*}{Theorem}
\newtheorem{proposition}[theorem]{Proposition}

\newtheorem{remark}[theorem]{Remark}
\newtheorem{example}[theorem]{Example}
\newtheorem{definition}[theorem]{Definition}
\newtheorem{corollary}[theorem]{Corollary}
\newtheorem{lemma}[theorem]{Lemma}

\newtheorem{propanition}[theorem]{Proposition/Definition}

\newtheorem{conjecture}[theorem]{Conjecture}
\newtheorem{convention}[theorem]{Convention}

\newcommand{\alxydim}[2]{\begin{aligned}\xymatrix#1{#2}\end{aligned}}

\newcommand{\brem}{\begin{remark}}
	\newcommand{\erem}{\end{remark}}
\newcommand{\beg}{\begin{example}}
	\newcommand{\eeg}{\end{example}}
\newcommand{\bedef}{\begin{definition}}
	\newcommand{\exdef}{\end{definition}}
\newcommand{\berop}{\begin{proposition}}
	\newcommand{\eerop}{\end{proposition}}
\newcommand{\belem}{\begin{lemma}}
	\newcommand{\elem}{\end{lemma}}
\newcommand{\bethe}{\begin{theorem}}
	\newcommand{\ethe}{\end{theorem}}
\newcommand{\becor}{\begin{corollary}}
	\newcommand{\ecor}{\end{corollary}}
\newcommand{\beroof}{\noindent\begin{proof}}
	\newcommand{\eroof}{\end{proof}}
\newcommand{\becon}{\begin{convention}}
	\newcommand{\econ}{\end{convention}}
	\newcommand{\efact}{\begin{flushright}$\checkmark$\end{flushright}\end{Fact}}
\newcommand{\becj}{\begin{conjecture}}
	\newcommand{\ecj}{\begin{flushright}$\boxtimes$\end{flushright}\end{conjecture}}

\newcommand{\barr}{\begin{array}}
	\newcommand{\earr}{\end{array}}
\newcommand{\ben}{\begin{enumerate}}
	\newcommand{\een}{\end{enumerate}}
\newcommand{\bit}{\begin{itemize}}
	\newcommand{\eit}{\end{itemize}}

\newcommand{\qq}{\begin{eqnarray}}
	\newcommand{\qqq}{\end{eqnarray}}

\newcommand{\nn}{\nonumber}

\newcommand{\ovl}[1]{\overline{#1}}
\newcommand{\unl}[1]{\underline{#1}}

\newcommand{\Reqref}[1]{Eq.\,\eqref{#1}}

\newcommand\void[1]{}

\newcommand{\tx}[1]{\textrm{#1}} 

\newcommand{\gt}[1]{\mathfrak{#1}}

\def\cA{\mathcal{A}}

\def\cC{\mathcal{C}}
\def\cD{\mathcal{D}}
\def\cE{\mathcal{E}}

\def\cG{\mathcal{G}}

\def\cI{\mathcal{I}}
\def\cJ{\mathcal{J}}
\def\cK{\mathcal{K}}
\def\ceL{\mathcal{L}}

\def\cO{\mathcal{O}}
\def\xcP{\mathcal{P}}

\def\cR{\mathcal{R}}
\def\cS{\mathcal{S}}
\def\cT{\mathcal{T}}
\def\cU{\mathcal{U}}
\def\cV{\mathcal{V}}
\def\cW{\mathcal{W}}

\def\xcC{\mathscr{C}}
\def\xcD{\mathscr{D}}
\def\xcE{\mathscr{E}}
\def\xcF{\mathscr{F}}
\def\xcG{\mathscr{G}}

\def\xcL{\mathscr{L}}
\def\xcM{\mathscr{M}}

\def\xcP{\mathscr{P}}

\def\xcV{\mathscr{V}}
\def\xcW{\mathscr{W}}

\def\bbB{\mathbf{B}}

\def\t{\mathbf{t}}

\def\bB{{\mathbb{B}}}
\def\bC{{\mathbb{C}}}

\def\bH{{\mathbb{H}}}

\def\bN{{\mathbb{N}}}

\def\bR{{\mathbb{R}}}

\def\bZ{{\mathbb{Z}}}

\def\a{\alpha}
\def\b{\beta}
\def\g{\gamma}
\def\G{\Gamma}
\def\d{\delta}
\def\D{\Delta}

\def\vep{\varepsilon}

\def\la{\lambda}

\def\om{\omega}
\def\Om{\Omega}

\def\si{\sigma}
\def\Si{\Sigma}

\def\t{\tau}

\def\Ups{\Upsilon}
\def\z{\zeta}

\def\ggt{\gt{g}}

\newcommand{\sfd}{{\mathsf d}}

\newcommand{\sfi}{{\mathsf i}}
\newcommand{\sfI}{{\mathsf I}}

\newcommand{\sfJ}{{\mathsf J}}
\newcommand{\sfk}{{\mathsf k}}

\newcommand{\sfM}{{\mathsf M}}

\newcommand{\sfP}{{\mathsf P}}

\newcommand{\sfS}{{\mathsf S}}

\newcommand{\sfT}{{\mathsf T}}

\newcommand{\txa}{{\rm a}}
\newcommand{\txA}{{\rm A}}

\newcommand{\txB}{{\rm B}}

\newcommand{\txd}{{\rm d}}

\newcommand{\txE}{{\rm E}}

\newcommand{\txg}{{\rm g}}
\newcommand{\txG}{{\rm G}}

\newcommand{\txH}{{\rm H}}

\newcommand{\txm}{{\rm m}}

\newcommand{\txT}{{\rm T}}

\newcommand{\txV}{{\rm V}}

\def\vC{\check{C}}
\def\Cv{\v{C}}

\def\vd{\check{\d}}

\def\exp{{\rm exp}}
\def\id{{\rm id}}
\newcommand{\pr}{{\rm pr}}

\def\too{\longrightarrow}
\def\ev{{\rm ev}}

\newcommand{\grpd}[2]{\hspace{-5pt}\alxydim{@C=.5cm@R=1.cm}{ #1 \ar@<.25ex>[r] \ar@<-.25ex>[r] & #2}\hspace{-3pt}}

\def\obj{{\rm Ob}}

\def\Hom{{\rm Hom}}
\def\morf{{\rm Mor}}
\def\1morf{1{\rm -Mor}}
\def\2morf{2{\rm -Mor}}
\def\dim{{\rm dim}}
\def\im{{\rm im}}
\def\ker{{\rm ker}}

\def\End{{\rm End}}

\def\Ext{{\rm Ext}}
\newcommand{\Id}{{\rm Id}}
\def\Inv{{\rm Inv}}

\def\bgrb{\gt{BGrb}}

\newcommand{\Gr}{{\rm {\bf Gr}}}
\newcommand{\BisGr}{{\rm Bisec({\bf Gr})}}

\newcommand{\Set}{{\rm {\bf Set}}}

\newcommand{\Top}{{\rm {\bf Top}}}
\newcommand{\Man}{{\rm {\bf Man}}}

\def\Vol{{\rm Vol}}

\newcommand{\pLie}[1]{\,{-\hspace{-8pt}\xcL}_{#1}}
\def\p{\partial}

\newcommand{\Diff}{{\rm Diff}}

\def\curv{{\rm curv}}
\def\Hol{{\rm Hol}}
\def\Bun{{\rm {\bf Bun}}}
\newcommand{\breP}{{\rm \breve{P}}}

\def\bd1{{\boldsymbol{1}}}
\def\brd0{{\boldsymbol{0}}}

\def\det{{\rm det}}
\def\tr{{\rm tr}}

\def\Ad{{\rm Ad}}

\def\Lie{{\rm Lie}}

\newcommand{\uj}{{\rm U}(1)}

\newcommand\MCR{\theta_{\rm R}}

\def\x{\times}
\def\ox{\otimes}

\def\lx{{\hspace{-0.04cm}\ltimes\hspace{-0.05cm}}}
\def\rx{\rtimes}
\newcommand{\fibx}[2]{\hspace{-2pt}{}_{#1}\hspace{-3pt}\x_{#2}\hspace{-2pt}}
\def\ract{\vartriangleleft}
\def\lact{\vartriangleright}
\def\mact{\blacktriangleleft}
\def\mlact{\blacktriangleright}

\def\must{\stackrel{!}{=}}

\def\rstr{\big\vert}

\newcommand{\corr}[1]{\left\langle #1 \right\rangle}

\newcommand{\GBra}[2]{\lsem\,#1\,,\,#2\,\rsem}

\newcommand\colo{\hspace{-2pt}:}
\newenvironment{lemproof}{
	\noindent{\em Proof of Lemma:}\ }{\hfill $\blacksquare$ \newline\newline}

\numberwithin{equation}{section} 

\newcount\hour\newcount\minute
        \hour=\time \divide\hour by60 \minute=\time
        {\multiply\hour by60 \global\advance\minute by-\hour}
        \edef\militarytime{\number\hour:\ifnum\minute<10 0\fi\number\minute}

\begin{document}

\title{Gaugings of Groupoids,\ Strings in Shadows,\ \\ and Emergent Poisson $\si$-Models}

\author[R.R. Suszek]{\vspace{45pt}Rafa\l ~R.\ ~Suszek}
\address{Katedra Metod Matematycznych Fizyki,\ Wydzia\l ~Fizyki
	Uniwersytetu Warszawskiego,\ ul.\ Pasteura 5,\ PL-02-093 Warszawa,
	Poland} 
\email{suszek@fuw.edu.pl}

\begin{abstract}
The gauge principle is proposed for rigid Lie-groupoidal symmetries $\grpd{\xcG}{M}$ of the Polyakov--Alvarez--Gaw\c{e}dzki 2$d$ non-linear $\si$-model with metric target $(M,\txg_M)$ and the Wess--Zumino term given by a Cheeger--Simons differential character coming from an abelian gerbe $\cG$.\ The principle bases on the notion of principaloid bundle with connection $(\xcP,\Theta)$,\ introduced by Strobl and the Author.\ The descent of the model to the shadow of $\xcP$ is demonstrated to require a twisted $\xcG$-equivariant structure on $\cG$,\ prequantising a multiplicative extension $(H_M,\rho,0)$ of the gerbe's curvature $H_M$ to a 3-cocycle in the Bott--Schulman--Stasheff cohomology of the groupoid's nerve.\ The descent is accompanied by a combined $\txg_M$-isometric/$\rho$-holonomic reduction of the structure group of $\xcP$,\ and uses an augmentation of the original gerbe by a trivial one depending quadratically on $\Theta$.\ The latter couples to the field of the $\si$-model---in an extension of the scheme worked out for the action groupoid by Gaw\c{e}dzki,\ Waldorf and the Author---through the comomentum component of the Spencer pair of $\rho$,\ as described by Crainic {\it et al}.\ A fully fledged cohomological analysis of the relevant gauge anomaly and of inequivalent gaugings is presented.\ In the symplectic setting,\ the augmentation procedure is shown to lead to the emergence of the standard Poisson $\si$-model.\ Conversely,\ a coaugmentation of a (local) Poisson $\si$-model by a flat equivariant gerbe yields a novel field theory with the shadow of the underlying symplectic principaloid bundle as the configuration bundle,\ and with manifest lagrangean gauge symmetry.\ A simple Cartan-type associating mechanism is proposed to account for the reduction of the structure group of the principaloid bundle.\ The mechanism allows for the coupling of an arbitrary number of distinct charged matter-field species to a given gauge field. 
\end{abstract}

\maketitle

\begin{flushright}\vspace{-255pt}
{\em Dedicated to Thomas Strobl,\ a dear Friend,\\ on the happy occasion of His 60(+1).\ birthday.\\~\\ }\vspace{225pt}
\end{flushright}

\tableofcontents

\section{Introduction}

The concept of gauge symmetry and the construct of gauge field theory have long enjoyed the status of constitutive elements of modern physical thinking and modelling.\ On one hand,\ they serve to explain deviations of classical trajectories of charged probes---such as those observed in the Wilson cloud chamber---and interference patterns for their Feynman histories---such as those measured in the Aharonov--Bohm experiment---from those of neutral objects propagating in a given spacetime in the presence of external gauge fields.\ In this case,\ the coupling between the gauge field and the probe resp.\ the said interference are typically represented by a suitable transport operator resp.\ the corresponding differential character (holonomy) on Ehresmann's bundle of gauges with compatible connection,\ which geometrises the gauge field \cite{Alvarez:1984es,Gawedzki:1987ak}.\ The last 40 years have seen tremendous progress in the study of gauge theory thus understood for arbitrary $p$-form gauge fields in terms of higher homological algebra and category theory,\ see:\ \cite{Borsten:2024gox} for a recent review.\ On the other hand,\ the two---the concept and the construct---provide us with a coherent framework in which to formulate dynamics of fields with configuration spaces modelled on orbispaces---singular in general---for various symmetry models,\ {\it e.g.},\ orbifolds,\ quotients of Lie-group actions,\ orientifolds {\it etc.}\ \cite{Gawedzki:2003pm,Gawedzki:2007uz,Gawedzki:2008um,Gawedzki:2010rn,Gawedzki:2012fu}.\ Here,\ it is the gaug{\em ing} of a rigid symmetry of a given field theory,\ and the resulting gauge field-mediated---{\it e.g.},\ through the minimal coupling---descent of dynamics to Cartan's associated bundles over the spacetime of arbitrary topology that play a key r\^ole,\ and equivariant cohomology becomes an indispensable tool.\ Also in this model-building context,\ there has recently been much progress---the systematic gerbe-theoretic construction and cohomological classification of the Gaw\c{e}dzki--Kupiainen coset $\si$-models of the 2$d$ RCFT,\ worked out by Gaw\c{e}dzki {\it et al.} \cite{Gawedzki:2010rn,Gawedzki:2012fu,deFromont:2013iy},\ is an important example.\ We make the latter aspect of gauge theory the general theme of the present work.\smallskip

The classic formulation of gauge theory,\ referred to above,\ bases on a particular choice of the symmetry model,\ which by now has become essentially synonymous with the notion of symmetry,\ to wit:\ a group $\txG$ acting on the configuration fibre $M$ of the field theory as $\la\colo\txG\to\Diff(M)$.\ The inherent rigidity of this model  and the ensuing limitations of its applicability in realistic configurational geometries were demonstrated convincingly by Weinstein \cite{Weinstein:1996grpd}:\ Given a group $\txG$ acting on a manifold $M$ by automorphisms of the structure present on $M$ ({\it e.g.},\ by diffeomorphisms in the generic situation),\ the removal of {\em some but not all} points from some of the $\txG$-orbits---which clearly does {\em not} affect the {\em local} geometric equivalences,\ away from the excisions---leads to the disappearance of $\txG$ as a {\em global} symmetry model.\ Recovering the local symmetry within the smooth paradigm calls for a renewed conceptualisation of the notion of symmetry,\ which allows for a fibration of the symmetry model over the space whose symmetries we seek to describe,\ but now nontrivial and without a typical fibre in general.\ Thus,\ we arrive at a {\em specialised} internalisation in the smooth category $\Man$ of the concept of equivalence relation between points in $M$,\ forming a set $\xcG$ :
\bit
\item[$\bullet$] {\it reflexivity}:\ identity maps are among (self-)equivalences,
\qq\nn
\{\xy *{\underset{m}{\bullet}} \ar@(ur,ul)_{\Id_m}
\endxy\}_{m\in M}\subset\xcG\,;
\qqq
\item[$\bullet$] {\it transitivity}:\ there exists an associative $M$-fibred composition of ~~~equivalences,
\qq\nn
\forall\alxydim{@C=.75cm@R=.5cm}{ \underset{t(g)}{\bullet} & \underset{s(g)}{\bullet} \ar@/_1.pc/[l]_g }\underset{=}{,}\alxydim{@C=1.15cm@R=.5cm}{ \underset{t(h)}{\bullet} & \underset{s(h)}{\bullet} \ar@/_1.pc/[l]_h }\in\xcG\quad\Longrightarrow\quad\exists\alxydim{@C=.75cm@R=.5cm}{ \underset{{\tiny \barr{c} t(g.h)\\ =t(g) \earr}}{\bullet} & \underset{{\tiny \barr{c} s(g.h)\\ =s(h) \earr} }{\bullet} \ar@/_1.pc/[l]_{g.h} }\hspace{-5pt}\in\xcG\,;
\qqq
\item[$\bullet$] {\it symmetry}:\ all equivalences admit inverses,
\qq\nn
&&\forall\alxydim{@C=1.15cm@R=.5cm}{ \underset{t(g)}{\bullet} & \underset{s(g)}{\bullet} \ar@/_1.pc/[l]_g }\in\xcG\quad\Longrightarrow\quad\exists \alxydim{@C=.45cm@R=.5cm}{ \underset{{\tiny \barr{c} t(g)\\ =s(g^{-1}) \earr}}{\bullet} \ar@/^1.pc/[r]^{g^{-1}} & \underset{{\tiny \barr{c} s(g)\\ =t(g^{-1}) \earr} }{\bullet} }\hspace{-10pt}\in\xcG\ :\ \xy *{\underset{t(g)}{\bullet}} \ar@(ur,ul)_{g.g^{-1}}
\endxy\equiv\xy *{\underset{t(g)}{\bullet}} \ar@(ur,ul)_{\Id_{t(g)}}
\endxy\ \land\ \xy *{\underset{s(g)}{\bullet}} \ar@(ur,ul)_{g^{-1}.g}
\endxy\equiv\xy *{\underset{s(g)}{\bullet}} \ar@(ur,ul)_{\Id_{s(g)}}
\endxy\,;
\qqq
\item[$\bullet$] {\it smoothness}:\ $\xcG$ carries a smooth structure,\ {\em and} equivalences `chart' $M$ in that for every equivalence $g\in\xcG$ there is a submanifold $O_g\ni g$ in $\xcG$ such that $t(O_g)\cong O_g\cong s(O_g)$ (diffeomorphisms).\vskip-10pt
\begin{figure}[h]
	\centering
	\includegraphics[width=0.25\textwidth]{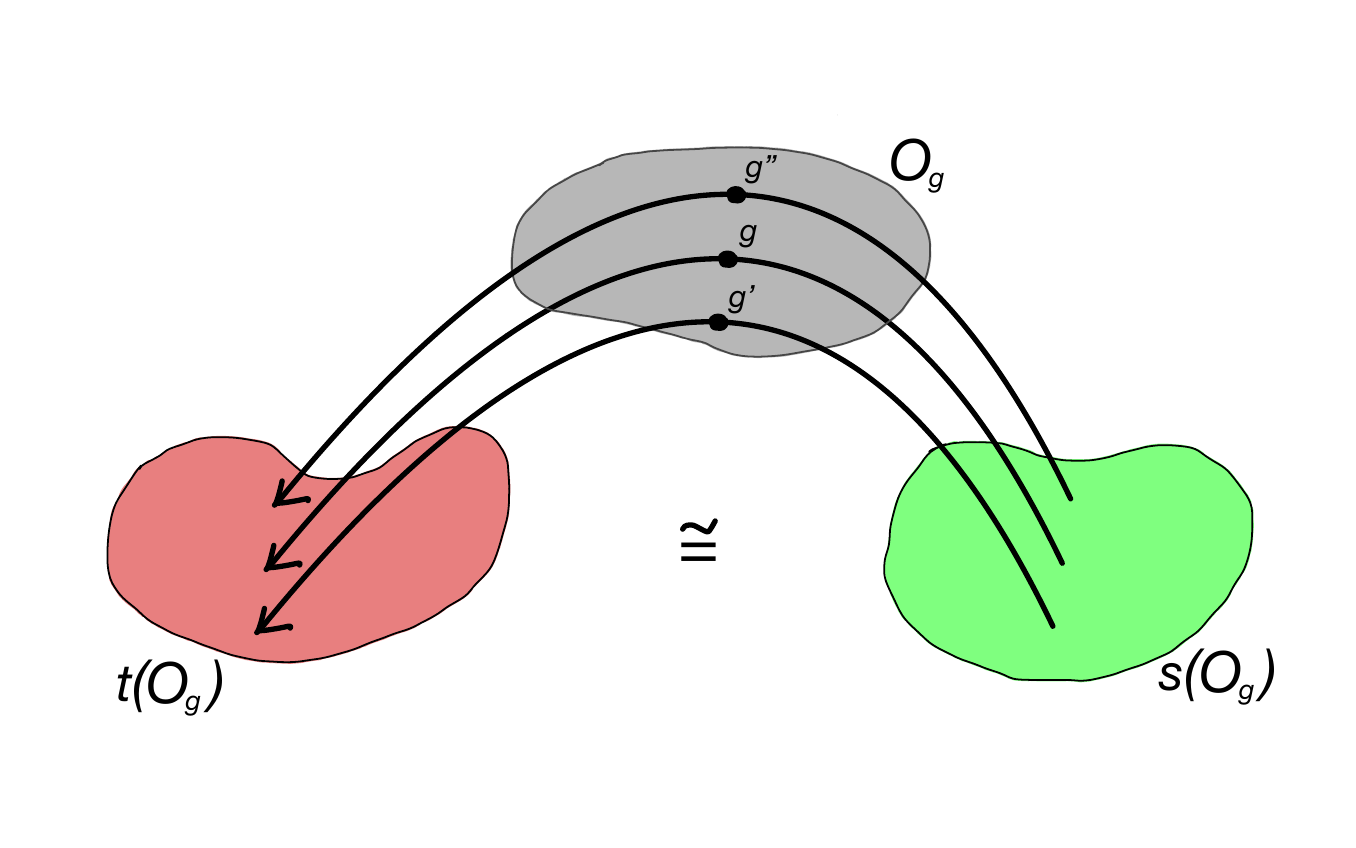}
\end{figure}
\eit
Altogether,\ these elementary requirements determine a {\em Lie groupoid} $\grpd{\xcG}{M}$,\ a small category with all morphisms invertible,\ composed of smooth manifolds:\ that of objects $M$,\ and that of morphisms $\xcG$,\ with all structure maps smooth,\ and with the source $s$ and target $t$ submersively surjective.\ The definition clearly subsumes the classic situation in the form of the action groupoid $\txG\,\lx_\la M\equiv\grpd{\txG\x M}{M}$ with $s=\pr_2$ and $t=\la$.\ It is,\ then,\ for this more comprehensive model of symmetry that we want to formulate a gauge principle.\ Besides the conceptual motivation to do so,\ outlined above and paving the way towards field theories with configuration fibres given by generic foliations,\ there is an important practical and concrete one grounded in the recent theoretical developments from the interface between topological field theory and Poisson/symplectic geometry,\ of fundamental relevance to Kontsevich's theory of quantisation \cite{Kontsevich:1997vb}:\ There is a class of two-dimensional field theories with Poisson configuration fibres $(M,\Pi)$---known as Poisson $\si$-models\footnote{Not unrelatedly,\ these topological field theories have also played an important r\^ole in the elucidation of the classical origin of quantum-group symmetries in the WZNW model of string theory with a Lie-group target \cite{Gawedzki:1990jc,Falceto:1992bf,Falceto:2001eh}.} \cite{Ikeda:1993fh,Schaller:1994es}---which exhibit tangential (infinitesimal) gauge symmetries modelled by smooth maps from the field theory's spacetime to the cotangent Lie algebroid over $(M,\Pi)$ \cite{Bojowald:2004wu}.\ The latter integrates,\ under special circumstances,\ to Weinstein's symplectic Lie groupoid \cite{Weinstein:1987,Coste:1987}.\ Hence,\ it becomes desirable to work out a field-theoretic framework which would incorporate Lie-groupoidal and---through linearisation at the submanifold of units $\Id(M)$---Lie-algebroidal gauge symmetries consistently and naturally.\ The result of our endeavour is envisaged as a generalisation of the classic prototype,\ and so we begin our investigation by carefully reformulating the latter in Lie-groupoidal terms in Sec.\,\ref{sec:class-gau-princ}.\smallskip

The construction at the core of the classic gauge theory for symmetries with a Lie-group model is Ehresmann's principal bundle of gauges $P\to\Si$,\ which internalises the concept of reference frame (or observer) in the slice $\Man/\Si$ over the spacetime $\Si$ of the field theory upon a passive reinterpretation of the underlying symmetry,\ and in which the idea of identification of local field configurations in $M$ related by families of frame transformations varying smoothly over $\Si$---modelled on locally smooth maps $\Si\to\txG$---takes its root.\ The choice of the corresponding generalisation for symmetries modelled on arbitrary Lie groupoids calls for some caution,\ lest the principle of homogeneity of internal structures of the field theory---such as the reference frames,\ or internal degrees of freedom---should be violated,\ which we choose to avoid.\ This leads us to the notion of {\bf principaloid $\xcG$-bundle} $\xcP\xrightarrow{\pi_\xcP}\Si$,\ a fibre-bundle object in the category of $\xcG$-modules with typical fibre $\xcG$---and so with a fibration of the symmetry model over $M$ defined by the target (resp.\ source) map---and the structure group given by the group ${\rm Bisec}(\grpd{\xcG}{M})$ of global bisections\footnote{These are global sections $\b\colo M\to\xcG$ of the source map $s$,\ with the property that $t_*\b\equiv t\circ\b\in\Diff(M)$.} of the structure groupoid.\ These were introduced by Strobl and the Author in \cite{Strobl:2025}.\ They constitute the basis of our approach to the gauging of Lie-groupoidal symmetries,\ and a natural counterpart of Cartan's extension $P\x M$ of Ehresmann's prototype $P$ through the space $M$ of internal degrees of freedom,\ on which the typical fibre $\txG$ of $P$ acts by $\la$.\ The classic extension is endowed with a principal (right) action of the relevant Lie groupoid $\txG\,\lx_\la M$,\ and so there arises a smooth Godement quotient $(P\x M)/\txG\,\lx_\la M\equiv P\x_\la M\xrightarrow{\pi_{P\x_\la M}}\Si$ with typical fibre $M$,\ known as the associated bundle,\ and interpreted as the (charged-matter) configuration bundle of the ensuing gauge field theory.\ This fundamental mechanism generalises straightforwardly to principaloid bundles,\ producing candidate configuration bundles of field theories with Lie-groupoidal symmetries gauged---the {\bf shadow bundles} $\xcP/\xcG\equiv\xcF\xrightarrow{\pi_\xcF}\Si$.\ This is reviewed in Sec.\,\ref{sec:principaloid},\ with due emphasis on the modelling of gauge transformations of the principaloid $\xcG$-bundle $\xcP$ and its shadow $\xcF$ on the canonical $\xcF$-fibred actions of the associated Ehresmann--Atiyah(-type) groupoid $\grpd{{\rm At}(\xcP)}{\xcF}$,\ and on the accompanying phenomenon of induction of an adapted connection on $\xcF$ from a $\xcG$-invariant one on $\xcP$.\ The latter parallels the classic Kobayashi--Nomizu mechanism for extended and associated bundles,\ and leads to the all-important definition of a gauge-covariant derivative of sections of the charged-matter bundle $\xcF$.\ It is,\ in particular,\ recalled how a $\xcG$-equivariant connection 1-form $\Theta\in\G(\txT\xcP\ox\ker\,\txT\pi_\xcP)$ on $\xcP$ gives rise to a (local) gauge field,\ taking values in the tangent algebroid $\xcE\equiv\Lie(\xcG)$ of the structure Lie groupoid.\smallskip

The rigidity of the global group model of symmetry is compensated by the modality of the associated field-theoretic construction,\ in which multiple charged-matter fields can be coupled to a single gauge field---encoded by a principal connection on $P$---with the help of the respective associating actions $\la$ and the corresponding fundamental vector fields.\ The groupoidal generalisation seems to reverse this picture:\ Introducing a fibred model of symmetry freezes the choice of the geometric base $M$ of the symmetry model,\ and thus ties the gauge field to the charged-matter field with internal degrees of freedom $M$.\ This apparent shortcoming of the original proposal of \cite{Strobl:2025} is overcome in Sec.\,\ref{sec:red-ass},\ in which the groupoidal variant of Cartan's associating construction is worked out (Thm.\,\ref{thm:del-as-prince}),\ including a Crittenden-type induction mechanism for connections.\ Here,\ the point of departure of the relevant construction in $\Man/\Si$ is the notion of groupoid module $(X,\mu_X,\la_X)$ with moment map $\mu_X\colo X\to M$ and $M$-fibred action $\la_X\colo\xcG\,\fibx{s\,}{\mu_X}X\to X$.\ The derived construction of the action groupoid $\xcG\,\lx_{\la_X}X\equiv\grpd{\xcG\,\fibx{s\,}{\mu_X}X}{X}$ enables us to subsequently reinterpret the corresponding {\bf $\la_X$-associated bundles} $\xcP\,\lx_{\la_X}X$--- defined in structural analogy with Cartan's associated bundles $\,P\x_\la M$ for the delooping groupoid $\bbB\txG\equiv\grpd{\txG}{\bullet}$---as shadows of principaloid $\xcG\,\lx_{\la_X}X$-bundles with a reduced structure group ${\rm Im}\,\widetilde\mu{}^*_M\subset{\rm Bisec}(\xcG\,\lx_{\la_X}X)$,\ canonically induced by a group homomorphism $\widetilde\mu{}^*_M\colo{\rm Bisec}(\grpd{\xcG}{M})\to{\rm Bisec}(\xcG\,\lx_{\la_X}X)$ (Props.\,\ref{prop:ass-as-oidle} and \ref{prop:Ass-aut}).\ In this way,\ the associating construction is shown to answer the need for a systematic treatment of the reduction of the structure group,\ which arises naturally in the presence of extra geometric structure---such as,\ {\it e.g.},\ a metric tensor,\ or a higher-geometric object---over the configuration fibre of the field theory.\ This complements the attainment of the original goal---the coupling of multiple charged-matter fields,\ represented by sections of the associated bundles $\xcP\,\lx_{\la_X}X$,\ to a single algebroid-valued gauge field,\ as described at length in Sec.\,\ref{sub:ass-appl}.\smallskip

As for many physical ideas and the corresponding mathematical constructions before,\ an ideal testing ground for those introduced above is provided by the distinguished model of charged-loop dynamics---the two-dimensional non-linear closed $\si$-model with the topological Wess--Zumino term \cite{Friedan:1985phd},\ with configuration fibre (aka the target space) given by the object manifold $M$ of a Lie groupoid $\grpd{\xcG}{M}$.\ This field theory---in its fully fledged Polyakov--Alvarez--Gaw\c{e}dzki (PAG) formulation using a degree-3 Cheeger--Simons differential character as the definition of the Wess--Zumino term \cite{Alvarez:1984es,Gawedzki:1987ak},\ recalled in Sec.\,\ref{sec:PAGsi}---combines tensorial (a metric) and non-tensorial (a gerbe \cite{Murray:1994db,Murray:1999ew}) structures on the target space,\ responsible for different structure group--reduction mechanisms for $\xcP$,\ examined at length in Sec.\,\ref{sec:reduction},\ and leading to different schemes of configurational descent $M\searrow M//\xcG$,\ discussed in Sec.\,\ref{sec:EAS}.\ In either case,\ the necessary intuition comes from the exhaustive study of the gauging of group-modelled symmetries in the PAG $\si$-model,\ carried out by Gaw\c{e}dzki,\ Waldorf and the Author in \cite{Gawedzki:2010rn,Gawedzki:2012fu},\ which has been recapitulated in Sec.\,\ref{sec:G-gau-sigmod}. 

A straightforward adaptation of the classic gauging scheme for the metric term in the Dirac--Feynman (DF) amplitude of the PAG $\si$-model---which consists in the coupling of the metric $\txg_M\in\G(S^2\txT^*M)$ tensor to the gauge field through the evaluation of the former on the covariant derivative on $\G(\xcF)$ in local trivialisations $\xcF\t_i\colo\pi_\xcF^{-1}(O_i)\xrightarrow{\cong}O_i\x M,\ i\in I$---yields a smooth symmetric tensor over the pullback cover $\pi_\xcF^{-1}\cO\equiv\bigsqcup_{i\in I}\,\pi_\xcF^{-1}(O_i)$ of $\xcF$ induced by the trivialising cover $\cO\equiv\{O_i\}_{i\in I}$ of $\Si$.\ The tensor is readily seen to descend to a {\em gauge-invariant} global section of $S^2\txT^*\xcF$ (Prop.\,\ref{thm:gauge-desc-aid}) in the usual,\ sheaf-theoretic sense.\ This happens if the structure group of $\xcP$ (and so also of $\xcF$) is reduced to the preimage of the isometry subgroup ${\rm Isom}(M,\txg_M)\subset\Diff(M)$ along the canonical shadow action $t_*\colo{\rm Bisec}(\xcG)\to\Diff(M),\ \b\mapsto t\circ\b$.

The descent mechanism for the target gerbe $\cG$,\ on the other hand,\ is much subtler.\ Upon taking into account the interpretation of the Bott--Shulman--Stasheff (BSS) cohomology of the nerve $N_\bullet(\grpd{\xcG}{M})$ of the structure groupoid as a model of the cohomology of the orbispace $M//\xcG$ \cite{Bott:1976bss},\ and adducing the correspondence between gerbe bi-modules and prequantisable symmetries of the PAG $\si$-model \cite{Runkel:2008gr,Suszek:2011hg,Suszek:2011},\ we have based our considerations on the assumption of the existence of a {\bf multiplicative extension} $\rho\in\Om^2(\xcG)$ of the curvature $H_M\in Z^3(M)$ of $\cG$ to a BSS 3-cocycle $(H_M,\rho,0)$ on the simpicial manifold $N_\bullet(\grpd{\xcG}{M})$.\  Such extensions have been shown to canonically induce vertical vector-bundle morphisms $\kappa\colo\xcE\to\txT^*M$---dubbed (2-){\bf comomenta} in the present context---as part of Spencer data of $\rho$,\ discovered by Crainic {\it et al.} in \cite{Crainic:2015msp}.\ An immediate consequence of the assumption is the reduction of the structure group of $\xcP$ to the {\bf $\rho$-holonomic group} $\bB_\rho(\xcG)=\{\ \b\in{\rm Bisec}(\grpd{\xcG}{M})\ |\ \b^*\rho=0 \ \}$,\ suggested by the aforementioned bi-module/symmetry correspondence.\ Crucially,\ the comomentum has been proven equivariant with respect to canonical actions of $\bB_\rho(\xcG)$ on its domain and codomain.\ Whenever there exists a bicategorial geometrisation of the BSS 3-cocycle $(H_M,\rho,0)$ in the form of a {\bf $\rho$-twisted $\xcG$-equivariant structure on} $\cG$,\ the reduction is readily seen to ensure effective descent---in the sense of Breen's 2-stack structure on the smooth site $\Man$,\ realised by gerbes \cite{Stevenson:2000wj}---of a gerbe over $\pi_\xcF^{-1}\cO$ obtained from $\cG$ through its augmentation by (the pullback of) a trivial gerbe on $\bigsqcup_{i\in I}\,O_i\x M$ determined by the comomentum $\kappa$ coupled to the local gauge field in a way essentially dictated by the pioneering analysis of \cite{Gawedzki:2010rn}.\ The descended gerbe over $\xcF$ has been verified to define a {\em gauge-invariant} holonomy (Thm.\,\ref{thm:gauge-desc-aid}).

The existence of a twisted equivariant structure on a gerbe may,\ in general,\ be obstructed topologically,\ which---as argued extensively in \cite{Gawedzki:2010rn} in the case of group-modelled symmetry---leads to various pathologies upon quantisation of the PAG $\si$-model coupled to the gauge field in the manner described above.\ Hence,\ it is vital to identify and quantify the obstructions,\ also known as {\bf gauge anomalies}.\ This goal has been attained in the case under consideration in terms of a $\xcG$-equivariant---{\it i.e.},\ simplicial,\ over $N_\bullet(\grpd{\xcG}{M})$---extension of the standard Beilinson--Deligne hypercohomology describing 0-,\ 1- and 2-cells of the bicategory of gerbes over a given base.\ The same formalism has been employed to classify inequivalent gaugings,\ represented by non-cohomologous $\rho$-twisted $\xcG$-equivariant structures on a given target gerbe.\ Both results---the anomaly analysis (Props.\,\ref{prop:Ups-obstr},\ \ref{prop:gam-obstr} and \ref{prop:gam-coh-obstr}) and the classification (Prop.\,\ref{prop:class-cohom})---extend straightforwardly (and are motivated by) those derived in \cite{Gawedzki:2010rn}.

Whenever a multiplicative extension $\rho$ exists,\ and the gauge anomaly vanishes,\ a $\rho$-twisted $\xcG$-equivariant structure on the target gerbe from any one of the equivalence classes identified above can be used to define a {\bf gauged PAG $\si$-model} with the configuration bundle given by a shadow $\xcF$ of an arbitrary principaloid $\xcG$-bundle with the structure group fully reduced as ${\rm Bisec}(\grpd{\xcG}{M})\searrow t_*^{-1}({\rm Isom}(M,\txg_M))\cap\bB_\rho(\xcG)\equiv\bB_\si(\xcG)$ (Thm.\,\ref{thm:gausigmod}).\ The latter group can be viewed as composed by a family of distinguished submanifolds within the arrow manifold $\xcG$ which project diffeomorphically onto $M$ along both:\ the source map and the target map.\ In this picture,\ the structure group of the principaloid $\xcG$-bundle $\xcP$ prior to the background-enforced reduction,\ ${\rm Bisec}(\grpd{\xcG}{M})$,\ yields---for a class of Lie groupoids,\ and among them those canonically associated with integrable Lie algebroids (as well as arbitrary group-action groupoids)---the most complete realisation of the structure groupoid within $\Diff(M)$ that can be conceived,\ namely one with an element $\b\in{\rm Bisec}(\grpd{\xcG}{M})$ through {\em every} arrow.\ It is,\ then,\ in this sense that we may think of the construction reported in this paper as giving us a genuinely novel gauge field theory with a Lie-groupoidal gauge symmetry,\ modelled on $t_*(\bB_\si(\xcG))$.\smallskip

There is an important theoretical context in which the hitherto general considerations concretise and gain even more structure,\ to wit---the geometry of Weinstein's symplectic groupoids \cite{Weinstein:1987,Coste:1987},\ with its deep conceptual relation to Kontsevich's (deformation) quantisation \cite{Kontsevich:1997vb},\ mediated by (topological) field theory in the manner originally uncovered by Cattaneo and Felder \cite{Cattaneo:1999fm,Cattaneo:2000iw}.\ An attempt---made in Sec.\,\ref{sec:symplscen}---at internalising the former results in the symplectic category leads to an unexpected finding:\ Given a symplectic groupoid $(\grpd{\xcG}{M},\om)$,\ with its multiplicative symplectic 2-form $\om\in Z^2(\xcG)$ and its canonically Poisson object manifold $(M,\Pi)$,\ the holonomy of the previously mentioned (trivial-gerbe) augmentation of a gerbe $\cG$ over $M$ endowed with an $\om$-twisted $\xcG$-equivariant structure reproduces---upon specialisation to the trivial symplectic principaloid bundle,\ with the structure group reduced to the subgroup $\bB_\om(\xcG)$ of {\bf lagrangean bisections}---the DF amplitude of the Poisson $\si$-model on the cotangent algebroid over $(M,\Pi)$ \cite{Ikeda:1993fh,Schaller:1994es}.\ This provides us with an intricate,\ novel conceptualisation of the Poisson $\si$-model---as a flat trivial twist of the gerbe of a (dynamical) PAG $\si$-model on the Poisson target $(M,\Pi)$,\ untwisting that gerbe's $\xcG$-equivariant structure---and leads,\ through the reversal of the logic of the original gauging procedure,\ to an extension of the hitherto definition of the Poisson $\si$-model which promotes the latter to the rank of a (topological) field theory with a Lie-groupoidal gauge symmetry modelled on the group $t_*(\bB_\om(\xcG))$ of Poisson diffeomorphisms.\ The extension consists in co-augmenting the said trivial gerbe with an arbitrary $\om$-twisted $\xcG$-equivariant gerbe over $M$.\ By previous results,\ such a co-augmentation automatically allows to extend---through the (2-)stacky descent---the definition of the field theory further to Poisson shadow bundles of {\em arbitrary} topology (Thm.\,\ref{thm:coaugPSM-gauge}).\ In the light of the previous remarks,\ this permits us to think of the novel field theory as one with the effective configuration fibre $M//\xcG$.\ Given the structural interpretation of the Euler--Lagrange equations of the Poisson $\si$-model as defining Lie-algebroid morphisms \cite{Bojowald:2004wu},\ special attention has been given to the distinguished co-augmentations which leave these equations unchanged---the {\bf flat coaugmentations},\ characterised in natural cohomological terms in Prop.\,\ref{prop:coaugPSM-anomaly}.\medskip

The present work presents the first physical application of the abstract mathematical constructs introduced in the precursor paper \cite{Strobl:2025}.\ Bearing in mind its character,\ we have gone to great lengths ({\it sic}!) to make it self-contained,\ and thus more easily accessible---this is the rationale behind the many examples scattered in the main text,\ and the numerous appendices.\ Understandably,\ the work leaves a number of questions open,\ and prompts further research in several directions---both have been addressed in the concluding Sec.\,\ref{sec:summ-out}.\ We hope to return to them in a future study.\bigskip 

\noindent{\bf Acknowledgements:}\ The Author acknowledges many interesting and congenial discussions with Thomas Strobl over the last few years of the ongoing collaboration---revolving,\ {\it i.a.},\ around some of the problems discussed in the present work---and the joint intelectual adventure backed up with a very special friendship,\ which goes back in time as far as\ldots the happy period 2005--2007 shared in Lyon.

It is also the Author's pleasure to thank the organisers of conferences in La Rochelle,\ Prague,\ and Arpino,\ as well as participants of the Trans-Carpathian Seminar on Geometry and Physics in Warsaw and Bucharest for the opportunity to present and discuss,\ in the course of the last year,\ partial results of the research reported in full herein.

In the course of this work,\ the Author was supported by the Excellence Initiative -- Research University (IDUB) programme at the University of Warsaw.

\section{A groupoidal reformulation of the classic gauge principle}\label{sec:class-gau-princ}

The main goal of the present investigation is the formulation of a gauge principle for rigid configurational symmetries of a dynamical system modelled on a Lie groupoid,\ and its subsequent application in a field-theoretic setting of choice.\ The path towards the goal thus set is bound to lead through and,\ in so doing,\ subsume the classic gauge principle for Lie group-modelled symmetries,\ whose geometric foundations were laid by Ehresmann \cite{Ehresmann:1950} and Cartan \cite{Cartan:1950mix}.\ A careful reformulation of this prototypical construction in the language of the theory of Lie groupoids and their tangent algebroids affords us the opportunity to identify the key components of the classic principle and structural relations between them,\ which we may subsequently employ as the basis of the desired generalisation.\ This is,\ then,\ the course of action that we take below.\ (As the raw mathematical content of the present discussion,\ as well as its physical interpretation are completely standard by now,\ we choose a narrative,\ descriptive mode of exposition instead of the formal,\ definition-theorem-proof one,\ which we reserve for the remainder of the paper.)\smallskip

Ehresmann's classic construction begins from a choice of a manifold $M$ modelling the space of internal degrees of freedom of the field theory-to-be over a given spacetime $\Si$,\ accompanied by a choice of a Lie group $\txG$ of symmetries $\la\colo\txG\to\Diff(M)$ of the field's dynamics.\ The latter are interpreted passively as distinguished redefinitions of an internal reference frame (or {\em gauge}).\ There then follows an identification of a smooth homogeneous distribution,\ over $\Si$,\ of ``observers'' (as they are expressly called by Pradines in \cite{Pradines:2007}),\ each of which is represented by a $\txG$-torsor of equivalent choices of an internal reference frame for his or her observations/measurements of a physical (matter) field.\ The ensuing trivial local models $O_i\x\txG$ of the distribution of frames over an open cover $\cO\equiv\{O_i\}_{i\in I}$ of $\Si$ are glued together over non-empty intersections $O_{ij}\equiv O_i\cap O_j$ by smoothly $O_{ij}$-indexed families of diffeomorphisms of the typical fibre $\txG$ which commute with the right regular action $r\colo\txG\x\txG\to\txG,\ (g,h)\mapsto g\cdot h$ of the structure groupoid $\bbB\txG\equiv\grpd{\txG}{\bullet}$ (see:\ Ex.\,\ref{eg:deloop}) on $\txG$.\ The latter action represents a global (vertical) choice of frame.\ This yields a principal $\txG$-bundle of gauges over $\Si$:
\qq\label{eq:princ-G-bndl}\qquad\qquad
\alxydim{@C=.75cm@R=1.cm}{ \txG \ar@{^{(}~>}[d] & & \\ P \ar[d]_{\pi_P} \ar@{-->}[dr] & & \txG \ar@{=>}[dl] \ar@/_1.pc/[ll]_{r_P} \ar@/_1.pc/[ull]_r \\ \Si & \bullet & }\,,\qquad\qquad\qquad P\cong\bigsqcup_{i\in I}\,(O_i\x\txG)/\sim_{l_{g_{\cdot\cdot}}}\,,\qquad(\si,j,g)\sim\bigl(\si,i,g_{ij}(\si)\cdot g\bigr)\,,
\qqq
with a (principal fibrewise right) action $r_P$ of $\bbB\txG$ on its total space $P$,\ modelled---through $\txG$-equivariant local trivialisations $P\t_i\colo\pi_P^{-1}(O_i)\xrightarrow{\cong}O_i\x\txG$---on the aforementioned right regular action $r$ on the typical fibre $\txG$,\ and with the gluing realised in terms of the left regular action $l\colo\txG\to\Diff(\txG),\ h\mapsto h\cdot(\cdot)\equiv l_h(\cdot)$ of the transition 1-cocycle $g_{ij}\colo O_{ij}\to\txG$.

Next,\ we use the principle of $r_P(\txG)$-equivariance to distinguish those bundle automorphisms of $P$ which---similarly to the local gluings---commute with global redefinitions of frame,
\qq\nn
{\rm Aut}(P)={\rm Aut}_{\Bun(\Si)}(P)\cap\Diff_{r_P(\txG)}(P)\,.
\qqq
In the local (trivialised) picture,\ these are represented by smooth maps from the base $\Si$ into $\txG$ realised on the typical fibre by $l$,\ as before.\ As understood by Ehresmann already in \cite{Ehresmann:1950},\ their action on $P$ admits a straightforward Lie-groupoidal description:\ The distinguished automorphisms are in bijection with global bisections of the Ehresmann--Atiyah (or gauge) groupoid $(P\x P)/\txG\equiv{\rm At}(P)$ (see:\ Def.\,\ref{def:bisec}).\ The groupoid acts on $P$ along $\Si$ from the left as $\la_P\colo{\rm At}(P)\x_\Si P\to P,\ ([(p_2,p_1)],p_1)\mapsto p_2$,\ with moment map given by the base projection $\pi_\Si$ (see:\ Def.\,\ref{def:gr-mod}).\ The restriction of the action to any one of the bisections reproduces the corresponding automorphism.\ The action commutes with the right defining action $r_P$ of the structure groupoid $\bbB\txG$,\ and the latter preserves the moment map $\pi_\Si$ by definition.\ Moreover,\ $\la_P$ is manifestly principal relative to the canonical map $P\dashedrightarrow\bullet$ to the object manifold of the structure groupoid $\bbB\txG$,\ so that a prototypical bi-bundle arises in the sense of Def.\,\ref{def:bibndl}:
\qq\nn
\alxydim{@C=.75cm@R=1.cm}{ {\rm At}(P) \ar@{=>}[rd] \ar@/^1.pc/[rr]^{\la_P} & & P \ar@{-->}[dr] \ar[dl]_{\pi_P} & & \txG \ar@{=>}[ld] \ar@/_1.pc/[ll]_{r_P} \\ & \Si & & \bullet & }\,.
\qqq
Inspection of the anatomy of the gauge groupoid reveals additional structure behind the above picture:\ ${\rm At}(P)$ is a fibre-bundle object in the category of Lie groupoids,\ with base ${\rm Pair}(\Si)$ and typical fibre $\bbB\txG$.\ Its action $\la_P$ on $P$ covers the canonical left-fibred action of its base on that of $P$ (see:\ Ex.\,\ref{eg:fibred-act}),\ and is locally modelled on the canonical left-fibred action of its fibre on that of $P$.\ Thus,\ altogether,\ we discover the (left) trident (see:\ Def.\,\ref{def:Trident-oid})
\qq\label{diag:G-trident}
\alxydim{@C=.75cm@R=1.cm}{ \txG \ar@{=}[r] \ar@{^{(}~>}[dd] & \txG \ar@{^{(}~>}[dd] \ar@{=>}[dr] \ar@/^1.pc/[rr]^l & & \txG \ar@{^{(}~>}[dd] \ar@{-->}[dl] & & \\ & & \bullet \ar@{^{(}~>}[dd] & & & \\ \Ad(P) \ar[dd] \ar@{^{(}->}[r] & {\rm At}(P) \ar@{=>}[dr] \ar[dd]_{(\pi_1,\pi_2)} \ar@/^1.pc/[rr]^(.46){\la_P} |!{[ur];[dr]}\hole & & P \ar[dl]^{\pi_P} \ar[dd]^{\pi_P} \ar@{-->}[dr] & & \txG \ar@{=>}[dl] \ar@/_1.pc/[ll]_{r_P} \ar@/_1.pc/[uull]_r \\ & & \Si \ar@{=}[dd] & & \bullet & \\ \Si \ar@{^{(}->}[r] & \Si\x\Si \ar@{=>}[dr] \ar@/^1.pc/[rr]^(.39){\pr_1\circ\pr_1} |!{[ur];[dr]}\hole & & \Si \ar@{=}[dl] & \\ & & \Si & & & }\,.
\qqq
In it,\ additionally,\ there appears---over the identity bisection $\Id(\Si)$ of the base groupoid ${\rm Pair}(\Si)$---the adjoint bundle $\Ad(P)$,\ encoding {\em vertical} automorphisms of $P$:
\qq\nn
\alxydim{@C=.75cm@R=1.cm}{  P \ar[rr]^{\chi} \ar[dr]_{\pi_P} & & P \ar[dl]^{\pi_P} \\ & \Si & }\,,
\qqq
with a local presentation
\qq\nn
P\t_i\circ\chi\circ P\t_i^{-1}(\si,g)=\bigl(\si,h_i(\si)\cdot g\bigr)
\qqq
in terms of some smooth maps $h_i\colo O_i\to\txG$,\ related over intersections $O_{ij}\ni\si$ of the respective domains by the transition 1-cocycle of $P$ as 
\qq\nn
h_i(\si)=\Ad_{g_{ij}(\si)}\bigl(h_j(\si)\bigr)\,.
\qqq

The next stage of the classic construction,\ worked out by Cartan \cite{Cartan:1950mix} (see also:\ \cite{Tu:2020}),\ consists in extending Ehresmann's bundle of gauges $P$ by the $\txG$-manifold $(M,\la)$.\ As a result,\ another bundle:\ $\pi_P\circ\pr_1\colo P\x M\to\Si$ arises over $\Si$,\ with typical fibre given by the arrow manifold $\txG\x M$ of the action groupoid $\txG\,\lx_\la M$ associated with the smooth action $\la$ (see:\ Ex.\,\ref{eg:actgrpd}).\ The groupoid acts (fibrewise from the right) on the total space of the extended bundle in a manner which emulates the diagonal action of $\txG$ combining $r_P$ an $\la$,\ {\it i.e.},\ through
\qq\label{eq:Gact-ext}\qquad\qquad
\widetilde\la\colo(P\x M)\x_M(\txG\x M)\too P\x M,\ \bigl((p,m),\bigl(g,\la_{g^{-1}}(m)\bigr)\bigr)\longmapsto\bigl(r_P(p,g),\la_{g^{-1}}(m)\bigr)\,.
\qqq
The action being free and proper,\ there emerges a smooth quotient manifold $(P\x M)/\txG\equiv(P\x M)/(\txG\,\lx_\la M)\equiv P\x_\la M$,\ which fibres over $\Si$ with typical fibre given by the object manifold $M$ of the previously identified structure groupoid $\txG\,\lx_\la M$.\ The surjective submersion $\pi_\sim\colo P\x M\to P\x_\la M,\ (p,m)\mapsto[(p,m)]$ is modelled on the target map $\la$ between the respective typical fibres,\ and so we arrive at a non-trivial principal $\txG\lx_\la M$-bundle $P\x M\to P\x_\la M$ in the slice over $\Si$,\ modelled on the unit bundle of Ex.\,\ref{eg:triv-grpd-bndle}:
\qq\label{diag:princ-Gact-bndl}
\alxydim{@C=.75cm@R=1.cm}{ & \txG\x M \ar@{^{(}~>}[d] \ar[dl]_{t\equiv\la} & & \\ M \ar@{^{(}~>}[d] & P\x M \ar[d]_{\pr_1} \ar[dr]^{\pr_2} \ar[dl]_{\pi_\sim} & & \txG\x M \ar@{=>}[dl] \ar@/_1.pc/[ll]_{\widetilde\la} \ar@/_1.pc/[ull]_r \\ P\x_\la M \ar[dr]_{\pi_{P\x_\la M}} & P \ar[d]_{\pi_P} & M & \\ & \Si & & }\,.
\qqq
The quotient $\pi_\sim$ realises an important conceptual objective:\ It couples the choice of a reference frame to the outcome of an `observation' of a (local) physical field $\varphi\colo O\to M$ (over $O\subset\Si$) through the identity
\qq\nn
[(r_P(p,g),\varphi(\si))]\equiv[(p,\la_g(\varphi(\si)))]\,.
\qqq
Thus,\ `observations' of a matter field $\varphi\in\G(P\x_\la M)$,\ carried out in local trivialisations $\xcF\t_i\colo\pi_{P\x_\la M}^{-1}(O_i)\xrightarrow{\cong}O_i\x M,\ i\in I$ and represented by smooth maps $\varphi_i=\pr_2\circ\xcF\t_i\circ\varphi\rstr_{O_i}\colo O_i\to M$,\ are related over intersections $O_{ij}\ni\si$ of the respective domains by the transition 1-cocycle of $P$ as 
\qq\nn
\varphi_i(\si)=\la_{g_{ij}(\si)}\bigl(\varphi_j(\si)\bigr)\,.
\qqq
In fact,\ the latter structure is far from generic due to the cartesian factorisation of the total space of the fibre bundle $P\x M\to\Si$,\ implying,\ in particular,\ that the gluing of the bundle's local trivialisations is realised by the transition 1-cocycle of the sub-structure $P\to\Si$,\ taking values in $\txG$.\ This conforms with the identification of $\txG$ as the symmetry/equivalence structure,\ which relates local descriptions (trivialisations) of the field bundle $P\x_\la M$.\ Hence,\ effectively,\ the structure groupoid $\bbB\txG$ (whose action is emulated by that of $\txG\,\lx_\la M$) and the structure group of the extended bundle $P\x M$ coincide---we see the Lie group $\txG$ in both r\^oles.\ The resultant erasure of ontological/functional distinction between the two entities in the prototypical construction obscures the path to generalisation.\ Restoring the two functional types was the first step on that path undertaken in \cite{Strobl:2025},\ which we recapitulate in the next section.

The non-genericity noted above propagates to the next level,\ on which distinguished automorphisms of the associated bundle $P\x_\la M$ are induced from those of the extended bundle $P\x M$.\ The former being recognised as a configuration bundle of a field theory with the {\em group} symmetry $\txG$ gauged,\ we are naturally interested in automorphisms with a local model furnished by the symmetry equivalence $\la$.\ These are induced---by the very construction of the principal bundle $P\x M\to P\x_\la M$---from $\txG$-equivariant automorphisms of $P$ previously encoded in the construction of the Ehresmann--Atiyah groupoid of the bundle of gauges $P$.\ In particular,\ the vertical ones $\xcF_*(\chi)\in{\rm Aut}_{\Bun(\Si)/\Si}(P\x_\la M)$ among them---termed gauge transformations---are locally presented as 
\qq\nn
\xcF\t_i\circ\xcF_*(\chi)\circ\xcF\t_i^{-1}(\si,m)=\bigl(\si,\la_{h_i(\si)}(m)\bigr)\,,
\qqq
with the $h_i$ as before.\  Thus,\ we arrive at---non-generic from the point of view of the structure of the action groupoid---{\em extended} trident of the Cartan construction:
\qq\label{dig:ext-G-trident}
\alxydim{@C=.75cm@R=1.cm}{ & \txG\x M \ar@{^{(}~>}[dddd] \ar@{=>}[dr] \ar@/^1.pc/[rr]^l & & \txG\x M \ar@{^{(}~>}[dd] \ar[dl]_{\la} & & \\ & & M \ar@{^{(}~>}[dd] & & & \\ & & & P\x M \ar[dd]_{\pr_1} \ar[dl]_{\pi_\sim} \ar[dr]_{\pr_2} & & \txG\x M \ar@{=>}[dl] \ar@/_1.pc/[ll]_{\widetilde\la} \ar@/_1.pc/[uull]_r \ar[dd]^{\pr_1} \\ & & P\x_\la M \ar[dd]^(.25){\pi_{P\x_\la M}} & & M \ar@{-->}[dd] & \\ \Ad(P) \ar[dd] \ar@{^{(}->}[r] & {\rm At}(P) \ar@{=>}[dr] \ar[dd]_{(\pi_1,\pi_2)} \ar@/^1.pc/[rr]_(.33){\la_P} |!{[ur];[dr]}\hole \ar@/^1.pc/[ur]_{\la_{P\x_\la M}} \ar@/^2.5pc/[uurr]^{\widetilde\la{}_P} |!{[uur];[ur]}\hole & & P \ar[dl]^{\pi_P} \ar[dd]^{\pi_P} \ar@{-->}[dr] & & \txG \ar@{=>}[dl] \ar@/_1.pc/[ll]_(.33){r} \ar@/_1.75pc/[uull]_(.68){\widetilde r} |!{[ul];[dlll]}\hole \\ & & \Si \ar@{=}[dd] & & \bullet & \\ \Si \ar@{^{(}->}[r] & \Si\x\Si \ar@{=>}[dr] \ar@/^1.pc/[rr]^(.33){\pr_1\circ\pr_1} |!{[ur];[dr]}\hole & & \Si \ar@{=}[dl] & \\ & & \Si & & & }\,,
\qqq
whose left wing,\ with its lifted action
\qq\nn
\widetilde\la{}_P\colo{\rm At}(P)\x_\Si(P\x M)\too P\x M,\ \bigl([(p_2,p_1)],(p_1,m)\bigr)\longmapsto(p_2,m)
\qqq
descending to the quotient as
\qq\nn
\la_{P\x_\la M}\colo{\rm At}(P)\x_\Si(P\x_\la M)\too P\x_\la M,\ \bigl([(p_2,p_1)],[(p_1,m)]\bigr)\longmapsto[(p_2,m)]
\qqq
(due to the mutual commutativity of $\widetilde\la{}_P$ and $\widetilde\la$),\ captures the automorphisms of interest through restriction of the latter action to bisections within the arrow manifold ${\rm At}(P)$ of the gauge groupoid.

The Ehresmann--Cartan construction rephrased above readily lifts to the tangents of the three bundles:\ $P,P\x M$,\ and $P\x_\la M$,\ giving rise to distinguished Ehresmann connections,\ with the respective horizontal distributions:\ a $\txG$-invariant one $\txH P$ over $P$,\ its $\txG$-invariant pullback $\pr_1^*\txH P$ over $P\x M$,\ and one over $P\x_\la M$ induced from the latter as $\txT\pi_\sim(\pr_1^*\txH P)$ \`a la Kobayashi--Nomizu \cite{Kobayashi:1963}.\ The definition of connections admits the usual equivalent reformulation in terms of smooth families of projectors onto the respective verticals along the distinguished horizontals.\ It starts with a $\txG$-equivariant connection 1-form on $P$.\ In consequence of the global trivialisation\footnote{It implies the same property of $\txV(P\x M)\cong P\x(\ggt\x M)$,\ with the factorised---{\it i.e.},\ trivial as a vector bundle over $M$---fibre $\ggt\x M\cong\Lie(\txG\,\lx_\la M)$,\ another manifestation of non-genericity of the present construction.} of $\txV P\cong P\x\ggt,\ \ggt\equiv\Lie(\txG)$,\ realised by the fundamental vector field $\cK^P\colo P\x\ggt\to\txV P\subset\txT P$ of the free defining action $r_P$,\ the latter vector-valued 1-form is customarily presented as $\cA\in\Om^1(P)\ox\ggt$,\ with $\txH P=\ker\,\cA$.\ As such,\ it yields the standard model of a (local) gauge field $A_i=\si_i^*\cA\in\Om^1(O_i)\ox\ggt$,\ induced with the help of a flat unital section $\si_i=P\t_i^{-1}(\cdot,e)\colo O_i\to P$.\ In the simplest gauging scheme,\ applicable to field theories with lagrangean densities defined in terms of $\txG$-invariant tensors on the configuration fibre $M$,\  the gauge field is coupled to the charged-matter fields $\varphi\in\G(P\x_\la M)$ through the verticalised derivative
\qq\label{eq:cov-der}
\nabla^\cA\colo\G(P\x_\la M)\ni\varphi\longmapsto\Theta_{P\x_\la M}\circ\txT\varphi\equiv\nabla^\cA\varphi\in\G\bigl(\txT^*\Si\ox\varphi^*\txV(P\x_\la M)\bigr)\,,
\qqq
written in terms of the induced (Crittenden) connection 1-form $\Theta_{P\x_\la M}\in\G(\txT^*(P\x_\la M)\ox\txV(P\x_\la M))$,\ which satisfies $\ker\,\Theta_{P\x_\la M}=\txT\pi_\sim(\pr_1^*\ker\,\cA)$.\ The coupling is mediated by the fundamental vector field $\cK^M\colo M\x\ggt\to\txT M$ of the symmetry $\la$,\ a fact manifest in the local trivialisations $\xcF\t_i$,\ in which $\nabla^\cA\varphi$ takes the form
\qq\label{eq:cov-der-loc}
\xcD^{A_i}\varphi_i\equiv\txT\xcF\t_i\circ\nabla^\cA\varphi=\txT\varphi_i-\cK^M(\varphi_i)\circ A_i\,.
\qqq
These local objects exhibit simple (tensorial) behaviour:
\qq\nn
\xcD^{A^\chi_i}\bigl(\chi\circ\varphi\bigr)_i(\si)=\txT_{\varphi_i(\si)}\la_{h_i(\si)}\circ\xcD^{A_i}\varphi_i\,,\qquad\si\in O_i
\qqq
under a {\em simultaneous} gauge transformation of the (local) matter field:
\qq\nn
\varphi_i(\cdot)\longmapsto\la_{h_i(\cdot)}\bigl(\varphi_i(\cdot)\bigr)\equiv(\xcF_*(\chi)\circ\varphi)_i(\cdot)\,,
\qqq
and that of the (local) gauge field:
\qq\label{eq:gauge-trafo-loc-gfield}
A_i\longmapsto\txT_e\Ad_{h_i}\circ A_i+h_i^*\theta_{\rm R}\equiv A^\chi_i\,.
\qqq
The form of the latter is a consequence of having promoted $\chi$ to the rank of a connection-preserving automorphism of $P$,\ with
\qq\label{eq:GaugetrafoA}
\chi^*\cA^\chi\equiv\cA^\chi\circ\txT\chi=\cA\,.
\qqq
Accordingly,\ given a $\txG$-invariant tensor $\cT\in\G(\txT^*M^{\ox n})$ (for $n\in\bN^\x$) defining a density on $\Si$ through an evaluation\footnote{The evaluation may involve a nontrivial geometric structure on $\Si$,\ as is,\ {\it e.g.},\ the case for the metric term in the Polyakov--Alvarez--Gaw\c{e}dzki $\si$-model from Def.\,\ref{def:sigmod}.} 
\qq\label{eq:Tensev}
\corr{\cT,\txT X^{\ox n}}
\qqq
on the tangent of $X\in C^\infty(\Si,M)$,\ we form the corresponding $\Si$-{\em local} gauge invariant by replacing the derivative $\txT X$ in the above expression with the local presentation $\xcD^{A_i}\varphi_i$ of the covariant derivative of the charged-matter field $\varphi\in\G(P\x_\la M)$,\ whereby a {\em minimally coupled} term
\qq\label{eq:min-coupl}
\corr{\cT,\xcD^{A_i}\varphi_i^{\ox n}}
\qqq
arises over $O_i$.\ Resummation of these $\Si$-local gauge invariants over an arbitrary tessellation of the spacetime $\Si$ subordinate\footnote{The subordination of a tessellation $\triangle_\Si$ to the cover $\cO\equiv\{O_i\}_{i\in I}$ boils down to the existence of a map $\iota\colo\triangle_\Si\to I$ such that,\ for every cell $\t\in\triangle_\Si$,\ we have $\t\subset O_{\iota(\t)}$.} to the trivialising cover $\cO$ produces the anticipated {\em global} gauge invariant,\ independent of the choice of tessellation.\ This is a concrete realisation---in the spirit of a field theory with gauge-symmetry defects (see:\ \cite{Runkel:2008gr,Suszek:2012ddg,Suszek:2013})---of the following sheaf-theoretic construction:\ Replace $\cT$ with a family $\{\widetilde\cT[A_i]\}_{i\in I}$ of $\Si$-local tensors on $P\x_\la M$ (written in a self-explanatory shorthand notation,\ in which we identify tensors on $O_i$ resp.\ $M$ with distinguished tensors on $O_i\x M$)
\qq\nn
\widetilde\cT[A_i]:=\xcF\t_i^*\bigl(\cT\circ\bigl(\id_{\txT M}-\cK^M\circ A_i\bigr)^{\ox n}\bigr)\in\G\bigl(\txT^*\pi_{P\x_\la M}^{-1}(O_i)^{\ox n}\bigr)\,,
\qqq
see:\ \eqref{eq:cov-der-loc}.\ The family defines a smooth tensor $\widetilde\cT[\cA]\in\G(\txT^*\pi_{P\x_\la M}^{-1}\check{Y}_\cO^{\ox n})$ on the total space of the surjective submersion $\check{\pi}_{P\x_\la M}\colo\pi_{P\x_\la M}^{-1}\check{Y}_\cO\equiv\bigsqcup_{i\in I}\,\pi_{P\x_\la M}^{-1}(O_i)\to P\x_\la M,\ (\xcF\t_i^{-1}(\si,m),i)\mapsto\xcF\t_i^{-1}(\si,m)$,\ induced by the pullback cover $\pi_{P\x_\la M}^{-1}\cO$ of $P\x_\la M$,\ as {\it per}
\qq\nn
\widetilde\cT[\cA]\rstr_{\pi_{P\x_\la M}^{-1}(O_i)}=\widetilde\cT[A_i]\,.
\qqq
The assumed $\txG$-invariance of $\cT$,\ in conjunction with the affine form of the gluing law 
\qq\nn
A_i\rstr_{O_{ij}}=\txT_e\Ad_{g_{ij}}\circ A_j\rstr_{O_{ij}}+g_{ij}^*\theta_{\rm R}
\qqq
for the local gauge fields over the intersections $O_{ij}=O_i\cap O_j$,\ structurally identical with that of the formerly considered gauge transformation \eqref{eq:gauge-trafo-loc-gfield},\ ensure the descent of $\widetilde\cT[\cA]$ to a unique smooth tensor
\qq\label{eq:desc-min-coupl-T}
\unl{\cT[\cA]}\in\G(\txT^*(P\x_\la M)^{\ox n})
\qqq
with the property
\qq\nn
\check{\pi}_{P\x_\la M}^*\unl{\cT[\cA]}=\widetilde\cT[\cA]\,.
\qqq
This is a simple manifestation of the sheaf property of $\G(\txT^*(P\x_\la M)^{\ox n})$.\ The descended tensor $\unl{\cT[\cA]}$ can now be used in the definition of a coupling 
\qq\nn
\corr{\unl{\cT[\cA]},\txT\varphi^{\ox n}}
\qqq
in the gauge(d) field theory,\ in full analogy with the original one \eqref{eq:Tensev}.\ A specific example of this standard construction is presented in detail in Sec.\,\ref{sec:G-gau-sigmod},\ alongside an important illustration of a more general gauging scheme relevant in the presence of non-tensorial topological couplings.\medskip

We close this section with an answer to the natural question about the geometric significance of the choice of (the isoclass of) the bundle of gauges underlying the construction reviewed above.\ The answer reveals the actual meaning of the latter,\ and---in so doing---emphasises its usefulness as one of systematic procedures that serve to chart the universe of field theories.\ It bases on Steenrod's concept of universal principal $\txG$-bundle \cite{Steenrod:1951} (internal to the category of CW-complexes)
\qq\label{diag:univ-princ-G-bndl}
\alxydim{@C=.75cm@R=1.cm}{ \txG \ar@{^{(}~>}[d] & & \\ \txE\txG \ar[d]_{\pi_{\txE\txG}} \ar@{-->}[dr] & & \txG \ar@{=>}[dl] \ar@/_1.pc/[ll]_{r_{\txE\txG}} \ar@/_1.pc/[ull]_r \\ \txB\txG & \bullet & }\,,
\qqq
determined (up to homotopy equivalence) by its property:\ Given an arbitrary principal $\txG$-bundle \eqref{eq:princ-G-bndl} over a CW-complex $\Si$,\ there exists a continuous map $f\colo\Si\to\txB\txG$ such that $P\cong f^*\txE\txG$,\ and two such pullback bundles $f_a^*\txE\txG,\ a\in\{1,2\}$ are isomorphic iff the corresponding maps $f_a,\ a\in\{1,2\}$ are homotopic.\ Thus,\ inequivalent choices of the bundle of gauges can be viewed---in the spirit of the Yoneda Lemma---as corresponding to topologically inequivalent $\Si$-points in the base $\txB\txG$ of the universal bundle,\ called the classifying space of $\txG$.\ Both $\txB\txG$ and $\txE\txG$ admit concrete (if also formal) models (see:\ \cite{Segal:1968}):\ The former can be identified with the (fat) geometric realisation $\Vert N_\bullet(\bbB\txG)\Vert$ of the nerve of the delooping groupoid $\bbB\txG$,\ the latter -- with the (fat) geometric realisation $\Vert N_\bullet(\txG\,\lx_l\txG)\Vert$ of the nerve of the (self-)action groupoid $\txG\,\lx_l\txG$ (associated with the left regular action),\ see:\ Def.\,\ref{def:fat-geom-real}.\ The surjective submersion $\txE\txG\to\txB\txG$ can be viewed---tautologically,\ but in a conceptually broader context---as a principal $\bbB\txG$-bundle with the same universal property now internalised in the category of principal groupoid bundles of Def.\,\ref{def:princ-gr-bun}.\ This point of view proves very useful,\ as it leads to a meaningful generalisation---the notion of universal principal $\xcG$-bundle $\txE\xcG\to\txB\xcG$,\ with total space 
\qq\nn
\txE\xcG:=\Vert N_\bullet(\xcG\,\lx_l\xcG)\Vert\,,
\qqq
and base,\ termed the {\bf classifying space of} $\xcG$,\ 
\qq\label{eq:class-grpd}
\txB\xcG:=\Vert N_\bullet(\xcG)\Vert\,,
\qqq
for an {\em arbitrary} Lie groupoid $\grpd{\xcG}{M}$,\ see:\ \cite{Arias:2011}.\ The classifying space has an independent interpretation as a (regular) homotopy model of the orbispace $M//\xcG$ (aka the characteristic foliation of $\xcG$).\ In the case of immediate interest,\ that is for the action groupoid $\txG\,\lx_\la M$ encountered before,\ it is completely straightforward to identify:\ It coincides with the universal associated bundle 
\qq\nn
\txE\txG\x_\la M\cong\txB(\txG\,\lx_\la M)\,,
\qqq
the base of the relevant principal groupoid bundle $\txE\txG\x M\to\txE\txG\x_\la M$,\ see:\ \eqref{diag:princ-Gact-bndl}.\ Thus,\ we recognise in the universal prototype $\txE\txG\x_\la M$ of our field-theoretic structure $P\x_\la M$ a homotopy model of the orbispace $M//\txG\,\x_\la M\equiv M//\txG$.\ It also features in Cartan's mixing diagram \cite{Cartan:1950mix}
\qq\label{diag:univ-princ-Gact-bndl}
\alxydim{@C=1.75cm@R=1.5cm}{ \txE\txG \ar[d]_{\pi_{\txE\txG}} \ar@{^{(}~>}[dr] & \txE\txG\x M \ar[l] _{\pr_1} \ar[r]^{\pr_2} \ar[d]_{\pi_\sim} & M \ar@{..>}[d]^{\varpi} \ar@{_{(}~>}[dl] \\ \txB\txG & \txE\txG\x_\la M \ar[l]^{\pi_{\txE\txG\x_\la M}} \ar@{..>}[r]_{\xi} & M//\txG }\,,
\qqq
which emphasises the symmetry between the two fibrations:
\qq\nn
\alxydim{@C=1.5cm@R=1.5cm}{ M \ar@{^{(}~>}[r] & \txE\txG\x_\la M \ar[r]_{\quad\pi_{\txE\txG\x_\la M}} & \txB\txG}\,,\qquad\qquad\alxydim{@C=1.5cm@R=1.5cm}{ \txE\txG \ar@{^{(}~>}[r] & \txE\txG\x_\la M \ar@{..>}[r]_{\xi} & M//\txG }\,,
\qqq
with $\varpi\colo M\to M//\txG,\ m\mapsto[m]$ (the quotient map) and $\xi\colo\txE\txG\x_\la M\to M//\txG,\ [(p,m)]\mapsto[m]$,\ and thus---in the light of the (weak) contractibility of the typical fibre $\txE\txG$ of the second one \cite{Steenrod:1951}---provides an alternative way of establishing the status of $\txE\txG\x_\la M$ as the homotopy quotient,\ see:\ \cite{Tu:2020}.\ We conclude that it is perfectly legitimate to interpret the Ehresmann--Cartan gauging scheme (for group-modelled symmetry) as a construction of a smooth model of the configuration bundle of a field theory with the reduced space of internal degrees of freedom $M//\txG$ (even if---and especially if---the latter is {\em not} smooth).\ However,\ for the interpretation to be meaningful,\ that is for the smooth model to be effective,\ as measured by its affinity with the universal model,\ it is imperative that {\em all} inequivalent $\Si$-points $P\x_\la M$ of the universal model be incorporated in the ensuing gauge(d) field theory.\ Therefore,\ we always ought to look for and satisfy sufficient conditions for the existence of a formulation of that gauge(d) field theory on an {\em arbitrary} associated bundle,\ {\em independently} of its type,\ as reflected by (the cohomology class of) its transition 1-cocycle,\ or local data of the connection,\ for that matter.\ Additional corroboration of this conclusion,\ putting on display the relevance of the potential variability of the gauge field,\ shall be recalled,\ after \cite{Gawedzki:2010rn,Gawedzki:2012fu},\ in Sec.\,\ref{sec:G-gau-sigmod}.\ The Reader is also encouraged to consult \cite[Sec.\,4.4,6.7]{Gawedzki:2010rn} for a concretisation of the present argument through a detailed study of contributions of nontrivial gauge fields to partition functions of the Gaw\c{e}dzki--Kupiainen coset $\si$-models.

\section{Principaloid bundles with connection and their Godement quotients}\label{sec:principaloid}

In this section,\ we recapitulate the essential elements of the recent proposal---due to Strobl and the Author \cite{Strobl:2025}---of a mathematical construction underlying the gauge principle for a symmetry modelled on an arbitrary Lie groupoid,\ based on {\em the fundamental assumption of homogeneity of the distribution of internal structures} (field degrees of freedom,\ gauges,\ {\it etc.})\ {\em over the spacetime of the physical model}.\ The construction generalises,\ in a nontrivial manner,\ the one from the previous section,\ employed in the gauging of symmetries with the (Lie-)group model.\ (Proofs of propositions and theorems quoted in this section can be found in \cite{Strobl:2025}.)

\subsection{The bundles}

We begin with a generalisation of Ehresmann's concept of bundle of gauges,\ or reference frames under equivalence\footnote{These were also referred to by Pradines \cite{Pradines:2007} as `measures of events',\ or `observations',\ carried out by `observers' (represented by fibres of the bundle).}.\ We model the totality of these frames by the arrow manifold of a given Lie groupoid,\ which is then promoted to the rank of the typical fibre of the resultant fibre bundle,\ in keeping with the assumption of homogeneity.\ Furthermore,\ we take as the organising principle of our subsequent construction equivariance with respect to the canonical right-fibred action of the Lie groupoid on the latter manifold,\ admitting a smooth quotient.

\bedef\label{def:principaloid}
Adopt the notation of Def.\,\ref{def:gr-mod}.\ Let $\grpd{\xcG}{M}$ be a Lie groupoid,\ and let $\Si$ be a smooth manifold.\ A {\bf principaloid $\xcG$-bundle} is a fibre-bundle object in $\Man_\xcG$,\vspace{15pt}
\qq\nn
\alxydim{@C=1.25cm@R=1.cm}{ \cO\x\xcG & \cO\x_\Si\xcP \ar[l]^{\cong}_{\xcP\t} \ar[r] \ar[d] & \xcP \ar[d]_{\pi_\xcP} \ar[dr]^{\mu} & & \xcG \ar@{=>}[dl] \ar@/_1.pc/[ll]^{\varrho} \ar@/_2.25pc/[llll]_r \\ & \cO \ar[r]_{\jmath} & \Si & M & }
 \qqq
with a trivial $\xcG$-action on the base $\Si$,\ and that on the total space,\ $\varrho\colo \xcP\,\fibx{\mu}{t}\,\xcG\to\xcP$,\ defined for a moment map $\mu\colo\xcP\to M$ with local model $s$,\ and modelled---through $\xcG$-equivariant local trivialisations $\xcP\t$ over an open cover $\cO=\bigsqcup_{i\in I}\,O_i$---on the canonical right-fibred action $r$ of $\grpd{\xcG}{M}$ on its typical fibre $\xcG$ (see:\ Ex.\,\ref{eg:fibred-act}).
\exdef
\noindent In the light of the standard Clutching Theorem and Prop.\,\ref{prop:rGequiv-LB},\ we obtain
\becor\label{cor:loc-mode-P}
Adopt the notation of Def.\,\ref{def:bisec-act}.\ A principaloid $\xcG$-bundle $\xcP\to\Si$ from Def.\,\ref{def:principaloid} admits a model
\qq
\xcP\cong\bigsqcup_{i\in I}\,(O_i\x\xcG)/\sim_{L_{\b_{\cdot\cdot}}}\,,\qquad\qquad(\si_2,j,g_2)\sim(\si_1,i,g_1)\quad\Longleftrightarrow\quad(\si_1,g_1)=\bigl(\si_2,L_{\b_{ij}(\si_1)}(g_2)\bigr)\,,\cr \label{eq:clutch-xcP}
\qqq
relative to the open cover $\cO$,\ and for a {\bf transition 1-cocycle} 
\qq\nn
\b_{ij}\colo O_{ij}\too\bB(\xcG)\,,
\qqq
satisfying the {\bf 1-cocycle identities}
\qq\label{eq:cocycle}
\b_{ij}\cdot\b_{jk}\rstr_{O_{ijk}}=\b_{ik}\rstr_{O_{ijk}}\,,
\qqq
and realised by the left-multiplication $L\colo\bB(\xcG)\to\Diff(\xcG)$.
\ecor
\noindent It is worth noting that the {\em realisation} $\{L_{\b_{ij}}\}_{i,j\in I}$ carries exactly the same (sheaf-theoretic) information as the underlying 1-cocycle $\{\b_{ij}\}_{i,j\in I}$.\ Indeed,\ we recover the latter by evaluating the former on the identity bisection within $\xcG$.\ In this manner,\ the finite-dimensional principaloid bundle becomes {\em uniquely} associated with the generically infinite-dimensional principal $\bB(\xcG)$-bundle over its base,\ represented by the transition 1-cocycle.

The special character of the action of the structure groupoid on the total space of the principaloid bundle leads to the emergence of a daughter fibre bundle over the same base as a smooth quotient relative to that action.
\bethe\label{thm:duck-as-prince}
Adopt the notation of Defs.\,\ref{def:tstar} and \ref{def:princ-gr-bun},\ and Cor.\,\ref{cor:loc-mode-P}.\ A principaloid $\xcG$-bundle $\xcP$ from Def.\,\ref{def:principaloid} canonically induces a fibre bundle $\xcF$ with model
\qq
\xcF\cong\bigsqcup_{i\in I}\,\bigl( O_i\x M\bigr)/\sim_{  t_*\b_{\cdot\cdot}}\,,\qquad\qquad(\si_2,j,m_2)\sim(\si_1,i,m_1)\quad\Longleftrightarrow\quad(\si_1,m_1)=\bigl(\si_2,t_*(\b_{ij})(\si_1)(m_2)\bigr)\,,\cr\label{eq:clutch-xcF}
\qqq
written in terms of the transition 1-cocycle $\{ \b_{ij}\}_{i,j\in I}$ of $\xcP$,\ which we realise by the shadow action $t_*\colo\bB(\xcG)\to\Diff(M)$.\ It comes with a bundle map
\qq\label{diag:sitting-duck}
\alxydim{@C=.75cm@R=1.cm}{ \xcP \ar[rr]^{\xcD} \ar[dr]_{\pi_\xcP} & & \xcF \ar[dl]^{\pi_\xcF} \\ & \Si & }
\qqq
locally modelled on $ t\colo \xcG\to M$.\ The triple $(\xcP,\xcF,\xcD)$ carries a canonical structure of a principal-$\xcG$-bundle object in the category of fibre bundles over $\Si$, 
\qq\label{diag:duck-as-prince}
\alxydim{@C=.75cm@R=1.cm}{ \Si \ar@{=}[d] & & \xcP \ar[ll]_{\pi_\xcP} \ar@{->>}[d]_{\xcD} \ar[rd]^{\mu} & & \xcG \ar@{=>}[ld] \ar@/_1.pc/[ll]_{\varrho} \\ \Si & & \xcF \ar[ll]^{\pi_\xcF} & M & }\,,
\qqq
with a division map 
\qq\nn
\phi_\xcP\colo\xcP\x_\xcF\xcP\too\xcG
\qqq
with local restrictions
\qq\label{eq:div-maP-loc}\qquad
\phi_\xcP\rstr_{\pi_\xcP^{-1}(O_i)\x_\xcF\pi_{O_i}^{-1}(O_i)}\colo\pi_\xcP^{-1}(O_i)\x_\xcF\pi_{O_i}^{-1}(O_i)\ni\bigl(\xcP\t_i^{-1}(\si,g_1),\xcP\t_i^{-1}(\si,g_2)\bigr)\longmapsto g_1^{-1}.g_2\in\xcG\,,
\qqq
In particular,\ the bundle $\xcF$ can be identified with the smooth Godement quotient $\xcP/\xcG$ (see:\ Thm.\,\ref{thm:Godement}).
\ethe
\noindent By its very construction,\ generalising that of the classic group quotient $P\x_\la M\equiv(P\x M)/\txG$,\ the new bundle becomes a natural candidate for the configuration bundle of a gauge(d) field theory-to-be.\ Its global sections $\varphi\in\G(\xcF)$,\ to be interpreted as (charged-matter) fields of the said theory,\ admit---through local trivialisations $\xcF\t_i\colo\pi_\xcF^{-1}(O_i)\xrightarrow{\cong}O_i\x M$---the anticipated  presentation by locally smooth maps $\varphi_i:=\pr_2\circ\xcF\t_i\circ\varphi\rstr_{O_i}$ to the space $M$ of internal degrees of freedom of the theory,\ related by local gauge (or reference-frame) redefinitions $t_*(\b_{ij}(\si))$ over intersections $O_{ij}\ni\si$ of their respective domains. 
\bedef
Adopt the notation of Thm.\,\ref{thm:duck-as-prince}.\ The Godement quotient $\xcF\cong\xcP/\xcG$ is called the {\bf shadow of} $\xcP$,\ and the quotient map $\xcD$ is called the {\bf sitting-duck map}.
\exdef

The original identification of the principaloid bundle as a bundle of gauges immediately suggests a suitable generalisation of the classic definition of a fibre bundle over the spacetime $\Si$ encoding---through its fibred action---the distinguished $\xcG$-equivariant automorphisms of the principaloid bundle,\ commuting with the {\em global} (vertical) fibrewise action of the structure groupoid,\ and,\ as such,\ representing {\em local} changes of gauge.\ While a na\"ive immitation of the prototypical quotient construction $(P\x P)/\txG$,\ which works for the delooping groupoid $\bbB\txG$ of Ex.\,\ref{eg:deloop},\ is ruled out for a general Lie groupoid due to the fibred nature of the defining action $\varrho$,\ we nevertheless have
\bedef\label{def:AtP-bun}
Adopt the notation of Def.\,\ref{def:bisec-act},\ Cor.\,\ref{cor:loc-mode-P} and Prop.\,\ref{prop:G-in-B_act}.\ 
We call the fibre bundle $\pi_{{\rm At}(\xcP)}\colo{\rm At}(\xcP)\to\Si$ with typical fibre $\xcP$ and model
\qq\nn
\bigsqcup_{i\in I}\,\bigl(\xcP\x O_i\bigr)/\sim_{B\varrho_{\b_{\cdot\cdot}^{-1}}}\too\Si,\ [(p,\si)]\longmapsto\si\,,
\qqq
written in terms of the action $B\varrho\colo\bB(\xcG)\to\Diff(\xcP)$ induced from $\varrho$,\ the {\bf Atiyah bundle of} $\xcP$.\ We also denote,\ in the above model,
\qq\nn
\pi_1\colo\bigsqcup_{i\in I}\,\bigl(\xcP\x O_i\bigr)/\sim_{B\varrho_{\b_{\cdot\cdot}^{-1}}}\too\Si,\ [(p,\si)]\longmapsto\pi_\xcP(p)\,.
\qqq
The bundle admits further resolution:
\qq\label{eq:AtP-loc-res}
{\rm At}(\xcP)\cong\bigsqcup_{i,j\in I}\,\bigl(O_i^{(1)}\x\xcG\x O_j^{(2)}\bigr)/\sim_{L^{(1)}_{\b_{\cdot\cdot}}\circ R^{(2)}_{\b_{\cdot\cdot}^{-1}}}\,,\\ \cr
(\si_{21},g_2,\si_{22},k,l)\sim(\si_{11},g_1,\si_{12},i,j)\quad\Longleftrightarrow\quad(\si_{11},g_1,\si_{12})=\bigl(\si_{21},\bigl(L_{\b_{ik}(\si_{11})}\circ R_{\b_{jl}(\si_{12})^{-1}}\bigr)(g_2),\si_{22}\bigr)\,,\nn
\qqq
expressed in terms of the the left- and right-multiplication $L,R\colo\bB(\xcG)\to\Diff(\xcG)$.
\exdef
\noindent Just as in the classic setting,\ we recognise in the total space of the bundle a natural fibration over the pair groupoid of $\Si$ (encoding diffeomorphisms of the base),\ with the typical fibre given by $\grpd{\xcG}{M}$.
\bethe\label{thm:AtPoid}
Adopt the notation of Defs.\,\ref{def:AtP-bun} and \ref{def:bisec-act},\ and of Ex.\,\ref{eg:pairgrpd}.\ The pair $({\rm At}(\xcP),\xcF)$ carries a canonical structure of a Lie groupoid\vspace{-8pt}
\qq\label{diag:Grmatbndldiag}\qquad\qquad
\alxydim{@C=1.75cm@R=1.25cm}{ {\rm At}(\xcP)\x_{\xcF}{\rm At}(\xcP) \ar[r]^{\quad\sfM} & {\rm At}(\xcP) \ar[r]^\sfJ & {\rm At}(\xcP) \ar@<.25ex>[r]^{\sfS} \ar@<-.25ex>[r]_{\sfT} & \xcF \ar@/_1.5pc/[l]_{\sfI} }
\qqq
with structure maps covering the corresponding ones of ${\rm Pair}(\Si)$,\ and modelled fibrewise on the corresponding ones of $\xcG$ in local trivialisations, 
\qq
&\sfS\colo {\rm At}(\xcP)\too\xcF,\ [(\si_1,g,\si_2,i,j)]\longmapsto[(\si_2, s( g),j)]\,,&\cr\cr
&\sfT\colo {\rm At}(\xcP)\too\xcF,\ [(\si_1,g,\si_2,i,j)]\longmapsto[(\si_1, t( g),i)]\,,&\nn\cr\cr
&\sfI\colo \xcF\too{\rm At}(\xcP),\ [(\si,m,i)]\longmapsto[(\si,\Id_m,\si,i,i)]\,,&\cr\cr
&\sfJ\colo {\rm At}(\xcP)\too{\rm At}(\xcP),\ [(\si_1,g,\si_2,i,j)]\longmapsto[(\si_2,g^{-1},\si_1,j,i)]\,,&\cr\cr
&\sfM\colo {\rm At}(\xcP){}_{\sfS}\hspace{-1pt}\x_{\sfT}\hspace{-1pt}{\rm At}(\xcP)\too {\rm At}(\xcP),\ \bigl([(\si_1,g_1,\si_2,i,j)],[(\si_2,g_2,\si_3,j,l)]\bigr)\longmapsto[(\si_1,g_1.g_2,\si_3,i,l)]\,.&
\qqq
The Lie groupoid fits into the following short exact sequence of Lie groupoids:
\qq\label{ses:Atiyah}\qquad
\alxydim{@C=1.75cm@R=1.25cm}{ {\rm Ad}(\xcP) \ar@{^{(}->}[r]^{j_{{\rm Ad}(\xcP)}} \ar@<.5ex>[d]_{\sfT\vert\;\; } \ar@<-.5ex>[d]^{\;\; \sfS\vert} & {\rm At}(\xcP) \ar@{->>}[r]^{\pi} \ar@<.5ex>[d]_{\sfT \;\;} \ar@<-.5ex>[d]^{\;\; \sfS} & \Si\x\Si{} \ar@<.5ex>[d]_{\pr_1\;\; } \ar@<-.5ex>[d]^{\;\; \:\pr_2}\\ \xcF \ar@{=}[r] & \xcF \ar[r]_{\pi_\xcF} & \Si }\,,
\qqq
where $\pi=(\pi_1,\pi_{{\rm At}(\xcP)})$ and $j_{{\rm Ad}(\xcP)}$ is the embedding of ${\rm Ad}(\xcP) = \pi^{-1}(\Id(\Sigma))$.\ In particular,\ the quadruple $({\rm At}(\xcP), \mathrm{Pair}(\Sigma), \xcG, \pi)$ is a fibre-bundle object in the category of Lie groupoids: 
\qq\label{eq:EA-grpd-bndl}
\alxydim{@C=.5cm@R=1.cm}{\xcG \: \ar@{^{(}~>}[r] & {\rm At}(\xcP) \ar[d]^{\pi} \\ & {\rm Pair}(\Si)}\,.
\qqq

The sub-bundle ${\rm Ad}(\xcP)$ has a model
\qq\label{eq:clutch-xcA}
{\rm Ad}(\xcP)\cong\bigsqcup_{i\in I}\,\bigl( O_i\x \xcG\bigr)/\sim_{C_{\b_{\cdot\cdot}}}\,,
\qqq
written in terms of the transition 1-cocycle $\{\b_{ij}\}_{i,j\in I}$ of $\xcP$,\ realised by the conjugation $C\colo\bB(\xcG)\to\Diff(\xcG)$.
\ethe

\bedef\label{def:AtP-grpd}
Adopt the notation of Thm.\,\ref{thm:AtPoid}.\ We call $\grpd{{\rm At}(\xcP)}{\xcF}$ the {\bf Ehresmann--Atiyah groupoid},\ or the {\bf gauge groupoid} of $\xcP$.\ The short exact sequence \eqref{ses:Atiyah} is referred to as the {\bf Atiyah sequence} for $\xcP$.\ The bundle ${\rm Ad}(\xcP)$ is termed the {\bf adjoint bundle} of $\xcP$,\ and $\grpd{{\rm Ad}(\xcP)}{\xcF}$ the {\bf adjoint groupoid} of $\xcP$.
\exdef

The fundamental feature of the bundle object \eqref{eq:EA-grpd-bndl} is the existence of a coherent lift of the canonical action of its base ${\rm Pair}(\Si)$ on that of the principaloid bundle to the respective total spaces.\ The lift commutes with $\varrho$,\ and so it descends to the shadow.\ This is crucial for the induction of distinguished automorphisms of the latter bundle from the $\xcG$-equivariant ones of the principaloid bundle,\ as shall be recalled in the next (sub)section.
\bethe\label{thm:princ-as-At-mod} 
Adopt the notation of Defs.\,\ref{def:AtP-grpd} and \ref{def:Trident-oid}.\ The Ehresmann--Atiyah groupoid ${\rm At}(\xcP)$ acts on the principaloid bundle $\xcP$ and its shadow $\xcF$ in the following way:
\bit
\item[$\bullet$] On every principaloid bundle $\xcP$,\ there exists a canonical structure of a left ${\rm At}(\xcP)$-module,\ with momentum $\mu_\xcP=\xcD\colo\xcP\to\xcF$ and action
\qq\label{eq:lamPa}\qquad\qquad
\la_\xcP\colo {\rm At}(\xcP){}_{\sfS}\hspace{-2pt}\x_{\xcD}\hspace{-2pt}\xcP\too\xcP,\ \bigl([(\si_1,g,\si_2,i,j)],[(\si_2,h,k)]\bigr)\longmapsto[(\si_1,g.(\b_{jk}(\si_2)\lact h),i)]\,,
\qqq
written in the local models:\ \eqref{eq:AtP-loc-res} for ${\rm At}(\xcP)$ and \eqref{eq:clutch-xcP} for $\xcP$,\ for $\si_1\in O_i$ and $\si_2\in O_{jk}$.\
\item[$\bullet$] This structure covers the canonical structure of a left ${\rm Pair}(\Si)$-module on $\Si$,\ with momentum $\mu_\Si\equiv\id_\Si$ and action
\qq\nn
\la_\Si\colo(\Si\x\Si)\,{}_{\pr_2}\hspace{-3pt}\x_{\id_\Si}\hspace{-2pt}\Si\too\Si,\ \bigl((\si_1,\si_2),\si_2\bigr)\longmapsto\si_1\,.
\qqq
\item[$\bullet$] On the corresponding shadow $\xcF$,\ there exists a canonical structure of a left ${\rm At}(\xcP)$-module,\ with momentum $\mu_\xcF=\id_\xcF$ and action 
\qq\label{eq:lamFa}
\la_\xcF\colo{\rm At}(\xcP){}_{\sfS}\hspace{-2pt}\x_{\id_\xcF}\hspace{-2pt}\xcF\too\xcF,\ \bigl([(\si_1,g,\si_2,i,j)],[(\si_2,m,k)]\bigr)\longmapsto[(\si_1,t(g),i)]\,.
\qqq
\item[$\bullet$] The action $\la_\xcP$ is intertwined with $\la_\xcF$ by the sitting-duck map $\xcD$,\ as reflected in the identities
\qq\label{eq:lamPa-int-lamFa}
\mu_\xcP=\mu_\xcF\circ\xcD\,,\qquad\qquad\la_\xcF\circ\bigl(\id_{{\rm At}(\xcP)}\x\xcD\bigr)=\xcD\circ\la_\xcP\,.
\qqq
\eit
The quintuple $({\rm At}(\xcP),\xcP,\xcG;\Si,\xcG)$ is a left trident,\ captured by the following diagram
\qq\label{diag:Trident}
\alxydim{@C=.75cm@R=1.cm}{ & \xcG \ar@{^{(}.>}[dr] & & & \\ {\rm At}(\xcP) \ar@{=>}[rd] \ar@/^1.pc/[rr]^{\la_\xcP} & & \xcP \ar[dl]_{\xcD} \ar[rd]^{\mu} \ar[dd]^{\pi_\xcP} & & \xcG \ar@{=>}[ld] \ar@/_1.pc/[ll]_{\varrho} \\ & \xcF & & M & \\ & & \Si & &}\,.
\qqq
In particular,\ there exists a division map 
\qq\label{eq:divAt}
\psi_\xcP\colo\xcP\fibx{\mu}{\mu}\xcP\too{\rm At}(\xcP)
\qqq
with the property
\qq\label{eq:divAt-def-id}
(\psi_\xcP,\pr_2)=(\la_\xcP,\pr_2)^{-1}\,.
\qqq
\ethe

The above construction is readily seen to subsume the classic ones,\ long-known to capture the geometry of the gauge principle for group-modelled symmetries.
\beg\label{eg:deloop-oid}
For the delooping groupoid $\bbB\txG$ from Ex.\,\ref{eg:deloop},\ we recover as the principaloid bundle a principal $G$-bundle $P$ of gauges $\si\in\G_{\rm loc}(P)$:\vspace{15pt}
\qq\nn
\alxydim{@C=1.25cm@R=1.cm}{ \cO\x G & \cO\x_\Si P \ar[l]^{P\t}_{\cong} \ar[r] \ar[d] & P \ar@{->>}[d]_{\pi_P} \ar@{-->}[dr] & & G \ar@{=>}[dl] \ar@/_1pc/[ll]^{r_P} \ar@/_2.25pc/[llll]_r \\ & \cO \ar[r]_{\ \jmath} & \Si & \bullet & }\,.
\qqq
The shadow coincides with the base $\Si$ of $P$.\ The Ehresmann--Atiyah groupoid of this principaloid bundle is the classic one,\ ${\rm At}(P)$,\ and it gives rise to the trident \eqref{diag:G-trident}.
\eeg

\beg\label{eg:act-grp-oid-red}
For the action groupoid $G\lx_\la M$ from Ex.\,\ref{eg:actgrpd},\ and for the {\em reduced} structure group $\iota(G)$ composed of flat bisections from Ex.\,\ref{eg:flatbisec},\ we find the extended bundle\vspace{15pt}
\qq\nn
\alxydim{@C=1.25cm@R=1.cm}{ \cO\x(G\x M) & \cO\x_\Si(P\x M) \ar[l]^{P\t\x\id}_{\cong} \ar[r] \ar[d] & P\x M \ar@{->>}[d]_{\pi_P\circ\pr_1} \ar[dr]^{\pr_2} & & G\x M \ar@{=>}[dl] \ar@/_1pc/[ll]^{\widetilde\la} \ar@/_2.25pc/[llll]_r \\ & \cO \ar[r]_{\ \jmath} & \Si & M & }
\qqq
as the principaloid bundle,\ and the associated bundle
\qq\nn
\alxydim{@C=1.25cm@R=1.cm}{ \cO\x M & \cO\x_\Si\xcF \ar[l]^{\xcF\t}_{\cong} \ar[r] \ar[d] & \xcF\equiv(P\x M)/G\equiv P\x_\la M \ar@{->>}[d]_{\pi_\xcF} \\ & \cO \ar[r]_{\ \jmath} & \Si }
\qqq
as its shadow.\ As for every principaloid bundle,\ there exists an Ehresmann--Atiyah groupoid ${\rm At}(P\x M)$,\ which,\ however,\ has not been considered in the classic setting.\ The obvious reason for that is the existence of the abstract symmetry model $\txG$,\ which subsequently enters the definition of the gauge group,\ whence the more complex structure of the extended trident \eqref{dig:ext-G-trident}.\ The symmetry model $\txG$ is typically distinguished as a finite-dimensional subgroup within the group of diffeomorphisms of the configuration fibre $M$ of a given field theory consisting of those diffeomorphisms which preserve some additional geometric structure on $M$ determining the dynamics of that field theory,\ such as,\ {\it e.g.},\ a metric tensor,\ or a Maxwell field {\it etc.}\ As we shall see in Sec.\,\ref{sec:reduction} (and as has long been known,\ {\it e.g.},\ in the symplectic/Poisson setting of \cite{Weinstein:1987,Coste:1987},\ see:\ \cite{Rybicki:2001}),\ putting such extra structure on the object (resp.\ arrow) manifold of a given Lie groupoid also leads,\ in general,\ to a reduction of the group of bisections to a subgroup composed of those bisections which preserve the extra structure when realised through $t_*$ (resp.\ $L$) as diffeomorphisms of $M$ (resp.\ $\xcG$).\ Anticipating this eventuality,\ we shall consider the groupoidal variant of Cartan's associating construction in Sec.\,\ref{sec:red-ass}.
\eeg

\subsection{The automorphisms,\ and gauge transformations}

The principle of $\xcG$-equivariance,\ introduced in the previous section,\ leads directly to an adaptation of the notion of automorphism.
\bedef\label{def:princ-auts}
Adopt the notation of Def.\,\ref{def:principaloid}.\ We call a bundle automorphism of a principaloid $\xcG$-bundle $\xcP$ {\bf $\xcG$-principal},\ or simply an {\bf automorphisms of $\xcP$},\ if it is $\xcG$-equivariant.\ Whenever,\ in addition,\ it covers the identity on the base $\Si$,\ we term it {\bf vertical},\ or a {\bf gauge transformation}.\ The {\bf group of} {\bf automorphisms} of $\xcP$ shall be denoted as 
\qq\nn
{\rm Aut}(\xcP)\equiv {\rm Aut}_{{\rm {\bf Bun}}(\Si)}(\xcP)\cap\Diff_{\varrho(\xcG)}(\xcP)\,,
\qqq
and its subgroup composed of vertical automorphisms,\ to be referred to as the {\bf gauge group} of $\xcP$,\ as 
\qq\nn
{\rm Gauge}(\xcP) \equiv {\rm Aut}(\xcP)_{\rm vert}\,.
\qqq
\exdef
\noindent Upon invoking Prop.\,\ref{prop:rGequiv-LB} once more,\ we establish
\becor\label{cor:Requiv-principoidle-Auts}
Adopt the notation of Cor.\,\ref{cor:loc-mode-P} and Def.\,\ref{def:princ-auts}.\ A gauge transformation $\Phi\in{\rm Gauge}(\xcP)$ admits a local presentation by smooth maps valued in the structure group $L(\bB(\xcG))$ of $\xcP$,\ {\it i.e.},
\qq\nn
\Phi\rstr\colo \pi_\xcP^{-1}(O_i)\xrightarrow{\ \cong\ }\pi_\xcP^{-1}(O_i),\ \xcP\t_i^{-1}(\si,g)\longmapsto\xcP\t_i^{-1}\bigl(\si,L_{\g_i(\si)}(g)\bigr)
\qqq
for some family of smooth maps
\qq\nn
\g_i\colo O_i\too\bB(\xcG)\,,\quad i\in I\,,
\qqq
satisfying the {\bf gauge 1-coboundary identities}
\qq\label{eq:gau-coboundary}
\b_{ij}\cdot\g_j\rstr_{O_{ij}}=\g_i\cdot\b_{ij}\rstr_{O_{ij}}\,.
\qqq
\ecor

The equivariance of the automorphisms of the principaloid bundle distinguished above enables us to push them down to the shadow along the quotient map $\xcD$.
\berop\label{prop:Ups-ind}
Adopt the notation of Def.\,\ref{def:princ-auts} and Thm.\,\ref{thm:duck-as-prince}.\ There is a canonical group homomorphism
\qq\nn
\xcF_*\colo{\rm Aut}(\xcP)\too{\rm Aut}_{{\rm {\bf Bun}}(\Si)}(\xcF)\,,
\qqq
which induces an action of ${\rm Aut}(\xcP)$ on $\xcF$ intertwined by the sitting-duck map,\ as determined by the identity
\qq\nn
\xcD\circ\Phi=\xcF_*(\Phi)\circ\xcD\,.
\qqq
\eerop
\noindent In this manner,\ we arrive at a natural model of groupoidal gauge symmetry of the configuration bundle of the gauge(d) field theory-to-be.
\bedef
Adopt the notation of Prop.\,\ref{prop:Ups-ind}.\ We call the subgroup $\im\,\xcF_*\subset{\rm Aut}_{{\rm {\bf Bun}}(\Si)}(\xcF)$ the {\bf shadow automorphism group},\ and its elements -- {\bf shadow automorphisms}.\ Upon restriction to ${\rm Gauge}(\xcP)\subset{\rm Aut}(\xcP)$,\ we obtain the {\bf shadow gauge group}
\qq\nn
{\rm Gauge}(\xcF):=\xcF_*\bigl({\rm Gauge}(\xcP)\bigr)\,,
\qqq
whose elements are termed {\bf shadow gauge transformations}.
\exdef

A neat closure of the intrinsically (bi-)Lie-groupoidal model of gauge symmetry systematically reconstructed in our hitherto considerations,\ in the spirit of Ehresmann's original work \cite{Ehresmann:1950} (see also:\ \cite{Pradines:2007}),\ calls for one last
\bedef\label{def:Bis-piproj}
Adopt the notation of Thm.\,\ref{thm:AtPoid}.\ A bisection $\b$ of the Ehresmann--Atiyah groupoid $\grpd{\mathrm{At}(\xcP)}{\xcF}$  is called {\bf $\pi$-projectable} whenever there exists a bisection $\unl\b$ of ${\rm Pair}(\Si)$ such that the dotted rectangle at the right in the diagram
\qq\label{eq:Atiyah-bisec}\qquad
\alxydim{@C=1.5cm@R=1.cm}{ {\rm Ad}(\xcP) \ar@{^{(}->}[r]^{j_{{\rm Ad}(\xcP)}} \ar@<.5ex>[d]_{\sfT\;\; } \ar@<-.5ex>[d]^{\;\; \sfS} & {\rm At}(\xcP) \ar@{->>}[r]^{\pi} \ar@<.5ex>[d]_{\sfT \;\;} \ar@<-.5ex>[d]^{\;\; \sfS} & \Si\x\Si{} \ar@<.5ex>[d]_{\pr_1\;\; } \ar@<-.5ex>[d]^{\;\; \:\pr_2} \\ \xcF \ar@{=}[r] & \xcF \ar@{->>}[r]_{\pi_\xcF} \ar@/^2.5pc/@<-.25ex>@{..>}[u]^{\b} & \Si \ar@/^-2.5pc/@<-.25ex>@{..>}[u]_{\unl{\b}} }
\qqq
is commutative.\ Such bisections compose a subgroup in ${\rm Bisec}({\rm At}(\xcP))$,\ denoted as
\qq\nn
{\rm Bisec}_\pi\bigl({\rm At}(\xcP)\bigr)\,.
\qqq
\exdef
\noindent Comparison of local data of $\xcG$-equivariant automorphisms of the principaloid bundle with those of $\pi$-projectable bisections of the corresponding Ehresmann--Atiyah groupoid now yields 
\bethe\label{thm:autP-from-BisAt}
Adopt the notation of Def.\,\ref{def:Bis-piproj} and Prop.\,\ref{prop:Ups-ind}.\ There is a canonical group isomorphism
\qq\nn
\b_\cdot\colo{\rm Aut}(\xcP)\xrightarrow{\ \cong\ }{\rm Bisec}_\pi\bigl({\rm At}(\xcP)\bigr)\,,
\qqq
which restricts to a group isomorphism 
\qq\nn
{\rm Gauge}(\xcP)\cong{\rm Bisec}_\pi({\rm Ad}(\xcP))
\qqq
between the respective subgroups.\ The former isomorphism renders commutative the following diagram
\qq\label{diag:Tbeta}
\alxydim{@C=.5cm@R=1.cm}{ & {\rm Bisec}_\pi\bigl({\rm At}(\xcP)\bigr) \ar[dr]^{\sfT_*} & \\ {\rm Aut}(\xcP) \ar[rr]_{\xcF_*} \ar[ur]^{\b_\cdot} & & {\rm Aut}(\xcF) }\,,
\qqq
in which $\sfT_*$ is the standard pushforward along $\sfT$.
\ethe
\noindent The last theorem leads to a natural geometrisation of $\xcG$-equivariant automorphisms of the principaloid bundle,\ in which the full trident structure \eqref{diag:Trident} is seen to come together.
\berop
Adopt the notation of Thms.\,\ref{thm:princ-as-At-mod} and \ref{thm:autP-from-BisAt}.\ The structure of a left ${\rm At}(\xcP)$-module on $\xcP$ gives rise to a geometric implementation of automorphisms ${\rm Aut}(\xcP)\ni\Phi$ on $\xcP$ (resp.\ on $\xcF$) through restriction of $\la_\xcP$ (resp.\ of $\la_\xcF$) to the corresponding submanifolds $\b_\Phi(\xcF)\subset{\rm At}(\xcP)$,\ embedded by $\pi$-projectable bisections $\b_\Phi\in{\rm Bisec}_\pi({\rm At}(\xcP))$,
\qq\nn
\alxydim{@C=.75cm@R=1.cm}{ & & & \xcP \ar@/_1.25pc/[ddl]_{\mu_\xcP\equiv\xcD} \ar@{->>}[dr]^{\xcD} & \\ {\rm At}(\xcP) \ar@{=>}[drr] \ar@{->>}[dd]_{\pi} \ar@/^1.75pc/[urrr]^{\la_\xcP} \ar@/^2.25pc/[rrrr]^(.38){\la_\xcF} |!{[urrr];[dr]}\hole & \b_\Phi(\xcF) \ar@{_{(}->}[l] & & & \xcF \ar[dll]^{\mu_\xcF\equiv\id_\xcF} \ar@{->>}[dd]^{\pi_\xcF} \\ & & \xcF \ar@{->>}[dd]^{\pi_\xcF} \ar[ul]_{\b_\Phi} & & \\ \Si\x\Si \ar@{=>}[drr] \ar@/^2.25pc/[rrrr]^(.38){\pr_1\circ\pr_1} |!{[urr];[drr]}\hole & (f,\id_\Si)(\Si) \ar@{_{(}->}[l] & & & \Si \ar[dll]^{\mu_\Si} \\ & & \Si \ar[ul]_{(f,\id_\Si)} & & }\,.
\qqq
This is captured by the following commutative diagrams:
\qq\label{diag:betPhi-Phi}
\alxydim{@C=2.cm@R=1.cm}{ \xcF{}_{\id_\xcF}\hspace{-3pt}\x_{\mu_\xcP}\hspace{-1pt}\xcP \ar[r]^{\b_\Phi\x\id_\xcP} & {\rm At}(\xcP){}_{\sfS}\hspace{-2pt}\x_{\xcD}\hspace{-2pt}\xcP \ar[d]^{\la_\xcP} \\ \xcP \ar[u]^{(\mu_\xcP,\id_\xcP)} \ar[r]_{\Phi} & \xcP }\,,
\qqq
and
\qq\nn
\alxydim{@C=2.cm@R=1.cm}{ & {\rm At}(\xcP)\,\,\fibx{\sfS}{\id_\xcF}\xcF \ar[d]^{\la_\xcF} \\ \xcF \ar[ur]^{\b_\Phi\x\id_\xcF} \ar[r]_{\xcF_*(\Phi)} & \xcF }\,.
\qqq
\eerop

\beg
Automorphisms of the Ehresmann principaloid $\bbB\txG$-bundle from Ex.\,\ref{eg:deloop-oid} are the standard $\txG$-equivariant bundle automorphisms of $P$,\ in bijection with (all) bisections of the Ehresmann--Atiyah groupoid ${\rm At}(P)$.
\eeg

\beg\label{eg:red-class-EA-grpd}
Automorphisms of the Cartan principaloid $\txG\,\lx_\la M$-bundle $P\x M$ from Ex.\,\ref{eg:act-grp-oid-red} are usually not considered in all generality,\ for the same reason as that given for the absence of the Ehresmann--Atiyah groupoid ${\rm At}(P\x M)$ in the literature.\ Instead,\ distinguished automorphisms of $P\x M$ are induced from those of $P$,\ represented by the underlying Ehresmann--Atiyah groupoid ${\rm At}(P)$,\ which acts on the first cartesian factor in $P\x M$.\ The action automatically commutes with the right action of the structure groupoid $\txG\,\lx_\la M$,\ the latter reducing to the right defining action of the structure group $\txG$ on that factor,\ and so it descends to the shadow $P\x_\la M$.\ Once more,\ we end up with the reduced structure group $\iota(\txG)\subsetneq{\rm Bisec}(\txG\,\lx_\la M)$.
\eeg

\brem\label{rem:CB-4-grpd}
We may invoke,\ once more,\ the construction of the universal principal $\xcG$-bundle $\txE\xcG\to\txB\xcG$ from \cite{Arias:2011},\ recalled in sufficient generality at the end of Sec.\,\ref{sec:class-gau-princ},\ to justify the interpretation of the gauge principle employing {\em all} inequivalent principal $\xcG$-bundles $\xcD\colo\xcP\to\xcF$ in $\Man/\Si$ as a method of constructing an effective model of a field theory with configuration bundle $M//\xcG$ over a given spacetime $\Si$ {\em in keeping with the principle of homogeneity}.\ This last aspect of the construction advocated in \cite{Strobl:2025} and in the present paper deserves an extra comment:\ If we were to adopt as the prototype of the present generalisation Ehresmann's construction of the abstract bundle of gauges \eqref{eq:princ-G-bndl} (resp.\ \eqref{diag:univ-princ-G-bndl}) instead of Cartan's mixing construction \eqref{diag:princ-Gact-bndl} (resp.\ \eqref{diag:univ-princ-Gact-bndl}) of {\rm gauges-in-realisation,\ fibred over the space of internal degrees of freedom},\ we might be led to identify the base of the principal $\xcG$-bundle with the spacetime of the field theory under construction---this is the case for applications of the old construct,\ due to MacKenzie,\ Moerdijk and Mr\v cun,\ of a principal $\xcG$-bundle,\ introduced as a mathematical model underlying the gauge principle for groupoidal symmetries.\ The important feature of this construct which distinguishes it from the principaloid $\xcG$-bundle (and the attendant shadow bundle) is the violation of the aforementioned principle of homogeneity in general.\ Indeed,\ as demonstrated in \cite{Moerdijk:2003mm},\ every principal $\xcG$-bundle is locally trivialisable,\ with the local model given by the canonical principal $\xcG$-bundle---the unit bundle from Ex.\,\ref{eg:triv-grpd-bndle}.\ However,\ the latter is,\ in general,\ only a surjective submersion,\ without a typical fibre,\ as its local fibres are $t$-fibres of $\xcG$,\ and it is easy to find Lie groupoids with pairwise {\em non-}diffeomorphic $t$-fibres.\ A simple example is provided by a full subgroupoid of an action groupoid $\txG\,\lx_\la M$ obtained by the removal of some,\ but not all points from some $\txG$-orbits.
\erem

\subsection{The connections,\ and gauge fields}

The principle of $\xcG$-equivariance can be lifted coherently to the tangent of the principaloid bundle,\ whereby an adaptation of the notion of connection arises which completes Ehresmann's construction in the present Lie-groupoidal setting.
\bedef\label{def:principaloid-conn}
Adopt the notation of Def.\,\ref{def:principaloid}.\ A (compatible) {\bf connection} on a principaloid $\xcG$-bundle $\xcP$ is an Ehresmann connection $\txT\xcP=\txV\xcP\oplus\txH\xcP$ with a $\xcG$-invariant horizontal distribution $\txH\xcP\subset\ker\,\txT\mu$.\ In other words,\ for every $g\in\xcG$ and $p\in\mu^{-1}(\{t(g)\})$,\ the horizontal distribution satisfies
\qq \label{eq:conn} 
\txT_p\varrho_g (\txH_p\xcP)=\txH_{\varrho_g(p)}\xcP\,,
\qqq
where $\varrho_g(p)\equiv\varrho(p,g)$.
\exdef 
\noindent Equivalently,\ and conveniently from the field-theoretic point of view,\ we may work with smooth families of $\xcG$-equivariant projectors onto the vertical.
\bedef\label{def:Ginv-principal-conn} 
Adopt the notation of Def.\,\ref{def:principaloid}.\ A (compatible) {\bf connection 1-form} on $\xcP$ is a $\txV\xcP$-valued 1-form $\Theta\in\Om^1(\xcP,\txV\xcP)\equiv\G(\txT^*\xcP\ox\txV\xcP)$
such that,\ when viewed as an element in $\End{(\txT\xcP)}$:
\bit
\item[(GC1)] $\Theta\rstr_{\txV\xcP}=\id_{\txV\xcP}$\, (a projection onto $\txV\xcP$);
\item[(GC2)] $\txT\mu\circ\Theta=\txT\mu$\, (kernel inclusion);
\item[(GC3)] $\txT\varrho\circ(\Theta\x\id_{\txT\xcG})\rstr_{\txT(\xcP\,\fibx{\mu}{t}\,\xcG)}=\Theta\circ\txT\varrho$\, ($\xcG$-equivariance).
\eit
\exdef
\noindent We have the anticipated
\berop\label{prop:conn-1-Econn}
Connections on a principaloid bundle are in a one-to-one correspondence with connection 1-forms on it.
\eerop
\noindent With the help of the standard construction using a smooth partition of unity associated with the trivialising cover of the base of the principaloid bundle,\ we readily establish
\bethe
On every principaloid bundle,\ there exists a connection.
\ethe

Similarly as in the case of automorphisms,\ building the principle of $\xcG$-equivariance into the definition of the connection enables us to employ the classic Kobayashi--Nomizu mechanism \cite{Kobayashi:1963} to descend the connection to the shadow.
\begin{propanition}\label{prop:shad-conn}
Adopt the notation of Def.\,\ref{def:principaloid-conn}.\ A connection on a principaloid bundle $\xcP$ canonically induces an Ehresmann connection on $\xcF$ along the sitting-duck map $\xcD$ as
\qq\label{eq:Whitney-shadow}
\txT\xcF=\txV\xcF\oplus\txH\xcF\,,\qquad\qquad\txV\xcF=\txT\xcD(\txV\xcP)\,,\qquad\txH\xcF=\txT\xcD(\txH\xcP)\,.
\qqq
We call it the {\bf shadow connection} on $\xcF$.
\end{propanition}
\noindent Once more,\ we have the equivalent description in terms of a smooth family of projectors.
\bedef\label{def:shadow-conn}
Adopt the notation of Prop.\,\ref{prop:shad-conn} and Thm.\,\ref{thm:duck-as-prince}.\ Given a connection $\txT\xcP=\txV\xcP\oplus\txH\xcP$ on $\xcP$ and the corresponding shadow connection $\txT\xcF=\txV\xcF\oplus\txT\xcD(\txH\xcP)$ on $\xcF$,\ the {\bf shadow connection 1-form} on $\xcF$ is a $\txV\xcF$-valued 1-form $\Theta_\xcF\in\Om^1(\xcF,\txV\xcF)\equiv\G(\txT^*\xcF\ox\txV\xcF)$ with the property $\txT\xcD(\txH\xcP)=\ker\,\Theta_\xcF$.
\exdef
\noindent A simple structural relation between the two families of projectors---instrumental in establishing the physically all-important gauge-covariance property of the verticalised derivative on $\G(\xcF)$,\ to be reviewed shortly---is captured by
\berop\label{prop:shadconn-idef}
The shadow connection 1-form from Def.\,\ref{def:shadow-conn} is uniquely determined by the identity
\qq\nn
\Theta_\xcF\circ\txT\xcD=\txT\xcD\circ\Theta\,.
\qqq
\eerop

The key field-theoretic r\^ole of a principal connection is that of a source of a background gauge field coupling to various charged-matter fields.\ A suitable adaptation of the former notion,\ taking into account the fibre bundle behind the Lie algebra $\G_{\rm c}(\xcE)$ of the structure group $\bB(\xcG)$ under consideration (see:\ Thm.\,\ref{thm:LieBis}),\ is provided in
\bedef\label{def:principoidle-conn-cech}
Adopt the notation of Defs.\,\ref{def:tstar} and \ref{def:bisec-act},\ of Cor.\,\ref{cor:loc-mode-P},\ and of Ex.\,\ref{eg:MC}.\ Let $\grpd{\xcG}{M}$ be a Lie groupoid,\ and let $\xcE$ be its tangent algebroid.\ {\bf Local connection data} on a principaloid $\xcG$-bundle $\xcP$ over $\Si$ from Def.\,\ref{def:principaloid} are an assignment,\ to elements of a trivialising cover $\cO\equiv\{O_i\}_{i\in I}$ of $\Si$,\ of $\txT O_i$-foliated $\pr_2^*\xcE$-valued 1-forms $A_i$ on $O_i\x M$,\ $A_i\in\G(\pr_1^*\txT^*O_i\ox\pr_2^*\xcE)$,\ subject,\ over double intersections $O_{ij}\ni\si$,\ to the following {\bf gluing law}:
\qq\label{eq:glue-Aij}
A_i\bigl(\si,t_*\bigl(\b_{ij}(\si)\bigr)(m)\bigr)=\txT_{\Id_m}C_{\b_{ij}(\si)}\circ A_j(\si,m)+\theta_{\rm R}\circ\txT_\si\widetilde\b{}_{ij}(\cdot,m)\,,
\qqq
expressed in terms of smooth maps 
\qq\label{eq:bij-tilde}
\widetilde\b{}_{ij}\colo O_{ij}\x M\too\xcG,\ (\si,m)\longmapsto\b_{ij}(\si)(m)\,,
\qqq
using the transition 1-cocycle $\{ \b_{ij}\}_{i,j\in I}$ of $\xcP$,\ and the right-invariant Maurer--Cartan form $\theta_{\rm R}$ on $\xcG$ (see:\ Ex.\,\ref{eg:MC}).\ The 1-forms are termed {\bf local gauge fields}.
\exdef
\noindent A structural link between the two notions is established in
\bethe\label{thm:loc-data-conn}
Adopt the notation of Defs.\,\ref{def:Ginv-principal-conn} and \ref{def:principoidle-conn-cech}.\ For every principaloid $\xcG$-bundle $\xcP$,\ a connection 1-form determines local connection data,\ and {\it vice versa}.\ The relation between the two is given by the formula
\qq\label{eq:loc-data-conn}
\bigl(\xcP\t_i^{-1\,*}\Theta\bigr)(\si,g)=\id_{\txT\xcG}\rstr_g-\txT_{\Id_{t(g)}}r_g\circ A_i\bigl(\si,t(g)\bigr)\,,\qquad (\si,g)\in O_i\x\xcG\,.
\qqq
\ethe

The realisation of the geometric function of the connection---that is:\ covariantisation of derivatives of matter fields through their verticalisation---requires that automorphisms of the principaloid bundle be promoted to connection-preserving ones.\ This reasoning leads to
\bedef\label{def:gauge-trafo-conn}
Adopt the notation of Defs.\,\ref{def:princ-auts} and \ref{def:Ginv-principal-conn}.\ Let $\xcP$ be a principaloid bundle with a connection 1-form $\Theta\in\Om^1(\xcP,\txV\xcP)$.\ Given an automorphism $\Phi\in{\rm Aut}(\xcP)$,\ we denote 
\qq\label{eq:gauge-transform}
\Theta^\Phi:=\Phi^{-1\,*}\Theta\equiv\txT\Phi\circ\Theta\circ\txT\Phi^{-1}\in\Om^1(\xcP,\txV\xcP)\,.
\qqq
Whenever $\Phi\in{\rm Gauge}(\xcP)$,\ the 1-form $\Theta^\Phi$ is called a {\bf gauge transform} of $\Theta$.
\exdef
\noindent The definition implies a simple behaviour of the shadow connection under gauge transformations.
\berop\label{prop:covariance-of-shadow}
Adopt the notation of Defs.\,\ref{def:shadow-conn} and \ref{def:gauge-trafo-conn} and Prop.\,\ref{prop:Ups-ind}.\ For every gauge transformation $\Phi\in{\rm Gauge}(\xcP)$ of a principaloid bundle $\xcP$ with connection 1-form $\Theta$,\ the shadow connection 1-form $\Theta_\xcF\in\Om^1(\xcF,\txV\xcF)$ satisfies the identity
\qq\nn
\xcF_*(\Phi)^*\:\Theta_\xcF^\Phi=\Theta_\xcF\,,
\qqq
in which $\Theta_\xcF^\Phi$ is the shadow connection induced by the gauge transform $\Theta^\Phi$,\ and in which the pullback along $\xcF_*(\Phi)$ is explicitly given by
\qq\nn
\xcF_*(\Phi)^*\:\Theta_\xcF^\Phi\equiv\txT\bigl(\xcF_*(\Phi)\bigr)^{-1}\circ\Theta_\xcF^\Phi\circ\txT\bigl(\xcF_*(\Phi)\bigr)\,.
\qqq
\eerop
\noindent That the above effects the desired promotion is documented by the following
\berop\label{prop:principoidle-connPhi-cech}
Adopt the notation of Defs.\,\ref{def:gauge-trafo-conn},\ \ref{def:tstar} and \ref{def:bisec-act},\ of Cor.\,\ref{cor:loc-mode-P},\ of Thm.\,\ref{thm:loc-data-conn},\ and of Ex.\,\ref{eg:MC}.\ For every gauge transformation $\Phi\in{\rm Gauge}(\xcP)$,\ the corresponding gauge transform $\Theta^\Phi$ of the connection 1-form $\Theta$ is,\ again,\ a connection 1-form on $\xcP$.\ Its local data $\{A^\Phi_i\}_{i\in I}$ associated with a trivialising cover $\cO\equiv\{O_i\}_{i\in I}$ of $\Si$ obey,\ over double intersections $O_{ij}\ni\si$,\ the {\em same} gluing law as their counterparts $\{A_i\}_{i\in I}$ (see:\ \eqref{eq:glue-Aij}):
\qq\label{eq:glue-APhij}
A^\Phi_i\bigl(\si,t_*\bigl(\b_{ij}(\si)\bigr)(m)\bigr)=\txT_{\Id_m}C_{\b_{ij}(\si)}\circ A^\Phi_j(\si,m)+\theta_{\rm R}\circ\txT_\si\widetilde\b{}_{ij}(\cdot,m)\,.
\qqq
\eerop
\beroof
We begin by checking the defining properties of a connection 1-form.\ Thus,\ using $\txT\Phi^{-1}(\txV\xcP)=\txV\xcP$,\ we find
\qq\nn
\Theta^\Phi\rstr_{\txV\xcP}\equiv\txT\Phi\circ\Theta\circ \txT\Phi^{-1}\rstr_{\txV\xcP}=\txT\Phi\circ\id_{\txV\xcP}\circ \txT\Phi^{-1}\rstr_{\txV\xcP}=\id_{\txV\xcP}\,.
\qqq
Next,\ the $\la_\xcP$-invariance of the moment map $\mu$ yields
\qq\nn
\txT\mu\circ\Theta^\Phi\equiv\txT\bigl(\mu\circ\Phi\bigr)\circ\Theta\circ \txT\Phi^{-1}=\txT\mu\circ\Theta\circ\txT\Phi^{-1}=\txT\mu\circ \txT\Phi^{-1}=\txT\bigl(\mu\circ\Phi^{-1}\bigr)=\txT\mu\,.
\qqq
Finally,\ we invoke the commutativity of $\la_\xcP$ and $\varrho$ to compute
\qq\nn
\Theta^\Phi\circ\txT\varrho&=&\txT\Phi\circ\Theta\circ \txT\bigl(\Phi^{-1}\circ\varrho\bigr)=\txT\Phi\circ\bigl(\Theta\circ \txT\varrho\bigr)\circ\bigl(\txT\Phi^{-1}\x\id_{\txT\xcG}\bigr)\rstr_{\txT(\xcP\,\fibx{\mu}{t}\,\xcG)}\cr\cr 
&=&\txT\bigl(\Phi\circ\varrho\bigr)\circ\bigl(\bigl(\Theta\circ\txT\Phi^{-1}\bigr)\x\id_{\txT\xcG}\bigr)\rstr_{\txT(\xcP\,\fibx{\mu}{t}\,\xcG)}=\txT\varrho\circ\bigl(\Theta^\Phi\x\id_{\txT\xcG}\bigr)\rstr_{\txT(\xcP\,\fibx{\mu}{t}\,\xcG)}\,.
\qqq

The gluing law for the local data $\{A^\Phi_i\}_{i\in I}$ of the thus established connection 1-form $\Theta^\Phi$ is obtained from an analysis---conceptually identical to the one carried out for $\Theta$ in \cite[Sec.\,3.2]{Strobl:2025}---of the relations between the local presentations $\xcP\t_i^{-1\,*}\Theta^\Phi$ and $\xcP\t_j^{-1\,*}\Theta^\Phi$ over the intersections $O_{ij}\x\xcG$ of their respective domains.\ Indeed,\ it is relative to the trivialisations $\xcP\t_i$ that the local data of the gauge transform $\Theta^\Phi$ were derived {\it ibidem}.\ Thus,\ the dependence of the local presentation on the index of the cover comes solely from the pullback along the trivialisation,\ without any reference to $\Phi$.
\eroof

Putting the global content of Def.\,\ref{def:gauge-trafo-conn} in conjunction with the local content of Def.\,\ref{def:principoidle-conn-cech},\ we discover the familiar affine-like behaviour of the local gauge field under redefinitions of gauge.
\berop\label{prop:gautrafo-conn-is-conn}
Adopt the notation of Cor.\,\ref{cor:Requiv-principoidle-Auts} and Prop.\,\ref{prop:principoidle-connPhi-cech}.\ For every gauge transformation $\Phi\in{\rm Gauge}(\xcP)$ presented by local data $\g_i\colo O_i\to\bB(\xcG),\ i\in I$,\ local data $\{A_i^\Phi\}_{i\in I}$ of $\Theta^\Phi$ are related to those of $\Theta$ by the following transformation law:
\qq\label{eq:gaugetrafo}
A_i^\Phi\bigl(\si,t_*\bigl(\g_i(\si)\bigr)(m)\bigr)=\txT_{\Id_m}C_{\g_i(\si)}\circ A_i(\si,m)+\theta_{\rm R}\circ\txT_\si\widetilde\g{}_i(\cdot,m)\,,
\qqq
written,\ for arbitrary $\si\in O_i$ and $m\in M$,\ 
in terms of smooth maps 
\qq\label{eq:gammi-tilde}
\widetilde\g{}_i\colo O_i\x M\too\xcG,\ (\si,m)\longmapsto\g_i(\si)(m)\,.
\qqq
\eerop
\noindent The last proposition provides us with a convenient local restatement of the gauge covariance of the (minimal) coupling between charged-matter and gauge fields modelled by the verticalised derivative,\ to be recalled next.\ As such,\ it proves to be instrumental in the analysis of a connection-aided configurational descent that is at the core of the present work.

\beg
Compatible connections on the Ehresmann principaloid $\bbB\txG$-bundle from Ex.\,\ref{eg:deloop-oid} are standard principal connections.\ The corresponding $\txV P$-valued connection 1-forms are usually replaced by $\ggt$-valued 1-forms on $P$ with the help of the trivialisation $\txV P\cong P\x\ggt$ realised by the fundamental vector field $\cK^P$ of the defining action $r_P$. 
\eeg

\beg\label{eg:act-grp-oid-red-conn}
Compatible connections on the Cartan principaloid $\txG\,\lx_\la M$-bundle $P\x M$ from Ex.\,\ref{eg:act-grp-oid-red} are usually not considered in all generality for the same reason as before (see:\ Ex.\,\ref{eg:act-grp-oid-red}).\ Instead,\ those encountered in the literature are induced---through pullback---from compatible connections on the principaloid $\bbB\txG$-bundle $P$.\ As such,\ they automatically descend to the shadow $P\x_\la M$,\ see:\ \cite{Kobayashi:1963}.
\eeg

\subsection{The gauge-covariant derivative}

A prototypical source of a coupling between charged-matter and gauge fields in a generic gauge(d) field theory is the verticalised derivation of the former,\ introduced in
\bedef\label{def:cov-der}
Adopt the notation of Def.\,\ref{def:shadow-conn}.\ The {\bf covariant derivative} of a section $\varphi\in\G(\xcF)$ of the shadow bundle $\xcF$ of a principaloid bundle $\xcP$ {\bf relative to} a connection 1-form $\Theta$ on $\xcP$ is the $C^\infty(\Si,\bR)$-linear mapping
\qq\label{eq:cov-der-oid}
\nabla^\Theta_\cdot\varphi\colo \G(\txT\Si)\too\G(\varphi^*\txV\xcF),\ 
\cV\longmapsto\Theta_\xcF\circ\txT\varphi(\cV)\equiv\nabla^\Theta_\cV\varphi\,.
\qqq
\exdef
\noindent The prime r\^ole of the verticalisation,\ which consists in the covariantisation of the derivative relative to the action of the shadow gauge group,\ is documented in
\berop
Adopt the notation of Defs.\,\ref{def:gauge-trafo-conn} and \ref{def:cov-der},\ and of Prop.\,\ref{prop:Ups-ind}.\ 
For a section $\varphi\in\G(\xcF)$ of the shadow bundle $\xcF$ of a principaloid bundle $\xcP$,\ and a gauge transformation $\Phi\in{\rm Gauge}(\xcP)$ of $\xcP$,\ let $\varphi^\Phi$ denote the corresponding gauge transform of $\varphi$: 
\qq\nn
\varphi^\Phi\equiv\xcF_*(\Phi)\circ\varphi\,.
\qqq
The covariant derivative of $\varphi$ relative to a connection 1-form $\Theta$ on $\xcP$ transforms \emph{covariantly} as 
\qq\nn
\nabla^{\Theta^\Phi}_\cdot\bigl(\varphi^\Phi\bigr)=\txT\bigl(\xcF_*(\Phi)\bigr)\circ\nabla^\Theta_\cdot\varphi\,.
\qqq
\eerop
\brem 
The last result illustrates an important structural feature of the gauge principle that emerges from our considerations:\ The simple transformational behaviour of the geometric coupling between the charged-matter and gauge fields hinges on the simultaneous redefinition of {\em both}:\ the differentiated matter field {\em and} the verticalising gauge field.\ We shall encounter certain complex variations on this universal theme in what follows.
\erem

The coupling referred to above becomes apparent in the local picture,\ in which,\ moreover,\ the tensorial transformation law under redefinitions of gauge acquires a form accessible to a field-theoretic construction which starts---as is usually the case ({\it e.g.},\ in the approach to the gauging that employs gauge-symmetry defects modelling the twisted sector of the gauged field theory,\ see:\ \cite{Runkel:2008gr,Suszek:2012ddg,Suszek:2013})---with geometric structures (tensors,\ higher-geometric objects {\it etc.}) on the configuration fibre $M$ to be reduced.
\bedef\label{def:cov-der-loc-pres}
Adopt the notation of Def.\,\ref{def:cov-der} and Thm.\,\ref{thm:loc-data-conn}.\ For every global section $\varphi\in\G(\xcF)$ of the shadow $\xcF$ of $\xcP$,\ with local data $\varphi_i:=\pr_2\circ\xcF\t_i\circ\varphi\rstr_{O_i}\colo O_i\to M,\ i\in I$ in local trivialisations $\xcF\t_i\colo\pi_\xcF^{-1}(O_i)\xrightarrow{\cong}O_i\x M$ induced by the $\xcP\t_i$ (see:\ the proof of Thm.\,\ref{thm:duck-as-prince}),\ we call the expression
\qq\nn
\xcD^{A_i}\varphi_i:=\pr_2\circ\txT\xcF\t_i\circ\nabla^\Theta_\cdot\varphi\rstr_{O_i}
\qqq
the {\bf local presentation of the covariant derivative} of $\varphi$ {\bf relative to} $\Theta$ (in trivialisation $\xcF\t_i$).
\exdef
\noindent Explicitly,\ we find
\berop\label{prop:loc-pres-cov-der}
Adopt the notation of Def.\,\ref{def:cov-der-loc-pres},\ Cor.\,\ref{cor:Requiv-principoidle-Auts} and Prop.\,\ref{prop:principoidle-connPhi-cech}.\ The local presentation of the covariant derivative of $\varphi\in\G(\xcF)$ relative to $\Theta$ takes the form
\qq\label{eq:cov-der-loc-oid}
\xcD^{A_i}\varphi_i(\si)=\txT_\si\varphi_i-\a_\xcE\circ A_i\bigl(\si,\varphi_i(\si)\bigr)\,,\quad\si\in O_i\,.
\qqq
Its components are subject,\ over double intersections $O_{ij}\ni\si$,\ to the following gluing law:
\qq\nn
\xcD^{A_i}\varphi_i(\si)=T_{\varphi_j(\si)}t_*\bigl(\b_{ij}(\si)\bigr)\circ\xcD^{A_j}\varphi_j(\si)\,,
\qqq
and its tensorial behaviour under a gauge transformation $\Phi\in{\rm Gauge}(\xcP)$ with local presentation $\g_i\colo O_i\to\bB(\xcG),\ i\in I$ is captured by the formula
\qq\nn
\xcD^{A^\Phi_i}\varphi^\Phi_i(\si)=T_{\varphi_i(\si)}t_*\bigl(\g_i(\si)\bigr)\circ\xcD^{A_i}\varphi_i(\si)\,.
\qqq
\eerop
\beroof
Obvious.
\eroof

\beg
The covariant derivative on sections of the shadow $P\x_\la M$ of the Cartan principaloid $\txG\,\lx_\la M$-bundle $P\x M$ from Ex.\,\ref{eg:act-grp-oid-red} obtained from a distinguished connection 1-form on $P\x M$ induced from a principal connection 1-form $\cA\in\Om^1(P)\ox\ggt$ on the principaloid $\bbB\txG$-bundle $P$ is the standard Crittenden(-type) derivative,\ see:\ \cite{Crittenden:1962cd}.
\eeg

\section{Association:\ Reduction of the gauge group \& proliferation of field species}\label{sec:red-ass}

The construction of a principaloid bundle,\ as reviewed in the previous section,\ exhibits a certain intrinsic rigidity,\ which may prove inconvenient from the perspective of both physical and mathematical model building:\ First of all,\ the definition of a principaloid bundle for a {\em given} Lie groupoid $\grpd{\xcG}{M}$,\ in which objects and morphisms are tied once and for all,\ does not come with a natural mechanism of reduction of the canonical structure group given by the group of (all) bisections of the structure groupoid,\ other than {\it ad hoc} and by hand.\ An obvious potential source of such reduction is the presence of an extra geometric structure on the groupoid,\ {\it e.g.},\ a metric structure (see:\ \cite{delHoyo:2018r}),\ a symplectic structure (see:\ \cite{Weinstein:1987,Coste:1987}),\ or an $n$-gerbe geometrising a (higher) gauge field that couples to the charge carried by a physical field (as in a $\si$-model),\ see:\ Sec.\,\ref{sec:reduction} below.\ One might wish to increase the modularity of the construction by separating the structure {\rm groupoid} of the bundle of gauges from its structure {\rm group},\ encoded by the relevant gauge groupoid.\ This is,\ in fact,\ the standard situation encountered in the classic formulation of the gauge principle,\ in which---as observed in Ex.\,\ref{eg:red-class-EA-grpd}---the gauge groupoid ${\rm At}(P)$ provides a proper subset of the set of ($\pi$-projectable) bisections of the Ehresmann--Atiyah groupoid ${\rm At}(P\x M)$ of the principaloid bundle under consideration.\ Another structural problem ensues from the identification of the principaloid bundle as a source of the gauge field:\ Based on standard physical experience (see,\ {\it e.g.}:\ the Standard Model of particle physics),\ one would expect to be able to couple {\em the same} gauge field to multiple {\em distinct} charged-matter fields,\ each coming with its own configuration fibre.\ In the present setting,\ the type of the charged-matter field is frozen as $M$.

Below,\ we propose a geometric mechanism which resolves both issues without ever sacrificing the key advantage of the novel construction of \cite{Strobl:2025},\ which is the existence of a nontrivial fibration of symmetry over the configuration space of the field theory,\ modelled on the source resp.\ target map of the structure groupoid.\ The mechanism is inspired by and,\ in particular,\ subsumes the original Cartan(--Borel) mixing construction reviewed in Sec.\,\ref{sec:class-gau-princ}.\ It is gratifying to conclude,\ once it has been formulated in full detail,\ that it actually fits within the broad conceptual framework of the original definition of a principaloid bundle,\ which thus retains its status of an overarching construction,\ but now acquires some `fine' structure,\ and,\ with it,\ greater modularity.

\subsection{The geometry of the associating construction}

We begin by adapting the Cartan(--Borel) mixing construction to the present setting,\ in which the bundle of gauges has been chosen in the categorified form of a principaloid bundle.
\bedef\label{def:biasbndl}
Adopt the notation of Def.\,\ref{def:bibndl}.\ Let $\,\grpd{\xcG_a}{M_a},\ a\in\{1,2\}$ be Lie groupoids,\ and let $(\widehat P;\mu_1,\la_1;$ $\mu_2,\varrho_2)$ be a biprincipal $(\xcG_1,\xcG_2)$-bibundle.\ Furthermore,\ let $(X_1,\mu_{X_1},\la_{X_1})$ be a left $\xcG_1$-module and let $(X_2,\mu_{X_2},\varrho_{X_2})$ be a right $\xcG_2$-module.\ A {\bf $(\xcG_1,\xcG_2)$-bibundle $(\la_{X_1},\varrho_{X_2})$-associated to} $\widehat P$ is a sextuple $(\check P,\check\pi;\check\mu{}_1,\check{\unl\la}{}_1;\check\mu{}_2,\check{\unl\varrho}{}_2)$ composed of
\bit
\item a manifold $\check P$,\ termed a {\bf $(\mu_{X_1},\mu_{X_2})$-biextension} of $\widehat P$;
\item a map $\check\pi\colo\check P\to\widehat P$,\ termed a {\bf principal projection};
\item surjective submersions $\check\mu{}_a\colo\check P\to X_a,\ a\in\{1,2\}$,\ termed the {\bf left} ($a=1$) and {\bf right extending moment map} ($a=2$),\ respectively;
\item a map $\check\la{}_1\colo\xcG_1{}_{s_1}\hspace{-3pt}\x_{\check{\unl\mu}{}_1}\hspace{-1pt}\check P\to\check P$,\ written for $\check{\unl\mu}{}_1=\mu_{X_1}\circ\check\mu{}_1$ and termed the {\bf left associating action};  
\item a map $\check\varrho{}_2\colo\check P{}_{\check{\unl\mu}{}_2}\hspace{-3pt}\x_{t_2}\hspace{-1pt}\xcG_2\to\check P$,\ written for $\check{\unl\mu}{}_2=\mu_{X_2}\circ\check\mu{}_2$ and termed the {\bf right associating action}\,,
\eit
such that
\bit
\item $(\check P,X_2,\check\mu{}_2,\check{\unl\mu}{}_1,\check{\unl\la}{}_1)$ is a left principal $\xcG_1$-bundle;
\item $(\check P,X_1,\check\mu{}_1,\check{\unl\mu}{}_2,\check{\unl\varrho}{}_2)$ is a right principal $\xcG_2$-bundle;
\item $\check\pi$ is a morphism of $\xcG_a$-modules for each $a\in\{1,2\}$;
\item $\check\mu{}_a$ is $\xcG_a$-equivariant for each $a\in\{1,2\}$.
\eit 

We represent a $(\xcG_1,\xcG_2)$-bibundle $\check P$ $(\la_{X_1},\varrho_{X_2})$-associated to $\widehat P$ by the following (commutative) extended $W$-shaped diagram:
\qq\label{diag:ext-W-diagram}
\alxydim{@C=.5cm@R=.5cm}{ & & & & \check P \ar[ddd]_{\check\pi} \ar[drr]^{\check\mu{}_2} \ar[dll]_{\check\mu{}_1} & & & & \\ & & X_1 \ar[ddd]_(.5){\mu_{X_1}} & & & & X_2 \ar[ddd]^(.5){\mu_{X_2}} & & \\ \xcG_1 \ar@{=>}[ddrr] \ar@/^1.pc/[urr]_{\la_{X_1}} \ar@/^2.pc/[uurrrr]^{\check{\unl\la}{}_1} \ar@/^1.5pc/[drrrr]^(.6){\la_1} |!{[urr];[drr]}\hole & & & & & & & & \xcG_2 \ar@{=>}[ddll] \ar@/_1.pc/[ull]^{\varrho_{X_2}} \ar@/_2.pc/[uullll]_{\check{\unl\varrho}{}_2} \ar@/_2.pc/[dllll]_(.6){\varrho_2} |!{[ull];[dll]}\hole \\ & & & & \widehat P \ar[drr]^(.4){\mu_2} \ar[dll]_(.4){\mu_1} & & & & \\ & & M_1 & & & & M_2 & &}\,.
\qqq
\exdef

\beg
For every Lie groupoid $\grpd{\xcG}{M}$ and every left $\xcG$-module space $(X,\mu_X,\la_X)$ (viewed simultaneously as a right $\xcG$-module space with moment map $\mu_X$ and action $\varrho_X=\la_X\circ(\Inv\circ\pr_2,\pr_1))$),\ the arrow manifold $\xcG\,\fibx{s}{\mu_X}X$ of the corresponding action groupoid $\xcG\,\lx_{\la_X}X$ gives rise to a canonical structure of a $(\xcG,\xcG)$-bibundle $(\la_X,\varrho_X)$-associated to the biprincipal $(\xcG,\xcG)$-bibundle from Ex.\,\ref{eg:can-biprinc},\ for the associating actions 
\qq\nn
&\check{\unl\la}{}_1\equiv l\x\id_X\colo\xcG\,\fibx{s}{\mu_X\circ\la_X}\bigl(\xcG\,\fibx{s}{\mu_X}X\bigr)\too\xcG\,\fibx{s}{\mu_X}X,\ \bigl(h,(g,x)\bigr)\longmapsto(h.g,x)\,,&\cr\cr
&\check{\unl\varrho}{}_2\colo\bigl(\xcG\,\fibx{s}{\mu_X}X\bigr)\,\fibx{\mu_X\circ\pr_2}{t}\xcG
\too\xcG\,\fibx{s}{\mu_X}X,\ \bigl((g,x),h\bigr)\longmapsto\bigl(g.h,\varrho_X(x,h)\bigr)\,.&
\qqq 
The structure is captured by the extended $W$-diagram
\qq\nn
\alxydim{@C=.5cm@R=.5cm}{ & & & & \xcG\,\fibx{s}{\mu_X}X \ar[ddd]_{\pr_1} \ar[drr]^{\pr_2} \ar[dll]_{\la_X} & & & & \\ & & X \ar[ddd]_(.5){\mu_X} & & & & X \ar[ddd]^(.5){\mu_X} & & \\ \xcG \ar@{=>}[ddrr] \ar@/^1.pc/[urr]_{\la_X} \ar@/^2.5pc/[uurrrr]^{\check{\unl\la}{}_1} \ar@/^1.5pc/[drrrr]^(.6){l} |!{[urr];[drr]}\hole & & & & & & & & \xcG \ar@{=>}[ddll] \ar@/_1.pc/[ull]^{\varrho_X} \ar@/_2.5pc/[uullll]_{\check{\unl\varrho}{}_2} \ar@/_1.5pc/[dllll]_(.6){r} |!{[ull];[dll]}\hole \\ & & & & \xcG \ar[drr]^{s} \ar[dll]_{t} & & & & \\ & & M & & & & M & &}\,.
\qqq
\eeg
\noindent The next definition provides us with a useful refinement of the `raw' definition above.
\bedef\label{def:mom-extrid}
Adopt the notation of Defs.\,\ref{def:biasbndl} and \ref{def:Trident-oid}.\ Let $\grpd{\xcG_a}{M_a},\ a\in\{1,2\}$ be Lie groupoids,\ let $(X_1,\mu_{X_1},\la_{X_1})$ be a left $\xcG_1$-module and let $(X_2,\mu_{X_2},\varrho_{X_2})$ be a right $\xcG_2$-module.\ Furthermore,\ let $(\check P,\check\pi;\check\mu{}_1,\check{\unl\la}{}_1;\check\mu{}_2,\check{\unl\varrho}{}_2)$ be a $(\xcG_1,\xcG_2)$-bibundle $(\la_{X_1},\varrho_{X_2})$-associated to a biprincipal $(\xcG_1,\xcG_2)$-bibundle $(\widehat P;\mu_1,\la_1;\mu_2,\varrho_2)$.\ We shall call the septuple 
\qq\nn
\bigl(\grpd{\xcG_1}{M_1};(\check P,\check\pi;\check\mu{}_1,\check{\unl\la}{}_1;\check\mu{}_2,\check{\unl\varrho}{}_2),(\widehat P;\mu_1,\la_1;\mu_2,\varrho_2);\grpd{\xcG_2}{M_2};\pi_{\widehat P},\Si;\grpd{\xcG_3}{M_3})
\qqq
a ({\bf left}) {\bf moment-extended trident} if
\bit
\item $(\xcG_1,\widehat P,\xcG_2;\pi_{\widehat P},\Si;\xcG_3)$ is a (left) trident relative to the two commuting principal-bundle structures:\ $(\widehat P,\mu_1,\la_1)$ (left) and $(\widehat P,\mu_2,\varrho_2)$ (right);
\item the (left-)$\xcG_1$-module structure on $\check P$ is modelled on the canonical left $\xcG_3$-module structure on $\xcG_3\,\,\fibx{s_3}{\mu_{X_3}}X_3$ (in a local trivialisation),\ as captured by the diagram:
\qq\nn
\alxydim{@C=.75cm@R=.75cm}{ & & & & & \xcG_3\,\,\fibx{s_3}{\mu_{X_3}}X_3 \ar@{^{(}.>}[ddrrr] \ar[dl]^{\la_{X_3}} \ar[dddd]^(.4){\pr_1} & & & \\ & & & & X_3 \ar@{^{(}.>}[ddrr] \ar[dddd]_(.4){\mu_{X_3}} & & & & \\ & & & & & & & & \check P \ar[dddd]_{\check\pi} \ar[dll]_{\check\mu{}_1} \\ & & & & & & X_1 \ar[dddd]^(.4){\mu_{X_1}} & & \\ \xcG_3 \ar@{^{(}.>}[ddrrrr] \ar@{=>}[drrrr] \ar@/^1.25pc/[rrrrr]^{l_3} |!{[urrrr];[drrrr]}\hole \ar@/^1.25pc/[uuurrrr]_{\la_{X_3}} \ar@/^3.pc/[uuuurrrrr]^{l_3\x\id_{X_3}} & & & & & \xcG_3 \ar@{^{(}.>}[ddrrr] \ar[dl]_{t_3} & & & \\ & & & & M_3 \ar@{^{(}.>}[ddrr] & & & & \\ & & & & \xcG_1 \ar@{=>}[drr] \ar[dd]_{\pi_{\xcG_1}} \ar@/^1.25pc/[rrrr]^{\la_1} |!{[urr];[drr]}\hole \ar@/_1.25pc/[uuuurrrr]_(.75){\check{\unl\la}{}_1} |!{[urr];[drr]}\hole \ar@/_1.25pc/[uuurr]^(.8){\la_{X_1}} & & & & \widehat P \ar[dll]_(.4){\mu_1} \ar[dd]^{\pi_{\widehat P}} \\ & & & & & & M_1 \ar[dd]_(.6){\pi_{M_1}} & & \\ & & & & \Si\x\Si \ar@{=>}[drr] \ar@/^1.25pc/[rrrr]^{\pr_1\circ\pr_1} |!{[urr];[drr]}\hole & & & & \Si \ar@{=}[dll] \\ & & & & & & \Si & & }\,.
\qqq
\eit
A moment-extended trident shall be represented by the following diagram:
\qq\label{diag:Ass-trident}\qquad
\alxydim{@C=.75cm@R=.5cm}{ & \xcG_3\,\,\fibx{s_3}{\mu_{X_3}}X_3 \ar@{^{(}.>}[ddrrr] \ar[dl]_{\la_{X_3}} & & & & & & & & & \\ X_3 \ar@{^{(}.>}[ddrr] & & & & & & & & & & \\ & & & & \check P \ar[ddd]_{\check\pi} \ar[drr]^{\check\mu{}_2} \ar[dll]_{\check\mu{}_1} & & & & & & \\ & & X_1 \ar[ddd]_(.65){\mu_{X_1}} & & & & X_2 \ar[ddd]^(.65){\mu_{X_2}} & & & & \\ \xcG_1 \ar@{=>}[ddrr] \ar@/^2.pc/[uurrrr]^(.6){\check{\unl\la}{}_1} \ar@/^1.25pc/[urr]_(.4){\la_{X_1}} \ar@/^1.25pc/[drrrr]^(.4){\la_1} |!{[urr];[drr]}\hole & & & \xcG_3 \ar@{^{(}.>}[dr] & & & & & & & \xcG_2 \ar@{=>}[ddllll] \ar@/_2.pc/[uullllll]_(.6){\check{\unl\varrho}{}_2} \ar@/_1.25pc/[ullll]^(.4){\varrho_{X_2}} \ar@/_1.25pc/[dllllll]_(.4){\varrho_2} |!{[ulllll];[dlll]}\hole \\ & & & & \widehat P \ar[drr]^(.4){\mu_2} \ar[dll]_(.4){\mu_1} \ar[dd]^{\pi_{\widehat P}} & & & & & & \\ & & M_1 & & & & M_2 & & & & \\ & & & & \Si & & & & & & }\,.
\qqq
\exdef

We are finally ready to discuss the physically motivated mixing construction featuring the principaloid bundle of \cite{Strobl:2025} as the bundle of gauges. 

\bethe\label{thm:del-as-prince}
Adopt the notation of Def.\,\ref{def:princ-gr-bun} and Cor.\,\ref{cor:loc-mode-P}.\ Let $(X,\mu_X,\la_X)$ be a left $\xcG$-module.\ Every principaloid $\xcG$-bundle $\xcP$ canonically induces a fibre bundle $(\Xi\equiv\xcP\,\lx_{\la_X}X,\Si,X,\pi_{\xcP\,\lx_{\la_X}X})$ with model
\qq\label{eq:clutch-XassP}
\Xi\cong\bigsqcup_{i\in I}\,\bigl( O_i\x X\bigr)/\sim_{ B\la_{X\,\b_{\cdot\cdot}}}\xrightarrow{\ \pi_\Xi\ }\Si,\ [(\si,x,i)]\longmapsto\si\,,\\ \cr (\si_2,j,x_2)\sim(\si_1,i,x_1)\quad\Longleftrightarrow\quad(\si_1,x_1)=\bigl(\si_2,B\la_X\bigl(\b_{ij}(\si_1),x_2\bigr)\bigr)\,,\nn
\qqq
written in terms of the transition 1-cocycle $\{ \b_{ij}\}_{i,j\in I}$ of $\xcP$,\ which we realise by the action of $\bB(\xcG)$ on $X$ induced from $\la_X$ along the lines of Prop.\,\ref{prop:G-in-B_act}.\ It comes with a vertical bundle map
\qq\label{diag:del-bun-map}
\alxydim{@C=.75cm@R=1.cm}{ \xcP\,\fibx{\mu}{\mu_X} X\ar[rr]^{\quad\d_X} \ar[dr]_{\pi_\xcP\circ\pr_1} & & \Xi \ar[dl]^{\pi_\Xi} \\ & \Si & }
\qqq
locally modelled on $\la_X\colo\xcG\,\fibx{s}{\mu_X}X\to X$.\
The triple $(\xcP\,\fibx{\mu}{\mu_X}X,\Xi,\d_X)$ carries a canonical structure of a principal-$\xcG$-bundle object in the category of fibre bundles over $\Si$, 
\qq\label{diag:duck-as-prince-ass}
\alxydim{@C=.75cm@R=1cm}{ \Si \ar@{=}[d] & & \xcP\,\fibx{\mu}{\mu_X}X \ar[ll]_{\pi_\xcP\circ\pr_1\quad} \ar@{->>}[d]_{\d_X} \ar[rd]^{\widetilde\mu} & & \xcG \ar@{=>}[ld] \ar@/_1.pc/[ll]_{\widetilde\varrho} \\ \Si & & \Xi \ar[ll]^{\pi_\Xi} & M & }\,,
\qqq
with moment map 
\qq\nn
\widetilde\mu=\mu\circ\pr_1\equiv\mu_X\circ\pr_2\colo\xcP\,{}_\mu\hspace{-3pt}\x_{\mu_X}\hspace{-1pt}X\too M 
\qqq
and right action 
\qq\label{eq:varrho-def}
\widetilde\varrho\colo (\xcP\,\fibx{\mu}{\mu_X}X)\,\,\fibx{\widetilde\mu}{t}\xcG\too\xcP\,\fibx{\mu}{\mu_X}X\,,\ \bigl((p,x),g\bigr)\longmapsto\bigl(\varrho(p,g),\la_X\bigl(g^{-1},x\bigr)\bigr)\,.
\qqq
The above implies a canonical identification
\qq\nn
\Xi\cong(\xcP\,\fibx{\mu}{\mu_X}X)/\xcG
\qqq
\ethe
\beroof
Consider smooth local maps
\qq\label{eq:del-quot-loc}\qquad
\d_{X\,i}\colo\pi_\xcP^{-1}(O_i)\,\,\fibx{\mu}{\mu_X}X\too\Xi,\ \bigl(\xcP\t_i^{-1}(\si,g),x\bigr)\longmapsto[(\si,\la_X(g,x),i)]\,,\quad i\in I\,,
\qqq
written in the convenient model \eqref{eq:clutch-XassP} of $\Xi$.\ At an arbitrary point $\si'\in O_{ij}$,\ we find
\qq\nn
&&\d_{X\,i}\bigl(\xcP\t_j^{-1}(\si',g),x\bigr)=\d_{X\,i}\bigl(\xcP\t_i^{-1}\bigl(\si',\b_{ij}(\si')\lact g\bigr),x\bigr)\equiv[(\si,\la_X(\b_{ij}(\si')\lact g,x),i)]\cr\cr
&=&[(\si,(\la_X((\b_{ij}(\si'))(t(g)),\la_X(g,x)),i)]\equiv[(\si,(\la_X((\b_{ij}(\si'))(\mu_X(\la_X(g,x))),\la_X(g,x)),i)]\cr\cr
&\equiv&[(\si,B\la_X(\b_{ij}(\si'),\la_X(g,x)),i)]=[(\si,\la_X(g,x),j)]\equiv\d_{X\,j}\bigl(\xcP\t_j^{-1}(\si',g),x\bigr)\,,
\qqq 
meaning that the $\d_{X\,i}$ are restrictions of a globally smooth map
\qq\nn
\d_X\colo\xcP\,\,\fibx{\mu}{\mu_X}X\too\Xi\,,\qquad\d_X\rstr_{\pi_\xcP^{-1}(O_i)\,\,\fibx{\mu\ }{\mu_X}X}=\d_{X\,i}\,.
\qqq

In the next step,\ we check,\ for arbitrary $\si\in O_i,\ (g,h)\in\xcG\fibx{s}{t}\xcG$ and $x\in\mu_X^{-1}(\{s(g)\}))$,\ the identity
\qq\nn
&&\bigl(\d_X\circ\widetilde\varrho_h\bigr)\bigl(\xcP\t_i^{-1}(\si,g),x\bigr)\equiv\d_X\bigl(\varrho\bigl(\xcP\t_i^{-1}(\si,g),h\bigr),\la_X\bigl(h^{-1},x\bigr)\bigr)=\d_X\bigl(\xcP\t_i^{-1}(\si,g.h),\la_X\bigl(h^{-1},x\bigr)\bigr)\cr\cr
&=&[(\si,\la_X(g.h,\la_X(h^{-1},x)))]=[(\si,\la_X(g.h.h^{-1},x))]=[(\si,\la_X(g,x))]\equiv\d_X\bigl(\xcP\t_i^{-1}(\si,g),x\bigr)\,,
\qqq
which implies that $(\pr_1,\widetilde\varrho)$ maps as
\qq\nn
(\pr_{1,2},\widetilde\varrho)\colo (\xcP\,\,\fibx{\mu}{\mu_X}X)\,\,\fibx{\widetilde\mu\,}{t}\xcG\too(\xcP\,\,\fibx{\mu}{\mu_X}X)\,\,\fibx{\d_X}{\d_X}(\xcP\,\,\fibx{\mu}{\mu_X}X)
\qqq

At this stage,\ it suffices to prove the existence of a (smooth) division map
\qq\nn
\widetilde\phi\colo(\xcP\,\,\fibx{\mu}{\mu_X}X)\,\,\fibx{\d_X}{\d_X}(\xcP\,\,\fibx{\mu}{\mu_X}X)\too\xcG
\qqq
with the property $(\pr_1,\widetilde\phi)=(\pr_1,\widetilde\varrho)^{-1}$.\ As $\d_X$ is a vertical bundle map,\ we may define the map sought after as
\qq\nn
\widetilde\phi:=\phi_\xcP\circ\pr_{1,3}
\qqq
in terms of the division map $\phi_\xcP$ of $\xcP$.\ We readily verify,\ for $((p,x),g)\in(\xcP\,\,\fibx{\mu}{\mu_X}X)\,\,\fibx{\widetilde\mu}{t}\xcG$,
\qq\nn
\bigl((\pr_1,\widetilde\phi)\circ(\pr_1,\widetilde\varrho)\bigr)\bigl((p,x),g\bigr)\equiv(\pr_1,\widetilde\phi)\bigl((p,x),\bigl(\varrho(p,g),\la_X\bigl(g^{-1},x\bigr)\bigr)\bigr)\equiv\bigl((p,x),\phi_\xcP\bigl(p,\varrho(p,g)\bigr)\bigr)=\bigl((p,x),g\bigr)\,,
\qqq
and,\ for $((p_1,x_1),(p_2,x_2))\in(\xcP\,\,\fibx{\mu\ }{\mu_X}\,X)\,\,\fibx{\d_X\   }{\d_X}\,(\xcP\,\,\fibx{\mu\ }{\mu_X}\,X)$ such that $(p_a,x_a)=(\xcP\t_i^{-1}(\si,g_a),x_a),$ $a\in\{1,2\}$,
\qq\nn
&&\bigl((\pr_1,\widetilde\varrho)\circ(\pr_1,\widetilde\phi)\bigr)\bigl((p_1,x_1),(p_2,x_2)\bigr)\equiv(\pr_1,\widetilde\varrho)\bigl((p_1,x_1),\phi_\xcP(p_1,p_2)\bigr)\cr\cr &=&\bigl((p_1,x_1),\widetilde\varrho\bigl((p_1,x_1),\phi_\xcP(p_1,p_2)\bigr)\bigr)=\bigl((p_1,x_1),\bigl(\varrho\bigl(p_1,\phi_\xcP(p_1,p_2)\bigr),\la_X\bigl(\phi_\xcP(p_1,p_2)^{-1},x_1\bigr)\bigr)\bigr)\cr\cr 
&=&\bigl((p_1,x_1),\bigl(p_2,\la_X\bigl(\phi_\xcP(p_1,p_2)^{-1},x_1\bigr)\bigr)\bigr)=\bigl((p_1,x_1),(p_2,x_2)\bigr)\,, 
\qqq
where the last equality follows from
\qq\nn
[(\si,\la_X(g_1,x_1),i)]\equiv\d_X(p_1,x_1)\must\d_X(p_2,x_2)\equiv[(\si,\la_X(g_2,x_2),i)]\,,
\qqq
the latter implying 
\qq\nn
\la_X\bigl(\phi_\xcP(p_1,p_2)^{-1},x_1\bigr)=\la_X\bigl((g_1^{-1}.g_2)^{-1},x_1\bigr)=\la_X\bigl(g_2^{-1}.g_1,x_1\bigr)=x_2\,.
\qqq

\eroof

\bedef\label{def:assoc-bndl-oid}
Adopt the notation of Thm.\,\ref{thm:del-as-prince}.\ We shall call the fibre bundle $(\xcP\,\,\fibx{\mu\,\,}{\mu_X} X,\Si,\xcG\,\,\fibx{s\,\,}{\mu_X}X,\pi_\xcP\circ\pr_1)$ a {\bf $\mu_X$-extended bundle},\ and the fibre bundle $(\Xi,\Si,X,\pi_\Xi)$ a {\bf bundle associated to} $\xcP$ {\bf by $\la_X$},\ or a {\bf $\la_X$-associated bundle} for short.\ The map $\la_X$ shall be termed,\ in this context,\ the {\bf associating $\xcG$-action},\ and the map $\d_X$ the {\bf $\xcG$-quotient map}.
\exdef

We may readily reinterpret the construction of the associated bundle $(\xcP\,\,\fibx{\mu}{\mu_X}X)/\xcG$ as a distinguished specialisation of the construction of the shadow bundle of a principaloid bundle with the structure groupoid given by the action groupoid $\xcG\,\lx_{\la_X}X$ from Prop.\,\ref{prop:grpd-act-grpd}.\ The specialisation hinges upon
\berop\label{prop:Bis-Bis}
Adopt the notation of Defs.\,\ref{def:bisec} and \ref{def:gr-mod},\ and of Prop.\,\ref{prop:grpd-act-grpd}.\ 
Let $(X,\mu_X,\la_X)$ be a left $\xcG$-module.\ The moment map $\mu_X\colo X\to M$ canonically induces a group homomorphism
\qq\nn
\widetilde\mu{}_X^*=\bigl(\mu_X^*(\cdot),\id_X\bigr)\colo{\rm Bisec}\bigl(\grpd{\xcG}{M}\bigr)\too{\rm Bisec}\bigl(\xcG\,\lx_{\la_X}X\bigr),\ \b\longmapsto\bigl(\b\circ\mu_X,\id_X\bigr)\,.
\qqq
Whenever $\mu_X$ is surjective,\ the homomorphism is injective.
\eerop
\beroof
Consider an arbitrary bisection $\b\in{\rm Bisec}(\grpd{\xcG}{M})$.\ We then have
\qq\nn
\widetilde s\circ\widetilde\mu{}_X^*(\b)\equiv\pr_2\circ\bigl(\b\circ\mu_X,\id_X\bigr)=\id_X\,,
\qqq
and so,\ indeed,\ $\widetilde\mu{}_X^*(\b)$ is a section of the surjective submersion $\widetilde s$.\ In order to verify that its manifestly smooth post-composition with $\widetilde t$, 
\qq\nn
\widetilde t\circ\widetilde\mu{}_X^*(\b)\equiv\la_X\circ\bigl(\b\circ\mu_X,\id_X\bigr)\,,
\qqq
yields a diffeomorphism of $X$,\ we check directly that another smooth map:
\qq\nn
\widetilde t\circ\widetilde\mu{}_X^*\bigl(\b^{-1}\bigr)\equiv\la_X\circ\bigl(\Inv\circ\b\circ(t_*\b)^{-1}\circ\mu_X,\id_X\bigr)
\qqq
is the inverse of $\widetilde t\circ\widetilde\mu{}_X^*(\b)$,
\qq\nn
\widetilde t\circ\widetilde\mu{}_X^*\bigl(\b^{-1}\bigr)=\bigl(\widetilde t\circ\widetilde\mu{}_X^*(\b)\bigr)^{-1}\,.
\qqq
Hence,
\qq\nn
\widetilde t\circ\widetilde\mu{}_X^*(\b)\in\Diff(X)\,,
\qqq
and,\ consequently, 
\qq\nn
\widetilde\mu{}_X^*(\b)\in{\rm Bisec}\bigl(\xcG\,\lx_{\la_X}X\bigr)\,,
\qqq
as desired.\ The homomorphicity of $\widetilde\mu{}_X^*$ is readily verified.

Finally,\ the injectivity of $\widetilde\mu{}_X^*$ in the case of a surjective moment map $\mu_X$ is obvious.
\eroof
\noindent With the last result in hand,\ we may state the anticipated 
\berop\label{prop:ass-as-oidle}
Adopt the notation of Prop.\,\ref{prop:Bis-Bis} and Thms.\,\ref{thm:duck-as-prince} and \ref{thm:del-as-prince}.\ The fibre bundle $(\xcP\,\,\fibx{\mu\,\,}{\mu_X}X,\Si,\xcG\,\,\fibx{s\,\,}{\mu_X}X,\pi_\xcP\circ\pr_1)$ is a principaloid $\xcG\,\lx_{\la_X}X$-bundle with the structure group reduced to $L({\rm Im}\,\widetilde\mu{}_X^*)\subset L({\rm Bisec}(\xcG\,\lx_{\la_X}X))$.\ The fibre bundle $(\Xi,\Si,X,\pi_\Xi)$ associated to $\xcP$ by $\la_X$ is the corresponding shadow bundle,\ and the $\xcG$-quotient map $\d_X$ is the sitting-duck map between the two.\ This is captured by the diagram
\qq\label{diag:ass-as-prince}
\alxydim{@C=.75cm@R=1.cm}{ \Si \ar@{=}[d] & & \xcP\,\,\fibx{\mu}{\mu_X}X \ar[ll]_{\pi_\xcP\circ\pr_1\quad} \ar@{->>}[d]_{\d_X} \ar[rd]^{\widehat\mu{}_X} & & \xcG\,\,\fibx{s}{\mu_X}X \ar@{=>}[ld] \ar@/_1.pc/[ll]_{\widehat\varrho{}_X} \\ \Si & & \Xi \ar[ll]^{\pi_\Xi} & X & }\,,
\qqq
written for the moment map 
\qq\nn
\widehat\mu{}_X=\pr_2\colo\xcP\,\,\fibx{\mu}{\mu_X}X\too X\,,
\qqq 
and for the action map
\qq\nn
\widehat\varrho{}_X=\widetilde\varrho\circ\pr_{1,2,3}&\colo&(\xcP\,\,\fibx{\mu}{\mu_X}X)\,\,\fibx{\widehat\mu}{\la_X}(\xcG\,\,\fibx{s}{\mu_X}X)\too\xcP\,\,\fibx{\mu}{\mu_X}X\cr\cr  &\colo&\bigl((p,x),\bigl(g,\la_X\bigl(g^{-1},x\bigr)\bigr)\bigr)\longmapsto\bigl(\varrho(p,g),\la_X\bigl(g^{-1},x\bigr)\bigr)\,.
\qqq
\eerop
\beroof
The model \eqref{eq:clutch-xcP} of $\xcP$ gives us a convenient presentation of the bundle of interest:
\qq\nn
\xcP\,\,\fibx{\mu}{\mu_X}X\cong\bigsqcup_{i\in I}\,\bigl(O_i\x\bigl(\xcG\,\,\fibx{s}{\mu_X}X\bigr)\bigr)/\sim_{\widetilde L{}_{\b_{\cdot\cdot}}}\,,
\qqq
in which the identification of fibres over $O_{ij}\ni\si$ is realised by the diffeomorphisms
\qq\nn
(\si,g,x,j)\sim\bigl(\si,\b_{ij}(\si)\lact g,x,i\bigr)\,,
\qqq
consistently with the fibration over $M$,
\qq\nn
s\bigl(\b_{ij}(\si)\lact g\bigr)=s(g)=\mu_X(x)\,.
\qqq
The diffeomorphisms can be rewritten as
\qq\nn
&&\bigl(\b_{ij}(\si)\lact g,x\bigr)\equiv\bigl(\b_{ij}(\si)\bigl(t(g)\bigr).g,x\bigr)\equiv\bigl(\b_{ij}(\si)\bigl(t(g)\bigr),\la_X(g,x)\bigr).(g,x)\cr\cr
&\equiv&\bigl(\b_{ij}(\si)\bigl(\bigl(\mu_X\circ\la_X\bigr)(g,x)\bigr),\la_X(g,x)\bigr).(g,x)\equiv\widetilde\mu{}_X^*\bigl(\b_{ij}(\si)\bigr)\bigl(\widetilde t(g,x)\bigr).(g,x)\equiv\widetilde\mu{}_X^*\bigl(\b_{ij}(\si)\bigr)\lact(g,x)\,,
\qqq
and so we arrive at
\qq\nn
\xcP\,\,\fibx{\mu}{\mu_X}X\cong\bigsqcup_{i\in I}\,\bigl(O_i\x\bigl(\xcG\,\,\fibx{s}{\mu_X}X\bigr)\bigr)/\sim_{ L_{\widetilde\mu{}_X^*\circ\b_{\cdot\cdot}}}\,.
\qqq

The identification of the $\xcG$-quotient (bundle) map $\d_X\colo\xcP\,\,\fibx{\mu}{\mu_X}X\to\Xi$ as the principal $\xcG\,\lx_{\la_X}X$-bundle now follows from the local formul\ae ~\eqref{eq:del-quot-loc},\ from which the desired local model $\la_X\equiv\widetilde t$ of the map emerges.\ Equivalently,\ we note that,\ for $\si\in O_{ij}$ and $x\in X$,
\qq\nn
\widetilde t{}_*\bigl(\bigl(\widetilde\mu{}_X^*\circ\b_{ij}\bigr)(\si)\bigr)(x)=\bigl(\la_X\circ\bigl(\b_{ij}(\si)\circ\mu_X,\id_X\bigr)\bigr)(x)=\la_X\bigl(\b_{ij}(\si)\bigl(\mu_X(x)\bigr),x\bigr)\equiv B\la_X\bigl(\b_{ij}(\si),x\bigr)\,.
\qqq

The remaining statements of the proposition are straightforward.
\eroof

The two principal groupoid-bundle structures on $(\xcP\,\,\fibx{\mu}{\mu_X}X,\Xi,\d_X)$ discussed above are related by the Lie-groupoid morphism of Prop.\,\ref{prop:actgrpd-morph-grpd} between the two structure groupoids,\ as captured by the commutative diagram
\qq\nn
\alxydim{@C=.75cm@R=1.cm}{ \Si \ar@{=}[d] & & \xcP\,\,\fibx{\mu}{\mu_X}X \ar[ll]_{\pi_\xcP\circ\pr_1\quad} \ar@{->>}[d]_{\d_X} \ar[rd]^{\widehat\mu{}_X} \ar[rddd]_{\widetilde\mu} & & \xcG\,\,\fibx{s}{\mu_X}X \ar@/_1.pc/[ll]_{\widehat\varrho{}_X} \ar@{=>}[ld] \ar[dd]^{\pr_1} \\ \Si & & \Xi \ar[ll]^{\pi_\Xi} & X \ar[dd]^{\mu_X} & \\ & & & & \xcG \ar@/_2.5pc/[uull]_(.6){\widetilde\varrho} |!{[ul];[dlll]}\hole \ar@/_1.25pc/[ul]^{\varrho_X} \ar@{=>}[ld] \\ & & & M & }\,.
\qqq
The morphism extends readily to the corresponding principal groupoid bundles under consideration.
\berop\label{prop:Ups-intro}
Adopt the notation of Prop.\,\ref{prop:ass-as-oidle}.\ There exists a canonical morphism of principal groupoid bundles $(\pr_1,\Upsilon_X,\Theta)$
\qq\label{diag:ass-P-morph}
\alxydim{@C=1.5cm@R=1.cm}{ \xcP\,\,\fibx{\mu}{\mu_X}X \ar[r]^{\qquad\pr_1} \ar[d]_{\d_X} & \xcP \ar[d]^{\xcD} \\ \Xi \ar[r]_{\Upsilon_X} & \xcF }
\qqq
locally modelled on the Lie-groupoid morphism $\Theta=(\pr_1,\mu_X)$ from Prop.\,\ref{prop:actgrpd-morph-grpd} as
\qq\label{eq:Ups}
\Upsilon_X\colo\Xi\too\xcF\,,\ \Xi\t_i^{-1}(\si,x)\longmapsto\xcF\t_i^{-1}\bigl(\si,\mu_X(x)\bigr)\,,
\qqq
where $\xcF\t_i\colo\pi_\xcF^{-1}(O_i)\xrightarrow{\cong}O_i\x M$ resp.\ $\Xi\t_i\colo\pi_\Xi^{-1}(O_i)\xrightarrow{\cong}O_i\x X,\ i\in I$ are 
local trivialisations of $\xcF$ resp.\ $\Xi$ induced by those of $\xcP$ (see:\ the proof of Thm.\,\ref{thm:duck-as-prince} and \Reqref{eq:clutch-XassP}).\ Equivalently,\ $(\pr_1,\Upsilon_X,\Id_{\Gr})$ is a morphism of principal $\xcG$-bundles (for $\Gr\equiv\grpd{\xcG}{M}$).
\eerop
\beroof
First of all,\ we need to check the global smoothness of $\Upsilon_X$ as defined (locally) in \eqref{eq:Ups}.\ To this end,\ consider $\si\in O_{ij}$ and calculate
\qq\nn
&&\Upsilon_X\bigl(\Xi\t_j^{-1}\bigl(\si,B\la_X\bigl(\b_{ji}(\si),x\bigr)\bigr)\bigr)\equiv\Upsilon_X\bigl(\Xi\t_j^{-1}\bigl(\si,\la_X\bigl(\b_{ji}(\si)\bigl(\mu_X(x)\bigr),x\bigr)\bigr)\bigr)\cr\cr 
&=&\xcF\t_j^{-1}\bigl(\si,\mu_X\bigl(\la_X\bigl(\b_{ji}(\si)\bigl(\mu_X(x)\bigr),x\bigr)\bigr)=\xcF\t_j^{-1}\bigl(\si,t_*\bigl(\b_{ji}(\si)\bigr)\bigl(\mu_X(x)\bigr)\bigr)=\xcF\t_i^{-1}\bigl(\si,\mu_X(x)\bigr)\equiv\Upsilon_X\bigl(\Xi\t_i^{-1}(\si,x)\bigr)\,.
\qqq

Having verified the well-definedness of $\Upsilon_X$,\ we may,\ next,\ inspect diagram \eqref{diag:ass-P-morph}.\ Thus,\ at an arbitrary point $(p,x)\in\xcP\,\,\fibx{\mu}{\mu_X}X$ with $p=\xcP\t_i^{-1}(\si,g)$,\ we obtain
\qq\nn
&&(\xcD\circ\pr_1)(p,x)=\xcD\bigl(\xcP\t_i^{-1}(\si,g)\bigr)=\xcF\t_i^{-1}\bigl(\si,t(g)\bigr)\equiv\xcF\t_i^{-1}\bigl(\si,\bigl(\mu_X\circ\la_X\bigr)(g,x)\bigr)\cr\cr 
&\equiv&\Upsilon_X\bigl(\Xi\t_i^{-1}\bigl(\si,\la_X(g,x)\bigr)\bigr)\equiv\bigl(\Upsilon_X\circ\d_X\bigr)\bigl(\xcP\t_i^{-1}(\si,g),x\bigr)\equiv\bigl(\Upsilon_X\circ\d_X\bigr)(p,x)\,,
\qqq
which establishes the stated commutativity of the diagram.

That the fibres of the respective moment maps are preserved by the above morphism follows from the tautological commutativity of the diagram
\qq\nn
\alxydim{@C=1.5cm@R=1.cm}{ \xcP\,\,\fibx{\mu}{\mu_X}X \ar[r]^{\qquad\pr_1} \ar[d]_{\widehat\mu{}_X\equiv\pr_2} & \xcP \ar[d]^{\mu} \\ X \ar[r]_{\mu_X} & M }\,.
\qqq 

Finally,\ the equivariance of the morphism,\ as reflected in the commutativity of the diagram
\qq\nn
\alxydim{@C=2.cm@R=1.cm}{ \bigl(\xcP\,\,\fibx{\mu}{\mu_X}X\bigr)\,\,\fibx{\widehat\mu{}_X}{\la_X}(\xcG\,\,\fibx{s}{\mu_X}X) \ar[r]^{\qquad\qquad\pr_1\x\pr_1} \ar[d]_{\widehat\varrho{}_X} & \xcP\,\,\fibx{\mu}{t}\xcG \ar[d]^{\varrho} \\ \xcP\,\,\fibx{\mu}{\mu_X}X \ar[r]_{\qquad\pr_1} & \xcP }\,,
\qqq
is readily checked in a direct calculation.
\eroof

\beg\label{eg:CB-from-BG-ass}
For the principaloid bundle $P$ of the delooping groupoid $\bbB\txG$ from Ex.\,\ref{eg:deloop-oid},\ and a left $\bbB\txG$-module $(M,\bullet,\la)$,\ we recover the extended bundle $P\x M$ as the $\bullet$-extended bundle,\ and Cartan's associated bundle $P\x_\la M$ as the corresponding bundle associated to $P$ by $\la$.
\eeg

\beg 
The shadow $\xcF$ of an arbitrary principaloid $\xcG$-bundle $\xcP$ can be viewed as a bundle associated to $\xcP$ by the canonical target action of $\xcG$ on the object manifold $M$ of its structure groupoid $\grpd{\xcG}{M}$ from Ex.\,\ref{eg:st-act}.\ This is best seen in local models as $B(t\circ\pr_1)\equiv t_*$. 
\eeg

\subsection{Associated gauge transformations}

In the next step,\ we study another manifestation of the reduction of the structure group through association,\ to wit:\ the reduction of the gauge group of the associated principaloid bundle.\ As anticipated,\ the reduction is encoded in the action of the Ehresmann--Atiyah groupoid of the underlying principaloid bundle $\xcP$.

\bethe\label{thm:ass-as-AtP-mod}
Adopt the notation of Prop.\,\ref{prop:Ups-intro} and Thm.\,\ref{thm:princ-as-At-mod}.\ The Ehresmann--Atiyah groupoid ${\rm At}(\xcP)$ of a principaloid $\xcG$-bundle $\xcP$ acts on the $\mu_X$-extended bundle $\xcP\,\,\fibx{\mu}{\mu_X}X$ and on the corresponding $\la_X$-associated bundle $\Xi$ in the following way:
\bit
\item[$\bullet$] On every $\mu_X$-extended bundle $\xcP\,\,\fibx{\mu}{\mu_X}X$,\ there exists a canonical structure of a left ${\rm At}(\xcP)$-module,\ with moment map $\widetilde\mu{}_\xcP=\mu_\xcP\circ\pr_1\colo\xcP\,\,\fibx{\mu}{\mu_X}X\to\xcF$ and action
\qq\label{eq:tilamPa}\qquad\qquad
\widetilde\la{}_\xcP=\la_\xcP\x\id_X\colo {\rm At}(\xcP)\,\,\fibx{\sfS}{\widetilde\mu{}_\xcP}(\xcP\,\,\fibx{\mu}{\mu_X}X)\too\xcP\,\,\fibx{\mu}{\mu_X}X,\ \bigl(a,(p,x)\bigr)\longmapsto\bigl(\la_\xcP(a,p),x\bigr)\,,
\qqq
written in terms of the left-${\rm At}(\xcP)$-module structure $(\xcP,\mu_\xcP,\la_\xcP)$.
\item[$\bullet$] This structure covers the canonical structure of a left ${\rm Pair}(\Si)$-module on $\Si$,\ with moment map $\mu_\Si\equiv\id_\Si$ and action
\qq\nn
\la_\Si\colo(\Si\x\Si)\,\,\fibx{\pr_2}{\id_\Si}\Si\too\Si,\ \bigl((\si_1,\si_2),\si_2\bigr)\longmapsto\si_1\,.
\qqq
\item[$\bullet$] On the corresponding $\la_X$-associated bundle $\Xi$,\ there exists a canonical structure of a left ${\rm At}(\xcP)$-module,\ with moment map $\mu_\Xi=\Upsilon_X$ and action 
\qq\label{eq:tilamXi}
\la_\Xi\colo{\rm At}(\xcP)\,\,\fibx{\sfS}{\Upsilon_X}\Xi\too\Xi,\ \bigl(a,\d_X(p,x)\bigr)\longmapsto\d_X\bigl(\la_\xcP(a,p),x\bigr)\,.
 \qqq
\item[$\bullet$] The action $\widetilde\la{}_\xcP$ is intertwined with $\la_\Xi$ by the $\xcG$-quotient map $\d_X$,\ as reflected in the identities
\qq\label{eq:tilamPa-int-tilamXi}
\widetilde\mu{}_\xcP=\mu_\Xi\circ\d_X\,,\qquad\qquad\la_\Xi\circ\bigl(\id_{{\rm At}(\xcP)}\x\d_X\bigr)=\d_X\circ\widetilde\la{}_\xcP\,.
\qqq
\eit
\ethe
\beroof
A trivial direct check.
\eroof
\noindent The relation between the two groupoid-module structures on the $\mu_X$-extended bundle is established in
\bethe
Adopt the notation of Def.\,\ref{def:mom-extrid} and Thm.\,\ref{thm:ass-as-AtP-mod}.\ For every principaloid $\xcG$-bundle $\xcP$ and every left $\xcG$-module $(X,\mu_X,\la_X)$,\ the $\mu_X$-extended bundle $\xcP\,\,\fibx{\mu\,}{\mu_X}X$ carries a canonical structure of an $({\rm At}(\xcP),\xcG)$-bibundle $(\la_\Xi,\varrho_X)$-associated to $\xcP$,\ which turns the septuple
\qq\nn
\bigl(\grpd{{\rm At}(\xcP)}{\xcF};(\xcP\,\,\fibx{\mu}{\mu_X}X,\pr_1;\d_X,\widetilde\la{}_\xcP;\pr_2,\widetilde\varrho),(\xcP;\xcD,\la_\xcP;\mu,\varrho);\grpd{\xcG}{M};\pi_\xcP,\Si;\grpd{\xcG}{M}\bigr)
\qqq
into a (left) moment-extended trident,\ represented by the following diagram:
\qq\nn
\alxydim{@C=.75cm@R=.5cm}{ & \xcG\,\,\fibx{s}{\mu_X}X \ar@{^{(}.>}[ddrrr] \ar[dl]_{\la_X} & & & & & & & & & \\ X \ar@{^{(}.>}[ddrr] & & & & & & & & & & \\ & & & & \xcP\,\,\fibx{\mu}{\mu_X}X \ar[ddd]_{\pr_1} \ar[drr]^{\pr_2} \ar[dll]_{\d_X} & & & & & & \\ & & \Xi \ar[ddd]_(.65){\Upsilon_X} & & & & X \ar[ddd]^(.65){\mu_X} & & & & \\ {\rm At}(\xcP) \ar@{=>}[ddrr] \ar@/^2.pc/[uurrrr]^(.6){\widetilde\la{}_\xcP} \ar@/^1.25pc/[urr]_(.4){\la_\Xi}  \ar@/^1.25pc/[drrrr]^(.4){\la_\xcP} |!{[urr];[drr]}\hole & & & \xcG \ar@{^{(}.>}[dr] & & & & & & & \xcG \ar@{=>}[ddllll] \ar@/_2.pc/[uullllll]_(.6){\widetilde\varrho} \ar@/_1.25pc/[ullll]^(.4){\varrho_X} \ar@/_1.25pc/[dllllll]_(.4){\varrho} |!{[ulllll];[dlll]}\hole \\ & & & & \xcP \ar[drr]^(.4){\mu} \ar[dll]_(.4){\xcD} \ar[dd]^{\pi_\xcP} & & & & & & \\ & & \xcF & & & & M & & & & \\ & & & & \Si & & & & & & }\,.
\qqq
\ethe
\beroof
The commutativity of the extended $W$-diagram inscribed into the trident follows directly from the very definition of the fibred product $\xcP\,\,\fibx{\mu}{\mu_X}X$ and from Prop.\,\ref{prop:Ups-intro},\ see Diag.\,\ref{diag:ass-P-morph}.\ It identifies $(\xcP\,\,\fibx{\mu}{\mu_X}X,\Xi,\d_X,\mu_X\circ\pr_2,\widetilde\varrho)$ as the (right) principal $\xcG$-bundle described in Thm.\,\ref{thm:del-as-prince}.

Next,\ we consider the left ${\rm At}(\xcP)$-module $(\xcP\,\,\fibx{\mu}{\mu_X}X,X,\Upsilon_X\circ\d_X\equiv\widetilde\mu{}_\xcP,\widetilde\la{}_\xcP)$ of Thm.\,\ref{thm:ass-as-AtP-mod},\ the form of the moment map following,\ once more,\ from the commutativity of the diagram above.\ The action $\widetilde\la{}_\xcP$ trivially preserves $\pr_2$-fibres,\ see \eqref{eq:tilamPa},\ and so we have a well-defined map
\qq\nn
\bigl(\widetilde\la{}_\xcP,\pr_{2,3}\bigr)\colo{\rm At}(\xcP)\,\,\fibx{\sfS}{\widetilde\mu{}_\xcP}\bigl(\xcP\,\,\fibx{\mu}{\mu_X}X\bigr)\too\bigl(\xcP\,\,\fibx{\mu}{\mu_X}X\bigr)\,\,\fibx{\pr_2}{\pr_2}\bigl(\xcP\,\,\fibx{\mu}{\mu_X}X\bigr)\,.
\qqq
Define a (smooth) map
\qq\nn
\widetilde\psi{}_\xcP:=\psi_\xcP\circ\pr_{1,3}\colo\bigl(\xcP\,\,\fibx{\mu}{\mu_X}X\bigr)\,\,\fibx{\pr_2}{\pr_2}\bigl(\xcP\,\,\fibx{\mu}{\mu_X}X\bigr)\too{\rm At}(\xcP),\ \bigl((p_1,x),(p_2,x)\bigr)\longmapsto\psi_\xcP(p_1,p_2)\,,
\qqq
taking into account the identity $\mu(p_2)=\mu_X(x)=\mu(p_1)$,\ and then invoking Thm.\,\ref{thm:AtPoid} (see:\ \eqref{eq:divAt}).\ This is readily verified to be the division map for $\widetilde\la{}_\xcP$.\ Thus,\ the principality of $(\xcP\,\,\fibx{\mu}{\mu_X}X,X,\pr_2,\Upsilon_X\circ\d_X,\widetilde\la{}_\xcP)$ is established,\ whence,\ in particular,\ the canonical identification
\qq\nn
X\cong\bigl(\xcP\,\,\fibx{\mu}{\mu_X}X\bigr)/{\rm At}(\xcP)
\qqq
ensues by the Godement Criterion,\ see:\ Thm.\,\ref{thm:Godement}.

We may,\ now,\ investigate properties of the two associating actions:\ $\widetilde\la{}_\xcP$ and $\widetilde\varrho$ relative to each other,\ to the principal projection $\pr_1$,\ and to the respective extending moment maps $\d_X$ and $\pr_2$.\ We start by noting that the two associating actions commute because so do the underlying actions $\la_\xcP$ and $\varrho$ on $\xcP$.\ Moreover,\ the first component in each pair of chiral actions:\ $(\widetilde\la{}_\xcP,\la_\xcP)$ and $(\widetilde\varrho,\varrho)$ being a lift of the second one along $\pr_1$,\ the latter map automatically intertwines the two components.\ Finally,\ we note that $\d_X$ is ${\rm At}(\xcP)$-equivariant by definition,\ and,\ likewise,\ $\pr_2$ intertwines $\widetilde\varrho$ with $\varrho_X\equiv\la_X\circ(\Inv\circ\pr_2,\pr_1)$,\ see \eqref{eq:varrho-def}.\ Thus,\ by now,\ we have the structure of an $({\rm At}(\xcP),\xcG)$-bibundle $(\la_\Xi,\varrho_X)$-associated to the principaloid bundle $\xcP$ established on the $\mu_X$-extended bundle $\xcP\,\,\fibx{\mu}{\mu_X}X$,\ and captured neatly by the extended $W$-diagram 
\qq\nn
\alxydim{@C=.75cm@R=.5cm}{ & & & & \xcP\,\,\fibx{\mu}{\mu_X}X \ar[ddd]_{\pr_1} \ar[drr]^{\pr_2} \ar[dll]_{\d_X} & & & & \\ & & \Xi \ar[ddd]_(.4){\Upsilon_X} & & & & X \ar[ddd]^(.4){\mu_X} & & \\ {\rm At}(\xcP) \ar@{=>}[ddrr] \ar@/^1.pc/[urr]_{\la_\Xi} \ar@/^2.5pc/[uurrrr]^{\widetilde\la{}_\xcP} \ar@/^1.75pc/[drrrr]^(.6){\la_\xcP} |!{[urr];[drr]}\hole & & & & & & & & \xcG \ar@{=>}[ddll] \ar@/_1.pc/[ull]^{\varrho_X} \ar@/_2.5pc/[uullll]_{\widetilde\varrho} \ar@/_1.75pc/[dllll]_(.6){\varrho} |!{[ull];[dll]}\hole \\ & & & & \xcP \ar[drr]^(.4){\mu} \ar[dll]_(.4){\xcD} & & & & \\ & & \xcF & & & & M & &}\,.
\qqq

That the left wing of the above diagram fibres as claimed over the pair groupoid ${\rm Pair}(\Si)$ of the common base $\Si$ of the bundles $\xcP\,\,\fibx{\mu}{\mu_X}X,\Xi,\xcP$ and $\xcF$ follows directly from the definitions of the four bundles,\ put in conjunction with the fact that all these maps are $\Si$-vertical,\ {\it i.e.},\ they are morphisms in the slice category $\Bun(\Si)/\Si$. 
\eroof

The induction of the structure of an ${\rm At}(\xcP)$-module space on the $\mu_X$-extended bundle $\xcP\,\,\fibx{\mu}{\mu_X}X$ and its descent to the quotient $(\xcP\,\,\fibx{\mu}{\mu_X}X)/\xcG\cong\Xi$ pave the way to the introduction of distinguished automorphisms of the $\la_X$-associated bundle $\Xi$,\ including its gauge transformations.
\berop\label{prop:Ass-aut}
Adopt the notation of Props.\,\ref{prop:Ups-ind} and \ref{prop:Ups-intro}.\ There exist canonical group homomorphisms
\qq\nn
\Ext_X\colo{\rm Aut}(\xcP)\too{\rm Aut}\bigl(\xcP\,\,\fibx{\mu}{\mu_X}X\bigr)\,,\qquad\qquad\Xi_*\colo {\rm Aut}(\xcP)\too{\rm Aut}(\Xi)\,,
\qqq
where ${\rm Aut}(\xcP\,\,\fibx{\mu}{\mu_X}X)$ is the group of automorphisms of the $\mu_X$-extended bundle $\xcP\,\,\fibx{\mu}{\mu_X}X$ as a principaloid $\xcG\,\lx_{\la_X}X$-bundle (see:\ Prop.\,\ref{prop:ass-as-oidle}),\ whereas ${\rm Aut}(\Xi)$ is the group of automorphisms of $\Xi$ as a fibre bundle.\ These homomorphisms satisfy,\ for all $\Phi\in{\rm Aut}(\xcP)$,\ the equivariance identities
\qq
&\d_X\circ\Ext_X(\Phi)=\Xi_*(\Phi)\circ\d_X\,,\label{eq:Ups-coin-delX}&\\ \cr
&\Upsilon_X\circ\Xi_*(\Phi)=\xcF_*(\Phi)\circ\Upsilon_X\,.\label{eq:Ups-coin-Duck}&
\qqq
\eerop
\beroof
The existence of $\Ext_X$ and $\Xi_*$ is ensured by Thm.\,\ref{thm:autP-from-BisAt} and Prop.\,\ref{prop:G-in-B_act}.\ When put in conjunction,\ these yield,\ for $\Phi\in{\rm Aut}(\xcP)$ arbitrary and all $(p,x)\in\xcP\,\,\fibx{\mu}{\mu_X}X$,
\qq\nn
\Ext_X(\Phi)(p,x)&=&B\widetilde\la{}_\xcP\bigl(\b_\Phi,(p,x)\bigr)\equiv\widetilde\la{}_\xcP\bigl(\b_\Phi\bigl(\bigl(\Upsilon_X\circ\d_X\bigr)(p,x)\bigr),(p,x)\bigr)=\bigl(\la_\xcP\x\id_X\bigr)\bigl(\bigl(\b_\Phi\circ\xcD\bigr)(p)\bigr),(p,x)\bigr)\cr\cr
&\equiv&\bigl(\la_\xcP\bigl(\bigl(\b_\Phi\circ\mu_\xcP\bigr)(p),p\bigr),x\bigr)=\bigl(\Phi(p),x\bigr)\,,
\qqq
where the third and last equalities use the commutativity of Diags.\,\eqref{diag:ass-P-morph} and \eqref{diag:betPhi-Phi},\ respectively,\ and,\ for all $\xi\in\Xi$,
\qq\nn
\Xi_*(\Phi)(\xi)=B\la_\Xi(\b_\Phi,\xi)\equiv\la_\Xi\bigl(\b_\Phi\bigl(\Upsilon_X(\xi)\bigr),\xi\bigr)\,.
\qqq
Hence,\ we may write
\qq\nn
\Ext_X(\Phi)=\Phi\x\id_X\,,
\qqq
and
\qq\nn
\Xi_*(\Phi)=\la_\Xi\circ(\b_\Phi\circ\Upsilon_X,\id_\Xi)\,.
\qqq
It is,\ in particular,\ clear that the $\varrho$-equivariance of $\Phi$ ensures the $\widetilde\varrho$-equivariance of the induced bundle automorphism $\Ext_X(\Phi)$.

Identity \eqref{eq:Ups-coin-delX} now follows from a direct calculation,\ invoking the second of identities \eqref{eq:tilamPa-int-tilamXi} and,\ once more,\ Diags.\,\eqref{diag:ass-P-morph} and \eqref{diag:betPhi-Phi},
\qq\nn
\bigl(\Xi_*(\Phi)\circ\d_X\bigr)(p,x)&=&\la_\Xi\bigl(\b_\Phi\bigl(\bigl(\Upsilon_X\circ\d_X\bigr)(p,x)\bigr),\d_X(p,x)\bigr)\equiv\bigl(\la_\Xi\circ\bigl(\id_{{\rm At}(\xcP)}\x\d_X\bigr)\bigr)\bigl(\bigl(\b_\Phi\circ\xcD\bigr)(p),(p,x)\bigr)\cr\cr
&=&\bigl(\d_X\circ\widetilde\la{}_\xcP\bigr)\bigl(\bigl(\b_\Phi\circ\xcD\bigr)(p),(p,x)\bigr)\equiv\d_X\bigl(\la_\xcP\bigl(\bigl(\b_\Phi\circ\mu_\xcP\bigr)(p),p\bigr),x\bigr)=\d_X\bigl(\Phi(p),x\bigr)\,.
\qqq
Similarly,\ for \eqref{eq:Ups-coin-Duck},\ we compute,\ using the commutativity of Diag.\,\ref{diag:Tbeta} and the identification of $\Upsilon_X$ as the moment map for the action $\la_\Xi$ along the way,
\qq\nn
\bigl(\Upsilon_X\circ\Xi_*(\Phi)\bigr)(\xi)\equiv\bigl(\Upsilon_X\circ\la_\Xi\bigr)\bigl(\b_\Phi\bigl(\Upsilon_X(\xi)\bigr),\xi\bigr)=\sfT\bigl(\b_\Phi\bigl(\Upsilon_X(\xi)\bigr)\bigr)\equiv\bigl(\sfT\circ\b_\Phi\bigr)\bigl(\Upsilon_X(\xi)\bigr)=\xcF_*(\Phi)\bigl(\Upsilon_X(\xi)\bigr)\,.
\qqq
\eroof

\beg 
In the case of the bundle $P\x M$ associated with the principaloid $\bbB\txG$-bundle $P$ by the action $\la$,\ as in Ex.\,\ref{eg:CB-from-BG-ass},\ we recover the prototypical moment-extended trident \eqref{dig:ext-G-trident},\ which we may now rewrite concisely as
\qq\nn
\alxydim{@C=.75cm@R=.5cm}{ & \txG\x M \ar@{^{(}.>}[ddrrr] \ar[dl]_{\la} & & & & & & & & & \\ M \ar@{^{(}.>}[ddrr] & & & & & & & & & & \\ & & & & P\x M \ar[ddd]_{\pr_1} \ar[drr]^{\pr_2} \ar[dll]_{\pi_\sim} & & & & & & \\ & & P\x_\la M \ar[ddd]_(.65){\pi_{P\x_\la M}} & & & & M \ar@{-->}[ddd] & & & & \\ {\rm At}(P) \ar@{=>}[ddrr] \ar@/^1.pc/[urr]_{\la_{P\x_\la M}} \ar@/^2.5pc/[uurrrr]^{\widetilde\la{}_P} \ar@/^1.25pc/[drrrr]^(.4){\la_P} |!{[urr];[drr]}\hole & & & \txG \ar@{^{(}.>}[dr] & & & & & & & \txG \ar@{=>}[ddllll] \ar@/_2.pc/[uullllll]_(.6){\widetilde r{}_P} \ar@/_1.25pc/[ullll]^(.45){\la\circ(\Inv_\txG\circ\pr_2,\pr_1)\qquad} \ar@/_1.25pc/[dllllll]_(.4){r_P} |!{[ulllll];[dlll]}\hole \\ & & & & P \ar@{-->}[drr] \ar[dll]_(.4){\pi_P} \ar[dd]^{\pi_P} & & & & & & \\ & & \Si & & & & \bullet & & & & \\ & & & & \Si & & & & & & }\,.
\qqq
\eeg

\subsection{Induced connections on associated bundles}

The last element of the associating construction is the induction of distinguished connections on the $\mu_X$-extended bundle and on the $\la_X$-associated bundle from a $\xcG$-invariant connection on the underlying principaloid bundle $\xcP$.\ This is described in
\berop\label{prop:ind-assoc-conn}
Adopt the notation of Def.\,\ref{def:principaloid-conn} and Thm.\,\ref{thm:del-as-prince}.\ Every (compatible) Ehresmann connection $\txT\xcP\cong\txH\xcP\oplus\txV\xcP$ on a principaloid $\xcG$-bundle $\xcP$ canonically induces a $\xcG$-invariant,\ and hence---equivalently---$\xcG\,\lx_{\la_X}X$-invariant Ehresmann connection on the $\mu_X$-extended bundle $(\xcP\,\,\fibx{\mu\,}{\mu_X} X,\Si,\xcG\,\,\fibx{s\,}{\mu_X}X,\pi_\xcP\circ\pr_1)$ as
\qq\nn
&\txT(\xcP\,\,\fibx{\mu}{\mu_X}X)=\txV(\xcP\,\,\fibx{\mu}{\mu_X}X)\oplus\txH(\xcP\,\,\fibx{\mu}{\mu_X}X)\,,&\cr\cr
&\txV(\xcP\,\,\fibx{\mu}{\mu_X}X)=\pr_1^*\txV\xcP\,{}_{\txT\mu}\hspace{-3pt}\oplus_{\txT\mu_X}\hspace{-1pt}\pr_2^*\txT X\,,\qquad\txH(\xcP\,\,\fibx{\mu}{\mu_X}X)=\pr_1^*\txH\xcP\,,&
\qqq
where the Whitney sum is $\txT M$-fibred as
\qq\nn
\pr_1^*\txV\xcP\,{}_{\txT\mu}\hspace{-3pt}\oplus_{\txT\mu_X}\hspace{-1pt}\pr_2^*\txT X=\bigl\{\ (V,\xi)\in\txV_p\xcP\oplus\txT_x X\quad\vert\quad\txT_p\mu(V)=\txT_x\mu_X(\xi)\,,\quad(p,x)\in\xcP\,\,\fibx{\mu}{\mu_X}X\ \bigr\}\,.
\qqq
Consequently,\ $\txH\xcP$ also induces---along the $\xcG$-quotient map $\d_X$---a connection on the $\la_X$-associated bundle $(\Xi\cong(\xcP\,\,\fibx{\mu}{\mu_X}X)/\xcG,\Si,X,\pi_\Xi)$ as
\qq\label{eq:Whitney-G-quotient}\qquad\qquad
\txT\Xi=\txV\Xi\oplus\txH\Xi\,,\qquad\qquad\txV\Xi=\txT\d_X(\pr_1^*\txV\xcP\,{}_{\txT\mu}\hspace{-3pt}\oplus_{\txT\mu_X}\hspace{-1pt}\pr_2^*\txT X)\,,\qquad\txH\Xi=\txT\d_X(\pr_1^*\txH\xcP)\,.
\qqq
\eerop
\beroof
The tangent bundle of the $\mu_X$-extended bundle can be obtained from that of the cartesian product
\qq\nn
\txT(\xcP\x X)\cong\pr_1^*\txT\xcP\oplus\pr_2^*\txT X
\qqq
by defining the relevant $\txT M$-fibred subbundle\footnote{Note that $\txT\mu\circ\pr_1-\txT\mu_X\circ\pr_2$ has maximal and hence constant rank owing to the submersivity of $\mu$.}
\qq\nn
&&\ker\,(\txT\mu\circ\pr_1-\txT\mu_X\circ\pr_2)=\pr_1^*\txT\xcP\,{}_{\txT\mu}\hspace{-3pt}\oplus_{\txT\mu_X}\hspace{-1pt}\pr_2^*\txT X\cr\cr
&\equiv&\bigl\{\ (V,\xi)\in\txT_p\xcP\oplus\txT_x X\quad\vert\quad\txT_p\mu(V)=\txT_x\mu_X(\xi)\,,\quad(p,x)\in\xcP\x X\ \bigr\}
\qqq
and subsequently restricting the latter to the submanifold $\xcP\,\,\fibx{\mu}{\mu_X}X\subset\xcP\x X$,\ 
whereby the tangent of that submanifold is recovered,
\qq\nn
&&\txT(\xcP\,\,\fibx{\mu}{\mu_X}\hspace{-1pt} X)=\pr_1^*\txT\xcP\,{}_{\txT\mu}\hspace{-3pt}\oplus_{\txT\mu_X}\hspace{-1pt}\pr_2^*\txT X\,.
\qqq
Taking into account Prop.\,\ref{prop:ass-as-oidle} in conjunction with Def.\,\ref{def:principaloid-conn},\ we see that the horizontal distribution of a $\xcG\,\lx_{\la_X}X$-invariant connection on $\xcP\,\,\fibx{\mu}{\mu_X}X$---which is the same as a $\xcG$-invariant connection on that bundle---necessarily lies in 
\qq\nn
\ker\,\txT\widehat\mu{}_X\equiv\ker\,\txT\pr_2=\pr_1^*\ker\,\txT\mu\supset\pr_1^*\txH\xcP\,.
\qqq
Upon further noting that $\widehat\varrho$ restricts to $\varrho$ in the first cartesian factor in $\xcP\,\,\fibx{\mu\,}{\mu_X}X$,\ we conclude that the pullback along $\pr_1$ of a $\xcG$-invariant distribution $\txH\xcP$ within $\ker\,\txT\mu$ is automatically $\xcG\,\lx_{\la_X}X$-invariant,\ and so does,\ indeed,\ determine a $\xcG\,\lx_{\la_X}X$-invariant connection on $\xcP\,\,\fibx{\mu}{\mu_X}X$ relative to the \emph{common} base $\Si$ of $\xcP$ and $\xcP\,\,\fibx{\mu\,}{\mu_X}X$.

The identification of the vertical subbundle in $\txT(\xcP\,\,\fibx{\mu}{\mu_X}X)$,\ on the other hand,\ follows straightforwardly from the structure of the base projection,
\qq\nn
\pi_{\xcP\,\,\fibx{\mu\,\,}{\mu_X}X}=\pi_\xcP\circ\pr_1\qquad\Longrightarrow\qquad\txV(\xcP\,\,\fibx{\mu}{\mu_X}X)\equiv\ker\,\txT(\pi_\xcP\circ\pr_1)=\pr_1^*\ker\,\txT\pi_\xcP\,{}_{\txT\mu\,}\hspace{-3pt}\oplus_{\txT\mu^X}\hspace{-1pt}\pr_2^*\txT X\,.
\qqq

The last part of the proposition is now implied by Prop.\,\ref{prop:shad-conn} upon invoking Prop.\,\ref{prop:ass-as-oidle}.
\eroof

\bedef
Adopt the notation of Prop.\,\ref{prop:ind-assoc-conn}.\ We shall call the Ehresmann connection on $\Xi$ induced from one on $\xcP$ along the $\xcG$-quotient map $\d_X$ the {\bf $\xcG$-quotient connection}.
\exdef

The induced Ehresmann connections admit a standard (equivalent) definition in terms of smooth families of projectors,\ from which there emerges the definition of a coupling between the universal gauge field sourced by the connection 1-form on the principaloid bundle and (associated-)matter fields of interest.
\berop\label{prop:assoc-conn-form}
Adopt the notation of Prop.\,\ref{prop:ind-assoc-conn} and Thm.\,\ref{thm:loc-data-conn}.\ A (compatible) connection 1-form $\Theta\in\Om^1(\xcP,\txV\xcP)$ on a principaloid $\xcG$-bundle $\xcP$ induces a connection 1-form $\widehat\Theta{}_X\in\Om^1(\xcP\,\,\fibx{\mu\,}{\mu_X}X,\txV(\xcP\,\,\fibx{\mu\,}{\mu_X}X))$ on the $\mu_X$-extended bundle $\xcP\,\,\fibx{\mu\,}{\mu_X}X$ with the defining properties
\qq\nn
\widehat\Theta{}_X\rstr_{\txV(\xcP\,\,\fibx{\mu\ }{\mu_X}X)}=\id_{\txV(\xcP\,\,\fibx{\mu\ }{\mu_X}X)}\qquad{\rm and}\qquad\ker\,\widehat\Theta{}_X=\txH(\xcP\,\,\fibx{\mu}{\mu_X}X)\,.
\qqq
Let $\xcP\t_i\colo \pi_\xcP^{-1}(O_i)\xrightarrow{\ \cong\ }O_i\x\xcG$ be local trivialisations of $\xcP$ over an open cover $\{O_i\}_{i\in I}$ of $\Si$,\ and let $\{A_i\}_{i\in I}$ be the corresponding local data of $\Theta$.\ The induced connection 1-form is the $\txV(\xcP\,\,\fibx{\mu\,}{\mu_X}X)$-valued 1-form $\widehat\Theta{}_X$ with local presentations
\qq\nn
\bigl(\widehat{\xcP\t}{}_i^{-1\,*}\widehat\Theta{}_X\bigr)(\si,g,x)=\id_{\txT\xcG}\rstr_g+\id_{\txT X}\rstr_x-\txT_{\Id_{t(g)}}r_g\circ A_i\bigl(\si,t(g)\bigr)\,,\qquad (\si,g,x)\in O_i\x\bigl(\xcG\,\,\fibx{s\,}{\mu_X}X\bigr)\,,\qquad i\in I\,,
\qqq
written in terms of the local trivialisations 
\qq\nn
\widehat{\xcP\t}{}_i\colo(\pi_\xcP\circ\pr_1)^{-1}(O_i)\xrightarrow{\ \cong\ }O_i\x\bigl(\xcG\,\,\fibx{s\,}{\mu_X}X\bigr)\,,\ \bigl(\xcP\t_i^{-1}(\si,g),x\bigr)\longmapsto(\si,g,x)
\qqq
of $\xcP\,\,\fibx{\mu}{\mu_X}X$ induced by the $\xcP\t_i$.

The connection 1-form $\widehat\Theta{}_X$ is a compatible connection 1-form on the principaloid $\xcG\,\lx_{\la_X}X$-bundle $\xcP\,\,\fibx{\mu}{\mu_X}X$,\ in conformity with Prop.\,\ref{prop:ass-as-oidle}.
\eerop
\beroof
Obvious.
\eroof

\bedef\label{def:G-quot-conn-form}
Adopt the notation of Prop.\,\ref{prop:ind-assoc-conn}.\ Given a connection $\txT\xcP=\txV\xcP\oplus\txH\xcP$ on $\xcP$ and the corresponding connection $\txT(\xcP\,\,\fibx{\mu}{\mu_X}X)=\txV(\xcP\,\,\fibx{\mu}{\mu_X}X)\oplus\txH(\xcP\,\,\fibx{\mu}{\mu_X}X)$ on the $\mu_X$-extended bundle $\xcP\,\,\fibx{\mu}{\mu_X}X$,\ the {\bf $\xcG$-quotient connection 1-form} on the $\la_X$-associated bundle $\Xi$ is a $\txV\Xi$-valued 1-form $\Theta_\Xi\in\Om^1(\Xi,\txV\Xi)\equiv\G(\txT^*\Xi\ox\txV\Xi)$ with the property $\txT\d_X(\txH(\xcP\,\,\fibx{\mu}{\mu_X}X))=\ker\,\Theta_\Xi$.
\exdef
\noindent In analogy with Prop.\,\ref{prop:shadconn-idef},\ we readily establish
\berop\label{prop:Gquotconn-idef}
Adopt the notation of Def.\,\ref{def:G-quot-conn-form} and Prop.\,\ref{prop:assoc-conn-form}.\ The $\xcG$-quotient connection 1-form is uniquely determined by the identity
\qq\label{eq:Gquotconn-idef}
\Theta_\Xi\circ\txT\d_X=\txT\d_X\circ\widehat\Theta{}_X\,.
\qqq
It is,\ in particular,\ a shadow of the compatible connection 1-form $\widehat\Theta{}_X$.
\eerop
\beroof
Follows directly from Prop.\,\ref{prop:shadconn-idef} upon taking into account Props.\,\ref{prop:ass-as-oidle} and \ref{prop:assoc-conn-form}.
\eroof
\noindent Finally,\ upon descending to the $\la_X$-associated bundle as above,\ and passing to the local picture,\ we recover the aforementioned gauge-matter field coupling.
\berop\label{prop:ass-gauge-coupl}
Adopt the notation of Def.\,\ref{def:G-quot-conn-form},\ Prop.\,\ref{prop:Ups-intro} and Thm.\,\ref{thm:loc-data-conn}.\ Let $\{A_i\}_{i\in I}$ be local data of a connection 1-form $\Theta\in\Om^1(\xcP,\txV\xcP)$ on $\xcP$ associated to a trivialising cover $\{O_i\}_{i\in I}$ of $\Si$.\ The $\xcG$-quotient connection 1-form induced by $\Theta$ on the $\xcG$-quotient bundle $\Xi$ is the $\txV\Xi$-valued 1-form $\Theta_\Xi\in\Om^1(\Xi,\txV\Xi)$ with local presentations
\qq\nn
\bigl(\Xi\t_i^{-1\,*}\Theta_\Xi\bigr)(\si,x)=\id_{\txT X}\rstr_x-\a_{\xcE\,\lx_{\la_X} X}(x,\Id_{\mu_X(x)})\circ A_i\bigl(\si,\mu_X(x)\bigr)\,,\qquad\qquad (\si,x)\in O_i\x X\,,\qquad i\in I\,,
\qqq
written in terms of the local trivialisations $\Xi\t_i\colo\pi_\Xi^{-1}(O_i)\xrightarrow{\cong}O_i\x X$ induced by those of $\xcP$,\ and of the fundamental vector field $\a_{\xcE\,\lx_{\la_X}X}$ for the action $\la_X$ introduced in Def.\,\ref{prop:algbd-act-algbd} (see:\ Rem.\,\ref{rem:algbd-act-algbd}).
\eerop
\beroof
A simple corollary of Props.\,\ref{prop:ass-as-oidle} and \ref{prop:assoc-conn-form},\ and Thm.\,\ref{thm:loc-data-conn}.
\eroof

\subsection{Covariant derivatives of associated matter fields}
The hitherto constructions provide us with a gauge-covariant differentiation of sections of the $\la_X$-associated bundle,\ through which a coupling between the universal gauge field and the corresponding species of charged matter arises functionally.
\bedef\label{def:cov-def-ass}
Adopt the notation of Def.\,\ref{def:G-quot-conn-form}.\ The {\bf covariant derivative} of a section $\xi\in\G(\Xi)$ of the $\la_X$-associated bundle $\Xi$ of $\xcP$ {\bf relative to} a $\xcG$-quotient connection 1-form $\Theta_\Xi$ on $\Xi$ is the $C^\infty(\Si,\bR)$-linear mapping
\qq\label{eq:ass-cov-der-oid}
\nabla^{\Theta_\Xi}_\cdot\xi\colo \G(\txT\Si)\too\G(\xi^*\txV\Xi),\cV\longmapsto\Theta_\Xi\circ\txT\xi(\cV)\equiv\nabla^{\Theta_\Xi}_\cV\xi\,.
\qqq
\exdef
\noindent The verticalised differentiation thus defined exhibits the desired covariance properties.
\berop
Adopt the notation of Defs.\,\ref{def:gauge-trafo-conn} and \ref{def:cov-def-ass} and Prop.\,\ref{prop:Ass-aut}.\ Given an arbitrary gauge transformation $\Phi\in{\rm Gauge}(\xcP)$ of $\xcP$,\ let $\Theta_\Xi^\Phi$ denote the $\xcG$-quotient connection on the $\la_X$-associated bundle $\Xi$ induced from the gauge transform $\Theta^\Phi$ of the connection 1-form $\Theta$.\ Furthermore,\ for an arbitrary section $\xi\in\G(\Xi)$ of $\Xi$,\ let $\xi^\Phi$ denote the corresponding gauge transform of $\xi$: 
\qq\nn
\xi^\Phi\equiv\Xi_*(\Phi)\circ\xi\,.
\qqq
The covariant derivative of $\xi$ relative to $\Theta_\Xi$ transforms \emph{covariantly} as 
\qq\nn
\nabla^{\Theta^\Phi_\Xi}_\cdot\bigl(\xi^\Phi\bigr)=\txT\bigl(\Xi_*(\Phi)\bigr)\circ\nabla^{\Theta_\Xi}_\cdot\xi\,.
\qqq
\eerop
\beroof
Begin by noting that $\Theta_\Xi^\Phi$,\ being induced from $\Theta^\Phi$,\ can be understood as a shadow of the gauge transform $\widehat\Theta{}^\Phi_X$ of $\widehat\Theta{}_X$,\ the transform being given by the formula 
\qq\nn
\widehat\Theta{}^\Phi_X=\txT\bigl(\Ext_X(\Phi)\bigr)\circ\widehat\Theta{}_X\circ\txT\bigl(\Ext_X(\Phi)\bigr)^{-1}\,,
\qqq
see \eqref{eq:gauge-transform} and Prop.\,\ref{prop:Ass-aut}.\ Accordingly,\ it satisfies---in the light of Eqs.\,\eqref{eq:Gquotconn-idef} and \eqref{eq:Ups-coin-delX}---the identity 
\qq\nn
&&\Theta_\Xi^\Phi\circ\txT\d_X=\txT\d_X\circ\widehat\Theta{}^\Phi_X=\txT\bigl(\d_X\circ\Ext_X(\Phi)\bigr)\circ\widehat\Theta{}_X\circ\txT\bigl(\Ext_X(\Phi)\bigr)^{-1}=\txT\bigl(\Xi_*(\Phi)\circ\d_X\bigr)\circ\widehat\Theta{}_X\circ\txT\bigl(\Ext_X(\Phi)\bigr)^{-1}\cr\cr
&=&\txT\bigl(\Xi_*(\Phi)\bigr)\circ\Theta_\Xi\circ\txT\bigl(\d_X\circ\Ext_X(\Phi)^{-1}\bigr)=\txT\bigl(\Xi_*(\Phi)\bigr)\circ\Theta_\Xi\circ\txT\bigl(\Xi_*(\Phi)\bigr)^{-1}\circ\txT\d_X\,,
\qqq
from which the simple formula for the gauge-transform of $\Theta_\Xi$ follows by the surjectivity of $\txT\d_X$:
\qq\nn
\Theta_\Xi^\Phi=\txT\bigl(\Xi_*(\Phi)\bigr)\circ\Theta_\Xi\circ\txT\bigl(\Xi_*(\Phi)\bigr)^{-1}\,.
\qqq
Taking the latter into account,\ we verify the desired covariance in an explicit calculation:
\qq\nn
&&\nabla^{\Theta^\Phi_\Xi}_\cdot\bigl(\xi^\Phi\bigr)\equiv\Theta^\Phi_\Xi\circ\txT\xi^\Phi=\txT\bigl(\Xi_*(\Phi)\bigr)\circ\Theta_\Xi\circ\txT\bigl(\Xi_*(\Phi)\bigr)^{-1}\circ\txT\bigl(\Xi_*(\Phi)\circ\xi\bigr)=\txT\bigl(\Xi_*(\Phi)\bigr)\circ\nabla^{\Theta_\Xi}_\cdot\xi\,.
\qqq
\eroof

At long last,\ we may write out the local presentation of the verticalised derivative,\ in which the gauge-matter field coupling becomes manifest.
\bedef\label{def:ass-cov-der-loc-pres}
Adopt the notation of Def.\,\ref{def:cov-def-ass} and Prop.\,\ref{prop:Ups-intro}.\ For every global section $\xi\in\G(\Xi)$,\ with local data $\xi_i:=\pr_2\circ\Xi\t_i\circ\xi\rstr_{O_i}\colo O_i\to X,\ i\in I$,\ we call the expression
\qq\nn
\xcD^{\txA_i}_\cdot\xi_i:=\pr_2\circ\txT(\Xi\t_i)\circ\nabla^{\Theta_\Xi}_\cdot\xi\rstr_{O_i}
\qqq
the {\bf local presentation of the covariant derivative} of $\xi$ {\bf relative to} $\Theta_\Xi$ (in trivialisation $\Xi\t_i$).
\exdef
\noindent The presentation takes the form explicited in the following
\berop
Adopt the notation of Def.\,\ref{def:ass-cov-der-loc-pres} and Prop.\,\ref{prop:ass-gauge-coupl}.\ The local presentation of the covariant derivative relative to $\Theta_\Xi$ of a global section $\xi\in\G(\Xi)$ with local data $\xi_i,\ i\in I$ takes the form
\qq\label{eq:ass-cov-der-loc-oid}
\xcD^{A_i}_\cdot\xi_i(\si)=\txT_\si\xi_i-\a_{\xcE\,\lx_{\la_X} X}\bigl(\xi_i(\si),\Id_{(\mu_X\circ\xi_i)(\si)}\bigr)\circ A_i\bigl(\si,\bigl(\mu_X\circ\xi_i\bigr)(\si)\bigr)\,,\quad\si\in O_i\,.
\qqq
\eerop
\beroof
We calculate directly,\ with the help of Prop.\,\ref{prop:ass-gauge-coupl}:
\qq\nn
&&\xcD^{A_i}_\cdot\xi_i(\si)\equiv\pr_2\circ\txT_{\xi(\si)}(\Xi\t_i)\circ\nabla^{\Theta_\Xi}_\cdot\xi(\si)\equiv\pr_2\circ\txT_{\xi(\si)}(\Xi\t_i)\circ\Theta_\Xi\bigl(\xi(\si)\bigr)\circ\txT_\si\xi\cr\cr
&=&\pr_2\circ\bigl(\Xi\t_i^{-1\,*}\Theta_\Xi\bigr)\bigl(\si,\xi_i(\si)\bigr)\circ\txT_\si\bigl(\Xi\t_i\circ\xi\bigr)\cr\cr
&=&\bigl(\id_{\txT_{\xi_i(\si)}X}-\a_{\xcE\,\lx_{\la_X} X}\bigl(\xi_i(\si),\Id_{(\mu_X\circ\xi_i)(\si)})\circ A_i\bigl(\si,\bigl(\mu_X\circ\xi_i\bigr)(\si)\bigr)\bigr)\bigl(\id_{\txT_\si\Si},\txT_\si\xi_i\bigr)\cr\cr
&=&\txT_\si\xi_i-\a_{\xcE\,\lx_{\la_X} X}\bigl(\xi_i(\si),\Id_{(\mu_X\circ\xi_i)(\si)}\bigr)\circ\txA_i\bigl(\si,\bigl(\mu_X\circ\xi_i\bigr)(\si)\bigr)\,.
\qqq
\eroof

\subsection{Applications of the associating construction}\label{sub:ass-appl}

The findings presented in the previous (sub)sec\-tions provide us with a solid mathematical framework for addressing the issues raised in the opening paragraph,\ which we now discuss briefly in that framework.

The possibility of using associated principaloid bundles in the modelling of structure-group reduction suggests itself readily:\ Whenever,\ for a given Lie groupoid $\grpd{\xcG}{M}$ and a subgroup $\bB_{\rm red}(\xcG)\subset\bB(\xcG)$,\ identified as the reduced structure group in a field-theoretic setting of interest ({\it e.g.},\ preserving an extra differential-geometric structure on $M$ or $\xcG$),\ there exists a Lie groupoid $\grpd{\xcG_{\rm red}}{M_{\rm red}}$ and a left $\xcG_{\rm red}$-module structure $(M,\mu_M,\la_M)$ on $M$ such that $\xcG_{\rm red}\,\lx_{\la_M}M\cong\xcG$ and $\bB_{\rm red}(\xcG)={\rm Im}\,\widetilde\mu{}^*_M$ for the group homomorphism $\widetilde\mu{}^*_M:\bB(\xcG_{\rm red})\to\bB(\xcG)$ from Prop.\,\ref{prop:Bis-Bis},\ then we may first erect a principaloid $\xcG_{\rm red}$-bundle $\xcP_{\rm red}\to\Si$ with (unreduced) structure group $\bB(\xcG_{\rm red})$,\ and subsequently associate with it a bundle $\xcP_{\rm red}\,\lx_{\la_M}M\to\Si$ by the action $\la_M$.\ In virtue of Prop.\,\ref{prop:ass-as-oidle},\ we thus obtain a principaloid $\xcG$-bundle with a structure group reduced to (an isomorphic image of) $\bB_{\rm red}(\xcG)$,\ as desired.

The other issue is of a more conceptual nature,\ as it concerns the modelling of multiple (different) charged-matter fields to which a single gauge field couples,\ the latter coming from a given principaloid bundle.\ Clearly,\ the induction mechanism for $\xcG$-quotient connections described in Prop.\,\ref{prop:ind-assoc-conn} paves the way to a systematic construction of invariants of the induced gauge groups $\Xi_*({\rm Gauge}(\xcP))$,\ {\it e.g.},\ in terms of $B\la_X(\bB(\xcG))$-invariant tensors\footnote{Going beyond the minimal-coupling scheme is also possible,\ see:\ Sec.\,\ref{sec:gau-sigmod}.} on $X$.\ Indeed,\ a straightforward adaptation of the minimal-coupling scheme from Sec.\,\ref{sec:class-gau-princ} (see:\ the passage between \eqref{eq:min-coupl} and \eqref{eq:desc-min-coupl-T}) associates to every such tensor field $\cT\in\G(\txT^*X^{\ox n})^{B\la_X(\bB(\xcG))}$ a smooth tensor field $\cT[\Theta_\Xi]\in\G((\txT^*\pi_\Xi^{-1}\check{Y}_\cO)^{\ox n})$ on the total space of the surjective submersion $\check{\pi}_\Xi\colo\pi_\Xi^{-1}\check{Y}_\cO\equiv\bigsqcup_{i\in I}\,\pi_\Xi^{-1}(O_i)\to\Xi,\ (\Xi\t_i^{-1}(\si,x),i)\mapsto\Xi\t_i^{-1}(\si,x)$ with local restrictions
\qq\nn
\widetilde\cT[\Theta_\Xi]\rstr_{\pi_\Xi^{-1}(O_i)}:=\Xi\t_i^*\bigl(\cT\circ\pr_2\circ\bigl(\Xi\t_i^{-1\,*}\Theta_\Xi\bigr)^{\ox n}\bigr)\equiv\cT\circ\pr_2\circ\bigl(\txT\Xi\t_i\circ\Theta_\Xi\bigr)^{\ox n}\rstr_{\pi_\Xi^{-1}(O_i)}\in\G\bigl(\txT^*\pi_\Xi^{-1}(O_i)^{\ox n}\bigr)
\qqq
(see:\ Prop.\,\ref{prop:ass-gauge-coupl}),\ obtained from $\cT$ through the minimal coupling of the local gauge field which is mediated by the verticalising $\xcG$-quotient connection 1-form from the covariant derivative \eqref{eq:ass-cov-der-oid} in a local presentation \eqref{eq:ass-cov-der-loc-oid}.\ The family manifestly descends---in the sheaf-theoretic sense---to a smooth tensor field 
\qq\nn
\unl{\cT[\Theta_\Xi]}\in\G\bigl(\txT^*\Xi^{\ox n}\bigr)\,,\qquad\qquad\check{\pi}_\Xi^*\unl{\cT[\Theta_\Xi]}=\widetilde\cT[\Theta_\Xi]\,.
\qqq
The latter is gauge invariant,\ as captured by the following identity,\ written for an arbitrary gauge transformation $\Phi\in{\rm Gauge}(\xcP)$,
\qq\nn
\Xi_*(\Phi)^*\unl{\cT[\Theta_\Xi^\Phi]}=\unl{\cT[\Theta_\Xi]}\,.
\qqq
This implies the invariance of the pullback 
\qq\nn
\cT[\xi,\Theta_\Xi]\equiv\xi^*\unl{\cT[\Theta_\Xi]}
\qqq
of the descended tensor along an arbitrary matter field $\xi\in\G(\Xi)$,
\qq\nn
\cT[\xi^\Phi,\Theta_\Xi^\Phi]=\cT[\xi,\Theta_\Xi]\,,\quad\Phi\in{\rm Gauge}(\xcP)\,.
\qqq
Thus,\ the initial geometric problem is solved,\ but there arises a new {\em dynamical} one,\ which follows from the fact that the local gauge field $A_i$,\ {\it i.e.},\ the local datum of the inducing connection 1-form $\Theta$,\ `lives' not only on the spacetime $\Si\supset O_i$ of the field theory,\ but also on the generically non-terminal manifold of units $M$ of the structure groupoid of $\xcP$,\ and so requires a \emph{choice} of a graph of a map $\varphi_i\colo O_i\to M$ to couple to a charged-matter field as
\qq\label{eq:gfield-on-Higgs}
A_i\bigl(\si,\varphi_i(\si)\bigr)\,,\quad\si\in O_i\,.
\qqq
A natural model of such a graph is a local presentation $\xcF\t_i\circ\varphi=(\id_{O_i},\varphi_i)$ of a section $\varphi\in\G(\xcF)$ of the distinguished shadow bundle.\ In view of its special r\^ole in the construction,\ we refer to $\xcF$ and $\varphi$ in what follows as the {\bf Higgs bundle} and a {\bf Higgs field},\ respectively.\ If we are interested in the dynamics of \emph{probe} matter fields,\ which do not affect a given gauge-field \emph{background},\ the above poses no additional technical difficulty,\ because we may simply fix the values of the gauge field on \emph{all} allowed Higgs fields by declaring its \emph{global} form $\Theta$ once and for all.\ The problem arises whenever we choose to treat the gauge field \emph{dynamically},\ {\it i.e.},\ when its preferred values are determined by a variational principle \emph{jointly with} and \emph{in relation to} those of the Higgs and (other) charged matter \emph{sources}.\ We then encounter two natural scenarios in the setting under consideration:
\bit
\item \emph{The hybrid scenario}:\ The Higgs-restricted gauge field \eqref{eq:gfield-on-Higgs} is determined by a variational principle for a Yang--Mills--Higgs(-type) (YMH) theory,\ which couples the gauge field to the distinguished Higgs matter,\ and is then extended smoothly to an arbitrary background for charged-matter fields coming from the associated bundles $\Xi$.\ The most natural and elementary choice of the YMH dynamics is one in which the Higgs field $\varphi$ becomes the dynamical source of the {\em curvature} of the gauge field,\ the latter being modelled in a standard manner---through local trivialisations (see:\  \cite[Sec.\,9.7]{Kolar:1993})---on the {\bf curvature of the connection} $\Theta$,
\qq\nn
R[\Theta]:=\Theta\circ[\cdot,\cdot]_{\G(\txT\xcP)}\circ\bigl((\id_{\txT\xcP}-\Theta)\wedge(\id_{\txT\xcP}-\Theta)\bigr)\colo\bigwedge\hspace{-3pt}{}^2\hspace{1pt}\txT\xcP\too\txV\xcP\,.
\qqq
The curvature measures an obstruction against integrability of the horizontal distribution $\txH\xcP\equiv\ker\,\Theta$,\ see:\ \cite[Sec.\,9.4]{Kolar:1993}.\ It is not difficult to write down simple gauge invariants involving local presentations of $R[\Theta]$,\ further restricted to Higgs graphs in the YMH action functional,\ at least as long as there exists a metric $\kappa_\xcE\in\G(S^2\xcE^*)$ on the tangent algebroid $\xcE$ of $\xcG$ invariant under the action of the group $\unl{\txT C}(\bB(\xcG))$ from Prop.\,\ref{prop:Bis-on-E}.\ Lie-algebroidal prototypes of such invariants were considered by Strobl and Kotov in \cite{Strobl:2004im,Kotov:2015},\ and we shall provide their generalisations for arbitrary principaloid bundles in an upcoming work.
\item \emph{The multi-field scenario}:\ The gauge field is determined on the collection of images of {\em all} charged-matter fields in $\xcF$ (obtained with the help of the respective bundle maps $\Upsilon_X$) by a variational principle for a (lagrangean) field theory in which every species of charged matter contributes \emph{independently} to the curvature of the gauged field ({\it e.g.},\ through a replica of the invariant from the previous point,\ but now restricted to local presentations of $\Upsilon_X\circ\xi$ in sector $\G(\Xi)\ni\xi$).
\eit
It is easy to see that both scenarios have serious shortcomings.\ In the former one,\ it is the fundamentally different ontological status of the matter fields---the Higgs field acts as a source for the gauge field,\ whereas the remaining matter fields are mere probes---that may seem unnatural,\ and even if it be viewed as a proposal for an \emph{effective} field theory of a Higgs field `strongly' coupled to the gauge field and remaining matter fields `weakly' coupled to that field,\ it is still technically cumbersome as the background gauge field would have to be allowed to vary at intersections of the dynamical Higgs section with the $\Upsilon_X$-images of the other dynamical matter sections in $\xcF$,\ in a way dictated by the variational principle for the YMH sector.\ In the latter one,\ on the other hand,\ independent determination of the gauge field in each matter sector might lead to contradictions at the aforementioned intersections of critical matter sections,\ and it is far from obvious how to exclude such contradictions other than by manually overriding the principle of least action through imposition of a suitable `superselection rule'.

Luckily,\ there is a simple method to circumnavigate the above-mentioned problems in harmony with the distinguished status of the mother principaloid bundle $\xcP$ and its Higgs shadow $\xcF$.\ It consists in the imposition,\ on all classical configurations of charged-matter fields $\xi\in\Xi$,\ of the {\bf Higgs alignment constraint}
\qq\nn
\Upsilon_{X\,*}(\xi)\must\varphi\,,
\qqq
the latter being written in terms of the map
\qq\nn
\Upsilon_{X\,*}\colo\G(\Xi)\too\G(\xcF),\ \xi\longmapsto\Upsilon_X\circ\xi\,,
\qqq
and hence consistent with the gauge symmetry present,
\qq\nn
\Upsilon_{X\,*}\bigl(\xi^\Phi\bigr)\equiv\bigl(\Upsilon_X\circ\Xi_*(\Phi)\bigr)(\xi)=\bigl(\xcF_*(\Phi)\circ\Upsilon_X\bigr)(\xi)\equiv\xcF_*(\Phi)\bigl(\Upsilon_{X\,*}(\xi)\bigr)\must\xcF_*(\Phi)(\varphi)\equiv\varphi^\Phi\,,
\qqq
see:\ \eqref{eq:Ups-coin-Duck}.\ Indeed,\ it yields the identity (for $\xi_i$ and $\varphi_i$ as before)
\qq\nn
\cT[\xi,\Theta_\Xi](\si)&\equiv&\cT\circ\xcD^{A_i}_\cdot\xi_i(\si)^{\ox n}=\cT\circ\bigl(\txT_\si\xi_i-\a_{\xcE\,\lx_{\la_X} X}\bigl(\xi_i(\si),\Id_{(\mu_X\circ\xi_i)(\si)}\bigr)\circ A_i\bigl(\si,\bigl(\mu_X\circ\xi_i\bigr)(\si)\bigr)\bigr)^{\ox n}\cr\cr
&\must&\cT\circ\bigl(\txT_\si\xi_i-\a_{\xcE\,\lx_{\la_X} X}\bigl(\xi_i(\si),\Id_{\varphi_i(\si)}\bigr)\circ A_i\bigl(\si,\varphi_i(\si)\bigr)\bigr)^{\ox n}\,,
\qqq
and so it ensures that the coupling between the matter field $\xi$ and the gauge field is fully determined by the restriction of the latter to the underlying Higgs section.\ We shall study physical consequences of the Higgs alignment in a future work.

\section{The charged loop and its rigid symmetries}\label{sec:PAGsi}

Below,\ we introduce a class of field theories whose inherently geometric functional definition,\ involving both:\ tensorial and cohomological objects on the configuration fibre,\ sets the stage for a nontrivial and structurally varied discussion of the gauge principle for groupoidal symmetries.\ The discussion features two conceptually related types of descent to the configuration bundle $\xcF$ (in the spirit of Sec.\,\ref{sec:class-gau-princ} and Rem.\,\ref{rem:CB-4-grpd}):\ the sheaf-theoretic descent of (minimally coupled) background tensors and the (2-)stacky descent of (connection-augmented) cohomological objects.\ Our choice of the field theories to be analysed can be defended independently on the grounds of their relevance in the context of---on one hand---the critical bosonic string theory on curved target spaces \cite{Friedan:1985phd},\ and---on the other hand,\ and for a specific choice of the target space---the effective field theory of (slow) spinons in certain quantum spin chains \cite{Affleck:1985crit}.

\bedef\label{def:sigmod}
A {\bf Polyakov--Alvarez--Gaw\c{e}dzki} ({\bf PAG}) {\bf $\si$-model} is a triple 
\qq\nn
\bigl((\Si,\eta_\Si),(M,\txg_M,H_M ),\cA^{\si(M)}_{\rm DF}\bigr)\equiv\xcM_\si
\qqq 
which consists of
\bit
\item a compact oriented two-dimensional smooth metric\footnote{The signature of $\eta_\Si$ does not play a r\^ole in the present considerations,\ and so shall be left unspecified.} manifold $(\Si,\eta_\Si)$ without boundary,\ $\p\Si=\emptyset$,\ termed the {\bf worldsheet};
\item a smooth manifold $M$,\ termed the {\bf target space},\ endowed with a metric $\txg_M\in\G(S^2\txT^*M)$ and a closed differential 3-form $H_M \in Z^3_{\rm dR}(M)$ with $2\pi$-integral periods,\ ${\rm Per}([H_M ])\subset 2\pi\bZ$;
\item a map
\qq\nn
\cA^{\si(M)}_{\rm DF}\colo C^\infty(\Si,M)\too\uj,\ \varphi\longmapsto\cA^{\rm metr}_{\rm DF}[\varphi]\cdot\cA^{\rm top}_{\rm DF}[\varphi]\equiv\cA^{\si(M)}_{\rm DF}[\varphi]\,,
\qqq
termed the {\bf Dirac--Feynman} ({\bf DF}) {\bf amplitude}\footnote{We stick to the formulation of the dynamics and the  attendant nomenclature consistently used by Gaw\c{e}dzki {\it et al.} (see,\ {\it e.g.},\ \cite{Runkel:2008gr,Gawedzki:2010rn,Gawedzki:2012fu}) since the pioneering paper \cite{Gawedzki:1987ak},\ in which Dirac's quantum-mechanical interpretation of the classical action functional \cite{Dirac:1933pi}---subsequently employed by Feynman \cite{Feynman:1948pi}---was emphasised and put to work in a rigorous modelling of the dynamics and (Aharonov--Bohm-type) interference phases of extended distributions of topological charge in external gravitational and (higher) charge/gauge fields.} with components:
\bit
\item the ({\bf Polyakov}) {\bf metric amplitude}
\qq\nn
\cA^{\rm metr}_{\rm DF}[\varphi]:=\exp\left(-\tfrac{\sfi}{2}\,\int_\Si\,\txg_M\bigl(\txT\varphi\,\overset{\wedge}{,}\star_{\eta_\Si}\txT\varphi\bigr)\right)\,,
\qqq
in which $\txT\varphi$ is regarded as a section of $\txT^*\Si\ox\varphi^*\txT M$,\  and then the Hodge operator $\star_{\eta_\Si}$ of $\eta_\Si$ and the exterior product $\wedge$ affect the first tensor factors in $\txT^*\Si$,\ whereas the target-space metric $\txg_M$ contracts the other tensor factors in $\varphi^*\txT M$,\ so that the exponent acquires the interpretation of a metric norm of $\txT\varphi$;
\item the ({\bf Wess--Zumino}) {\bf topological amplitude}
\qq\nn
\cA^{\rm top}_{\rm DF}[\varphi]:=\chi_{H_M}\bigl(\varphi(\Si)\bigr)\,,
\qqq
determined by a degree-3 Cheeger--Simons differential character $\chi_{H_M }\in\widehat H{}^3(M,\uj)$ of curvature $H_M $ (see:\ Def.\,\ref{def:CS-char}),\ or---equivalently (see:\ Thm.\,\ref{thm:CH-BD})---by a gerbe $\cG$ of the same curvature,\ with $\chi_{H_M}\equiv\Hol_\cG$  (see:\ Def.\,\ref{def:holG}).
\eit 
\eit
\exdef

\brem\label{rem:Lorentz}
In local coordinates $\{\si^a\}^{a\in\{1,2\}}$ near $\si\in\Si$ and $\{\varphi^\mu\}^{\mu\in\ovl{1,\dim M}}$ near $\varphi(\si)\in M$,\ the integrand in the metric amplitude takes the familiar form
\qq\nn
\txg_M\bigl(\txT\varphi\,\overset{\wedge}{,}\star_{\eta_\Si}\txT\varphi\bigr)=_{\rm loc}\Vol(\Si)\,\sqrt{|\det\,\eta_\Si|}\,\bigl(\eta_\Si^{-1}\bigr)^{ab}\,\txg_{M\,\mu\nu}(\varphi)\,\p_a\varphi^\mu\,\p_b\varphi^\nu\,.
\qqq
This,\ in conjunction with the co-defining property of the characters:
\qq\nn
\forall\,c\in C_3(M)\colo\chi_{H_M }(\p c)=\exp\left(\sfi\,\int_c\,H_M \right)\,,
\qqq
leads to the interpretation of the above field theory as that of (metric-)minimal embeddings $\varphi\colo\Si\to M$ perturbed by Lorentz-type forces
\qq\nn
\imath_{\wedge^2\txT\varphi(\Vol(\Si)^*)}H_M \,,
\qqq
written in terms of (a representative of) the orientation $\Vol(\Si)^*\in\G(\wedge^2\txT\Si)\setminus\{\brd0\}$ on $\Si$.
\erem
\noindent The point of departure of any application of the gauge principle is the identification of rigid symmetries of the field theory of interest.\ This we provide in
\bedef\label{def:rig-symm-sigmod}
A ({\bf large}) {\bf rigid symmetry} of a PAG $\si$-model $\xcM_\si$ from Def.\,\ref{def:sigmod} is a diffeomorphism $f\in\Diff(M)$ with the following properties:
\qq\nn
f^*\txg_M=\txg_M\qquad\Longleftrightarrow\qquad f\in{\rm Isom}(M,\txg_M)\,,
\qqq
and 
\qq\nn
\forall\varphi\in C^\infty(M)\colo\cA_{\rm DF}^{\rm top}[f\circ\varphi]=\cA_{\rm DF}^{\rm top}[\varphi]\qquad\Longleftrightarrow\qquad f^*\cG\cong\cG\,.
\qqq

A {\bf tangential rigid symmetry} of $\xcM_\si$ is a vector field $\cV\in\G(\txT M)$ with the following properties:
\qq\nn
\pLie{\cV}\txg_M=0\,,
\qqq
and
\qq\nn
\forall\,\varphi\in C^\infty(\Si,M)\colo\int_{\varphi(\Si)}\,\imath_\cV H_M=0\,,
\qqq
{\it i.e.},
\qq\nn
\exists\,\kappa_\cV\in\Om^1(M)\colo \imath_\cV H_M =-\sfd\kappa_\cV\,.
\qqq
We shall call the assignment $\cV\mapsto\kappa_\cV$ the {\bf comomentum map}.
\exdef

\brem
A derivation and an in-depth discussion of the above notion of symmetry,\ consistent with the definition of the PAG $\si$-model in terms of the differential character,\ can be found in \cite[Sec.\,2.2]{Gawedzki:2010rn} (see also:\ \cite{Runkel:2008gr}).\ Note,\ in particular,\ that we demand the invariance under $f$ (resp.\ $\cV$) of {\em each} of the two terms in the DF amplitude {\em independently}.\ This is justified by the (physically motivated) requirement that the invariance hold true for an {\em arbitrary} metric $\eta_\Si$,\ the latter being regarded as an independent field in the theory under consideration (see,\ {\it e.g.},\ \cite{Polyakov:1981rd}).\ Alternative approaches to the concept of (tangential) symmetry of a $\si$-model with a {\em gauge-fixed} worldsheet metric have also been considered,\ see,\ {\it e.g.},\ \cite{Chatzistavrakidis:2017tpk}.\ Note also,\ in this context,\ that there exist long-known {\em dualities} of charged-loop mechanics,\ such as T-duality \cite{Buscher:1987qj,Buscher:1987sk} and the Hughes--Polchinsky duality \cite{Hughes:1986dn,Gauntlett:1989qe} (see also \cite{Suszek:2019cum,Suszek:2020xcu,Suszek:2020rev,Suszek:2023ldu} for a recent application in the higher-(super)geometric setting),\ which mix the metric and topological degrees of freedom.
\erem
\noindent The above general discussion is placed in the groupoidal context of interest through the following
\bedef\label{def:rig-symm-grpd}
Adopt the notation of Defs.\,\ref{def:rig-symm-sigmod} and \ref{def:tstar}.\ Let $\xcM_\si$ be a PAG $\si$-model.\ A Lie groupoid $\grpd{\xcG}{M}$ shall be called a {\bf rigid-symmetry model} for $\xcM_\si$ if there exists a (nontrivial) subgroup $\bB_\si(\xcG)\subseteq\bB(\xcG)$ such that each diffeomorphism in $t_*(\bB_\si(\xcG))\subset\Diff(M)$ is a rigid symmetry of $\xcM_\si$.\ The subgroup $\bB_\si(\xcG)$ shall be referred to as a {\bf rigid groupoidal symmetry group of} $\xcM_\si$.
\exdef

\beg\label{eg:AMM-0}
Consider a compact simple and simply connected Lie group $\txG$ with tangent Lie algebra $\ggt\equiv\Lie(\txG)$.\ On the group manifold,\ we find a family of Cartan--Killing metrics
\qq\nn
\txg_{\rm CK}=-\tfrac{\sfk}{4\pi}\,\tr_\ggt\bigl(\theta_{\rm L}\ox\theta_{\rm L}\bigr)\,,\quad\sfk\in\bR\,,
\qqq
and a family of Cartan 3-forms
\qq\nn
H_{\rm C}=\tfrac{\sfk}{12\pi}\,\tr_\ggt\bigl(\theta_{\rm L}\wedge[\cdot,\cdot]\circ\bigl(\theta_{\rm L}\wedge\theta_{\rm L}\bigr)\bigr)\,,
\qqq
both written in terms of the left-invariant Maurer--Cartan form $\theta_{\rm L}=\d_{\rm L}\log(\id_\txG)\in\Om^1(\txG)\ox_\bR\ggt$ (the left logarithmic derivative of $\id_\txG$,\ see:\ \cite{Hilgert:2012slg}),\ with the understanding that the (antisymmetrised) tensoring takes place in the differential-form factor.\ Here,\ the trace $\tr_\ggt$ over $\ggt$ is normalised so that 
\qq\nn
{\rm Per}(H_{\rm C})\subset 2\pi\bZ\qquad\Longleftrightarrow\qquad\sfk\in\bZ\,,
\qqq
see,\ {\it e.g.},\ \cite{Gawedzki:1999bq}.\ We further constrain the {\bf level} $\sfk$ as $\sfk\in\bN$ in order to obtain the standard sign in front of the metric factor after the Wick rotation.\ The 3-form $H_{\rm C}$ now geometrises as the $\sfk{}^{\rm th}$ tensor power $\cG_{\rm b}^{\ox\sfk}\equiv\cG_\sfk$ of Meinrenkren's basic gerbe $\cG_{\rm b}$ over $\txG$ \cite{Meinrenken:2002} (see:\ Defs.\,\ref{def:Grb} and \ref{def:holG}).\ The ensuing PAG $\si$-model is termed the Wess--Zumino--Novikov--Witten (WZNW) model \cite{Witten:1983ar}. 

The group $\txG$ is the object manifold of the action groupoid $\txG\,\lx_\Ad\txG\equiv\grpd{\txG\x\txG}{\txG}$ with $(s,t,\Id,\Inv)=(\pr_2,\Ad,(e,\cdot),(\Inv_\txG\circ\pr_1,\Ad))$,\ which furnishes a rigid-symmetry model for this field theory,\ of particular relevance to the study of its (maximally symmetric) defects \cite{Fuchs:2007fw,Runkel:2009sp,Suszek:2022CSimpl}.\ The large symmetry induces a $\ggt$-indexed family of tangential symmetries:
\qq\nn
\cK^\txG\colo\txG\x\ggt\too\txT\txG,\ (g,X)\longmapsto (L_X-R_X)(g)\equiv\cK^\txG_X(g)\,,
\qqq
{\it i.e.},\ the fundamental vector field for the action $\Ad$,\ with the left-invariant $L_X\in\G(\txT\txG)$ and the right-invariant $R_X\in\G(\txT\txG)$ vector fields on $\txG$,
\qq\nn
L_X(g)=\tfrac{\sfd\ }{\sfd t}\rstr_{t=0}\bigl(g\cdot\exp(tX)\bigr)\,,\qquad\qquad R_X(g)=\tfrac{\sfd\ }{\sfd t}\rstr_{t=0}\bigl(\exp(tX)\cdot g\bigr)\,.
\qqq
The corresponding comomentum map reads
\qq\nn
\kappa_{\rm C}\colo\txG\x\ggt\too\txT^*\txG,\ (g,X)\longmapsto-\tfrac{\tx{$\sfk$}}{4\pi}\,\tr_{\tx{$\ggt$}}\bigl(X\bigl(\theta_{\rm L}(g)+\theta_{\rm R}(g)\bigr)\bigr)\equiv-\tfrac{\sfk}{4\pi}\,\tr_\ggt\bigl(X\,t_a\bigr)\,\bigl(\theta_{\rm L}^a(g)+\theta_{\rm R}^a(g)\bigr)\,.
\qqq
Above,\ $\{t_a\}_{a\in\ovl{1,\dim\,\txG}}$ is an arbitrary basis of $\ggt$,\ and $\theta_{\rm H}=\theta_{\rm H}^a\ox t_a,\ H\in\{{\rm L},{\rm R}\}$,\ with $\theta_{\rm R}=\d_{\rm R}\log(\id_\txG)$ the right-invariant Maurer--Cartan form on $\txG$.\ Note the properties:
\qq\nn
\imath_{\cK^\txG_X}\kappa_{\rm C}(Y)=-\imath_{\cK^\txG_Y}\kappa_{\rm C}(X)\,,\qquad\qquad\kappa_{\rm C}\bigl(\txT_e\Ad_h(X)\bigr)=\Ad_{h^{-1}}^*\bigl(\kappa_{\rm C}(X)\bigr)\,,
\qqq
written for arbitrary $X,Y\in\ggt$ and $h\in\txG$.
\eeg

\section{The classic pathway to groupoidal configurational descent}\label{sec:G-gau-sigmod}

We now proceed to briefly review the gauge principle for group-modelled symmetries in the setting of the PAG $\si$-model $\xcM_\si$ from Def.\,\ref{def:sigmod},\ as worked out by Gaw\c{e}dzki,\ Waldorf and the Author in \cite{Gawedzki:2010rn,Gawedzki:2012fu}.\ Our goal is to identify its key structural elements and to put them in a form and context which render the whole construction amenable to subsequent natural generalisation.\ In so doing,\ we turn,\ once more,\ to the narrative,\ descriptive mode of exposition,\ referring the Reader to the original works for the relevant propositions and proofs.

The point of departure of our considerations is a choice of subgroup $\txG\subset{\rm Isom}(M,\txg_M)$ of isometries of a given $\si$-model target $M$,\ with smooth action $\la\colo\txG\x M\to M$,\ and with the additional property
\qq\label{eq:comG}
\forall\ X\in\ggt\ \exists\ \kappa_X\in\Om^1(M)\colo\imath_{\cK_X}H_M=-\sfd\kappa_X\,,
\qqq
stated for the tangent Lie algebra $\ggt\equiv\Lie(\txG)$ of $\txG$,\ in terms of the fundamental vector field\footnote{The sign `-' in the definition ensures that $\cK$ is a Lie-algebra homomorphism rather than an {\em anti}-homomorphism.} of $\la$,
\qq\nn
\cK\colo M\x\ggt\too\txT M,\ (m,X)\longmapsto\txT_{(e,m)}\la\bigl(-X,\brd0_{\txT M}(m)\bigr)\equiv\cK_X(m)\,.
\qqq
Note that we may view the ensuing comomentum map as a vertical vector-bundle morphism
\qq\nn
\alxydim{@C=.75cm@R=1.cm}{ M\x\ggt \ar[rr]^{\kappa} \ar[dr]_{\pr_1} & & \txT^*M \ar[dl]^{\pi_{\txT^*M}} \\ & M & }\,,\qquad\qquad\kappa(m,X)\equiv\kappa_X(m)\,,
\qqq
with the domain canonically identified with the total space of the tangent algebroid $\ggt\,\lx_\la M$ of the underlying action groupoid $\txG\,\lx_\la M$,\ see:\ Ex.\,\ref{eg:actalgbrd}.

In view of the following identity (see:\ \cite[Prop.\,2.11]{Gawedzki:2012fu}),\ valid for all forms $\eta,\z\in\Om^\bullet(M)$,\ all $(g,m)\in\txG\x M$,\ and in a basis $\{t_a\}_{a\in\ovl{1,\dim\,\txG}}$ of $\ggt$,\ and written in the shorthand notation $\cK_{t_a}\equiv\cK_a$,
\qq\nn
\la^*\eta(g,m)=\left(\sum_{k=0}^\infty\,\tfrac{(-1)^n}{n!}\,\widetilde\imath{}_{\cK\circ\theta_{\rm L}(g)}^n\la_g^*\eta\right)(m)\,,\qquad\qquad\widetilde\imath{}_{\cK\circ\theta_{\rm L}(g)}\z(m):=\theta^a_{\rm L}(g)\wedge\imath_{\cK_a}\z(m)\,,
\qqq
the existence of the comomentum implies the relation
\qq\label{eq:kappath-HM}
\D^{(3)}_{(0)}H_M=\sfd\kappa[\theta_{\rm L}]\,,
\qqq
expressed in terms of the $0^{\rm th}$ Dupont operator $\D^{(\bullet)}_{(0)}=\la^*-\pr_2^*$ of the simplicial nerve $N_\bullet(\txG\,\lx_\la M)$ of the action groupoid,\ see:\ Def.\,\ref{def:simpl-obj} and Ex.\,\ref{eg:face-nerve-LieGrpd},\ and of a distinguished 2-form on the latter's arrow manifold:
\qq\label{eq:kappa-G}
\kappa[\theta_{\rm L}]=\pr_2^*\kappa_a\wedge\pr_1^*\theta_{\rm L}^a-\tfrac{1}{2}\,\pr_2^*\bigl(\imath_{\cK_a}\kappa_b\bigr)\,\pr_1^*\bigl(\theta_{\rm L}^a\wedge\theta_{\rm L} ^b\bigr)\in\Om^2(\txG\x M)\,,
\qqq
with $\kappa_{t_a}\equiv\kappa_a$.\ Invoking Thm.\,\ref{thm:W3-tors},\ we infer that there is a flat gerbe $\cD_0\in\cW^3(\txG\x M;0)$ and a 1-isomorphism
\qq\nn
\Upsilon\colo\la^*\cG\xrightarrow{\ \cong\ }\pr_2^*\cG\ox\cI_{\kappa[\theta_{\rm L}]}\ox\cD_0\,.
\qqq
These observations prepare the ground for the formulation of the gauging procedure,\ whose objective is to induce a $\si$-model background---{\it i.e.},\ a metric tensor and a gerbe---on the total space of the associated bundle $P\x_\la M$ from the original one $(\txg_M,\cG)$ on $M$.

First,\ we consider the topological component of the background---the gerbe $\cG$.\ We begin by pulling it back to the total space of the extended bundle $P\x M$ along the moment map $\pr_2\colo P\x M\to M$ of the principal $\txG\,\lx_\la M$-bundle from Ex.\,\ref{eg:CB-from-BG-ass},\ and subsequently ask about the necessary conditions for the descent of the pullback to the bundle's base $P\x_\la M$.\ By the previously reviewed construction,\ the surjective submersion $\pi_\sim\colo P\x M\to P\x_\la M$ is,\ in fact,\ a principal $\txG$-bundle,\ and so the conditions of descent are determined by the following specialisation of a more general result,\ due essentially to Stevenson \cite{Stevenson:2000wj},\ which we review and employ in Sec.\,\ref{sec:EAS}.
\bethe\cite[Thm.\,5.3]{Gawedzki:2010rn}
Let $\txG$ be a Lie group,\ and let $X$ be a $\txG$-manifold with smooth action $\la_X\colo\txG\x X\to X$.\ Whenever the orbispace $X//\txG$ carries a smooth structure,\ and $X$ fibres principally over it ({\it e.g.},\ when $\la_X$ is free and proper),\ there exists a canonical equivalence
\qq\nn
\bgrb_\nabla(X//\txG)\cong\bgrb_\nabla(X)^{\txG\,\lx_{\la_X}\hspace{-1pt}X}
\qqq
between the bicategory of gerbes over $X//\txG$ and that of $\txG\,\lx_\la X$-equivariant gerbes over $X$,\ see:\ Def.\,\ref{def:Gequiv-grb}.
\ethe
\noindent As a corollary to the last result,\ we infer that the isoclass of the gerbe $\cI_{\kappa[\theta_{\rm L}]}\ox\cD_0$ encountered earlier constitutes an obstruction to a {\em direct} descent of $\pr_2^*\cG$ to the associated bundle $P\x_\la M$.\ As demonstrated in \cite{Gawedzki:2010rn},\ the component of this anomaly sourced by a non-vanishing comomentum,\ $\cI_{\kappa[\theta_{\rm L}]}$,\ can be neutralised though a procedure which employs the connection 1-form  $\cA$ on the first cartesian factor of the base $P\x M$ of the pullback gerbe $\pr_2^*\cG$.\ Indeed,\ assuming $\cG$ trivial,\ $\cG=\cI_B$,\ and endowed with a $\txG$-invariant curving $B\in\Om^2(M)^\txG$,\ and further restricting to the topologically trivial case $P\cong\Si\x\txG$ with a global gauge field $A=A^a\ox t_a\in\Om^1(\Si)\ox\ggt$ for the sake of simplicity,\ we find ourselves in the r\'egime of applicability of the minimal-coupling scheme of \Reqref{eq:min-coupl}.\ The latter takes the form
\qq\nn
\Hol_{\widetilde\cG[A]}\bigl((\id_\Si,\varphi)(\Si)\bigr)\,,\qquad\varphi\in C^\infty(\Si,M)
\qqq
for an $A$-augmented gerbe
\qq\nn
\widetilde\cG[A]=\pr_2^*\cG\ox\cI_{\kappa[A]}\,,
\qqq
expressed in terms of the 2-form
\qq\label{eq:kap-on-ttriv}
\kappa[A]=\pr_2^*\kappa_a\wedge\pr_1^*A^a-\tfrac{1}{2}\,\pr_2^*\bigl(\imath_{\cK_a}\kappa_b\bigr)\,\pr_1^*\bigl(A^a\wedge A^b\bigr)\in\Om^2(\Si\x M)\,,
\qqq
to be compared with \eqref{eq:kappa-G}.\ For the holonomy of the $A$-augmented gerbe to be gauge-invariant,\ we need to ensure that the comomentum satisfy additional conditions,\ to wit,\ that it 
\bit
\item[(CG1)] be $\txG$-equivariant
\qq\nn
\kappa\circ\bigl(\la_g\x\txT_e\Ad_g\bigr)=\la_{g^{-1}}^*\circ\kappa\,,\quad g\in\txG\,;
\qqq
\item[(CG2)] define an isotropic distribution 
\qq\nn
\xcC_\si\equiv(\cK,\kappa)(M\x\ggt)\subset\txT M\oplus\txT^*M\equiv\txE^{1,1}M
\qqq
relative to the non-degenerate pairing
\qq\nn
\corr{\cdot,\cdot}\colo\txE^{1,1}M\x\txE^{1,1}M\too\bR,\ \bigl((v_1,\om_1),(v_2,\om_2)\bigr)\longmapsto\tfrac{1}{2}\,\bigl(\om_2(v_1)+\om_1(v_2)\bigr)\,.
\qqq
\eit
It is worth noting that Hitchin's generalised tangent bundle $\txE^{1,1}M$,\ thus arising naturally in the present setting,\ carries a richer structure \cite{Hitchin:2004ut},\ which helps to clarify the geometric meaning of the above conditions.\ Namely,\ it carries a Courant bracket
\qq\label{eq:SWC-bracket}
\GBra{\cdot}{\cdot}_{\rm C}^{H_M}&\colo&\G(\txE^{1,1}M)\x\G(\txE^{1,1}M)\too\G(\txE^{1,1}M)\cr\cr  &&\bigl((\cV_1,\om_1),(\cV_2,\om_2)\bigr)\longmapsto\bigl([\cV_1,\cV_2]_{TM},\pLie{\cV_1}\om_2-\pLie{\cV_2}\om_1-\tfrac{1}{2}\,\sfd\bigl(\imath_{\cV_1}\om_2-\imath_{\cV_2}\om_1\bigr)+\imath_{\cV_1}\imath_{\cV_2}H_M\bigr)
\qqq
twisted by $H_M$ \`a la \v{S}evera--Weinstein \cite{Severa:2001qm},\ which endowes $\txE^{1,1}M$ with the structure of a Courant algebroid $(\txE^{1,1}M,\GBra{\cdot}{\cdot}_{\rm C}^{H_M},\corr{\cdot,\cdot},\pr_1)$ (see:\ \cite{Dorfman:1987,Courant:1988}).\ The bracket is readily shown to close on the symmetry distribution $\xcC_\si$.\ The conditions for the gaugeability of the rigid symmetry $\txG$ in the simplified scenario considered above now translate---upon linearisation of the group action---into a statement of homomorphicity of the map 
\qq\nn
(\cK,\kappa)\colo\bigl(M\x \ggt,\cK,[\cdot,\cdot]_{\ggt\,\lx_\la M}\bigr)\too\bigl(\txE^{1,1}M,\pr_1,\GBra{\cdot}{\cdot}_{\rm C}^{H_M}\bigr)\,,
\qqq 
This fact was first observed in the context of gauging in \cite{Suszek:2012ddg} (see also:\ \cite{Gawedzki:2012fu}),\ with a field-theoretic inspiration coming from \cite{Alekseev:2004np}.\ It reflects the deeper structural relation between Hitchin's generalised geometry and the higher geometry of gerbes \cite{Hitchin:2004ut,Gualtieri:2003dx,Suszek:2012ddg}.

The above observation,\ derived for trivial gerbes with a $\txG$-invariant curving,\ may now---and was---turned into a postulate for gerbes of {\em arbitrary} topology:\ Any such gerbe $\cG$ is to be pulled back to $P\x M$ along the moment map $\pr_2$ and subsequently augmented by a trivial gerbe with curving determined by a comomentum $\kappa$,\ satisfying the conditions listed above,\ and with the obvious replacement of the trivial gauge field by the connection 1-form $\cA=\cA^a\ox t_a\in\Om^1(P)\ox\ggt$.\ Thus,\ we obtain an $\cA$-augmented gerbe
\qq\label{eq:Augrb}
\widetilde\cG[\cA]=\pr_2^*\cG\ox\cI_{\kappa[\cA]}
\qqq
over the extended bundle $P\x M$,\ with
\qq\label{eq:Augmntn}
\kappa[\cA]=\pr_2^*\kappa_a\wedge\pr_1^*\cA^a-\tfrac{1}{2}\,\pr_2^*\bigl(\imath_{\cK_a}\kappa_b\bigr)\,\pr_1^*\bigl(\cA^a\wedge \cA^b\bigr)\in\Om^2(P\x M)\,.
\qqq
Its properties relevant to the gauging procedure under consideration are summarised in
\bethe\cite[Prop.\,5.5]{Gawedzki:2010rn}
Adopt the notation of Def.\,\ref{def:Gequiv-grb}.\ Let $\txG$ be a Lie group with tangent Lie algebra $\ggt$,\ and let $(M,\la)$ be a $\txG$-manifold with a gerbe $\cG\in\bgrb_\nabla(M)$ over it,\ and with a comomentum $\kappa$ determined by the curvature $H_M$ of $\cG$ as in \eqref{eq:comG},\ and satisfying conditions {\rm (CG1)} and {\rm (CG2)}.\ Moreover,\ let $P$ be a principal $\txG$-bundle over a compact closed orientable two-dimensional base $\Si$,\ equipped with a connection 1-form $\cA\in\Om^1(P)\ox\ggt$.\ If the gerbe admits a $\kappa[\theta_{\rm L}]$-twisted $\txG\,\lx_\la M$-equivariant structure $(\Upsilon,\g)$,\ with the 2-form $\kappa[\theta_{\rm L}]$ from \Reqref{eq:kappa-G},\ then the corresponding $\cA$-augmented gerbe $\widetilde\cG[\cA]$ from \Reqref{eq:Augrb} carries a canonical $\txG\,\lx_\la M$-equivariant structure,\ and so it canonically induces a gerbe $\unl{\cG[\cA]}\in\bgrb_\nabla(P\x_\la M)$ over the associated bundle $P\x_\la M$,\ unique up to 1-isomorphism,\ whose pullback along $\pi_\sim$ is ($\txG\,\lx_\la M$-equivariantly) 1-isomorphic to $\widetilde\cG[\cA]$.\ Furthermore,\ the holonomy of $\unl{\cG[\cA]}$ is invariant under gauge transformations in the sense captured by the identity (see:\ \Reqref{eq:GaugetrafoA})
\qq\nn
\forall\ (\varphi,\chi)\in\G(P\x_\la M)\x\G(\Ad\,P)\colo\Hol_{\unl{\cG[\cA^\chi]}}\bigl((\chi\circ\varphi)(\Si)\bigr)=\Hol_{\unl{\cG[\cA]}}\bigl(\varphi(\Si)\bigr)\,.
\qqq
\ethe
\noindent Two comments are in order at this point,\ which shall play an important r\^ole when we come to discuss a generalisation of the present scenario:\ First of all,\ the existence of a $\kappa[\theta_{\rm L}]$-twisted $\txG\,\lx_\la M$-equivariant structure on $\cG$ necessitates that the identity 
\qq\label{eq:multipl-kappaG}
\D^{(2)}_{(1)}\kappa[\theta_{\rm L}]=0\,,
\qqq
expressed in terms of the $1^{\rm st}$ Dupont operator $\D^{(\bullet)}_{(1)}$ of $N_\bullet(\txG\,\lx_\la M)$,\ hold true.\ Secondly,\ the induction of an {\em un}twisted $\txG\,\lx_\la M$-equivariant structure on $\widetilde\cG[\cA]$ hinges on the identity
\qq\nn
\widetilde\la{}^*\bigl(\kappa[\cA]\bigr)+\bigl(\Inv\circ\pr_2\bigr)^*\bigl(\kappa[\theta_{\rm L}]\bigr)=\pr_1^*\bigl(\kappa[\cA]\bigr)
\qqq
over the domain $(P\x M)\x_M(\txG\x M)$ of the $\txG\,\lx_\la M$-action \eqref{eq:Gact-ext}.\ The identity uses {\em both} properties of the comomentum,\ (CG1) and (CG2),\ assumed in the definition of the $\cA$-augmented gerbe.\ Finally,\ note that the existence of a coherent pair $(\Upsilon,\gamma)$ is,\ in general,\ obstructed topologically (in particular,\ it calls for the vanishing of $[\cD_0]$),\ and the obstruction,\ as well as a classification of inequivalent such pairs whenever the obstruction vanishes,\ is amenable to a cohomological analysis in the natural setting of $\txG$-equivariant cohomology \cite{Tu:2020},\ see:\ \cite{Gawedzki:2010rn}.\ We provide its suitable generalisation to the groupoidal case and a detailed exposition in Sec.\,\ref{sec:anomaly}.

We finish the construction of the gauged $\si$-model by defining its metric term.\ The point of departure is,\ in this case,\ the symmetric tensor $\unl{\txg_M[\cA]}$ on $P\x_\la M$ descended from the family of $\Si$-local minimally coupled ones $\{\widetilde\txg{}_M[A_i]\}_{i\in I}$ as described in the passage between \eqref{eq:min-coupl} and \eqref{eq:desc-min-coupl-T}.\ We evaluate it on the tangent morphism $\txT\varphi$ for $\varphi\in\G(P\x_\la M)$,\ viewed as a section of $\txT^*\Si\ox\varphi^*\txT(P\x_\la M)$ and,\ as such,\ suitably acted upon in the domain (as before),\ whereby we obtain the metric coupling
\qq\nn
\corr{\unl{\txg_M[\cA]},\txT\varphi^{\ox 2}}=\unl{\txg_M[\cA]}\bigl(\txT\varphi\,\overset{\wedge}{,}\star_{\eta_\Si}\txT\varphi\bigr)\,.
\qqq 
Altogether,\ then,\ in the notation used hitherto,\ we arrive at the definition of the PAG $\si$-model with the symmetry $\la$ gauged:
\qq\nn
\cA^{\si(P\x_\la M)}_{\rm DF}\colo\G(P\x_\la M)\too\uj,\ \varphi\longmapsto \exp\left(-\tfrac{\sfi}{2}\,\int_\Si\,\unl{\txg_M[\cA]}\bigl(\txT\varphi\,\overset{\wedge}{,}\star_{\eta_\Si}\txT\varphi\bigr)\right)\cdot\Hol_{\unl{\cG[\cA]}}\bigl(\varphi(\Si)\bigr)\,.
\qqq
We shall next generalise the above reasoning to the case of a symmetry modelled on an arbitrary Lie groupoid,\ and of the shadow of the corresponding principaloid bundle.

\brem
It deserves to be noted that the idea of configurational reduction $M\searrow M//\txG$,\ discussed abstractly at the end of Sec.\,\ref{sec:class-gau-princ},\ admits an interesting concretisation in the setting of the gauged PAG $\si$-model whenever the gauge field trivialises ({\it i.e.},\ {\it e.g.},\ locally) {\em and} $M$ fibres principally over the smooth quotient $M/\txG$ in such a way that $\txg_M$ induces a non-degenerate metric on the fibres.\ As demonstrated directly in \cite{Gawedzki:2012fu},\ the variable---in conformity with the discussion in Sec.\,\ref{sec:class-gau-princ}---background gauge field may then be integrated out,\ producing an emergent $\si$-model background:\ a $\txG$-{\em basic} metric 
\qq\nn
\widetilde\txg{}_M=\txg_M+X^{ab}\,\bigl(\kappa_a\ox\kappa_b-\cK^\flat_a\ox\cK^\flat_b\bigr)+Y^{ab}\,\bigl(\kappa_a\ox\cK^\flat_b+\cK^\flat_a\ox\kappa_b\bigr)\,,
\qqq
and an {\em untwisted} $\txG$-equivariant gerbe
\qq\nn
\widetilde\cG=\cG\ox\cI_B\,,\qquad\qquad B=\tfrac{1}{2}\,Y^{ab}\,\bigl(\kappa_a\wedge\kappa_b-\cK^\flat_a\ox\cK^\flat_b\bigr)+X^{ab}\,\kappa_a\wedge\cK^\flat_b\,,
\qqq
both written in terms of the $\txg_M$-contracted fundamental vector field $\cK^\flat_a=\txg_M(\cK^M_a,\cdot)$,\ and of the matrices $(X^{ab})\equiv X=E^{-1}$ and $(Y^{ab})\equiv Y=h^{-1}\,c\,E^{-1}$ determined by the background matrix $E=h-c\,h^{-1}\,c$,\ with $h=(\txg_M(\cK^M_a,\cK^M_b))$ and $c=(\imath_{\cK_a}\kappa_b)$.\ Thus,\ in the special situation just described,\ the $\si$-model descends to $M/\txG$ along $\pi\colo M\to M/\txG$,\ with the induced background $(\unl\txg,\unl\cG)$ such that $\pi^*\unl\txg=\widetilde\txg{}_M$ and $\pi^*\unl\cG\cong\widetilde\cG$.
\erem

\section{Reduction:\ Holonomicity and isometricity}\label{sec:reduction}

In this section,\ we take the first step towards a generalisation of the gauge principle of Sec.\,\ref{sec:G-gau-sigmod} to arbitrary structure Lie groupoids $\grpd{\xcG}{M}$,\ and so also to arbitrary principal $\xcG$-bundles $(\xcP,\xcF)$ in a given slice $\Bun(\Si)/\Si$.\ In this first step,\ we consider the relevant tensorial background on the simplicial nerve $N_\bullet(\grpd{\xcG}{M})$ of $\grpd{\xcG}{M}$,\ and the way in which it constrains our choice of the structure {\rm group}.\ The latter now becomes functionally separated from the structure groupoid in a pronounced manner.\ We shall lift our findings to the higher geometry behind the tensors in Sec.\,\ref{sec:EAS}.\smallskip

The highly structured nature of the prototype reviewed in the previous section suggests an immitation,\ which,\ however,\ becomes obscured by the various non-generic simplifications present in the prototypical construction:
\bit 
\item an {\it a-priori} choice of the symmetry model $\txG$,\ tailored to the extra geometric structure on $M$ (the metric $\txg_M$ and the 3-form field $H_M$) and hence controlling the reduction of the structure group to the subgroup $\iota(\txG)$ from Ex.\,\ref{eg:flatbisec},\ the latter coinciding with the {\em de-facto} structure groupoid $\bbB\txG$;
\item the product form of the arrow manifold $\txG\x M$,\ and the resultant simple factorisation of the twist $\kappa[\theta_{\rm L}]$,\ with the $M$-factor coming directly from the tangential-symmetry analysis of the DF amplitude prior to the gauging;
\item the global triviality of the underlying Lie algebroid $\ggt\x M\to M$,\ and the resulting non-generic structure of the standard connection 1-form on the principaloid bundle $\xcP=P\x M$,\ determined by a symmetry Lie algebra-valued 1-form $\cA\in\Om^1(P)\ox\ggt$,\ to which the comomentum may couple directly in the augmentation procedure \eqref{eq:Augrb};
\eit
\noindent In the light of the discussion of Sec.\,\ref{sec:red-ass},\ we could follow the pattern delineated in the previous section directly by starting with a given (higher-)geometric structure on a manifold $M$ and fixing an {\em external} Lie groupoid that stabilises that structure (or integrates its tangential rigid symmetries).\ The groupoid would then become the basic ingredient of the associating construction.\ This is,\ most certainly,\ a legitimate bottom-up approach to the question of the gauging of categorified rigid symmetries.\ In what follows,\ though,\ we take a reverse approach,\ in which the structure groupoid $\grpd{\xcG}{M}$ is given {\em a priori} ({\it e.g.},\ as a symmetry of some initial structure on the manifold $M$),\ and a reduction of the attendant group $\bB(\xcG)$ is enforced by the subsequent adjunction of extra geometric structure on its object manifold.\ As it happens,\ this scheme,\ emphasising the reduction mechanism (akin to classic ones,\ such as,\ {\it e.g.},\ an isometric structure-group reduction on a vector bundle that accompanies the introduction of a metric),\ enables us to make contact with long-known and important constructions,\ such as that of a symplectic groupoid,\ which subsequently leads to the emergence of lagrangean bisections,\ see:\ \cite{Weinstein:1987,Coste:1987,Rybicki:2001}.

In this top-down approach,\ and upon relinquishing the non-genericity of the motivating construction,\ we need to carefully define the point of departure in our considerations.\ While it is completely straightforward in the case of tensorial components of the extra geometric structure on $M$,\ for which the right notion is that of {\bf invariance} under the shadow action $t_*$ of the (reduced) structure group within $\bB(\xcG)$,\ dealing with charge fields,\ such as the 3-form field $H_M$,\ and the associated higher-geometric constructs,\ such as the gerbe $\cG$,\ calls for a more refined organising principle.\ This we read off from the discussion at the end of Sec.\,\ref{sec:class-gau-princ},\ taken in conjunction with the main result of the seminal work \cite{Bott:1976bss} of Bott,\ Schulman and Stasheff (see also:\ \cite{Dupont:1976,Dupont:1988}),\ which is the extended de Rham Theorem in the relevant adaptation (see:\ \Reqref{eq:class-grpd} and Def.\,\ref{def:BSS-complex})
\qq\nn
H^\bullet(\txB\xcG)=H^\bullet_{\rm BSS}\bigl(\grpd{\xcG}{M}\bigr)\,.
\qqq
Thus,\ viewing the gauging implemented by principaloid bundles as a field-theoretic mechanism of configurational descent,\ we are led to assume the existence of a {\bf $\xcG$-equivariant extension} of the tensorial background $H_M$ as the basis of our considerations in the present section.\ That extension shall subsequently be geometrised coherently in Sec.\,\ref{sec:EAS}.

An extension postulated above takes the general form 
\qq\nn
(H_M,\rho,\a,\b)\in\Om^3(M)\oplus\Om^2(\xcG)\oplus\Om^1(\xcG_2)\oplus\Om^0(\xcG_3)\,,
\qqq
with 
\qq\nn
D_{(3)}(H_M,\rho,\a,\b)=0
\qqq
for the BSS coboundary operator $D_{(3)}$ from Def.\,\ref{def:BSS-complex}.\ The first of the three identities mixing components of the extension of different form-degrees reads
\qq\label{eq:rho-rel-HM}
\txd\rho=\D^{(3)}_{(0)}H_M\,,
\qqq
and so it reproduces the familiar identity \eqref{eq:kappath-HM} upon specialisation.\ In conjunction with the anticipated geometrisation
\qq\label{eq:Ups-first}
\Upsilon\colo t^*\cG\xrightarrow{\ \cong\ }s^*\cG\ox\cI_\rho\,,
\qqq
it acquires a meaningful physical interpretation of the (hyper)cohomological datum of a $\cG$-bi-brane,\ which yields---through a variant of Gaw\c{e}dzki's transgression map from \cite{Gawedzki:1987ak} discussed at length in \cite{Suszek:2011hg}---a prequantisation of the classical rigid symmetry modelled by $\grpd{\xcG}{M}$,\ see:\ \cite{Fuchs:2007fw,Runkel:2008gr} and \cite{Suszek:2011hg,Suszek:2022CSimpl}.\ It is then natural to enquire about conditions under which the corresponding worldsheet defect admits self-intersections,\ which are bound to arise in the quantised field theory $\xcM_\si$,\ {\it e.g.},\ through factorisation of path integrals,\ see:\ \cite{Runkel:2008gr}.\ The detailed analysis of \cite{Runkel:2008gr} yields the necessary condition,\ to be satisfied by the extending 2-form $\rho$---it reads:
\qq\label{eq:rho-multi}
\D^{(2)}_{(1)}\rho=0\,,
\qqq
and so it yields the formerly encountered identity \eqref{eq:multipl-kappaG} upon specialisation.\ Altogether,\ then,\ we are led---by our physical argument---to assume the very peculiar form of the extension (see:\ Def.\,\ref{def:BSS-complex}):
\qq\nn
(H_M,\rho,0,0)\in{\rm Tot}^3\bigl(\Om^{\bullet_1}(\xcG_{\bullet_2}),\sfd^{(\bullet_1)}_{(\bullet_2)},\D^{(\bullet_1)}_{(\bullet_2)}\bigr)\,,
\qqq
subject to the constraints \eqref{eq:rho-rel-HM} and \eqref{eq:rho-multi}.\ Such BSS extensions have long been known and studied in the literature in various contexts,\ see,\ {\it e.g.},\ \cite{Xu:2004,Bursztyn:2004,Laurent-Gengoux:2007,Bursztyn:2012,Gawedzki:2010rn,Suszek:2012ddg,Gawedzki:2012fu,Crainic:2015msp,Suszek:2022CSimpl}. 
\bedef\label{def:BSS-xt}
Adopt the notation of Def.\,\ref{def:mult-k-form}.\ Let $H\in\Om^{k+1}(M)$ be a de Rham cocycle on the object manifold of a Lie groupoid $\grpd{\xcG}{M}$.\ A BSS $(k+1)$-cocycle $(H,\rho,0,0,\ldots,0)$ with $\rho\in\Om^k(\cG)$ multiplicative shall be called a {\bf multiplicative BSS extension} of $H$.\ We shall also say,\ in this case,\ that  $H$ is {\bf BSS-extended by} $\rho$.
\exdef
\noindent The intimate relation between such BSS extensions and (vertical) vector-bundle morphisms $\xcE\to\txT^*M$ with rather tractable equivariance properties with respect to a natural action of (jets of) bisections,\ which we are about to explore after \cite{Crainic:2015msp},\ shall be seen to play an absolutely crucial r\^ole in the generalisation of the gauge principle under consideration.\ This is yet another---this time {\it a posteriori}---rationale for focusing our attention on them in what follows.

Having fixed the form of the extension of $H_M$,\ co-defining a $\si$-model background,\ we return to the basic question of the reduction of the structure group in the presence of that background.\ The question may now be rendered more concrete with the help of identity \eqref{eq:rho-rel-HM}.\ Indeed,\ upon pulling back the latter along a candidate rigid-symmetry bisection $\b\in\bB(\xcG)$,\ we obtain the equality
\qq\nn
\bigl(t_*\b\bigr)^*H_M=H_M+\sfd\bigl(\b^*\rho\bigr)\,,
\qqq
from which the necessary restriction ensues:
\qq\nn
\b^*\rho\overset{!}{\in}\ker\,\sfd\,.
\qqq
Indeed,\ for the induced diffeomorphism $t_*\b\in\Diff(M)$ to preserve the holonomy of $\cG$ (in the definition of the DF amplitude of the PAG $\si$-model),\ it has to map its curvature $H_M$ to itself under pullback (see:\ Thm.\,\ref{thm:CH-BD}).\ With hindsight,\ we impose on $\b$ the stronger condition of {\bf $\rho$-holonomicity} (see:\ \cite{Crainic:2015msp}): 
\qq\label{eq:holo-must}
\b^*\rho\must0\,,
\qqq
which is readily seen to ensure---{\it via} \eqref{eq:Ups-first}---the desired prequantisation of the rigid symmetry:
\qq\label{eq:rig-symm-ens}
(t_*\b)^*\cG\equiv\b^*t^*\cG\cong\b^*s^*\cG\ox\cI_{\b^*\rho}=\cG\ox\cI_0\equiv\cG\,,
\qqq
see:\ Def.\,\ref{def:rig-symm-sigmod}.\ The above gives rise to the following subgroup of the original structure group.
\begin{propanition}\label{prop:rholo-grp}
Adopt the notation of Defs.\,\ref{def:bisec} and \ref{def:mult-k-form}.\ Let $\rho\in\Om^k(\xcG)$ be multiplicative.\ The set 
\qq\nn
\bB_\rho(\xcG):=\{\ \b\in\bB(\xcG) \quad\vert\quad \b^*\rho=0\ \}
\qqq
is a subgroup of the group $\bB(\xcG)$ of bisections of $\xcG$.\ We shall call it the {\bf $\rho$-holonomic group} of $\xcG$.
\end{propanition}
\beroof
Let $\b_1,\b_2\in\bB_\rho(\xcG)$.\ Then,\ in virtue of \eqref{eq:rho-multi},
\qq\nn
(\b_1\cdot\b_2)^*\rho\equiv\bigl(\txm\circ\bigl(\b_2\circ t_*\b_1,\b_1\bigr)\bigr)^*\rho=\bigl(\b_2\circ t_*\b_1,\b_1\bigr)^*\bigl(\pr_1^*+\pr_2^*\bigr)\rho=(t_*\b_1)^*\b_2^*\rho+\b_1^*\rho=0\,,
\qqq
and so $\b_1\cdot\b_2\in\bB_\rho(\xcG)$.\ Moreover,\ by the same token,
\qq\nn
\Id^*\rho\equiv\bigl(\txm\circ(\Id,\Id)\bigr)^*\rho=(\Id,\Id)^*\bigl(\pr_1^*+\pr_2^*\bigr)\rho=2\Id^*\rho\,,
\qqq
whence also $\Id\in\bB_\rho(\xcG)$.\ Finally,\ using the above,\ we obtain,\ for every $\b\in\bB_\rho(\xcG)$,
\qq\nn
0=\Id^*\rho\equiv\bigl(\b^{-1}\cdot\b\bigr)^*\rho=\bigl(t_*\b^{-1}\bigr)^*\b^*\rho+\bigl(\b^{-1}\bigr)^*\rho=\bigl(\b^{-1}\bigr)^*\rho\,,
\qqq
which implies $\b^{-1}\in\bB_\rho(\xcG)$,\ and completes the proof.
\eroof

\brem\label{rem:weak-rho-holo}
The rigid-symmetry identity \eqref{eq:rig-symm-ens} does not necessitate the stringent condition of $\rho$-holonomicity.\ Instead,\ its cohomology variant:
\qq\nn
[\b^*\rho]\must 0\in H^k(M)
\qqq 
is sufficient (see:\ Def.\,\ref{def:rig-symm-sigmod}).\ It is clear that the larger set of {\bf weakly $\rho$-holonomic bisections},\ which obey this less stringent constraint,\ also forms a group.\ Taking into account the increased level of complexity of the analysis of stacky descent---to be carried out in the next section---in this more general case,\ we restrict our attention to $\rho$-holonomic bisections in what follows.\ We hope to return to this issue in a future study.
\erem

In the prototypical construction of Sec.\,\ref{sec:G-gau-sigmod},\ the BSS extension is determined by the comomentum,\ itself derived from the rigid-symmetry analysis and subsequently coupled to the Lie algebra-valued gauge field,\ under the additional assumption of $\txG$-equivariance.\ In the present situation,\ we have to partially reverse this reasoning,\ and seek to obtain the comomentum with the desired property from the multiplicative 2-form $\rho$.\ That this is feasible yet far from trivial in the general case is demonstrated in the following remarkable
\bethe\cite[Thm.\,1 \& Prop.\,4.1]{Crainic:2015msp}\label{thm:CSS}
Adopt the notation of Defs.\,\ref{def:J1B} and \ref{def:mult-k-form},\ and of Ex.\,\ref{eg:tanLiealgbrd}.\ Let $\grpd{\xcG}{M}$ be a Lie groupoid with tangent algebroid $\xcE$,\ and let $\grpd{J^1\xcG}{M}$ be the first-jet groupoid of $\xcG$.\ For every multiplicative $k$-form $\rho\in\Om^k(\xcG)$,\ there exists a pair $(\kappa,S)$ composed of a vector-bundle morphism 
\qq\label{eq:comom-expl}
\alxydim{@C=0.75cm@R=1.cm}{ \xcE \ar[rr]^{\kappa\qquad} \ar[dr]_{\pi_\xcE} & & \wedge^{k-1}\txT^*M \ar[dl]^{\pi_{\wedge^{k-1}\txT^*M}} \\ & M & }\,,\qquad\qquad \kappa(\vep)=\Id^*\bigl(\imath_\vep\rho\bigr)\,,
\qqq
and a map 
\qq\label{eq:Spencer-expl}
\alxydim{@C=0.75cm@R=1.cm}{ J^1\xcG \ar[rr]^{S\quad} \ar[dr]_{\t} & & \wedge^k\txT^*M \ar[dl]^{\pi_{\wedge^k\txT^*M}} \\ & M & }\,,\qquad\qquad S\bigl(j^1_m\b\bigr)=\rho_{\b(m)}\circ\wedge^k\bigl(\txT_m\b\circ\bigl(\widetilde t{}_{j^1_m\b}\bigr)^{-1}\bigr)\,,
\qqq
with the following properties,\ written---in terms of the canonical actions of $J^1\xcG$:\ $\widetilde t{}_\cdot$ on $\txT M$ and $\widetilde C$ on $\xcE$ described in Prop.\,\ref{prop:J1GonMandE}---for all $g\in\xcG$ with $s(g)=m$ and arbitrary $j^1_m\b\in J^1\xcG$ with $\b(m)=g$,\ $\vep,\vep_1,\vep_2\in\xcE_m$,\ and $v_1,v_2,\ldots,v_{k-1}\in\txT_m M$:
\qq\label{eq:comom-equiv}
&&\kappa\bigl(\widetilde C{}_{j^1_m\b}(\vep)\bigr)\bigl(\widetilde t{}_{j^1_m\b}(v_1),\widetilde t{}_{j^1_m\b}(v_2),\ldots,\widetilde t{}_{j^1_m\b}(v_{k-1})\bigr)-\kappa(\vep)(v_1,v_2,\ldots,v_{k-1})\\ \cr
&=&S\bigl(j^1_m\b\bigr)\bigl(\widetilde t{}_{j^1_m\b}\bigl(\a_\xcE(\vep)\bigr),\widetilde t{}_{j^1_m\b}(v_1),\widetilde t{}_{j^1_m\b}(v_2),\ldots,\widetilde t{}_{j^1_m\b}(v_{k-1})\bigr)\,,\nn
\qqq
and
\qq\label{eq:comom-asymm}
\imath_{\a_\xcE(\vep_1)}\kappa(\vep_2)+\imath_{\a_\xcE(\vep_2)}\kappa(\vep_1)=0\,.
\qqq
The $k$-form $\rho$ evaluates on an arbitrary $k$-tuple $(X_1,X_2,\ldots,X_k)\in\txT_g\xcG$ at $g\in s^{-1}(\{m\})$ as
\qq\label{eq:rho-as-Skappa}
\rho(X_1,X_2,\ldots,X_k)=S\bigl(j^1_m\b\bigr)\bigl(\widetilde t{}_{j^1_m\b}(\unl{X}_1),\widetilde t{}_{j^1_m\b}(\unl{X}_2),\ldots,\widetilde t{}_{j^1_m\b}(\unl{X}_k)\bigr)\\ \cr 
+\sum_{p=1}^k\,\sum_{\sigma\in{\rm Shuffle}(p,k-p)}\,(-1)^{|\si|}\,\bigl(\kappa\circ\theta_{\rm R}\bigr)(\vep_{\si(1)})\bigl(\a_\xcE(\vep_{\si(2)}),\ldots,\a_\xcE(\vep_{\si(p)}),\widetilde t{}_{j^1_m\b}(\unl{X}_{\si(p+1)}),\ldots,\widetilde t{}_{j^1_m\b}(\unl{X}_{\si(k)})\bigr)\,,\nn
\qqq
where $\unl{X}_i:=\txT_g s(X_i),\ i\in\ovl{1,k}$,\ and---for $j^1_m\b\in J^1\xcG$ with $\b(m)=g$ chosen arbitrarily---
\qq\nn
\vep_j:=\bigl(\id_{\txT_g\xcG}-\txT_m\b\circ\txT_g s\bigr)(X_j)\,,\quad j\in\ovl{1,k}\,.
\qqq
Above,\ ${\rm Shuffle}(p,k-p)\subset S_k$ denotes the set of $(p,k-p)$-shuffles of $\ovl{1,k}$,\ and $(-1)^{|\cdot|}\colo S_k\to\{-1,1\}$ is the sign function.
\ethe

\brem\label{rem:more-on-Spencer}
The statement of the last theorem has been extracted from the constructive proof of the beautiful result \cite[Thm.\,1]{Crainic:2015msp} of Crainic,\ Salazar and Struchiner on the one-to-one correspondence between multiplicative forms and (higher) Spencer operators\footnote{The operators generalise,\ in an appropriate sense,\ the classical Spencer operators on (first-jet bundles of) vector bundles.} on Lie groupoids (with values in vector-bundle modules of the latter).\ Crucially,\ the pair $(\kappa,S)$ can be demonstrated to possess further properties,\ which ensure the independence of presentation \eqref{eq:rho-as-Skappa} of the {\em arbitrary} choice of $j^1_m\b$.\ We choose to omit the discussion of these additional properties for the sake of brevity,\ but keep in mind the implication emphasised.

Note,\ in passing,\ the convenient decomposition 
\qq\nn
X_j=\txT_m\b(\unl{X}_j)+\vep_j
\qqq
of an arbitrary vector $X_j\in\txT_g\xcG$ used above.\ It corresponds to the transverse splitting 
\qq\nn
\txT_{\b(m)}\xcG\cong\txT_m\b\bigl(\txT_m M\bigr)\oplus\bigl(\ker\,\txT s\bigr)_{\b(m)}
\qqq
induced by (an arbitrary representative $\b$ of) the first jet $j^1_m\b$,\ in conformity with Prop.\,\ref{prop:beta-split},
\qq\nn
\txT_m\b(\unl{X}_j)\equiv\chi_\b(X_j)\,,\qquad\qquad\vep_j\equiv\upsilon_\b(X_j)\,.
\qqq
\erem

\bedef\label{def:Spencer-pair}
Adopt the notation of Thm.\,\ref{thm:CSS}.\ The bundle map $\kappa$ and the map $S$ shall be called $k${\bf -co\-mo\-men\-tum} and---after \cite{Crainic:2015msp}---the {\bf scalar} $k${\bf -Spencer operator} of $\rho$,\ respectively.\ The pair $(\kappa,S)$ shall be referred to as the {\bf Spencer pair} of $\rho$.
\exdef
\noindent Now that the (2-)comomentum has been induced from a BSS extension of the curvature of the $\si$-model's gerbe,\ we may try to imitate the augmentation procedure from Sec.\,\ref{sec:G-gau-sigmod}.\ The success of such an imitation hinges on the concurrence and subtle constructive interplay of three elements:
\bit
\item the existence of an $\xcE$-valued differential form over the spacetime of the field theory transforming in an affine manner under redefinitions of the Lie-groupoidal gauge (in a local presentation as in Cor.\,\ref{cor:Requiv-principoidle-Auts});
\item the equivariance of the comomentum under the action---defined through a specialisation of Prop.\,\ref{prop:J1GonMandE}---of global bisections representing the redefinitions locally;
\item a reduction of the underlying multiplicative 2-form to an expression fully determined by the comomentum upon restriction to points in the arrow manifold $\xcG$ which belong to the latter bisections (understood as submanifolds diffeomorphically mapped to $M$ by $s$ and $t$,\ see:\ Def.\,\ref{def:bisec}).
\eit
The first element is straightforward to conceive,\ at least as long as we stay within a domain $O\subset\Si$ of trivialisation of the shadow bundle $\xcF\to\Si$,\ in which we have at our disposal a local gauge field $A\in\G(\pr_1^*\txT^*O\ox\pr_2^*\xcE)$,\ obtained from $\Theta$ as in Thm.\,\ref{thm:loc-data-conn},\ and exhibiting the desired behaviour,\ see:\ Prop.\,\ref{prop:gautrafo-conn-is-conn}.\ The gauge field may then be coupled to the comomentum through a natural generalisation of \Reqref{eq:kap-on-ttriv},
\qq\label{eq:kappA}
\kappa[A]=\kappa\overset{\wedge}{\circ}A+\tfrac{1}{2}\,\corr{\kappa\circ A\,\overset{\wedge}{,}\,\a_\xcE\circ A}\,,
\qqq
in which
\bit
\item in the first term:\ $\kappa$ is viewed as a section of $\xcE^*\ox \txT^*M$,\ the contraction $\circ$ refers to the first tensor factor,\ and the exterior product indicates antisymmetrisation with respect to the second tensor factor,\ regarded as a direct summand in $\txT^*(O\x M)$;
\item in the second term:\ $\kappa\circ A$ is viewed as a section of $\txT^*O\ox\txT^*M$,\ and the exterior product indicates antisymmetrisation with respect to the first tensor factor,\ regarded as a direct summand in $\txT^*(O\x M)$,\ and present also in $\a_\xcE\circ A\colo\txT O(\subset\txT(O\x M))\to\xcE\to\txT M$;
\item the pairing $\corr{\cdot,\cdot}$ is that between $\txT^*M$ (the codomain of $\kappa$) and $\txT M$ (the codomain of $\a_\xcE$).
\eit

\brem\label{rem:kappAi-expl}
In view of the key r\^ole played by the 2-forms \eqref{eq:kappA} in the remainder of our discussion,\ we pause to write out their evaluation on pairs $((v_1,V_1),(v_2,V_2))\in(\txT_\si O\oplus\txT_m M)^{\x 2}\equiv \txT_{(\si,m)}(O\x M)^{\x 2}$.\ This is given by the formul\ae:
\qq\nn
\bigl(\kappa\overset{\wedge}{\circ}A_i\bigr)\bigl((v_1,V_1),(v_2,V_2)\bigr)=\bigl(\kappa\bigl(A_i(v_2)\bigr)\bigr)(V_1)-\bigl(\kappa\bigl( A_i(v_1)\bigr)\bigr)(V_2)\,,
\qqq
and
\qq\nn
\corr{\kappa\circ A_i\,\overset{\wedge}{,}\,\a_\xcE\circ A_i}\bigl((v_1,V_1),(v_2,V_2)\bigr)=\bigl(\kappa\bigl(A_i(v_1)\bigr)\bigr)\bigl(\a_\xcE\bigl(A_i(v_2)\bigr)\bigr)-\bigl(\kappa\bigl(A_i(v_2)\bigr)\bigr)\bigl(\a_\xcE\bigl(A_i(v_1)\bigr)\bigr)\,.
\qqq
\erem
\noindent The second element is captured by the following proposition,\ in which the naturality of the $\rho$-holonomicity constraint \eqref{eq:holo-must} (see:\ Prop.\,\ref{prop:rholo-grp}) becomes apparent in the light of the $S$-twisted $J^1\xcG$-equivariance \eqref{eq:comom-equiv} of the comomentum.
\berop\label{prop:Spencer-on-holo}
Adopt the notation of Defs.\,\ref{prop:rholo-grp},\ \ref{def:Spencer-pair} and \ref{def:J1B}.\ Let $(\kappa,S)$ be the Spencer pair of a multiplicative $k$-form $\rho\in\Om^k(\xcG)$ on a Lie groupoid $\grpd{\xcG}{M}$,\ and let
\qq\nn
\bB_\rho(J^1\xcG):=j^1\bigl(\bB_\rho(\xcG)\bigr)
\qqq
be the homomorphic image of the $\rho$-holonomic group of $\xcG$ in the group $\bB(J^1\xcG)$ of bisections of the first-jet groupoid $\grpd{J^1\xcG}{M}$ of $\xcG$ along the group homomorphism $j^1$ of \Reqref{eq:j1-hom}.\ The restriction of the scalar $k$-Spencer operator $S$ to $\bB_\rho(J^1\xcG)$ vanishes identically,\ {\it i.e.},
\qq\nn
\forall\ \b\in\bB_\rho(\xcG)\ \forall\ m\in M\colo S\bigl(j^1_m\b\bigr)=0\,.
\qqq
Hence,\ the $k$-comomentum $\kappa$ is equivariant with respect to the natural action of $\bB_\rho(J^1\xcG)$ on the tangent algebroid $\xcE$ of $\xcG$ and the exterior bundle $\wedge^\bullet\txT^*M$ over its object manifold $M$.
\eerop
\beroof
Let $\b\in\bB_\rho(\xcG)$,\ and $m\in M$.\ Upon taking into account the first identity in \eqref{eq:tildetC-4-glob},\ definition \eqref{eq:Spencer-expl} yields the desired result:
\qq
S\bigl(j^1_m\b\bigr)&=&\rho_{\b(m)}\circ\wedge^k\bigl(\txT_m\b\circ\bigl(\widetilde t{}_{j^1_m\b}\bigr)^{-1}\bigr)=\rho_{\b(m)}\circ\wedge^k\bigl(\txT_m\b\circ\txT_{t_*\b(m)}\bigl(t_*\b^{-1}\bigr)\bigr)\cr\cr
&\equiv&\bigl(\b^*\rho\bigr)_m\circ\wedge^k\bigl(\txT_{t_*\b(m)}\bigl(t_*\b^{-1}\bigr)\bigr)=0\,.\label{eq:S-0-onholo}
\qqq

Passing to the second statement of the proposition,\ we note that the two actions to be intertwined by the $k$-comomentum can be read off from the (commutative) diagram
\qq\label{eq:Bis-on-E-over-TM}
\alxydim{@C=1.5cm@R=1.cm}{ \xcE \ar[r]^{\unl{\txT C}_\b} \ar[d]_{\a_\xcE} & \xcE \ar[d]^{\a_\xcE}\\ \txT M \ar[d]_{\pi_{\txT M}} \ar[r]_{\txT(t_*\b)} & \txT M \ar[d]^{\pi_{\txT M}} \\ M \ar[r]_{t_*\b} & M}\,,\qquad\b\in\bB_\rho(\xcG)
\qqq
which `resolves' \eqref{eq:Bis-on-E}.\ The claim now follows directly from \eqref{eq:comom-equiv} by a reasoning fully analogous to the one employed above,\ with the help of \emph{both} identities \eqref{eq:tildetC-4-glob} taken in conjunction with the former result \eqref{eq:S-0-onholo}.
\eroof
\noindent The final argument in favour of our choice of the reduction scheme comes from the investigation of the third element.
\berop\label{prop:mult-on-bisec}
Adopt the notation of Defs.\,\ref{prop:rholo-grp} and \ref{def:Spencer-pair},\ and of Prop.\,\ref{prop:beta-split}.\ Let $\rho\in\Om^k(\xcG)$ be multiplicative,\ and let $\kappa$ be its $k$-comomentum.\ Consider an arbitrary $\b\in\bB_\rho(\xcG)$,\ and the splitting $(\chi_\b,\upsilon_\b)\colo\txT_\b\xcG\equiv\txT\xcG\rstr_{\b(M)}\cong\txT\b(M)\oplus\ker\,\txT s\rstr_{\b(M)}$ of $\txT_\b\xcG$ induced by it.\ The restriction of $\rho$ to $\txT_\b\xcG$ is determined by $\kappa$ as
\qq
\rho\rstr_{\b(M)}=\sum_{p=1}^k\,\sum_{\si\in{\rm Shuffle}(p,k-p)}\,(-1)^{|\si|}\,t^*\bigl(\kappa_\b\circ\widehat\si{}_1\bigr)\bigl(\upsilon_\b\circ\widehat\si{}_2,\upsilon_\b\circ\widehat\si{}_3,\ldots,\upsilon_\b\circ\widehat\si{}_p,\chi_\b\circ\widehat\si{}_{p+1},\chi_\b\circ\widehat\si{}_{p+2},\ldots,\chi_\b\circ\widehat\si{}_k\bigr)\,,\cr\cr
\label{eq:rho-on-bisec}
\qqq
where 
\qq\nn
\kappa_\b=\kappa\circ\theta_{\rm R}\circ\upsilon_\b\colo \txT_\b\xcG\too\wedge^{k-1}\txT^*M\,,
\qqq
and where we have used the shorthand notation
\qq\nn
\widehat\si{}_i\colo\txT_\b\xcG^{\oplus k}\too\txT_\b\xcG,\  (X_1,X_2,\ldots,X_k)\longmapsto X_{\si(i)}\,,\quad i\in\ovl{1,k}\,.
\qqq
\eerop
\beroof
The starting point of our discussion is formula \eqref{eq:rho-as-Skappa},\ which expresses the multiplicative $k$-form in terms of it Spencer pair $(\kappa,S)$ at a {\em generic} point in $\xcG$.\ When putting the formula on the bisection $\b(M)$,\ a global choice of a bisection $j^1(\b)$ is available such that,\ at every point $m\in M$,\ the identity $j^1_m\b=j^1(\b)(m)$ obtains,\ which ensures the vanishing of the contribution of the $k$-Spencer operator to $\rho\rstr_{\b(M)}$ in virtue of Prop.\,\ref{prop:Spencer-on-holo}.\ This settles the question conceptually:\ The restriction $\rho\rstr_{\b(M)}$ is fully determined by the $k$-comomentum $\kappa$.\ What remains,\ at this stage,\ is a direct calculation,\ in which \eqref{eq:tildetC-4-glob} is invoked once more (for the aforementioned global choice of the splitting bisection $j^1_\cdot\b$) alongside Rem.\,\ref{rem:more-on-Spencer}:
\qq\nn
&&\bigl(\kappa\circ\theta_{\rm R}\bigr)(\vep_{\si(1)})\bigl(\a_\xcE(\vep_{\si(2)}),\ldots,\a_\xcE(\vep_{\si(p)}),\widetilde t{}_{j^1_m\b}(\unl{X}_{\si(p+1)}),\ldots,\widetilde t{}_{j^1_m\b}(\unl{X}_{\si(k)})\bigr)\cr\cr 
&=&\bigl(\kappa\circ\theta_{\rm R}\bigr)(\vep_{\si(1)})\bigl(\a_\xcE(\vep_{\si(2)}),\ldots,\a_\xcE(\vep_{\si(p)}),\txT_m(t_*\b)(\unl{X}_{\si(p+1)}),\ldots,\txT_m(t_*\b)(\unl{X}_{\si(k)})\bigr)\cr\cr 
&=&t^*\bigl(\kappa\circ\theta_{\rm R}\bigr)(\vep_{\si(1)})\bigl(\vep_{\si(2)},\ldots,\vep_{\si(p)},\txT_m\b(\unl{X}_{\si(p+1)}),\ldots,\txT_m\b(\unl{X}_{\si(k)})\bigr)\cr\cr 
&\equiv&t^*\bigl(\kappa_\b(X_{\si(1)})\bigr)\bigl(\upsilon_\b(X_{\si(2)}),\ldots,\upsilon_\b(X_{\si(p)}),\chi_\b(X_{\si(p+1)}),\ldots,\chi_\b(X_{\si(k)})\bigr)\,,
\qqq
and which thus yields the anticipated result.
\eroof
\noindent The foregoing analysis amply demonstrates the functional compatibility of the two assumptions:\ that of the $\xcG$-equivariance of the 3-form component $H_M$ of the tensorial background of the $\si$-model,\ and that of the $\rho$-holonomic reduction of the structure group,\ and thus emphasises the adequacy of the latter in the present context.

In the next step,\ we address the issue of the invariance of the metric structure.
\bedef\label{def:isom-subgrp}
Adopt the notation of Def.\,\ref{def:bisec}.\ Let $\grpd{\xcG}{M}$ be a Lie groupoid with a metric object manifold $(M,\txg_M)$.\ The {\bf $\txg_M$-isometric group} of $\xcG$ is the subgroup
\qq\nn
\bB_{\txg_M}(\xcG):=\{\ \b\in\bB(\xcG) \quad\vert\quad (t_*\b)^*\txg_M=\txg_M \ \}
\qqq
of the group $\bB(\xcG)$ of (global) bisections of $\xcG$.
\exdef
\noindent The subtlety concealed by the seemingly benign concept is exposed in
\berop\label{prop:iso-2-much}
Adopt the notation of Def.\,\ref{def:isom-subgrp} and Ex.\,\ref{eg:st-act}.\ Let $\grpd{\xcG}{M}$ be a $\bB$-complete Lie groupoid,\ in the sense of Def.\,\ref{def:B-compl},\ with a metric object manifold $(M,\txg_M)$.\ If $\bB_{\txg_M}(\xcG)=\bB(\xcG)$,\ then all orbits $\xcG\mlact m,\ m\in M$ of the target action $\mlact$ of $\xcG$ on $M$ are zero-dimensional,
\qq\nn
\forall\ m\in M\colo\dim(\xcG\mlact m)=0\,.
\qqq
\eerop
\beroof
Write $[m]\equiv\xcG\mlact m$ for the sake of brevity.\ Assume $\dim\,[m]>0$,\ and subsequently choose two distinct points $x_a\in[m],\ a\in\{1,2\}$ alongside the respective non-zero but otherwise {\em arbitrary} tangent vectors $v_a\in\txT_{x_a}[m]$.\ In virtue of Thm.\,\ref{thm:orb-sub},\ there exists an arrow $g\in\xcG_{[m]}^{[m]}$ and a tangent vector $V\in\txT_g\xcG_{[m]}^{[m]}$ with the properties:\ $(t,s)(g)=(x_2,x_1)$ and $\txT_g(t,s)(V)=(v_2,v_1)$.\ By the assumption of $\bB$-completeness,\ we can find $\b\in\bB(\xcG)$ such that $\b(x_1)=g$,\ and then also the tangent vector $\xi:=\txT_{x_1}\b(v_1)\in \txT_g\b(M)$ such that $\txT_g s(\xi)=v_1$.\ We may have $\txT_g t(\xi)\neq v_2$,\ but,\ clearly,\ $V-\xi\in \txT_{\Id_{t(g)}}r_g(\xcE_{\Id_{t(g)}})$,\ and so---upon recalling Thm.\,\ref{thm:LieBis}---we deform $\b$ within $\bB(\xcG)$ (by flowing it along a suitably chosen compactly supported section of $\xcE$) to a new bisection $\widetilde\b\in\bB(\xcG)$ through $\widetilde\b(x_1)=\b(x_1)=g$ with $\txT_g\widetilde\b(v_1)=V$ (see:\ \cite{Filipek:2025MSc} for an explicit construction).\ We then obtain $v_2=\txT_g t(V)=\txT_g(t_*\widetilde\b)(v_1)$,\ which contradicts the original assumption about the arbitrariness of the two vectors as it implies that $v_2$ is an {\rm isometric} image of $v_1$,\ a property manifestly non-invariant under rescalings.
\eroof

While certainly very natural from the point of view of the abstract considerations concerning the (homotopy) modelling of the symmetry quotient $M//\xcG$,\ our treatment of the topological sector of the $\si$-model background has not yet been linked {\em directly} to rigid symmetries of the field theory under consideration.\ This last conceptual barrier is overcome in the concluding statement of the present section,\ in which we also combine the two hitherto independent reduction schemes.
\berop\label{prop:rig-sym-der}
Adopt the notation of Defs.\,\ref{def:sigmod},\ \ref{prop:rholo-grp} and \ref{def:isom-subgrp},\ and of Ex.\,\ref{eg:tanLiealgbrd}.\ Let $\grpd{\xcG}{M}$ be a Lie groupoid with tangent algebroid $\xcE$,\ and let $\rho\in\Om^2(\xcG)$ be a multiplicative BSS extension of the de Rham 3-cocycle $H_M\in Z^3_{\rm dR}(M)$.\ Consider $\vep\in\G_{\rm c}(\xcE)$.\ Whenever the flow of the right-invariant vector field $\iota_{\rm R}^{-1}(\vep)\in\G(\txT\xcG)_{\rm R}$ through the identity bisection $\Id(M)$ remains in
\qq\label{eq:sigma-symm-Bisec}
\bB_\si(\xcG):=\bB_{\txg_M}(\xcG)\cap\bB_\rho(\xcG)\subset\bB(\xcG)
\qqq
in a parameter (`time') range $I_\d\equiv]-\d,\d[\subset\bR$ for some $\d>0$,\ 
the corresponding fundamental vector field $\a_\xcE\circ\vep\in\G(\txT M)$ is a tangential rigid symmetry of the $\si$-model $\xcM_\si$ in the sense of Def.\,\ref{def:rig-symm-sigmod}.
\eerop
\beroof
The flow $\Phi_{\iota_{\rm R}^{-1}(\vep)}$ of the vector field $\iota_{\rm R}^{-1}(\vep)$ defines a path
\qq\nn
\Psi_\vep(\cdot;\cdot):=\Phi_{\iota_{\rm R}^{-1}(\vep)}(\cdot;\cdot)\circ(\id_\bR\x\Id)\colo I_\d\x M\too\xcG,\ (\t,m)\longmapsto\Phi_{\iota_{\rm R}^{-1}(\vep)}\bigl(\t;\Id_m\bigr)\,,\qquad\Psi_\vep\bigl(0;m\bigr)=\Id_m
\qqq
in $\bB_\si(\xcG)$ (see:\ \cite[Rem.\,2.11]{Crainic:2015msp}),\ that is,
\qq\nn
s\circ\Psi_\vep=\pr_2\,,
\qqq
and,\ for all $\tau\in I_\d$,
\qq\nn
\psi[\vep]_\t:=\bigl(t\circ\Psi_\vep\bigr)(\t;\cdot)\in{\rm Isom}(M,\txg_M)\,,\qquad\qquad\Psi_\vep(\t,\cdot)^*\rho=0\,.
\qqq
We have
\qq\nn
\tfrac{\sfd\ }{\sfd\t}\Psi_\vep(\t=0;\cdot)=\vep(\cdot)\,,
\qqq
and
\qq\nn
\tfrac{\sfd\ }{\sfd\t}\psi[\vep]_{\t=0}=\txT_{\Phi_{\iota_{\rm R}^{-1}(\vep)}(0;\cdot)}t\bigl(\tfrac{\sfd\ }{\sfd\t}\Psi_\vep(\t=0;\cdot)\bigr)\equiv\a_\xcE\circ\vep(\cdot)\,.
\qqq

Differentiation of the identity
\qq\nn
\psi[\vep]_\t^*\txg_M=\txg_M
\qqq
with respect to time $\t\in I_\d$ at $\t=0$ yields the first anticipated result:
\qq\nn
\pLie{\a_\xcE\circ\vep}\txg_M=0\,.
\qqq

We pull back identity \eqref{eq:rho-rel-HM} along the smooth map $\Psi_\vep$ and evaluate it on a triple $(\p_\t,\cV_1,\cV_2)$ of vector fields on its domain $I_\d\x M$,\ consisting of the global coordinate field $\p_\t\in\G(\txT\bR)$ and arbitrary $\cV_1,\cV_2\in\G(\txT M)$ (all three being viewed as distinguished sections of $\txT(I_\d\x M)\cong\pr_1^*\txT I_\d\oplus\pr_2^*\txT M$),\ to obtain---with the help of Cartan's formula---the equality
\qq\nn
&&H_M\bigl(\tfrac{\sfd\ }{\sfd\t}\psi[\vep],\txT_2\psi[\vep](\cV_1),\txT_2\psi[\vep](\cV_2)\bigr)=H_M\bigl(\tfrac{\sfd\ }{\sfd\t}\psi[\vep],\txT_2\psi[\vep](\cV_1),\txT_2\psi[\vep](\cV_2)\bigr)-H_M(0,\cV_1,\cV_2)\cr\cr 
&=&\bigl(H_M\circ\wedge^3 \txT(t\circ\Psi_\vep)\bigr)(\p_t,\cV_1,\cV_2)-\bigl(H_M\circ\wedge^3 \txT(s\circ\Psi_\vep)\bigr)(\p_t,\cV_1,\cV_2)=\sfd\bigl(\Psi_\vep^*\rho\bigr)(\p_t,\cV_1,\cV_2)\cr\cr 
&=&\p_\t\bigl(\bigl(\rho\circ\wedge^2 \txT_2\Psi_\vep\bigr)(\cV_1,\cV_2)\bigr)-\cV_1\bigl(\rho\bigl(\tfrac{\sfd\ }{\sfd\t}\Psi_\vep,\txT_2\Psi_\vep(\cV_2)\bigr)\bigr)+\cV_2\bigl(\rho\bigl(\tfrac{\sfd\ }{\sfd\t}\Psi_\vep,\txT_2\Psi_\vep(\cV_1)\bigr)\bigr)\cr\cr 
&&-\bigl(\Psi_\vep^*\rho\bigr)\bigl([\p_\t,\cV_1],\cV_2\bigr)+\bigl(\Psi_\vep^*\rho\bigr)\bigl([\p_\t,\cV_2],\cV_1\bigr)-\bigl(\Psi_\vep^*\rho\bigr)\bigl([\cV_1,\cV_2],\p_\t\bigr)\cr\cr 
&=&-\cV_1\bigl(\rho\bigl(\tfrac{\sfd\ }{\sfd\t}\Psi_\vep,\txT_2\Psi_\vep(\cV_2)\bigr)\bigr)+\cV_2\bigl(\rho\bigl(\tfrac{\sfd\ }{\sfd\t}\Psi_\vep,\txT_2\Psi_\vep(\cV_1)\bigr)\bigr)-\rho\bigl(\txT_2\Psi_\vep\bigl([\cV_1,\cV_2]\bigr),\tfrac{\sfd\ }{\sfd\t}\Psi_\vep\bigr)\,,
\qqq
where both $\txT_2\Psi_\vep$ and $\txT_2\psi[\vep]$ represent derivatives along $M$ with the time ($\t\in I_\d$) held fixed.\ We may,\ next,\ localise the above expression at $\t=0$,\ whereby it yields
\qq\nn H_M\bigl(\a_\xcE\circ\vep,\cV_1,\cV_2\bigr)&=&-\cV_1\bigl(\rho\bigl(\vep,\txT\Id(\cV_2)\bigr)\bigr)+\cV_2\bigl(\rho\bigl(\vep,\txT\Id(\cV_1)\bigr)\bigr)-\rho\bigl(\txT\Id\bigl([\cV_1,\cV_2]\bigr),\vep\bigr)\cr\cr 
&\equiv&-\bigl[\cV_1\bigl(\Id^*\bigl(\imath_\vep\rho\bigr)(\cV_2)\bigr)-\cV_2\bigl(\Id^*\bigl(\imath_\vep\rho\bigr)(\cV_1)\bigr)-\Id^*\bigl(\imath_\vep\rho\bigr)\bigl([\cV_1,\cV_2]\bigr)\bigr]\cr\cr 
&=&-\sfd\bigl(\Id^*\bigl(\imath_\vep\rho\bigr)\bigr)(\cV_1,\cV_2)\equiv-\sfd\bigl(\kappa(\vep)\bigr)(\cV_1,\cV_2)\,,
\qqq
see:\ \eqref{eq:comom-expl}.\ Thus,\ altogether,\ we obtain the desired result
\qq\label{eq:kapep-as-tansymm}
\imath_{\a_\xcE\circ\vep}H_M=-\sfd\kappa_{\a_\xcE\circ\vep}\,,\qquad\qquad\kappa_{\a_\xcE\circ\vep}\equiv\kappa(\vep)\,.
\qqq
\eroof

\beg\label{eg:AMM-1} 
The Cartan 3-form $H_{\rm C}$ on the object manifold $\txG$ of the action groupoid $\txG\,\lx_\Ad\txG$ from Ex.\,\ref{eg:AMM-0} admits a BSS extension $(H_{\rm C},\rho_{\rm AMM},0)$ determined by the multiplicative 2-form $\rho_{\rm AMM}\in\Om^2(\txG\x\txG)$ given by
\qq\nn
\rho_{\rm AMM}(h,g)&=&\kappa_{\rm C}(g,t_a)\wedge\theta_{\rm L}^a(h)-\tfrac{1}{2}\,\imath_{\cK_a(g)}\kappa_{\rm C}(g,t_b)\,\theta_{\rm L}^a(h)\wedge\theta_{\rm L}^b(h)\cr\cr 
&\equiv&\tfrac{\sfk}{4\pi}\,\tr_\ggt\bigl(\txT_e\Ad_g\circ\theta_{\rm L}(h)\wedge\theta_{\rm L}(h)+\theta_{\rm L}(h)\wedge\bigl(\theta_{\rm L}(g)+\theta_{\rm R}(g)\bigr)\bigr)\,,\quad(h,g)\in\txG\x\txG
\qqq
The quadruple $(D(\txG)\equiv\txG\x\txG,\la_{\rm AMM},\rho_{\rm AMM},\mu_{\rm AMM})$,\ with the extra components given by the left action
\qq\nn
\la_{\rm AMM}\colo(\txG\x\txG)\x D(\txG)\too D(\txG),\ \bigl((a,b),(h,g)\bigr)\longmapsto\bigl(a\cdot h\cdot b^{-1},\Ad_b(g)\bigr)\,,
\qqq
and the map 
\qq\nn
\mu_{\rm AMM}\colo D(\txG)\too\txG\x\txG,\ (h,g)\longmapsto\bigl(\Ad_h(g),g^{-1}\bigr)\,,
\qqq
is a prime example of an Alekseev--Malkin--Meinrenken (AMM) quasi-Hamiltonian $\txG\x\txG$-space \cite{Alekseev:1997}.\ The pair $(\txG\,\lx_\Ad\txG,(H_{\rm C},\rho_{\rm AMM}))$ is referred to as the {\bf AMM groupoid} over $\txG$. 

The group ${\rm Bisec}_{\rho_{\rm AMM}}(\txG\,\lx_\Ad\txG)$ of $\rho_{\rm AMM}$-holonomic bisections of the AMM groupoid contains a finite subgroup $\iota(\txG)\subset{\rm Bisec}_{\rho_{\rm AMM}}(\txG\,\lx_\Ad\txG)$ given by the image of the homomorphic embedding
\qq\nn
\iota\colo\txG\too{\rm Bisec}_{\rho_{\rm AMM}}(\txG\,\lx_\Ad\txG),\ h\longmapsto(h,\cdot)\,.
\qqq
Bisections from $\iota(\txG)\cong\txG$ are also manifestly $\txg_{\rm CK}$-isometric,\ which puts $\iota(\txG)$ in the r\^ole of the model of the gauge group of the gauged WZNW $\si$-model of \cite{Gawedzki:2010G,Gawedzki:2012fu}.
\eeg

\brem
It is easy to see that $\iota(\txG)$ is a {\em proper} subgroup of ${\rm Bisec}_{\rho_{\rm AMM}}(\txG\,\lx_\Ad\txG)$,\ 
\qq\nn
\iota(\txG)\subsetneq{\rm Bisec}_{\rho_{\rm AMM}}(\txG\,\lx_\Ad\txG)\,.
\qqq
Indeed,\ consider the (smooth) diagonal map $\D\colo\txG\to\txG\x\txG,\ g\longmapsto(g,g)$.\ The map satisfies the identities $s\circ\D=\id_\txG$ and $t\circ\D=\id_\txG\in\Diff(\txG)$,\ and so $\D\in{\rm Bisec}(\txG\,\lx_\Ad\txG)\setminus\iota(\txG)$.\ It is also trivially $\txg_{\rm CK}$-isometric.\ In order to see that $\D$ is $\rho_{\rm AMM}$-holonomic,\ note that $\txT\D=(\id_{\txT\txG},\id_{\txT\txG})$.\ Hence,
\qq\nn
\D^*\rho_{\rm AMM}(g)&=&\tfrac{\sfk}{4\pi}\,\tr_\ggt\bigl(\txT_e\Ad_g\circ\theta_{\rm L}(g)\wedge\theta_{\rm L}(g)+\theta_{\rm L}(g)\wedge\bigl(\theta_{\rm L}(g)+\theta_{\rm R}(g)\bigr)\bigr)\cr\cr 
&\equiv&\tfrac{\sfk}{4\pi}\,\tr_\ggt\bigl(\theta_{\rm L}(g)\wedge\theta_{\rm L}(g)+\theta_{\rm R}(g)\wedge\theta_{\rm L}(g)+\theta_{\rm L}(g)\wedge\theta_{\rm R}(g)\bigr)\,.
\qqq
However,\ $\tr_\ggt(\theta_{\rm H_1}\wedge\theta_{\rm H_2})=-\tr_\ggt(\theta_{\rm H_2}\wedge\theta_{\rm H_1})$ for all $H_1,H_2\in\{{\rm L},{\rm R}\}$,\ and so---indeed---
\qq\nn
\D^*\rho_{\rm AMM}(g)=0\,,
\qqq
{\it i.e.},
\qq\nn
\D\in{\rm Bisec}_{\rho_{\rm AMM}}(\txG\,\lx_\Ad\txG)\setminus\iota(\txG)\,.
\qqq
\erem

The tensorial analysis presented in this section provides a stepping stone for the higher-geometric considerations of the next section,\ ultimately leading to a fully fledged proposal for a gauge principle for the PAG $\si$-model with Lie-groupoidal rigid symmetry.\ We summarise our findings so far in
\begin{flushleft}
{\bf Assumption (R):}\ Adopt the notation of Prop.\,\ref{prop:rig-sym-der}.\ From now onwards,\ we presuppose the triple $(\grpd{\xcG}{M},\txg_M,(H_M,\rho,0))$ to exhibit a {\em nontrivial} combined {\bf $\txg_M$-isometric/$\rho$-holonomic reduction}\label{ass:red}
\qq\nn
\bB(\xcG)\supsetneq\bB_{\txg_M}(\xcG)\supseteq\bB_\si(\xcG)\,,
\qqq
which puts $\bB_\si(\xcG)=\bB_{\txg_M}(\xcG)\cap\bB_\rho(\xcG)$ in the r\^ole of the rigid groupoidal symmetry group of $\xcM_\si$ in the sense of Def.\,\ref{def:rig-symm-grpd}.
\end{flushleft}

\section{Gauge descent:\ Equivariance resp.\ invariance,\ augmentation,\ and stackiness}\label{sec:EAS}

The conceptual developments and technical results established in the previous section pave the way for a systematic descent of the field theory of interest to the total space of the shadow bundle $\xcF\to\Si$ over the theory's spacetime for an arbitrary symmetry model $\grpd{\xcG}{M}$.\ The exhaustive treatment of the classic gauging mechanism for group-modelled symmetry reported in \cite{Gawedzki:2010rn,Gawedzki:2012fu,Suszek:2012ddg} and recalled in Sec.\,\ref{sec:G-gau-sigmod} provides constructive insight as to how to proceed with the descent in the topologically trivial setting (or locally) in the presence of a multiplicative twist $\rho$ in the equivariant structure on the gerbe $\cG$ of the PAG $\si$-model.\ That said,\ an important and consequential departure from the very special structure present in the prototypical scenario,\ reviewed at length in the opening paragraphs of Sec.\,\ref{sec:reduction},\ is to be noted,\ to wit:\ On the total space of a principaloid bundle $\xcP$,\ there is,\ in general,\ no counterpart\footnote{Note that,\ whenever $\dim\,M>0$,\ the codomain $\txV\xcP$ of a connection 1-form $\Theta\in\Om^1(\xcP,\txV\xcP)$ from Def.\,\ref{def:Ginv-principal-conn} has a model $\txT\xcG$ of a rank {\em strictly larger} than that of the corresponding Lie algebroid $\xcE$ (which it contains over the identity bisection).} of the {\em globally smooth} Lie algebroid-valued 1-form given,\ in the special case $\xcP=P\x M$,\ by the connection 1-form $\pr_1^*\cA\in\Om^1(P\x M)\ox\ggt$.\ Hence,\ there is no {\em direct} lift of the obvious (structurally trivial) Lie-group{\em oidal} adaptation of the {\em local} augmentation mechanism \eqref{eq:Augrb} to the total space of the principaloid bundle,\ which---as emphasised in Sec.\,\ref{sec:reduction}---constitutes a natural point of departure for the descent,\ to the smooth quotient $\xcF\cong\xcP/\xcG$,\ of the (augmented) pullback $\mu^*\cG$ of the $\si$-model gerbe $\cG$ along the moment map $\mu\colo\xcP\to M$.\ Thus,\ we arrive at the disjunction:\ Either the multiplicative twist in the latter structure vanishes identically,\ $\rho\equiv 0$,\ so that we may try to descend the gerbe along $\xcD$ without involving the gauge field,\ or we have to work directly with the local augmentations over elements of a trivialising open cover $\cO=\{O_i\}_{i\in I}$ of $\Si$,\ where we have access to local models $O_i\x M$ of $\xcF$.\ As it turns out,\ the induction of a gerbe on $\xcF$ in {\em both} cases is governed by the same categorial principle of (2-)stacky descent,\ originally worked out by Stevenson in his Ph.D.\ thesis \cite{Stevenson:2000wj} in direct analogy with Brylinski's precursor result for principal $\bC^\x$-bundles \cite{Brylinski:1993ab}.\ It was subsequently abstracted and elaborated by Nikolaus and Schweigert in \cite{Nikolaus:2011ehg}.\ Although its application in the case of an arbitrary twist $\rho$ clearly subsumes the non-generic case $\rho=0$,\ we nevertheless discuss the latter,\ structurally `neater' scenario,\ thereby getting an opportunity to employ the said principle in its both (equivalent) formulations:\ the one for surjective submersions,\ and the one for open covers.\smallskip

The fundamental result of gerbe theory---concerning the existence of Breen's 2-stack structure on gerbes,\ which largely organises our subsequent considerations---is stated in
\bethe\cite{Stevenson:2000wj}\label{thm:descyk}
Given $Y,X\in\Man$,\ and a surjective submersion $\pi\colo Y\twoheadrightarrow X$,\ there exists a canonical equivalance
\qq\nn
\bgrb_\nabla(X)\cong\gt{Desc}(\pi)
\qqq
between the bicategory of (bundle) gerbes from Thm.\,\ref{thm:bgrb-bicat},\ and the descent bicategory from Def.\,\ref{def:desc-bicat}.
\ethe 
\noindent A practical consequence of the above,\ central to our field-theoretic construction,\ is extracted in 
\becor\label{cor:desc-ind}
Adopt the notation of Def.\,\ref{def:desc-bicat} and Thm.\,\ref{thm:bgrb-bicat}.\ Every 0-cell $(\cG,\Phi,\varphi)\in\gt{Desc}(\pi)$ canonically induces a gerbe $\unl\cG\in\bgrb_\nabla(X)$ for which there exists a 1-cell 
\qq\nn
(\Psi,\psi)\colo\bigl(\pi^*\unl\cG,\Id_{\pr_1^*\pi^*\unl\cG},\la_{\Id_{\pr_1^*\pi^*\unl\cG}}\bigr)\xrightarrow{\ \cong\ }\bigl(\cG,\Phi,\varphi\bigr)
\qqq
in $\gt{Desc}(\pi)$,\ written in terms of the left unitor $\la_{\Id_{\pr_1^*\pi^*\unl\cG}}\colo\Id_{\pr_1^*\pi^*\unl\cG}\circ\Id_{\pr_1^*\pi^*\unl\cG}\xrightarrow{\cong}\Id_{\pr_1^*\pi^*\unl\cG}$ of $\bgrb_\nabla(Y)$.
\ecor
\noindent The principle of (2-)stacky descent reflected in these results finds its most straightforward application in the non-generic situation in which the gerbe of the $\si$-model (on $M$) is endowed with an untwisted $\xcG$-equivariant structure,\ so that---in particular---its curvature obeys the strong condition $\D^{(3)}_{(0)}H_M=0$.\ We then also obtain the identity
\qq\nn
\bB_{\rho=0}(\xcG)\equiv\bB(\xcG)\,,
\qqq
which reflects the amenability of {\em all} bisections of the structure groupoid to non-anomalous gauging (in the topological sector).\ Moreover,\ the descent passes through the principaloid bundle,\ as in the prototypical construction of \cite{Gawedzki:2010rn}.
\bethe[Unaided gauge descent of gerbes]
Adopt the notation of Def.\,\ref{def:Gequiv-grb} and Prop.\,\ref{prop:Ups-ind}.\ Let $\grpd{\xcG}{M}$ be a Lie groupoid,\ and let $\xcF$ be the shadow of a principaloid $\xcG$-bundle $\xcP$ with structure group $\bB(\xcG)$.\ Every $\xcG$-equivariant gerbe $(\cG,\Upsilon,\g)$ canonically induces a gerbe $\unl\cG\in\bgrb_\nabla(\xcF)$ with the following property
\qq\label{eq:gauge-inv-indir}
\forall\ \Phi\in{\rm Gauge}(\xcP)\colo\xcF_*(\Phi)^*\unl\cG\cong\unl\cG\,.
\qqq
\ethe
\beroof
The theorem is a simple corollary to the more general Thm.\,\ref{thm:gauge-desc-aid},\ which shall be stated and proven presently.\ Therefore,\ here,\ we restrict to laying out the logic of the present proof,\ which is---essentially---that of (the proof of) \cite[Thm.\,5.9]{Gawedzki:2010rn},\ and,\ as such,\ is largely independent of that of the constructive proof of the general theorem.

In virtue of Prop.\,\ref{prop:simpl-ext-princmom},\ the principality of the defining action $\varrho$ of the structure groupoid $\xcG$ on $\xcP$ and the submersive surjectivity of the moment map $\mu\colo\xcP\to M$ imply the existence of the simplicial epimorphism $\vep_\xcP$ as in \eqref{eq:simpl-epi-ext}.\  The pullback of $(\xcG,\Upsilon,\g)$ along this epimorphism yields a descent datum 
\qq\nn
\vep_\xcP^*(\cG,\Upsilon,\g)\equiv\bigl(\mu^*\cG,\phi_\xcP^*\Upsilon,(\phi_\xcP\circ\pr_{1,2},\phi_\xcP\circ\pr_{2,3})^*\g\bigr)\in{\gt Desc}(\xcD)\,.
\qqq 
This is so due to the strict 2-functoriality of pullback (see:\ \cite{Waldorf:2007mm}),\ which implies
\qq\nn
\phi_\xcP^*\Upsilon\colo\pr_1^*\bigl(\mu^*\cG\bigr)=\phi_\xcP^*t^*\cG\xrightarrow{\ \cong\ }\phi_\xcP^*s^*\cG=\pr_2^*\bigl(\mu^*\cG\bigr)\,,
\qqq
and
\qq\nn
&&(\phi_\xcP\circ\pr_{1,2},\phi_\xcP\circ\pr_{2,3})^*\g\colo\pr_{2,3}^*\bigl(\phi_\xcP^*\Upsilon\bigr)\circ\pr_{1,2}^*\bigl(\phi_\xcP^*\Upsilon\bigr)\xLongrightarrow{\ \cong\ }\pr_{1,3}^*\bigl(\phi_\xcP^*\Upsilon\bigr)\,,
\qqq
the latter 2-isomorphism inheriting coherence from $\g$.\ Above,\ we have used the simpliciality of $\vep_\xcP$.\ The descent datum induces a gerbe $\unl\xcG\in\bgrb_\nabla(\xcF)$ by Thm.\,\ref{cor:desc-ind}.

The gauge invariance of the induced gerbe $\unl\xcG$ can now be proven along the lines of (the proof of) \cite[Thm.\,5.9]{Gawedzki:2010rn},\ given for the case of the action groupoid $\txG\,\lx_\la M$:\ In consequence of the $\xcG$-equivariance of $\Phi$,\ the existence of the 1-isomorphism \eqref{eq:gauge-inv-indir} is ensured by that of an isomorphism
\qq\nn
\Phi_\bullet^*\vep_\xcP^*(\cG,\Upsilon,\g)\cong\vep_\xcP^*(\cG,\Upsilon,\g)\,,
\qqq 
in which $\Phi_\bullet^*$ represents a natural simplicial lift of $\Phi^*$,\ with components $\Phi^*_n=(\Phi^{\x n+1})^*$.\ The meaningfulness of its application to the simplicial extension $\vep_\xcP^*(\cG,\Upsilon,\g)$ hinges,\ once more,\ on the strict 2-functoriality of pullback.

That the desired isomorphism (or,\ indeed,\ an equality) exists now follows from the $\la_\xcP$-invariance of the simplicial extension $\vep_\xcP$ of the manifestly $\la_\xcP$-invariant moment map $\mu$ (see:\ Thm.\,\ref{thm:princ-as-At-mod}).\ This ultimately stems from the $\la_\xcP$-invariance of the division map $\phi_\xcP$ of the principaloid bundle,\ which is readily demonstrated in its local presentation \eqref{eq:div-maP-loc} (see:\ Cor.\,\ref{cor:Requiv-principoidle-Auts}).
\eroof

In the presence of a non-vanishing twist,\ $\rho\neq 0$,\ the mechanism sketched above breaks down,\ and we need to couple the gerbe to the gauge field provided by the principaloid bundle $\xcP$ in order to remove the obstruction to its descent to $\xcF$.\ The stage for the connection-aided gauging of the gerbe's character is set in
\bedef\label{def:connaugerbe}
Adopt the notation of Defs.\,\ref{def:principoidle-conn-cech},\ \ref{prop:rholo-grp} and \ref{def:Spencer-pair},\ and of Thm.\,\ref{thm:bgrb-bicat}.\ Let $\grpd{\xcG}{M}$ be a Lie groupoid,\ and let $\rho\in\Om^2(\xcG)$ be a multiplicative 2-form with the corresponding $2$-comomentum $\kappa\colo\xcE\to\txT^* M$.\ Furthermore,\ let $\xcP\to\Si$ be a principaloid $\xcG$-bundle with connection 1-form $\Theta$ over a manifold $\Si$,\ with the structure group reduced to the $\rho$-holonomic group $\bB_\rho(\xcG)$,\ and let $\pi_\xcF\colo\xcF\to\Si$ be its shadow.\ Gauge transformations of $\xcP$ locally modelled on $L(\bB_\rho(\xcG))$ shall be called {\bf $\rho$-holonomic} and denoted as
\qq\nn
{\rm Gauge}_\rho(\xcP)\,.
\qqq

Fix a trivialising open cover $\cO=\{O_i\}_{i\in I}$ of $\Si$,\ together with the corresponding local trivialisations $\xcF\t_i\colo \pi_\xcF^{-1}(O_i)\xrightarrow{\cong}O_i\x M$ and local data $\{A_i\}_{i\in I}$ of $\Theta$.\ Given a gerbe $\cG\in\bgrb_\nabla(M)$ of curvature $H_M\in Z^3_{\rm dR}(M)$ which is BSS-extended by $\rho$,\ the {\bf connection-augmented gerbe} is the gerbe $\widetilde\cG[\Theta]\in\bgrb_\nabla(\pi_\xcF^{-1}\check Y{}_\cO)$ over the manifold $\pi_\xcF^{-1}\check Y{}_\cO\equiv\bigsqcup_{i\in I}\,\pi_\xcF^{-1}(O_i)$ (the pullback cover) with restrictions
\qq\nn
\widetilde\cG[A_i]\equiv\widetilde\cG[\Theta]\rstr_{\pi_\xcF^{-1}(O_i)}:=\xcF\t_i^*\bigl(\pr_2^*\cG\ox\cI_{\kappa[A_i]}\bigr)\,,
\qqq
in which the curving $\kappa[A_i]\in\Om^2(O_i\x M)$ of the trivial tensor factor is given by the formula (see:\ \eqref{eq:kappA})
\qq\label{eq:kappAi}
\kappa[A_i]=\kappa\overset{\wedge}{\circ}A_i+\tfrac{1}{2}\,\corr{\kappa\circ A_i\,\overset{\wedge}{,}\,\a_\xcE\circ A_i}\,.
\qqq
\exdef
\noindent Accordingly,\ the gauging circumnavigates,\ in this case,\ the total space of the principaloid bundle and starts directly over a covering family of trivialisations of the shadow.\ There,\ it calls for a different (albeit equivalent) formulation of the principle of stacky descent,\ which bases on the following specialisation of the formerly introduced notion of descent bicategory.
\bedef\cite{Stevenson:2000wj}\label{def:Cech-desc}
Adopt the notation of Def.\,\ref{eg:Cech-nerve} and Thm.\,\ref{thm:bgrb-bicat}.\ Let $X\in\Man$,\ and let $\cU=\{U_i\}_{i\in I}$ be an open cover of $X$.\ Over the manifold $\check{Y}_\cU:=\bigsqcup_{i\in I}\,U_i$,\ there arises the {\bf \v Cech descent bicategory} $\gt{Desc}(\check{\pi})$:\ Its 0-cells---{\bf \v Cech descent data}---are triples 
\qq\nn
\check{\cG}\equiv\bigl(\{\cG_i\}_{i\in I},\{\Phi_{ij}\}_{i,j\in I},\{\varphi_{ijk}\}_{i,j,k\in I}\bigr)
\qqq
composed of $I$-indexed families of
\bit
\item gerbes $\cG_i\in\bgrb_\nabla(U_i)$;
\item gerbe 1-isomorphisms $\Phi_{ij}\colo\cG_i\rstr_{U_{ij}}\xrightarrow{\cong}\cG_j\rstr_{U_{ij}}$;
\item gerbe 2-isomorphisms $\varphi_{ijk}\colo\Phi_{jk}\circ\Phi_{ij}\rstr_{U_{ijk}}\xLongrightarrow{\cong}\Phi_{ik}\rstr_{U_{ijk}}$
\eit 
satisfying the coherence identities
\qq\nn
\varphi_{ikl}\bullet\bigl(\id_{\Phi_{kl}}\circ\varphi_{ijk}\bigr)\rstr_{U_{ijkl}}=\varphi_{ijl}\bullet\bigl(\varphi_{jkl}\circ\id_{\Phi_{ij}}\bigr)\rstr_{U_{ijkl}}\,.
\qqq
Its 1-cells---{\bf coherent 1-isomorphisms between \Cv ech descent data} $\check{\cG}^a,\ a\in\{1,2\}$---are pairs 
\qq\nn
\check{\Psi}\equiv\bigl(\{\Psi_i\}_{i\in I},\{\psi_{ij}\}_{i,j\in I}\bigr)
\qqq
composed of $I$-indexed families of
\bit
\item gerbe 1-isomorphisms $\Psi_i\colo\cG_i^1\xrightarrow{\cong}\cG_i^2$;
\item gerbe 2-isomorphisms $\psi_{ij}\colo\Psi_j\circ\Phi_{ij}^1\rstr_{U_{ij}}\xLongrightarrow{\cong}\Phi_{ij}^2\circ\Psi_i\rstr_{U_{ij}}$,
\eit
satisfying the coherence identities
\qq\label{eq:psi-phi-coh}
\psi_{ik}\bullet\bigl(\id_{\Psi_k}\circ\varphi_{ijk}^1\bigr)\rstr_{U_{ijk}}=\bigl(\varphi_{ijk}^2\circ\id_{\Psi_i}\bigr)\bullet\bigl(\id_{\Phi_{jk}^2}\circ\psi_{ij}\bigr)\bullet\bigl(\psi_{jk}\circ\id_{\Phi_{ij}^1}\bigr)\rstr_{U_{ijk}}\,.
\qqq
Finally,\ its 2-cells---{\bf coherent 2-isomorphisms between} ({\bf coherent}) {\bf 1-isomorphisms} $\check{\xi}\colo\check{\Psi}{}^1\overset{\cong}{\Rightarrow}\check{\Psi}{}^2$ in $\gt{Desc}(\check\pi)(\check{\cG}{}^1,\check{\cG}{}^2)\ni\check{\Psi}{}^1,\check{\Psi}{}^2$---are $I$-indexed families gerbe 2-isomorphisms 
\qq\nn
\check{\xi}\equiv\{\ \xi_i\colo\Psi^1_i\xLongrightarrow{\ \cong\ }\Psi^2_i\ \}_{i\in I}
\qqq
satisfying the coherence identities
\qq\nn
\bigl(\id_{\Phi^2_{ij}}\circ\pr_1^*\xi_i\bigr)\bullet\psi^1_{ij}\rstr_{U_{ij}}=\psi^2_{ij}\bullet\bigl(\pr_2^*\xi_j\circ\id_{\Phi^1_{ij}}\bigr)\rstr_{U_{ij}}\,.
\qqq
\exdef
\noindent The last definition paves the way---{\it via} the corresponding specialisation of Thm.\,\ref{thm:descyk}---to the fundamental
\bethe[Connection-aided gauge descent of gerbes]\label{thm:gauge-desc-aid}
Adopt the assumptions and the notation of Defs.\,\ref{def:gauge-trafo-conn},\ \ref{def:connaugerbe},\ \ref{def:Cech-desc} and \ref{def:Gequiv-grb},\ and of Prop.\,\ref{prop:Ups-ind}.\ Consider the canonical surjective submersion 
\qq\label{eq:pull-bckover-ssubm}
\check{\pi}_\xcF\colo\pi_\xcF^{-1}\check Y{}_\cO\too\xcF,\ (f,i)\longmapsto f\,.
\qqq 
A $\rho$-twisted $\xcG$-equivariant structure $(\Upsilon,\g)$ on $\cG$ canonically defines an extension of the gerbe $\widetilde\cG[\Theta]\in\bgrb_\nabla(\pi_\xcF^{-1}\check Y{}_\cO)$ to a \v{C}ech descent datum
\qq\label{eq:desc-dat}
\bigl(\{\widetilde\cG[A_i]\}_{i\in I},\{\widetilde\Phi{}_{ij}\}_{i,j\in I},\{\widetilde\varphi{}_{ijk}\}_{i,j,k\in I}\bigr)\in\gt{Desc}(\check{\pi}_\xcF)\,.
\qqq
Hence,\ there exists a gerbe $\unl{\cG[\Theta]}\in\bgrb_\nabla(\xcF)$ such that $\unl{\cG[\Theta]}\rstr_{\pi_\xcF^{-1}(O_i)}\cong\widetilde\cG[A_i],\ i\in I$ (coherently).\ The descended gerbe has the property
\qq\label{eq:gauge-inv-descgrb}
\forall\ \Phi\in{\rm Gauge}_\rho(\xcP)\colo\xcF_*(\Phi)^*\unl{\cG[\Theta^\Phi]}\cong\unl{\cG[\Theta]}\,.
\qqq
\ethe
\beroof
See:\ App.\,\ref{app:proof-thm-aidesc}.
\eroof
\noindent Thus,\ through a subtle interplay between the affine transformation law for the $\xcE$-valued local gauge field on the principaloid bundle over the spacetime of the $\si$-model with an intrinsic bicategorial symmetry property of its gerbe---namely,\ its $\rho$-twisted equivariance---and the corresponding ($\rho$-holonomic) reduction of the group of diffeomorphisms,\ which ensures coherent preservation of the class of the gerbe in its presence and (restricted) equivariance of the induced comomentum map,\ we arrive at a definition of a gerbe over the configuration bundle of the $\si$-model with the groupoidal symmetry (implemented by) $\bB_\rho(\xcG)$ gauged.\ This is,\ clearly,\ a tightly knit geometric structure,\ whose naturality and indispensability in the present context is further corroborated and amplified by a refinement of Thm.\,\ref{thm:CSS},\ also due to Crainic {\it et al.} (see:\ \cite[Cor.\,3.18]{Crainic:2015msp},\ but also \cite[Thm.\,2.5]{Bursztyn:2004}),\ which establishes a one-to-one correspondence between multiplicative forms and their comomenta on source-simply connected Lie groupoids.\ Such groupoids are,\ in turn,\ uniquely associated with the underlying (integrable) Lie algebroids $\xcE$ in virtue of Lie's First Theorem for Lie algebroids as the corresponding {\bf monodromy groupoids}\label{quote:monodromy} (see:\ \cite[Prop.\,3.3]{Moerdijk:2002},\ and also \cite{Crainic:2003}),\ and so are,\ in particular,\ natural integrated/large-symmetry models for field theories with Lie-algebroidal tangential symmetries (as,\ {\it e.g.},\ the Poisson $\si$-model of \cite{Ikeda:1993fh,Schaller:1994es}).\ We shall invoke Thm.\,\ref{thm:gauge-desc-aid} in the formulation of the fully fledged gauged PAG $\si$-model in Sec.\,\ref{sec:gau-sigmod},\ after an in-depth analysis---carried out in the next section---of the cohomology behind the gauge principle emerging from our considerations.\smallskip

At this stage,\ we still need to deal with the metric term.\ In this case,\ we simply specialise (for $X=M$) the minimal-coupling recipe for invariant tensors on $M$.\ This yields
\bedef\label{def:connaumetr}
Adopt the notation of Defs.\,\ref{def:principoidle-conn-cech} and \ref{def:isom-subgrp}.\ Let $\grpd{\xcG}{M}$ be a Lie groupoid,\ and let $\txg_M\in\G(S^2\txT^*M)$ be a metric tensor on it object manifold.\ Furthermore,\ let $\xcP\to\Si$ be a principaloid $\xcG$-bundle with connection 1-form $\Theta$ over a manifold $\Si$,\ with the structure group reduced to the $\txg_M$-isometric group $\bB_{\txg_M}(\xcG)$,\ and let $\pi_\xcF\colo\xcF\to\Si$ be its shadow.\ Gauge transformations of $\xcP$ locally modelled on $L(\bB_{\txg_M}(\xcG))$ shall be called {\bf $\txg_M$-isometric} and denoted as
\qq\nn
{\rm Gauge}_{\txg_M}(\xcP)\,.
\qqq

Fix a trivialising open cover $\cO=\{O_i\}_{i\in I}$ of $\Si$,\ together with the corresponding local trivialisations $\xcF\t_i\colo \pi_\xcF^{-1}(O_i)\xrightarrow{\cong}O_i\x M$ and local data $\{A_i\}_{i\in I}$ of $\Theta$.\ The {\bf connection-augmented metric} is the tensor $\widetilde\txg{}_M[\Theta]\in\G(S^2\txT^*\pi_\xcF^{-1}\check Y{}_\cO)$ over the manifold $\pi_\xcF^{-1}\check Y{}_\cO\equiv\bigsqcup_{i\in I}\,\pi_\xcF^{-1}(O_i)$ with restrictions
\qq\nn
\widetilde\txg{}_M[\Theta]\rstr_{\pi_\xcF^{-1}(O_i)}=\xcF\t_i^*\bigl(\txg_M\circ\bigl(\xcF\t_i^{-1\,*}\Theta_\xcF\bigr)^{\ox 2}\bigr)\equiv\xcF\t_i^*\bigl(\txg_M\circ\bigl(\id_{\txT M}-\a_\xcE\circ A_i\bigr)^{\ox 2}\bigr)\in\G\bigl(S^2\txT^*\pi_\xcF^{-1}(O_i)\bigr)\,.
\qqq
\exdef
\noindent Thus prepared,\ we formulate the crucial
\berop\label{prop:desc-metr}
Adopt the notation of Defs.\,\ref{def:connaumetr} and \ref{def:gauge-trafo-conn},\ and of Prop.\,\ref{prop:Ups-ind},\ and consider the canonical surjective submersion $\check\pi{}_\xcF$,\ given in \Reqref{eq:pull-bckover-ssubm}.\ The connection-augmented metric $\widetilde\txg{}_M[\Theta]\in\G(S^2\txT^*\pi_\xcF^{-1}\check Y{}_\cO)$ canonically induces---in a sheaf-theoretic sense---a tensor $\unl{\txg_M[\Theta]}\in\G(S^2\txT^*\xcF)$ such that
\qq\nn
\check\pi{}_\xcF^*\unl{\txg_M[\Theta]}=\widetilde\txg{}_M[\Theta]\,.
\qqq
The descended tensor has the property
\qq\label{eq:gauge-inv-descmetr}
\forall\ \Phi\in{\rm Gauge}_{\txg_M}(\xcP)\colo\xcF_*(\Phi)^*\unl{\txg_M[\Theta^\Phi]}=\unl{\txg_M[\Theta]}\,.
\qqq
\eerop
\beroof 
A simple consequence of Prop.\,\ref{prop:loc-pres-cov-der}.
\eroof  

\beg\label{eg:AMM-3}
In the case of the AMM groupoid $(\txG\,\lx_\Ad\txG,(H_{\rm C},\rho_{\rm AMM}))$,\ and with the structure group reduced to the subgroup $\iota(\txG)\subset{\rm Bisec}_{\rho_{\rm AMM}}(\txG\,\lx_\Ad\txG)$ from Ex.\,\ref{eg:AMM-1},\ the augmentation of a $\rho_{\rm AMM}$-twisted $\txG\,\lx_\Ad\txG$-equivariant gerbe $\cG_\sfk$ can be defined over the total space $\sfP\x\txG$ of the surjective submersion $\pi_\sim\colo\sfP\x\txG\to\sfP\x_\Ad\txG$ determined by the principal $\txG$-bundle with connection $(\sfP,\cA)$ from Ex.\,(AMM-2),\ whereby the product gerbe $\widetilde\cG{}_\sfk[\cA]\equiv\pr_2^*\cG_\sfk\ox\cI_{\kappa_{\rm C}[\cA]}$ (see:\ \eqref{eq:Augrb}) arises,\ with augmentation $\kappa_{\rm C}[\cA]$ as in \eqref{eq:Augmntn},\ see:\ \cite{Gawedzki:2010rn}.\ A descent construction---fully analogous to the one performed in this section over the nerve of the surjective submersion $\pi_\xcF^{-1}\check Y{}_\cO$---can now be carried out over $N_\bullet({\rm Pair}_{\pi_\sim}(\sfP\x\txG))$ along the lines of the general Def.\,\ref{def:Gequiv-grb} and Thm.\,\ref{thm:descyk}.\ The construction yields a descended gerbe $\unl{\cG{}_\sfk[\cA]}$ over the total space of the adjoint bundle $\sfP\x_\Ad\txG\equiv\Ad\,\sfP$,\ on which gauge transformations are realised by 1-isomorphisms (preserving the holonomy).

Similarly,\ for the Cartan--Killing metric $\txg_{\rm CK}$,\ we readily establish $\iota(\txG)\subset{\rm Bisec}_{\txg_{\rm CK}}(\txG\,\lx_\Ad\txG)$,\ and so there arises a descended metric $\unl{\txg_{\rm CK}[\cA]}$ over $\Ad\,\sfP$,\ invariant under gauge transformations.
\eeg

\section{The gauge anomaly,\ and inequivalent gaugings}\label{sec:anomaly} 

The principle underlying our proposal for the connection-aided descent of the topological component of a prequantisable PAG $\si$-model to the shadow bundle is $\xcG$-equivariance.\ It was first assumed on the tensorial level,\ where it gave rise to a multiplicative BSS extension of the curvature of the Cheeger--Simons character in the DF amplitude,\ and was subsequently lifted to the higher-geometric object $\cG$ determining the latter as its holonomy.\ The requirement of the existence of the extra structure on $\cG$ is far from innocent,\ or structureless---indeed,\ in general,\ there is an obstruction to it,\ also known as the {\bf gauge anomaly},\ which admits a precise cohomological quantification.\ As usual in the theory of extensions of (algebro-)geometric structures,\ the cohomology containing the obstruction class also classifies---lower in the cohomology degree---inequivalent extensions,\ whose physical relevance hinges on the fact that they correspond to quantum-mechanically inequivalent gauged field theories.\ This justifies the attention devoted to a detailed study of the gauge anomaly in the present section,\ prior to formulating the definition of the gauged field theory in the case of a vanishing anomaly.\smallskip

We begin our discussion with the simpler,\ tensorial component of the anomaly.
\bedef\label{def:SGA}
Adopt the notation of Defs.\,\ref{def:rig-symm-grpd} and \ref{def:BSS-xt}.\ Let $\xcM_\si$ be a closed two-dimensional nonlinear $\si$-model with rigid-symmetry model $\,\grpd{\xcG}{M}$.\ The {\bf small gauge anomaly} of $\xcM_\si$ for $\,\grpd{\xcG}{M}$ is the obstruction to the existence of a multiplicative BSS extension $(H_M,\rho,0)$ of the curvature $H_M$.
\exdef
\noindent The small gauge anomaly controls the behaviour of the holonomy of the augmented gerbe $\widetilde\cG[A_i]$ under {\em linearised} (tangential) gauge transformations,\ represented by sections of the Lie algebroid $\xcE$ over a patch $O_i\subset\Si$ of the worldsheet embedded as $\varphi_i(O_i)$ in $M$ by a local presentation $\varphi_i=\pr_2\circ\xcF\t_i\circ\varphi\rstr_{O_i}\in C^\infty(O_i,M)$ of a matter field $\varphi\in\G(\xcF)$ in a local trivialisation $\xcF\t_i$ of $\xcF$,\ see:\ \cite[Prop.\,3.1]{Gawedzki:2010rn}.\ The sections in question are to be taken from the tangent Lie algebra---in the sense of \cite{Rybicki:2002ALG}---of the reduced structure group $\bB_\rho(\xcG)$,\ or---in the field-theoretic setting---of $\bB_\si(\xcG)$).\ Whenever they come from a Lie sub-algebroid $\xcE_\rho\subset\xcE$ (resp.\ $\xcE_\si\subset\xcE$)---which is a necessary condition for the application of the associating construction of Sec.\,\ref{sec:red-ass} in the modelling of the reduction---the small gauge anomaly can equivalently be understood as an obstruction to having a homomorphic realisation of $\xcE_\rho$ (resp.\ $\xcE_\si$),\ furnished by $\chi\equiv(\a_\xcE,\kappa)$ (the anchor $\a_\xcE$ of $\xcE$ together with the 2-comomentum $\kappa$ of the multiplicative form $\rho$),\ within the Courant algebroid $\gt{C}_\si\equiv(\txE^{1,1}M,\GBra{\cdot}{\cdot}_{\rm C}^{H_M},\corr{\cdot,\cdot},\pr_1)$ on Hitchin's generalised tangent bundle $\txE^{1,1}M=\txT M\oplus\txT^*M$ over $M$,\ equipped with the $H_M$-twisted Courant bracket \eqref{eq:SWC-bracket}.\ (This is closely related to an $H_M$-twisted Dirac structure over $M$ considered in \cite{Bursztyn:2004},\ with only the condition of the maximality of the isotropic structure relaxed.) Indeed,\ as shown in \cite[Prop.\,3.5]{Bursztyn:2004} (see also:\ \cite[Thm.\,3.1]{Bursztyn:2012}),\ the 2-comomentum $\kappa$ of a multiplicative 2-form $\rho$ extending $H_M$ satisfies
\qq\nn
\kappa\circ[\vep_1,\vep_2]_\xcE&=&\pLie{\a_\vep\circ\vep_1}(\kappa\circ\vep_2)-\imath_{\a_\vep\circ\vep_2}\sfd(\kappa\circ\vep_1)-\imath_{\a_\xcE\circ\vep_2}\imath_{\a_\xcE\circ\vep_1}H_M\,,
\qqq
and so the identity
\qq\nn
\GBra{\chi\circ\vep_1}{\chi\circ\vep_2}_{\rm C}^{H_M}=\chi\circ[\vep_1,\vep_2]_\xcE\,.
\qqq
follows.\ This algebroidal interpretation of the small gauge anomaly in the PAG $\si$-model was first given\footnote{The interpretation was inspired by the earlier identification---due to Alekseeev and Strobl \cite{Alekseev:2004np}---of the Courant algebroid $\gt{C}_\si$ as the target-space structure governing the Poisson algebra of Noether symmetry currents in the $\si$-model.} in \cite{Suszek:2012ddg},\ in the context of the long-known relation between Hitchin's generalised geometry and gerbes (see also:\ \cite{Gawedzki:2012fu}).

\subsection{The (large) gauge-anomaly classes}\label{sub:gauge-anomaly}

The second part of the proof of Thm.\,\ref{thm:gauge-desc-aid},\ given in App.\,\ref{app:proof-thm-aidesc},\ demonstrates convincingly that the stronger requirement of invariance of the holonomy of the augmented gerbe $\widetilde\cG[A_i]$ under arbitrary integrated ({\it i.e.},\ large) gauge transformations with model $\bB_\rho(\xcG)$ calls for the existence of the 1-isomorphism $\Upsilon$,\ and the structural demand of descendability of the family $\{\widetilde\cG[A_i]\}_{i\in I}$ to the shadow $\xcF$ entails---by the same theorem---a fully fledged geometrisation of the BSS extension $(H_M,\rho,0)$ in the form of a $\rho$-twisted $\xcG$-equivariant structure on $\cG$.\ This leads us to
\bedef\label{def:LGA}
Adopt the notation of Defs.\,\ref{def:SGA},\ \ref{def:holG} and \ref{def:Gequiv-grb}.\ Let $(H_M,\rho,0)$ be a multiplicative BSS extension of the curvature $H_M$ which lifts the small gauge anomaly of $\xcM_\si$ for $\,\grpd{\xcG}{M}$.\ Furthermore,\ let $\cG\in\bgrb_\nabla(M)$ be a gerbe of curvature $H_M$ which defines the topological term $\chi_{H_M}\equiv\Hol_\cG$ in $\cA_{\rm DF}$.\ The {\bf large gauge anomaly} of $\xcM_\si$ for $\grpd{\xcG}{M}$ is the (homological) obstruction to the existence of a $\rho$-twisted $\xcG$-equivariant structure $(\Upsilon,\g)$ on $\cG$.
\exdef
\noindent On the first level,\ relevant to both:\ the gauge invariance and the descent to $\xcF$,\ the anomaly admits a neat higher-geometric presentation given in
\berop\label{prop:Ups-obstr}
Adopt the notation of Def.\,\ref{def:LGA} and Thm.\,\ref{thm:W3-tors}.\ The obstruction to the existence of a 1-isomorphism $\Upsilon$ is given by the isomorphism class 
\qq\nn
[t^*\cG\ox s^*\cG^*\ox\cI_{-\rho}]\in\cW^3(\xcG,0)\cong H^2\bigl(\xcG,\uj\bigr)\,.
\qqq
\eerop
\beroof
The vanishing of the curvature of $t^*\cG\ox s^*\cG^*\ox\cI_{-\rho}$ is implied by the existence of an extension $[(H_M,\rho,0)]$ of the class $[H_M]$ of the curvature $H_M$ of $\cG$.\ The statement of the proposition now follows directly from Prop.\,\ref{prop:gerbes-as-H2} and Thm.\,\ref{thm:W3-tors}.
\eroof

\becor\label{cor:Ups-on-2conn}
Adopt the notation of Def.\,\ref{def:LGA}.\ Whenever the arrow manifold $\xcG$ has $H_2(\xcG)=\bd1$,\ the 1-isomorphism $\Upsilon$ exists.\ If also $H_1(\xcG)=\bd1$,\ then $\Upsilon$ is unique up to a 2-isomorphism.
\ecor
\beroof
The statement is a consequence of Props.\,\ref{prop:Ups-obstr},\ \ref{prop:gerbes-as-H2} and \ref{prop:1iso-H1},\ taken in conjunction with the Universal Coefficient Theorem.
\eroof
\noindent From the second level of the anomaly onwards,\ that is---in the context of the descent,\ it is convenient to base the analysis on the adapted \v Cech--Deligne--Dupont hypercohomological description (relative to a sufficiently fine simplicial cover of the nerve $N_\bullet(\grpd{\xcG}{M})$,\ in the sense of \cite{Tu:2006}),\ introduced in Def.\,\ref{def:CDD} and Prop.\,\ref{prop:ind-coc}.
\berop\label{prop:gam-obstr}
Adopt the notation of Def.\,\ref{def:LGA} and Prop.\,\ref{prop:ind-coc}.\ Assume given a 1-isomorphism $\Upsilon$,\ and let $p_\Ups\in A^1(\cO_{\xcG_1})$ represent the latter in the sense of Prop.\,\ref{prop:1iso-H1}.\ The obstruction to the existence of the 2-isomorphism $\g$ is---for $\cO_{\xcG_\bullet}$ sufficiently fine---given by the class
\qq\nn
[[\unl\D{}^{(1)}_{(1)}p_\Ups]]_{\widehat\d{}_\xcG}\in \widehat h{}^2(\xcG,\uj)\,.
\qqq
\eerop
\beroof
Let $b\in A^2(\cO_{\xcG_0})$ respresent the gerbe $\cG$ in the sense of Prop.\,\ref{prop:1iso-H1}.\ The existence of $\Upsilon$ then means that---for $\cO_{\xcG_\bullet}$ sufficiently fine---
\qq\nn
-\unl\D{}^{(2)}_{(0)}b+(\rho\rstr_{O^{(1)}_i},0,1)=D^{(1)}_{(1)}p_\Upsilon\,,
\qqq
with the 1-cochain $p_\Ups$ defined manifestly up to the addition $p_\Ups\mapsto p_\Ups+p$ of a 1-cocycle $p\in\ker\,D^{(1)}_{(1)}$.

In consequence of the homological nature of the Dupont operators,\ and of the multiplicativity of $\rho$,\ we then find
\qq\nn
D^{(1)}_{(2)}\unl\D{}^{(1)}_{(1)}(p_\Ups+p)=\unl\D{}^{(2)}_{(1)}D^{(1)}_{(1)}(p_\Ups+p)=-\unl\D{}^{(2)}_{(1)}\unl\D{}^{(2)}_{(0)}b+\bigl(\D^{(0,2)}_{(1)}\rho\rstr_{O_i^{(1)}},0,0\bigr)=0\,,
\qqq
and so $\unl\D{}^{(1)}_{(1)}(p_\Ups+p)$ defines a class
\qq\nn
[\unl\D{}^{(1)}_{(1)}(p_\Ups+p)]=[\unl\D{}^{(1)}_{(1)}p_\Ups]+[\unl\D{}^{(1)}_{(1)}][p]\in H^1\bigl(\xcG_2,\uj\bigr)\,.
\qqq
Every 1-cocycle $\unl\D{}^{(1)}_{(1)}(p_\Ups+p)$ represents the 1-isomorphism $(\txm^*\Upsilon^{-1}\ox\Id)\circ(\pr_2^*\Upsilon\ox\Id)\circ\pr_1^*\Upsilon$,\ and so the existence of $\g$ means that---for $\cO_{\xcG_\bullet}$ sufficiently fine---
\qq\nn
\unl\D{}^{(1)}_{(1)}(p_\Ups+p)=-D^{(0)}_{(2)}f_\g
\qqq
for some 0-cochain $f_\g\in A^0(\cO_{\xcG_2})$.\ Hence,\ it is the class $[\unl\D{}^{(1)}_{(1)}p_\Ups]\in P_\xcG^2$ up to corrections $[\unl\D{}^{(1)}_{(1)}][p]\equiv\widehat\d{}_\xcG^{(1)}[p]$ with $[p]\in H^1(\xcG_1,\uj)\equiv P_\xcG^1$ arbitrary that quantifies the obstruction to the existence of $\g$.\ By construction,\ the class is a 2-cocycle,
\qq\nn
\widehat\d{}_\xcG^{(2)}[\unl\D{}^{(1)}_{(1)}p_\Ups]\equiv[\unl{\D}{}^{(1)}_{(2)}\unl\D{}^{(1)}_{(1)}p_\Ups]=0\,.
\qqq
\eroof

\becor\label{cor:gamm-on-1conn}
Adopt the notation of Def.\,\ref{def:LGA}.\ If $\Upsilon$ exists,\ and the manifold $\xcG_2$ of pairs of composable arrows has $H_1(\xcG_2)=\bd1$,\ then the 2-isomorphism $\g$ exists.
\ecor
\beroof
The statement is a consequence of Props.\,\ref{prop:gam-obstr} and \ref{prop:1iso-H1},\ taken in conjunction with the Universal Coefficient Theorem.
\eroof
\noindent The last level of the obstruction---that to the coherence of a $\xcG$-equivariant structure---is quantified in
\berop\label{prop:gam-coh-obstr}
Adopt the assumptions and the notation of Def.\,\ref{def:LGA} and Prop.\,\ref{prop:ind-coc}.\ Assume given a 1-isomorphism $\Upsilon$ and a 2-isomorphism $\g$,\ and let $f_\g\in A^0(\cO_{\xcG_2})$ represent the latter in the sense of Prop.\,\ref{prop:1iso-H1}.\ The obstruction to the coherence of the 2-isomorphism $\g$,\ as expressed by identity \eqref{eq:gamma-coh},\ is---for $\cO_{\xcG_\bullet}$ sufficiently fine---given by the class
\qq\nn
[\unl\D{}^{(0)}_{(2)}f_\g]_{\d_\xcG}\in h^3\bigl(\xcG,\uj\bigr)\,.
\qqq
\eerop
\beroof
Let $p_\Ups\in A^1(\cO_{\xcG_1})$ and $f_\g\in A^0(\cO_{\xcG_2})$ represent the 1-isomorphism $\Upsilon$ and the 2-isomorphism $\g$,\ respectively,\ in the sense of Prop.\,\ref{prop:1iso-H1},\ so that equality 
\qq\label{eq:gam-BD}
\unl\D{}^{(1)}_{(1)}p_\Ups=-D^{(0)}_{(2)} f_\g
\qqq
holds true.\ Changing the representative of $p_\Upsilon$ within its cohomology class (or,\ equivalently,\ changing the representative of the 2-isomorphism class $[\Upsilon]$) as $p_\Upsilon\mapsto p_\Upsilon+D^{(0)}_{(1)}h,\ h\in A^0(\cO_{\xcG_1})$ becomes reflected in a correction $f_\g\mapsto f_\g+\unl\D{}^{(0)}_{(1)} h$.\ Taking into account the extra freedom of correction $f_\g\mapsto f_\g+k,\ k\in\ker\,D^{(0)}_{(2)}$,\ inscribed in \eqref{eq:gam-BD},\ we conclude that $f_\g$ is determined up to (and so can be corrected by) a replacement
\qq\nn
f_\g\mapsto f_\g+\unl\D{}^{(0)}_{(1)} h+k,\ (h,k)\in A^0(\cO_{\xcG_1})\x\ker\,D^{(0)}_{(2)}\,.
\qqq
We now obtain
\qq\nn
\unl\D{}^{(0)}_{(2)}\bigl(f_\g+\unl\D{}^{(0)}_{(1)} h+k\bigr)=\unl\D{}^{(0)}_{(2)}f_\g+\unl\D{}^{(0)}_{(2)}k\,,
\qqq
with
\qq\nn
D^{(0)}_{(3)}\unl\D{}^{(0)}_{(2)}f_\g=\unl\D{}^{(1)}_{(2)}D^{(0)}_{(2)}f_\g=-\unl\D{}^{(1)}_{(2)}\unl\D{}^{(1)}_{(1)}p_\Ups=0\,,
\qqq
which altogether infers that the quantitative measure of the obstruction to the coherence is $\unl\D{}^{(0)}_{(2)}f_\g\in\ker\,D^{(0)}_{(3)}$ up to corrections $\unl\D{}^{(0)}_{(2)}k$ with $k\in\ker\,D^{(0)}_{(2)}$ arbitrary.\ However,\ $\ker\,D^{(0)}_{(n)}\equiv\bH^0(\cO_{\xcG_n},\xcD^{\bullet}(2;\xcG_n))=\bH^0(\xcG_n,\xcD^{\bullet}(2;\xcG_n))$ for $\cO_{\xcG_\bullet}$ sufficiently fine,\ and so $\unl\D{}^{(0)}_{(2)}f_\g\in H^0(\xcG_3,\uj)\equiv\G^3_\xcG$ and $k\in\G^2_\xcG$ in virtue of Prop.\,\ref{prop:BD-vs-Huj}.\ Moreover,\ $\d_\xcG^{(3)}\unl\D{}^{(0)}_{(2)}f_\g\equiv\unl\D{}^{(0)}_{(3)}\unl\D{}^{(0)}_{(2)}f_\g=0$,\ that is $\unl\D{}^{(0)}_{(2)}f_\g$ is automatically a 3-cocycle,\ to be considered up to 3-coboundaries $\unl\D{}^{(0)}_{(2)}k\equiv\d_\xcG^{(2)}k$.
\eroof

\becor\label{cor:cohequiv-on-conn}
Adopt the notation of Def.\,\ref{def:LGA}.\ If both $\Upsilon$ and $\g$ exist,\ and the manifold $\xcG_3$ of triples of composable arrows is connected,\ then the 2-isomorphism $\g$ can be chosen coherent.
\ecor
\beroof
The statement is a consequence of Props.\,\ref{prop:gam-coh-obstr} and \ref{prop:1iso-H1},\ taken in conjunction with the Universal Coefficient Theorem.
\eroof

\subsection{The classifying cohomology group}

The interpretation of the gauging procedure emphasised in the concluding paragraphs of Sec.\,\ref{sec:class-gau-princ},\ dedicated to the universal construction,\ which presents the procedure as a model-building technique producing field theories with configuration fibres effectively reduced to orbispaces of Lie groupoids that model their rigid symmetries,\ prompts a classificatory question about inequivalent field theories obtained in this way for a {\em given} bundle of gauges $\xcP$.\ As the question trivialises on restriction to the metric sector,\ in which no extension (over $N_\bullet(\grpd{\xcG}{M})$) of the {\em invariant} tensorial background is needed,\ we end up with the following measure of the ambiguity.
\berop\label{prop:class-cohom}
Adopt the notation of Defs.\,\ref{def:LGA} and \ref{def:CDD},\ and assume the large gauge anomaly of $\xcM_\si$ for $\grpd{\xcG}{M}$ to vanish.\ The set of isomorphism classes of $\rho$-twisted $\xcG$-equivariant structures on a gerbe $\cG$ is---for $\cO_{\xcG_\bullet}$ sufficiently fine---a torsor of the hypercohomology group $\bH^2(\cJ)$ of the bounded cochain sub-bicomplex
\qq\nn
\cJ\qquad\colo\qquad\alxydim{@C=1.5cm@R=1.cm}{ \brd0 \ar[r] & A^0(\cO_{\xcG_3}) \ar[r]^{D^{(0)}_{(3)}} & \ker\,D^{(1)}_{(3)} \ar[r] & \brd0 \\ \brd0 \ar[r] & A^0(\cO_{\xcG_2}) \ar[r]^{D^{(0)}_{(2)}} \ar[u]^{\unl\D{}^{(0)}_{(2)}} & \ker\,D^{(1)}_{(2)} \ar[r] \ar[u]_{\unl\D{}^{(1)}_{(2)}} & \brd0 \\ \brd0 \ar[r] & A^0(\cO_{\xcG_1}) \ar[r]^{D^{(0)}_{(1)}} \ar[u]^{\unl\D{}^{(0)}_{(1)}} & \ker\,D^{(1)}_{(1)} \ar[r] \ar[u]_{\unl\D{}^{(1)}_{(1)}} & \brd0 \\ \brd0 \ar[r] & A^0(\cO_{\xcG_0}) \ar[r]^{D^{(0)}_{(0)}} \ar[u]^{\unl\D{}^{(0)}_{(0)}} & \ker\,D^{(1)}_{(0)} \ar[r] \ar[u]_{\unl\D{}^{(1)}_{(0)}} & \brd0 }
\qqq
of the \v{C}DD bicomplex of $\grpd{\xcG}{M}$ associated to $\cO_{\xcG_\bullet}$.
\eerop
\beroof
Let $(\Upsilon_a,\g_a),\ a\in\{1,2\}$ be two $\rho$-twisted $\xcG$-equivariant structures on $\cG\in\bgrb_\nabla(M)$,\ represented by the respective 2-cochains $(p_{\Ups_a},f_{\g_a})\in A^1(\cO_{\xcG_1})\oplus A^0(\cO_{\xcG_2})$.\ The difference 2-cochain
\qq\nn
(p_{1,2},f_{1,2}):=(p_{\Ups_2}-p_{\Ups_1},f_{\g_2}-f_{\g_1})
\qqq
satisfies the equations
\qq\nn
D^{(1)}_{(1)}p_{1,2}=0\,,\qquad\qquad\unl{\D}{}^{(1)}_{(1)}p_{1,2}=-D^{(0)}_{(2)}f_{1,2}\,,\qquad\qquad\unl{\D}{}^{(0)}_{(2)}f_{1,2}=0\,,
\qqq
which identify $(p_{1,2},f_{1,2})$ as a 2-cocycle in the total complex of $\cJ$,
\qq\nn
\nabla_{(2)}(0,p_{1,2},f_{1,2})=0\,.
\qqq

In the light of Def.\,\ref{def:Gequiv-grb},\ the $\rho$-twisted $\xcG$-equivariant structures represented by the 2-cochains $(p_{\Ups_a},f_{\g_a})$ are $\xcG$-equivariantly isomorphic if there exist
\bit
\item a 1-cochain $a_{1,2}\in A^1(\cO_{\xcG_0})$ representing a 1-automorphism $\Phi\colo\cG\xrightarrow{\cong}\cG$,\ and
\item a 0-cochain $h_{1,2}\in A^0(\cO_{\xcG_1})$ representing a (coherent) 2-isomorphism $\varphi\colo(d^{(1)*}_1\Phi\ox\Id_{\cI_{\rho}})\circ\Upsilon_1\overset{\cong}{\Longrightarrow}\Upsilon_2\circ d^{(1)*}_0\Phi$.
\eit
The two definitions translate into identities
\qq\nn
-D^{(1)}_{(0)}a_{1,2}=0\,,\qquad\qquad p_{\Ups_1}-d^{(1)*}_1 a_{1,2}+D^{(0)}_{(1)}h_{1,2}=p_{\Ups_2}-d^{(1)*}_0 a_{1,2}\,,
\qqq
which are to be augmented by an identity expressing the coherence condition \eqref{eq:Gequiv-iso-coh}:
\qq\nn
d^{(2)*}_1 h_{1,2}+f_{\g_1}=f_{\g_2}+d^{(2)*}_0 h_{1,2}+d^{(2)*}_2 h_{1,2}\,.
\qqq
Altogether,\ then,\ an equivalence between the two $\rho$-twisted $\xcG$-equivariant structures,\ whenever it exists,\ boils down to the triviality of the difference 2-cochain $(p_{1,2},f_{1,2})$ in the cohomology of $\cJ$,\ with the primitive given by the 1-cochain $(a_{1,2},h_{1,2})$,
\qq\nn
(0,p_{1,2},f_{1,2})=\nabla_{(1)}(a_{1,2},h_{1,2})\,.
\qqq
This completes the proof.
\eroof
\noindent While precise,\ the above quantification is also somewhat opaque and unwieldy,\ because---unlike that of the anomaly---it does not seem to be related to the topology of the simplicial manifold $N_\bullet(\grpd{\xcG}{M})$ in a tractable manner,\ through,\ say,\ the Universal Coefficient Theorem.\ This apparent shortcoming is at least partialy alleviated by the following observation,\ which refines and generalises the original,\ motivating one from \cite{Gawedzki:2010rn}.
\berop
Adopt the assumptions and the notation of Prop.\,\ref{prop:class-cohom}.\ There exists a canonical abelian-group homomorphism 
\qq\nn
\chi\colo\bH^2(\cJ)\too\widehat h{}^1\bigl(\xcG,\uj\bigr)
\qqq
with
\qq\nn
\ker\,\chi\cong h^2\bigl(\xcG,\uj\bigr)
\qqq
and
\qq\nn
\im\,\chi=\bigl\{ \ [[p]]_{\widehat\d{}_\xcG}\in\widehat h{}^1\bigl(\xcG,\uj\bigr) \quad\vert\quad \exists f\in A^0(\cO_{\xcG_2})\colo D^{(0)}_{(2)}f=-\unl{\D}{}^{(1)}_{(1)}p\quad \land\quad [\unl{\D}{}^{(0)}_{(2)}f]_{\d_\xcG}=0\ \bigr\}\,.
\qqq
\eerop
\beroof
Consider a 2-cochain $(p,f)\in\ker\,D^{(1)}_{(1)}\oplus A^0(\cO_{\xcG_2})$ in $\cJ$ which is closed,
\qq\nn
\nabla_{(2)}(0,p,f)=0\,,
\qqq
or,\ equivalently,
\qq\label{eq:HJ-coc}
D^{(1)}_{(1)}p=0\,,\qquad\qquad\unl{\D}{}^{(1)}_{(1)}p=-D^{(0)}_{(2)}f\,,\qquad\qquad\unl{\D}{}^{(0)}_{(2)}f=0\,.
\qqq
For a sufficiently fine cover $\cO_{\xcG_\bullet}$,\ the class $[p]\in\bH^1(\cO_{\xcG_1},\xcD^\bullet(2;\xcG_1))=\bH^1(\xcG_1,\xcD^\bullet(2;\xcG_1))\cong H^1(\xcG_1,\uj)\equiv P^1_\xcG$ (see:\ Prop.\,\ref{prop:BD-vs-Huj}) satisfies
\qq\nn
\widehat\d{}_\xcG^{(1)}[p]\equiv[\unl{\D}{}^{(1)}_{(1)}][p]=[\unl{\D}{}^{(1)}_{(1)}p]=0\,,
\qqq
and so the map
\qq\nn
\bZ^2(\cJ)\too\widehat h{}^1\bigl(\xcG,\uj\bigr),\ (p,f)\longmapsto[[p]]_{\widehat\d{}_\xcG}
\qqq
is well-defined.\ Whenever the 2-cocycle $(p,f)$ is a 2-coboundary,\ {\it i.e.},\ there exists $(a,h)\in\ker\,D^{(1)}_{(0)}\oplus A^0(\cO_{\xcG_1})$ such that
\qq\nn
(0,p,f)=\nabla_{(1)}(a,h)\,,
\qqq
or,\ in other words,\ 
\qq\nn
p=\unl\D{}^{(1)}_{(0)}a+D^{(0)}_{(1)}h\,,\qquad\qquad f=-\unl\D{}^{(0)}_{(1)}h\,,\qquad\qquad D^{(1)}_{(0)}a=0\,,
\qqq
we obtain
\qq\nn
[[p]]_{\widehat\d{}_\xcG}=[[\unl\D{}^{(1)}_{(0)}a]]_{\widehat\d{}_\xcG}=[[\unl\D{}^{(1)}_{(0)}][a]]_{\widehat\d{}_\xcG}=0\,.
\qqq
Hence,\ the above map induces the sought-after map in cohomology:
\qq\nn
\chi\colo\bH^2(\cJ)\too\widehat h{}^1\bigl(\xcG,\uj\bigr),\ [(p,f)]_\nabla\longmapsto[[p]]_{\widehat\d{}_\xcG}
\qqq

We begin the analysis of $\chi$ with the examination of its kernel.\ Given $\chi([(p,f)]_\nabla)=0$,\ we have $[p]=\widehat\d{}^{(0)}_\xcG[a]\equiv[\unl\D{}^{(1)}_{(0)}a]$ for some $a\in\ker\,D^{(1)}_{(0)}$,\ and hence there exists $h\in A^0(\xcG_1)$ such that $p-\unl\D{}^{(1)}_{(0)}a=D^{(0)}_{(1)}h$.\ This further implies $\unl\D{}^{(1)}_{(1)}p=\unl\D{}^{(1)}_{(1)}D^{(0)}_{(1)}h=D^{(0)}_{(2)}\unl\D{}^{(0)}_{(1)}h$,\ which in virtue of the cocyclicity \eqref{eq:HJ-coc} of $(p,f)$ yields (again,\ for a sufficiently fine cover) 
\qq\nn
k(p,f):=f+\unl\D{}^{(0)}_{(1)}h\in\ker\,D^{(0)}_{(2)}\equiv\bH^0\bigl(\cO_{\xcG_2},\xcD^\bullet(2;\xcG_2)\bigr)=\bH^0\bigl(\xcG_2,\xcD^\bullet(2;\xcG_2)\bigr)\cong H^0(\xcG_2,\uj)\equiv\G_\xcG^2\,,
\qqq
see:\ Prop.\,\ref{prop:BD-vs-Huj}.\ Thus,
\qq\nn
[(p,f)]_\nabla=[(\unl\D{}^{(1)}_{(0)}a+D^{(0)}_{(1)}h,-\unl\D{}^{(0)}_{(1)}h+k(p,f))]_\nabla\equiv[(0,k)+\nabla_{(1)}(a,h)]_\nabla=[(0,k(p,f))]_\nabla\,.
\qqq
Furthermore,\ $\d_\xcG^{(2)}k(p,f)=\unl{\D}{}^{(0)}_{(2)}f=0$
(again,\ in consequence of the cocyclicity of $(p,f)$),\ and allowed corrections $h\mapsto h+h_0$ with $h_0\in\ker\,D^{(0)}_{(2)}\equiv\G_\xcG^2\,$ induce shifts $k(p,f)\mapsto k(p,f)+\d_\xcG^{(1)}h_0$.\ Altogether,\ this indicates that we are dealing with a map $(p,f)\mapsto[k(p,f)]_{\d_\xcG}\in h^2(\xcG,\uj)$.\ The map is additive,\ and so upon checking that it sends $\nabla$-coboundries to $0$,
\qq\nn
[k(\nabla_{(1)}(a,h))]_{\d_\xcG}\equiv[k(\unl\D{}^{(1)}_{(0)}a+D^{(0)}_{(1)}h,-\unl\D{}^{(0)}_{(1)}h)]_{\d_\xcG}=[-\unl\D{}^{(0)}_{(1)}h+\unl\D{}^{(0)}_{(1)}(h+h_0)]_{\d_\xcG}\equiv 0\,,
\qqq
we thus obtain a well-defined homomorphism
\qq\nn
\widehat\iota\colo\ker\,\chi\too h^2\bigl(\xcG,\uj\bigr),\ [(p,f)]_\nabla\longmapsto[k(p,f)]_{\d_\xcG}\,.
\qqq
Note that $[k]_{\d_\xcG}=0$ implies the existence of $h\in\G_\xcG^1\equiv H^0(\xcG_1,\uj)\equiv\ker\,D^{(0)}_{(1)}$ such that $k=\d_\xcG^{(1)}h\equiv\unl{\D}{}^{(0)}_{(1)}h$,\ and then $[(0,k)]_\nabla=[(0,\unl{\D}{}^{(0)}_{(1)}h)]_\nabla\equiv[(-D^{(0)}_{(1)}h,\unl{\D}{}^{(0)}_{(1)}h)]_\nabla\equiv[\nabla_{(1)}(0,-h)]_\nabla=0$.\ Hence,\ $\widehat\iota$ is injective.\ Since it is also manifestly surjective,\ we conclude that it is the anticipated isomorphism
\qq\nn
\widehat\iota\colo\ker\,\chi\xrightarrow{\ \cong\ }h^2\bigl(\xcG,\uj\bigr)\,.
\qqq

We complete the proof by investigating the structure of $\im\,\chi$.\ By construction,
\qq\nn
\im\,\chi\subset\bigl\{ \ [[p]]_{\widehat\d{}_\xcG}\in\widehat h{}^1\bigl(\xcG,\uj\bigr) \quad\vert\quad \exists f\in A^0(\cO_{\xcG_2})\colo D^{(0)}_{(2)}f=-\unl{\D}{}^{(1)}_{(1)}p\quad \land\quad [\unl{\D}{}^{(0)}_{(2)}f]_{\d_\xcG}=0\ \bigr\}\,,
\qqq
see:\ \eqref{eq:HJ-coc},\ and so it remains to verify that every element of the set on the right-hand side of the above relation can be obtained from some 2-cocycle $(p,f)\in\bZ^2(\cJ)$.\ Consider such $[[p]]_{\widehat\d{}_\xcG}\in\widehat h{}^1(\xcG,\uj)$,\ and fix the corresponding $f\in A^0(\cO_{\xcG_2})$.\ The identity $[\unl{\D}{}^{(0)}_{(2)}f]_{\d_\xcG}=0$ implies the existence of $k\in\G^2_\xcG\equiv H^0(\xcG_2,\uj)\equiv\ker\,D^{(0)}_{(2)}$ such that $\unl{\D}{}^{(0)}_{(2)}f=\d_\xcG^{(2)}k\equiv\unl{\D}{}^{(0)}_{(2)}k$,\ and so we arrive at the complete list of equations
\qq\nn
D^{(1)}_{(1)}p=0\,,\qquad\qquad D^{(0)}_{(2)}f=-\unl{\D}{}^{(1)}_{(1)}p\,,\qquad\qquad \unl{\D}{}^{(0)}_{(2)}f=\unl{\D}{}^{(0)}_{(2)}k\,,\qquad\qquad D^{(0)}_{(2)}k=0
\qqq
describing $[[p]]_{\widehat\d{}_\xcG}$.\ The shifted 2-cochain $(p',f'):=(p,f)-(0,k)$ of $\cJ$ is closed,\ and yields $\chi([(p',f')])=[[p]]_{\widehat\d{}_\xcG}$,\ as desired.
\eroof

\beg\label{eg:AMM-4} 
The closed WZNW $\si$-model $\xcM_\si$ from Ex.\,\ref{eg:AMM-0} has a vanishing small gauge anomaly for the rigid-symmetry model given by the AMM groupoid from Ex.\,\ref{eg:AMM-1}.\ Moreover,\ the 2-connectedness of the group manifold $\txG$ ensures---in virtue of Cors.\,\ref{cor:Ups-on-2conn},\ \ref{cor:gamm-on-1conn},\ and \ref{cor:cohequiv-on-conn}---that there is no large gauge anomaly,\ either.\ The two cochain complexes from Prop.\,\ref{prop:ind-coc} trivialise in the present setting,\ and hence so does the classifying cohomology group $\bH^2(\cJ)\cong\ker\,\chi=\bd1$.
\eeg

\section{The gauged PAG $\si$-model}\label{sec:gau-sigmod}

In the present section,\ we put together the findings of Secs.\,\ref{sec:reduction} and \ref{sec:EAS}---under the assuption that the gauge anomalies described in Sec.\,\ref{sec:anomaly} vanish---and define a gauged field theory canonically induced from the PAG $\si$-model from Def.\,\ref{def:sigmod},\ with the configuration bundle given by the shadow of an arbitrary principaloid bundle with the structure group fully reduced according to Assumption (R) from p.\,\pageref{ass:red}. 

\bethe\label{thm:gausigmod}
Adopt the notation of Defs.\ \ref{def:rig-symm-grpd},\ \ref{def:connaugerbe},\ and \ref{def:connaumetr}.\ Consider a PAG $\si$-model $\xcM_\si$ with rigid-symmetry model $\grpd{\xcG}{M}$ for the rigid group\-oid\-al symmetry group $\bB_\si(\xcG)\subset\bB(\xcG)$ associated,\ as in Prop.\,\ref{prop:rig-sym-der},\ with the metric $\txg_M$ on the target space $M$ and a multiplicative BSS extension $(H_M,\rho,0)$ of the curvature $H_M$ of the $\si$-model gerbe.\ Fix a principaloid $\xcG$-bundle $\xcP\to\Si$ with connection 1-form $\Theta$ over the worldsheet $\Si$ of $\xcM_\si$,\ with the structure group reduced to $\bB_\si(\xcG)$,\ and let $\pi_\xcF\colo\xcF\to\Si$ be its shadow.\ Denote the subgroup of gauge transformations of $\xcP$ locally modelled on $L(\bB_\si(\xcG))$ as 
\qq\nn
{\rm Gauge}_\si(\xcP)\equiv{\rm Gauge}_\rho(\xcP)\cap{\rm Gauge}_{\txg_M}(\xcP)\subset{\rm Gauge}(\xcP)\,.
\qqq
The PAG $\si$-model canonically induces a field theory,\ with configuration bundle $\xcF$,\ determined by the DF amplitude
\qq\nn
\cA^{\si/\xcF,\bB_\si(\xcG)}_{\rm DF}&\colo&\G(\xcF)\too\uj\cr\cr  &\colo&\varphi\longmapsto\exp\left(-\tfrac{i}{2}\,\int_\Si\,\unl{\txg_M[\Theta]}\bigl(\txT\varphi\,\overset{\wedge}{,}\star_{\eta_\Si}\txT\varphi\bigr)\right)\cdot\Hol_{\unl{\cG[\Theta]}}\bigl(\varphi(\Si)\bigr)\equiv\widetilde\cA{}^{\si/\xcF,\bB_\si(\xcG)}_{\rm DF}[\varphi;\Theta]\,,
\qqq
in which $\unl{\txg_M[\Theta]}$ is the descended metric over $\xcF$ defined in Prop.\,\ref{prop:desc-metr},\ and $\unl{\cG[\Theta]}$ is the descended gerbe over $\xcF$ defined in Thm.\,\ref{thm:gauge-desc-aid}.\ The field theory has gauge symmetry ${\rm Gauge}_\si(\xcP)$:
\qq\nn
\forall\ (\varphi,\Phi)\in\G(\xcF)\x{\rm Gauge}_\si(\xcP)\colo\widetilde\cA{}_{\rm DF}^{\si/\xcF,\bB_\si(\xcG)}[\xcF_*(\Phi)\circ\varphi,\Theta^\Phi]=\widetilde\cA{}_{\rm DF}^{\si/\xcF,\bB_\si(\xcG)}[\varphi,\Theta]\,.
\qqq
\ethe
\beroof
The statement of the theorem combines those of Prop.\,\ref{prop:desc-metr} and Thm.\,\ref{thm:gauge-desc-aid}.
\eroof
\bedef
Adopt the notation of Thm.\,\ref{thm:gausigmod}.\ The subgroup ${\rm Gauge}_\si(\xcP)$ shall be referred to as the {\bf $\xcM_\si$-gauge group}.\ The induced $\si$-model determined by the DF amplitude $\cA^{\si/\xcF,\bB_\si(\xcG)}_{\rm DF}$ shall be called the {\bf $\bB_\si(\xcG)$-gauged PAG $\si$-model on} $\xcF$.
\exdef

\noindent We have thus obtained a genuinely novel\footnote{The construction of a prototypical gauged mechanical model (in dimension $0+1$),\ defined through (1-)stacky gauge descent to $\xcF$ for a CS differential character of degree 2,\ was first reported at the conference ``Rencontre Poisson \`a La Rochelle'' in May 2025,\ and subsequently at the 29${}^{\rm th}$ International Conference on Integrable Systems and Quantum Symmetries in Prague in July 2025 (see:\ \cite{Suszek:2025}).\ An elegant,\ equivalent approach to the gauging in the mechanical setting was subsequently worked out---in terms of the Cheeger--Simons cohomology in its axiomatic (functorial) Hopkins--Singer formulation,\ using the convenient Lerman--Malkin model---by Jakub Filipek in his M.Sci.\ Thesis \cite{Filipek:2025MSc}.} field theory (in spacetime dimension $1+1$) with a Lie-groupoidal symmetry gauged,\ the latter being represented by the action---restricted to isometrically and holonomically reduced bisections---of the Ehresmann--Atiyah groupoid of a principaloid bundle with the structure group likewise fully reduced,\ in conformity with Assumption (R).

\brem
The following general comment is well due at this stage:\ It might seem that we are not gauging (the target action of) $\xcG$ itself,\ but merely (the shadow action of) $\bB(\xcG)$---or,\ indeed,\ its reduction (which,\ however,\ might be viewed tentatively as coming from another structure groupoid through association).\ On the other hand,\ for $\bB$-complete groupoids the distinction becomes immaterial,\ and so---in view of the remark from p.\,\pageref{quote:monodromy}---we may safely regard the gauging as that of the monodromy groupoid of the tangent algebroid (and so of the linearised rigid symmetries) of the given Lie groupoid.
\erem

\beg\label{eg:AMM-5}
The unobstructed and unique descent of the closed WZNW $\si$-model $\xcM_\si$ from Ex.\,\ref{eg:AMM-0} yields the {\em maximally} gauged WZNW $\si$-model---the topological bulk $\txG/\txG$ coset theory of Gaw\c{e}dzki and Kupiainen \cite{Gawedzki:1988hq,Gawedzki:1988nj},\ whose correlators encode the structure of the Verlinde fusion ring of the original dynamical WZNW $\si$-model (see,\ {\it e.g.}:\ \cite{Gawedzki:1999bq}).
\eeg

\section{The symplectic scenario:\ An emergent Poisson $\si$-model}\label{sec:symplscen}

Among the multitude of Lie groupoids---regarded here as field-theoretic rigid-symmetry models--there is a class that has attracted considerable attention over the years,\ and can therefore provide us with a concrete realisation of the abstract construction contemplated in the present work,\ to wit:\ symplectic groupoids,\ as defined in \cite{Weinstein:1987,Coste:1987}.\ In this setting,\ the augmentation of Def.\,\ref{def:connaugerbe} acquires an intricate interpretation:\ For a symplectic Lie groupoid $\grpd{\xcG}{M}$ with the symplectic form $\om\in\Om^2(\xcG)$ on its arrow manifold (see:\ Def.\,\ref{eg:symplgrpd}),\ the comomentum assumes the canonical form 
\qq\label{eq:kappacan-id}
\kappa_{\rm can}=\id_{\txT^*M}
\qqq
under the isomorphism $\xcE\cong\txT^*M$ from Ex.\,\ref{eg:actalgbrd}.\ Indeed,\ formula \eqref{eq:comom-expl} now yields,\ in conjunction with the explicit identification \eqref{eq:smplalgbrd-secs-as-forms},
\qq\nn
\kappa(\a)\equiv\Id^*(\imath_\a\om)=\bigl(t\circ\Id\bigr)^*\eta[\a]=\eta[\a]\,.
\qqq
The closedness of $\om$ imposes severe constraints on gerbes over $M$ with the symplectic form as the twist of their $\xcG$-equivariant structure:\ The curvature $H_M$ of such a gerbe has to belong to the kernel of $\D^{(3)}_{(0)}$,\ see:\ \eqref{eq:rho-rel-HM}.\ Moreover,\ we may readily identify the $\om$-holonomic bisections of $\grpd{\xcG}{M}$:\ These are none other than the lagrangean bisections from Ex.\,\ref{eg:lagrbisec},
\qq\nn
\bB_\om(\xcG)\equiv\bB(\xcG,\om)\,.
\qqq  
Thus,\ upon picking up an arbitrary $\om$-twisted $\xcG$-equivariant gerbe $(\cG,\Upsilon,\g)$ with $\curv(\cG)=H_M\in\ker\,\D^{(3)}_{(0)}$,\ we may now seek to gauge the lagrangean group of $(\xcG,\om)$.\ In so doing,\ we arrive at the augmentation \eqref{eq:kappAi} over the local model $O_i\x M$ of $\xcF$,\ which pulls back to $O_i\ni\si$ along a local presentation $\widetilde\varphi{}_i\equiv(\id_{O_i},\varphi_i)=\xcF\t_i\circ\varphi\rstr_{O_i}$ of a section $\varphi\in\G(\xcF)$,\ and in local coordinates $\{x_i^\mu\}^{\mu\in\ovl{1,\dim\,M}}$ near $\varphi_i(\si)$,\ as
\qq\nn
\widetilde\varphi{}_i^*\kappa_{\rm can}[A_i]=\varphi_i^*\sfd  x_i^\mu\wedge A_{i\,\mu}\bigl(\widetilde\varphi{}_i(\cdot)\bigr)+\tfrac{1}{2}\,A_{i\,\nu}\bigl(\widetilde\varphi{}_i(\cdot)\bigr)\wedge\varphi_i^*\Pi^{\mu\nu}\,A_{i\,\mu}\bigl(\widetilde\varphi{}_i(\cdot)\bigr)\,,
\qqq
see:\ \eqref{eq:anchor-Poisson}.\ Denote $\txa_{i\,\mu}(\cdot):=A_{i\,\mu}(\widetilde\varphi_i(\cdot))$ to rewrite the above in the form
\qq\nn
\widetilde\varphi{}_i^*\kappa_{\rm can}[A_i]=-\txa_{i\,\mu}\wedge\varphi_i^*\sfd x_i^\mu-\tfrac{1}{2}\,\varphi_i^*\Pi^{\mu\nu}\,\txa_{i\,\mu}\wedge\txa_{i\,\nu}\,,
\qqq
in which we recognise the lagrangean density $\ceL_{\rm PSM}$ of the Poisson $\si$-model of \cite{Ikeda:1993fh,Schaller:1994es} with the Poisson target $(M,\Pi)$,\ calculated for the field configuration $(\varphi_i,\txa_i)$,
\qq\label{eq:PSM-as-aug}
\widetilde\varphi{}_i^*\kappa_{\rm can}[A_i]\equiv-\ceL_{\rm PSM}(\varphi_i,T\varphi_i,\txa_i)\,.
\qqq
Hence,\ gauging the lagrangean group of the symplectic groupoid $\grpd{\xcG}{M}$ in the PAG $\si$-model with a gerbe $\cG$ with a $\D^{(3)}_{(0)}$-closed curvature over the Poisson target space $(M,\Pi)$,\ equipped with an $\om$-twisted $\xcG$-equivariant structure $(\Upsilon,\g)$,\ can be viewed as coupling the $\si$-model (locally) to the Poisson $\si$-model.\ This sheds a new light on the latter topological field theory.\smallskip

The logic of the above argument can also be reversed to conclude that a correction of the local Poisson $\si$-model \eqref{eq:PSM-as-aug} by the holonomy of an $\om$-twisted $\xcG$-equivariant gerbe $(\cG,\Upsilon,\g)$ with a $\D^{(3)}_{(0)}$-closed curvature renders the model invariant under gauge transformations taking values in the structure group $\bB(\xcG,\om)$.\ (That the algebroidal symmetries of the theory prior to the coaugmentation do {\em not} integrate to the group(oid) level---as symmetries of the DF amplitude---seems to be signalled by the lack of  an {\em off-shell} (commutator) algebra structure on the vector space of these symmetries.)\ There is a distinguished class of gerbes for which such corrections additionally {\em do not affect the Euler--Lagrange equations of} $\ceL_{\rm PSM}$,\ namely\footnote{Recall that the contribution of the surface holonomy to the variation of the DF amplitude is proportional to the curvature of this differential character,\ see:\ Rem.\,\ref{rem:Lorentz}.}:\ {\it flat} $\om$-twisted $\xcG$-equivariant gerbes.\ Given the structural interpretation of solutions to the Euler--Lagrange equations ({\it i.e.},\ of classical field configurations of the original Poisson $\si$-model) as Lie-algebroid morphisms,\ see:\ \cite{Bojowald:2004wu},\ we give the coaugmentations with an unchanged space of classical configurations a special name -- that of {\bf flat coaugmentations},\ to be used below and in future studies.

\brem
Symplectic groupoids integrating---also in the formal sense (as singular spaces)---Poisson targets of the classic Poisson $\si$-model as their phase spaces were studied at length by Cattaneo and Felder in \cite{Cattaneo:2000iw}.\ These arise in connection to the canonical deformation quatisation of the Poisson structure $(M,\Pi)$ \`a la Kontsevich \cite{Kontsevich:1997vb},\ whose relation to the path-integral quantisation of the Poisson $\si$-model was discovered by the same authors in their seminal work \cite{Cattaneo:1999fm}.
\erem

\brem
It deserves to be noted that symplectic BSS extensions of flat 3-forms were mentioned parenthetically in \cite[Ex.\,3.19]{Crainic:2015msp} (for $\xcG$ $s$-simply connected).
\erem

While the conclusion from the previous paragraph holds true,\ in its present form,\ for a globally trivial principaloid bundle endowed with a global gauge field,\ it is completely clear---in the light of Thm.\,\ref{thm:gauge-desc-aid}---that the definition of the Poisson $\si$-model can now be extended,\ through a gerbe-theoretic twist and the subsequent (2-)stacky descent,\ to the total space of a shadow bundle of an arbitrary topology.\ We are thus led to
\bedef\label{def:coaug-PSM}
Adopt the notation of Defs.\,\ref{def:connaugerbe},\ \ref{def:simpl-obj} and \ref{def:Gequiv-grb},\ of Exs.\,\ref{eg:symplgrpd} and \ref{eg:lagrbisec},\ and of Thm.\,\ref{thm:gauge-desc-aid}.\ Let $\grpd{\xcG}{M}$ be a symplectic Lie groupoid with a multiplicative symplectic form $\om\in\Om^2(\xcG)$ on its arrow manifold,\ and the induced Poisson structure $(M,\Pi)$ on its object manifold.\ Let $\cG$ be a gerbe of curvature $H_M\in\ker\,\D^{(3)}_{(0)}$ over $M$ endowed with an $\om$-twisted $\xcG$-equivariant structure $(\Upsilon,\g)$.\ Given a closed compact orientable 2-dimensional manifold $\Si$,\ and a principaloid $\xcG$-bundle $\xcP$ with shadow $\xcF$,\ and with the structure group reduced to the lagrangean group $\bB(\xcG,\om)\subset\bB(\xcG)$ of $\grpd{\xcG}{M}$,\ denote the space of (compatible) connection 1-forms on $\xcP$ as ${\rm Conn}(\xcP)$.\ The {\bf $\cG$-coaugmented Poisson $\si$-model on} $\xcF$ is the field theory determined by the DF amplitude
\qq\nn
\cA_{\rm DF}^{P\si[\cG]/\xcF}\colo\G(\xcF)\x{\rm Conn}(\xcP)\too\uj,\ (\varphi,\Theta)\longmapsto\Hol_{\unl{\cG[\Theta]}}\bigl(\varphi(\Si)\bigr)^{-1}\,,
\qqq
in which $\unl{\cG[\Theta]}$ is the gerbe over $\xcF$ descended from the augmented gerbe $\widetilde\cG[\Theta]$ over $\pi_\xcF^{-1}\check Y{}_\cO$,\ with restrictions
\qq\nn
\widetilde\cG[\Theta]\rstr_{\pi_\xcF^{-1}(O_i)}=\xcF\t_i^*\bigl(\pr_2^*\cG\ox\cI_{\id_{T^*M}[A_i]}\bigr)
\qqq
\exdef
\noindent A fundamental symmetry property of the $\cG$-coaugmented Poisson $\si$-model is stated in
\bethe\label{thm:coaugPSM-gauge}
Adopt the notation and assumptions of Defs.\,\ref{def:gauge-trafo-conn} and \ref{def:coaug-PSM}.\ The $\cG$-coaugmented Poisson $\si$-model on $\xcF$ exhibits gauge symmetry ${\rm Gauge}_\om(\xcP)$,\ with local model $L(\bB(\xcG,\om))$,\ {\it i.e.},
\qq\nn
\forall\ (\varphi,\Theta,\Phi)\in\G(\xcF)\x{\rm Conn}(\xcP)\x{\rm Gauge}_\om(\xcP)\colo\cA_{\rm DF}^{P\si[\cG]/\xcF}[\xcF_*(\Phi)\circ\varphi,\Theta^\Phi]=\cA_{\rm DF}^{P\si[\cG]/\xcF}[\varphi,\Theta]\,.
\qqq
\ethe
\beroof
The statement of the theorem is a reformulation (and specialisation) of Thm.\,\ref{thm:gauge-desc-aid}.
\eroof
\noindent We might,\ at this stage,\ take a step back and enquire about the tangential {\em rigid} symmetry underlying the gauge symmetry of the $\cG$-coaugmented Poisson $\si$-model,\ which we took directly from the PAG $\si$-model analysis.\ The answer to a question thus posed can be read off from (an adaptation of) the proof of Prop.\,\ref{prop:rig-sym-der} (see,\ in particular:\ \Reqref{eq:kapep-as-tansymm}).\ Upon setting $\D^{(3)}_{(0)}H_M=0$,\ in keeping with the above discussion,\ we then find the defining identity 
\qq\nn
\sfd\kappa[\a]=0\,,
\qqq
to be satisfied by a tangential rigid symmetry represented by the section $\a\in\G_{\rm c}(\xcE)$ tangent to the $\om$-holonomically reduced structure group $\bB(\xcG,\om)$.\ In fact,\ the latter was shown to be a Lie subgroup of $\bB(\xcG)$ by Rybicki in \cite[Thm.\,4.5]{Rybicki:2001},\ and our analysis recovers---upon taking into account \eqref{eq:kappacan-id}---the identification of its Lie algebra:
\qq\nn
\Lie\bigl(\bB(\xcG,\om)\bigr)\cong Z^1_{\rm c}(M)\,,
\qqq
stated {\it ibidem}.\ Note also,\ in the passing,\ that in the special case of a flat coaugmentation,\ the defining property of the character ${\rm Hol}_\cG$ ensures that the contribution of the gerbe to the linearised gauge transformation of $\cA_{\rm DF}^{P\si[\cG]/\xcF}$ vanishes,\ and so we obtain the same Lie-algebroidal gauge symmetries as those considered in the original studies \cite{Bojowald:2004wu}.

\beg\label{eg:Lu-Weinstein}
Given a Poisson--Lie group $(\txG,\Pi)$ \cite{Drinfeld:1983,SemenovTianShansky:1985PL},\ there exists a canonical structure of a Lie algebroid $((\txT^*\txG\cong\txG\,\lx_{\Ad^*}\ggt^*,\txG,\bR^{\dim\,\txG},\pi_{\txT^*\txG}\equiv\pr_1),\Pi^\#,\{\cdot,\cdot\})$ on its cotangent bundle---the cotangent Lie algebroid,\ as defined in Ex.\,\ref{eg:cotang-Liealgd}.\ The corresponding classic Poisson $\si$-model was discussed at length in relation to the maximally gauged WZNW $\si$-model with the dual Poisson--Lie group $(\txG^*,\Pi^*)$ by Alekseev {\it et al.}\ in \cite{Alekseev:1995py},\ and subsequently by---{\it i.a.}---Falceto {\it et al.} in \cite{Falceto:2001eh} and \cite{Calvo:2003kv}.\ Here,\ the dual group $\txG^*$ has as its tangent Lie algebra the dual $\ggt^*$ of $\ggt\equiv\Lie(\txG)$,\ the latter being endowed with a Lie bracket $[\cdot,\cdot]^*$ induced from the Poisson bracket $\{\cdot,\cdot\}_\Pi$ on $C^\infty(\txG,\bR)\ni f_1,f_2$ through the identification $\ggt^*\equiv\txT^*_e\txG$ as
\qq\nn
[\txd f_1(e),\txd f_2(e)]^*:=\txd\{f_1,f_2\}_\Pi(e)\,.
\qqq 
The Poisson bracket on $C^\infty(\txG^*,\bR)$ is the unique one whose linearisation at the group unit $e^*\in\txG^*$ reproduces the Lie bracket $[\cdot,\cdot]$ on $\ggt\cong\ggt^*{}^*$.\ The Lie algebroid $\txG\,\lx_{\Ad^*}\ggt^*$ integrates to a symplectic Lie groupoid $\grpd{(\xcL\xcW}{\txG},\om_{\xcL\xcW})$ (see:\ Ex.\,\ref{eg:symplgrpd})---the {\bf Lu--Weinstein groupoid},\ introduced in \cite{Lu:1989sdg} (see also:\ \cite{Morgan:2019phd} for a detailed recent account).\ Hence,\ we may conceive symplectic principaloid $(\xcL\xcW,\om_{\xcL\xcW})$-bundles $\xcP(\xcL\xcW,\om_{\xcL\xcW})$ over a given closed compact orientable 2-dimensional manifold $\Si$,\ with the $\om_{\xcL\xcW}$-holonomically reduced structure group $\bB(\xcL\xcW,\om_{\xcL\xcW})$,\ and their shadows $\xcF(\xcL\xcW,\om_{\xcL\xcW})\cong\xcP(\xcL\xcW,\om_{\xcL\xcW})/\xcL\xcW$.\ The general discussion above opens an avenue to a generalisation of the classic Poisson $\si$-model with target $(\txG,\Pi)$---through (flat) augmentation---to shadows of an arbitrary topology.
\eeg

Finally,\ let us note---by way of a closing observation---that the obstruction analysis of Sec.\,\ref{sub:gauge-anomaly} calls for a reformulation and admits further refinement in the present setting {\em in the distinguished case of a flat coaugmentation}.\ Indeed,\ the point of departure is now a given multiplicative symplectic 2-form $\om$ on $\xcG$,\ or---equivalently---a trivial gerbe $\cI_\om\in\bgrb_\nabla(\xcG)$,\ and we look for a representative of a 1-isomorphism class $[\cG_0]\in\cW^3(M;0)\cong H^2(M,\uj)$ which,\ in particular,\ admits a 1-isomorphism 
\qq\nn
\Upsilon_0\colo t^*\cG_0\xrightarrow{\ \cong\ }s^*\cG_0\ox\cI_\om\,.
\qqq
The trivial gerbe determines a class
\qq\nn
[(\om\rstr_{O_i^{(1)}},0,1)]\in\cW^3(\xcG;0)\cong H^2\bigl(\xcG,\uj\bigr)
\qqq
in the kernel of the coboundary operator $\widetilde\d{}_\xcG^{(1)}$ (see:\ Prop.\,\ref{prop:ind-coc}),
\qq\nn
\widetilde\d{}_\xcG^{(1)}[(\om\rstr_{O_i},0,1)]\equiv\bigl[\bigl(\D^{(0,2)}_{(1)}\om\rstr_{O_i^{(1)}},0,1\bigr)\bigr]=0\,,
\qqq
and so we arrive at
\berop\label{prop:coaugPSM-anomaly}
Adopt the notation of Def.\,\ref{def:coaug-PSM} and \ref{def:mult-k-form},\ and of Prop.\,\ref{prop:ind-coc}.\ Given a multiplicative symplectic form $\om\in\Om^2(\xcG)$,\ the obstruction to the existence of a flat coaugmenting gerbe $\cG_0$ together with the 1-isomorphism $\Upsilon_0\colo t^*\cG_0\overset{\cong}{\to}s^*\cG_0\ox\cI_\om$ is---for $\cO_{\xcG_\bullet}$ sufficiently fine---captured by the class
\qq\nn
[[(\om\rstr_{O_i^{(1)}},0,1)]]_{\widetilde\d{}_\xcG}\in\widetilde h{}^1\bigl(\xcG,\uj\bigr)\,.
\qqq
If the pair $(\cG_0,\Upsilon_0)$ exists,\ with the 1-isomorphism $\Upsilon_0$ represented by $p_{\Ups_0}\in A^1(\cO_{\xcG_1})$,\ then the obstruction to the existence of the 2-isomorphism $\g_0$ is---for $\cO_{\xcG_\bullet}$ sufficiently fine---given by the class
\qq\nn
[[\unl\D{}^{(1)}_{(1)}p_{\Ups_0}]]_{\widehat\d{}_\xcG}\in \widehat h{}^2(\xcG,\uj)\,.
\qqq
Finally,\ if the triple $(\cG_0,\Upsilon_0,\g_0)$ exists,\ with the 2-isomorphism $\g_0$ represented by $f_{\g_0}\in A^0(\cO_{\xcG_2})$,\ then the obstruction to the coherence of the 2-isomorphism $\g_0$ is---for $\cO_{\xcG_\bullet}$ sufficiently fine---given by the class
\qq\nn
[\unl\D{}^{(0)}_{(2)}f_{\g_0}]_{\d_\xcG}\in h^3\bigl(\xcG,\uj\bigr)\,.
\qqq
\eerop
\beroof
In the light of Props.\,\ref{prop:gam-obstr} and \ref{prop:gam-coh-obstr},\ the only thing that needs to be verified is the statement concerning the first obstruction.\ A flat gerbe $\cG_0$ is represented by a 2-cocycle $b_0\in A^2(\cO_{\xcG_0})$ which defines a class
\qq\nn
[b_0]\in\cW^3(\xcG_0;0)\cong H^2\bigl(\xcG_0,\uj\bigr)\equiv W^0_\xcG\,.
\qqq
The existence of $\Upsilon_0$,\ which amounts to the triviality of $[t^*\cG_0\ox s^*\cG_0^*\ox\cI_{-\om}]\in H^2(\xcG,\uj)$,\ can now be understood as the vanishing of 
\qq\nn
[d^{(1)\,*}_0 b_0-d^{(1)\,*}_1 b_0-(\om\rstr_{O_i^{(1)}},0,1)]\equiv[\unl\D{}^{(2)}_{(0)}b_0-(\om\rstr_{O_i^{(1)}},0,1)]=\widetilde\d{}_\xcG^{(0)}[b_0]-[(\om\rstr_{O_i^{(1)}},0,1)]\,.
\qqq
This is the same as the triviality of $[[(\om\rstr_{O_i^{(1)}},0,1)]]_{\widetilde\d{}_\xcG}$.
\eroof

Our analysis of the symplectic scenario places the long-known topological field theory in an essentially new perspective:\ Depending on the preferred point of view,\ we may either view the original Poisson $\si$-model of \cite{Ikeda:1993fh,Schaller:1994es} as a topological gauge field theory with an intrinsically non-integrable (in general) Lie-algebroidal gauge symmetry,\ or we may regard the coaugmentation of Def.\,\ref{def:coaug-PSM} as an appropriate completion of the original field theory in which the linearised Lie-groupoidal symmetry has been integrated,\ and---at the same time---acces to configuration (shadow) bundles with nontrivial topology has been gained.\ In the former case,\ the $\cG$-coaugmented Poisson $\si$-model is to be seen as an extension of the original field theory,\ conceptually independent of the original one,\ endowed with a richer and topologically more intricate categorified gauge symmetry.\ In the latter case,\ it acquires the status of the complete and canonically prequantisable---through the gerbe-theoretic transgression of \cite{Gawedzki:1987ak}---formulation of the ancestor field theory,\ with {\em effectively Poisson-reduced} internal degrees of freedom $M//\xcG$.\ The latter point of view,\ which identifies the Poisson $\si$-model with the topological sector of a PAG $\si$-model with the target space given by the object manifold of a symplectic groupoid,\ and with the corresponding lagrangean bisections gauged,\ opens an avenue to the application of techniques of gerbe theory in the construction and cohomological classification of prequantisable boundary states and more general defects in the $\cG$-coaugmented Poisson $\si$-model.\ For a flat coaugmentation,\ we thus obtain a powerful (higher-)geometric criterion distinguishing branes and bi-branes (in the language of \cite{Fuchs:2007fw} and \cite{Runkel:2008gr}).\ In any event,\ the 2$d$ gauged field theory calls for an in-depth investigation,\ which is currently underway.

\section{Summary and Outlook}\label{sec:summ-out}.

We have investigated the gauging of rigid field-theoretic symmetries modelled on an arbitrary Lie groupoid $\grpd{\xcG}{M}$ with the object manifold $M$ given by the configuration fibre of the field theory of interest.\ In so doing,\ we have considered the recently proposed principaloid $\xcG$-bundles with (compatible) connection $(\xcP,\Theta)$ and their Godement shadows $(\xcF,\Theta_\xcF)$ (Sec.\,\ref{sec:principaloid}) as a natural point of departure and the main building block of the proposed construction---a source of a local Lie algebroid-valued gauge field (the former),\ and a model of the configuration bundle of the field theory with the groupoidal symmetry gauged (the latter).\ As argued in Sec.\,\ref{sec:class-gau-princ},\ the gauging procedure is to be understood as a way to model dynamics with a configuration fibre reduced to the characteristic foliation $M//\xcG$ of the groupoid,\ of particular relevance in the absence of a smooth structure on the latter orbispace.\ In our construction,\ grounded in the greater comprehensiveness of the categorified model of symmetry in contemporary field theory,\ we have drawn inspiration from a systematic reformulation of the classical gauge principle---originally worked out for action groupoids $\txG\,\lx_\la M$ associated with smooth (Lie-)group actions $\la$ on the configuration fibre---in the language of Lie groupoids,\ their associated Lie groups of global bisections,\ and principal bibundles (Sec.\,\ref{sec:class-gau-princ}).

As the first new result,\ we have worked out a groupoidal counterpart of Cartan's mixing construction,\ in which the fibering of the symmetry model over the space of internal degrees of freedom of the field theory---in the slice $\Bun(\Si)/\Si$ over a given spacetime $\Si$---has been generalised from the previously studied canonical one,\ given by the target map of the structure groupoid $\grpd{\xcG}{M}$,\ to its pullback along the moment map $\mu_X\colo X\to M$ of an arbitrary groupoid module $(X,\mu_X,\la_X)$ (Sec.\,\ref{sec:red-ass}).\ The thus obtained bundle associated to a given principaloid bundle $\xcP$ by the action $\la_X$ of $\grpd{\xcG}{M}$ has been interpreted as a special example of a principaloid $\xcG\,\lx_{\la_X}X$-bundle---one with a structure group reduced to the image of that of $\xcP$---{\it i.e.},\ of $\bB(\xcG)$---along a homomorphism canonically induced by $\mu_X$.\ The reduction is anticipated (and,\ indeed,\ has been demonstrated,\ in Sec.\,\ref{sec:reduction}) to be necessary in the presence of extra differential-geometric structure on the configuration fibre of the field theory---such as a metric tensor or a higher-geometric object (a bundle,\ a gerbe,\ {\it etc.})---and thus justifies the abstract construction.\ Another key rationale for the latter is the possibility it offers to couple multiple charged-matter fields,\ modelled on global sections of bundles associated to $\xcP$,\ to a single gauge field,\ encoded by the connection $\Theta$ on $\xcP$,\ through a groupoidal variant of the Crittenden construction of an associated connection.

The second key result of the present paper is the gauging prescription for Lie-groupoidal rigid symmetries in the two-dimensional Polyakov--Alvarez--Gaw\c{e}dzki $\si$-model with the topological Wess--Zumino term in the Dirac--Feynman amplitude given by a degree-3 Cheeger--Simons differential character---the surface holonomy of a gerbe $\cG$ over the metric target space $(M,\txg_M)$.\ The symmetry has been assumed to manifest itself as the invariance of the target metric $\txg_M$ and---independently---of the isomorphism class of the target gerbe $\cG$ under the target action $t_*$ of {\em distinguished} bisections of $\xcG$ on $M$.\ More specifically,\ a consistent implementation of the symmetry has been shown---in Sec.\,\ref{sec:reduction}---to naturally\footnote{See:\ Rem.\,\ref{rem:weak-rho-holo} and Prop.\,\ref{prop:iso-2-much},\ though.} call for a reduction of the structure group $\bB(\xcG)$ of $\xcP$ (and so also of $\xcF$) to the intersection $\bB_\si(\xcG)=\bB_{\txg_M}(\xcG)\cap\bB_\rho(\xcG)$ of its subgroup $\bB_{\txg_M}(\xcG)=t_*^{-1}({\rm Isom}(M,\txg_M))$,\ assumed proper,\ with another one,\ $\bB_\rho(\xcG)=\{\ \b\in\bB(\xcG)\ \vert\ \b^*\rho=0\ \}$,\ determined by a multiplicative extension $\rho\in\Om^2(\xcG)$ of the curvature $H_M$ of $\cG$ to a 3-cocycle $(H_M,\rho,0)$ in the Bott--Shulman--Stasheff cohomology of the simplicial nerve $N_\bullet(\grpd{\xcG}{M})$ of the structure groupoid.\ The reduction has been demonstrated to ensure descent of the target structures:\ the metric and the gerbe to the total space of the shadow bundle $\xcF$ from its trivialisation upon augmentation by the local gauge fields sourced by $\Theta$ (Sec.\,\ref{sec:EAS}),\ an idea largely inspired by and generalising the classic group-symmetry prototype worked out in \cite{Gawedzki:2010rn} (Sec.\,\ref{sec:G-gau-sigmod}).\ In the metric sector,\ a sheaf-theoretic descent has been attained through the usual minimal coupling,\ {\it i.e.},\ through the evaluation of the metric tensor on $\Theta$-verticalised (covariant) derivatives in a local presentation (Prop.\,\ref{prop:desc-metr}).\ In the topological sector,\ the augmentation requisite for a fully fledged 2-stacky descent has been found to be mediated by the 2-comomentum component of the Spencer pair of the multiplicative extension $\rho$,\ and based on the assumption of a coherent prequantisation of the latter in the form of a $\rho$-twisted $\xcG$-equivariant structure on the target gerbe $\cG$ (Thm.\,\ref{thm:gauge-desc-aid}).\ An exhaustive cohomological analysis of obstructions to the prequantisation,\ and a similar classification of inequivalent gaugings in the absence of such obstructions have been given in Sec.\,\ref{sec:anomaly}---once more drawing heavily on the methodology and findings of \cite{Gawedzki:2010rn}.\ This part of our study has been crowned---in Sec.\,\ref{sec:gau-sigmod}---with the definition and the statement of gauge symmetry of the gauged PAG $\si$-model with the configuration bundle given by the shadow of an arbitrary principaloid $\xcG$-bundle with the structure group reduced to $\bB_\si(\xcG)$ as above (Thm.\,\ref{thm:gausigmod}).

The last result of our investigation is a novel conceptualisation of the distinguished two-dimensional topological field theory---the Poisson $\si$-model of \cite{Ikeda:1993fh,Schaller:1994es}---obtained as an intricate spin-off of a specialisation of the previous discussion of the gauging to the symplectic setting of \cite{Weinstein:1987,Coste:1987},\ and a subsequent far-reaching generalisation of the original field theory proposed in Sec.\,\ref{sec:symplscen}.\ The specialisation---restricted to the topological sector of the PAG $\si$-model---consists in fixing the target space of the field theory in the form of a Poisson object manifold $(M,\Pi)$ of a symplectic groupoid $(\xcG,\om)$,\ the latter being equipped with a multiplicative symplectic form $\om\in Z^2_{\rm dR}(\xcG)$.\ Remarkably,\ the lagrangean Dirac--Feynman amplitude of the classical Poisson $\si$-model has been recovered as the trivial Cheeger--Simons differential character augmenting an $\om$-twisted $\xcG$-equivariant gerbe $\cG$ over $M$ in the manner proposed in Sec.\,\ref{sec:EAS}.\ In particular,\ the local gauge field coming from the connection $\Theta$ on the underlying {\em symplectic} principaloid $(\xcG,\om)$-bundle---with the structure group reduced to the lagrangean group $\bB(\xcG,\om)$ of the structure groupoid---has been suitably identified with the $\txT^*M$-valued gauge field of the Poisson $\si$-model.\ In the light of Thm.\,\ref{thm:gauge-desc-aid},\ this has opened the possibility of extending the original definition of the Poisson $\si$-model to shadows of symplectic principaloid $(\xcG,\om)$-bundles of arbitrary topology (Def.\,\ref{def:coaug-PSM}),\ and,\ simultaneously,\ of integrating their on-shell Lie-algebroidal gauge symmetries to off-shell Lie-groupoidal ones modelled on $\bB(\xcG,\om)$ (Thm.\,\ref{thm:coaugPSM-gauge}).\ Topological obstructions to the extension,\ contingent upon the existence of the coaugmenting $\om$-twisted $\xcG$-equivariant gerbe $\cG$,\ have been quantified in a specialisation of the former anomaly analysis (Prop.\,\ref{prop:coaugPSM-anomaly}).\ It has been observed that distinguished flat coaugmentations ({\it i.e.},\ those with $\cG$ flat) leave the form---and so also the Lie-algebroidal interpretation---of the field equations of the original model unchanged.\medskip

The findings reported in the present work provide a solid body of evidence for the applicability and relevance in the rich context of field theory of the abstract notion of principaloid bundle proposed in \cite{Strobl:2025}.\ They also prompt a number of follow-up questions,\ and chart directions for future research,\ of which we list a few in the closing paragraphs of this section.
\bit
\item[(F1)] A natural continuation of the study initited herein is a reinterpretation of the gauging of the PAG $\si$-model in terms of topological gauge-symmetry defect networks carrying the data of a $\rho$-twisted $\xcG$-equivariant structure on the target gerbe,\ along the lines of the pioneering work \cite{Fuchs:2007fw,Runkel:2008gr} and,\ in particular,\ \cite{Suszek:2012ddg,Suszek:2013} (see also:\ \cite{Suszek:2023ldu} for a recent application in the supergeometric setting).\ Results of a work thus oriented are currently being written up and shall be published soon.
\item[(F2)] Another obvious extension of our endeavour is a formulation of the gauging prescription in the presence of boundaries and more general defects in the field theory prior to the gauging,\ in structural analogy with the precursor project \cite{Gawedzki:2012fu} dedicated to Lie-group symmetries.\ This would entail,\ {\it i.a.},\ the identification and examination of natural conditions for the compatibility of defects with bulk groupoidal symmetries discussed in the present paper.\ In the light of the analyses presented in \cite{Runkel:2008gr,Runkel:2009sp,Suszek:2022CSimpl},\ the theory of double Lie groupoids,\ originally advanced by Ehresmann in \cite{Ehresmann:1963},\ and later developed,\ {\it i.a.},\ by Mackenzie in \cite{Mackenzie:1992dLa1,Mackenzie:2000dLa2},\ is expected to be of relevance.
\item[(F3)] The cohomological nature of the non-tensorial term in the DF amplitude of the PAG $\si$-model invites further examination of the amenability to gauging of {\em weakly} $\rho$-holonomic bisections:
\qq\nn
\widetilde\bB{}_\rho(\xcG):=\{\ \b\in\bB(\xcG) \quad\vert\quad [\b^*\rho]=0\in H^k(M) \ \}\supset\bB_\rho(\xcG)\,,
\qqq
see:\ Rem.\,\ref{rem:weak-rho-holo}.\ The findings of Crainic {\it et al.} on the relation between multiplicative $k$-forms on Lie groupoids and Spencer operators (see,\ in particular,\ \cite{Crainic:2015msp}) are certain to play an instrumental r\^ole in this context.
\item[(F4)] The definition of the PAG $\si$-model (and its higher-dimensional generalisations) uses Cheeger--Simons differential characters,\ which---while in a one-to-one correspondence with isoclasses of gerbes---provide us with a very particular,\ purely (homology-)algebraic description of the corresponding algebro-geometric objects.\ The latter may help to circumnavigate some technical complexities of the direct application of its higher-geometric counterpart developed by Murray {\it et al.} in the present work.\ Therefore,\ it seems apposite to inspect a reformulation of the hitherto work,\ as well as that proposed in the points (F1)-(F3) above,\ in terms of the axiomatic Cheeger--Simons cohomology,\ and in particular in (the Lerman--Malkin stacky rendering \cite{Lerman:2008,Lerman:2009} of) its functorial Hopkins--Singer model \cite{Hopkins:2002rd},\ which has been successfully employed in \cite{Filipek:2025MSc} in the gauge analysis of the prototypical 1$d$ $\si$-model of dynamics of a charged massive probe.
\item[(F5)] In all hitherto considerations,\ the Lie-algebroidal gauge field encoded by the connection 1-form on the principaloid $\xcG$-bundle has featured as a non-dynamical background in which (probe) charged-matter fields propagate.\ The formulation of a natural algebroidal Yang--Mills-type model of dynamics of the gauge field,\ inextricably entwined with the Higgs field (see:\ the discussion in Sec.\,\ref{sub:ass-appl}),\ is evidently required for the internal conistency of the theory,\ in which the gauge field should have sources of its curvature ({\it i.e.},\ charges).\ Proposals for such models,\ valid in the simplified scenario in which the gauge field is globally smooth and the gauge symmetry is modelled on (smooth functions from spacetime into) a Lie algebroid have been known for more than two decades,\ see:\ \cite{Strobl:2004im,Kotov:2015},\ and it is clear---with the concepts and techniques developed in \cite{Strobl:2025} and in the present paper---how to generalise them to principaloid (and shadow) bundles of arbitrary topology,\ and for the corresponding integrated gauge-symmetry models.\ The relevant constructions are currently under investigation and shall be presented in near future.
\item[(F6)] The previous point can be regarded as an ouverture to a more fundamental search for a universal gauge-field theory for principaloid bundles---{\it i.e.},\ for the corresponding variant of the Chern--Weil theory---building on,\ {\it i.a.},\ the work of Crainic {\it et al.},\ see:\ \cite{Arias:2011}.\ Establishing a field-theoretic manifestation of the fundamental concept of Morita equivalence of Lie groupoids (see:\ \cite{delHoyo:2013} for a nice introduction to the subject) would also be of great interest,\ all the more so in the light of the conceptualisation of the gauging as configurational descent to the characteristic foliation $M//\xcG$.
\item[(F7)] Besides modelling configurational reduction with respect to a Lie-groupoid action (as in the present paper),\ and exhibiting their autonomous Higgs-coupled dynamics (as in (F5)),\ Lie-algebroidal gauge fields are bound to enter models of dynamics of charged probes.\ Defining the latter calls for an understanding and explicit construction of holonomy and more general transport operators on principaloid bundles.\ It stands to reason that comomentum components of multiplicative forms,\ with their very tractable behaviour under the action of the (reduced) gauge group by the conjugation (in a local presentation),\ might be of use as candidate `trace' maps on Lie algebroids.\ The coaugmentation mechanism uncovered in Sec.\,\ref{sec:symplscen} in the setting of the Poisson $\si$-model also offers an interesting theoretical possibility.
\item[(F8)] In the symplectic context,\ one natural line of development would be to invoke the well-studied identification of defects (resp.\ boundary states) in the PAG $\si$-model---derived in \cite{Runkel:2008gr}---with gerbe bi-modules (resp.\ modules),\ and apply the bicategorial and higher-cohomological arsenal of gerbe theory \cite{Stevenson:2000wj,Waldorf:2007phd},\ in a novel approach to the reconstruction and classification of prequantisable (bi-)branes in the gerbe-coaugmented Poisson $\si$-model,\ building upon the hitherto lagrangean-TFT developments \cite{Cattaneo:2003dp,Calvo:2004mq} (see also:\ \cite{Falceto:2010rfp} for a review).\ Given the correspondence between $\si$-models defects and symmetries/dualities of the 2$d$ field theory,\ elaborated in \cite{Runkel:2008gr,Runkel:2009sp,Suszek:2011hg,Suszek:2022CSimpl},\ this opens an avenue to a higher-geometric and -categorial investigation of prequantisable dualities in this newly discovered environment.\ Its applications could be quite astounding,\ see,\ {\it e.g.},\  \cite{Nekrasov:2025egx}.
\item[(F9)] Remaining in the context of dualities of Poisson $\si$-models,\ one would be tempted to examine the Lu--Weinstein construction from Ex.\,\ref{eg:Lu-Weinstein} (or,\ potentially,\ also its formal generalisation from \cite{Cattaneo:2000iw}) from the new vantage point offered by the gauging of symplectic-groupoidal symmetries in the coaugmented Poisson $\si$-model.\ Indeed,\ the groupoid $(\grpd{\xcL\xcW(\txG,\Pi)}{\txG},\om_{\xcL\xcW(\txG,\Pi)})$ actually carries a richer structure than that implied by its status as the integration of the cotangent Lie algebroid over the Poisson--Lie group $(\txG,\Pi)$:\ It is a {\em double} symplectic Lie groupoid
\qq\nn
\alxydim{@C=1.25cm@R=1.cm}{ \xcL\xcW(\txG,\Pi) \ar@<.25ex>[r]^{\qquad t} \ar@<-.25ex>[r]_{\qquad s}  \ar@<.25ex>[d]_{s^*} \ar@<-.25ex>[d]^{t^*} & \txG \ar@{-->}@<.25ex>[d] \ar@{-->}@<-.25ex>[d] \\ \txG^* \ar@{-->}@<.25ex>[r] \ar@{-->}@<-.25ex>[r] & \bullet }\,,
\qqq
{\it i.e.},\ alongside the original structure of a symplectic Lie groupoid on $\xcL\xcW(\txG,\Pi)$ relative to the submanifold of units $(\txG,\Pi)$,\ there exists another one relative to the lagrangean submanifold of units $(\txG^*,\Pi^*)$---to be denoted as $\xcL\xcW^*(\txG,\Pi)\equiv\xcL\xcW(\txG,\Pi)$---and the two are mutually compatible in that both: 
\qq\nn
(t^*,\bullet),(s^*,\bullet)\colo\grpd{\xcL\xcW(\txG,\Pi)}{\txG}\to\bbB\txG^*
\qqq
and
\qq\nn
(t,\bullet),(s,\bullet)\colo\grpd{\xcL\xcW^*(\txG,\Pi)}{\txG^*}\to\bbB\txG
\qqq 
are (Lie-)groupoid epimorphisms.\ This implies that every symplectic principaloid bundle $\pi_\xcP\colo\xcP\to\Si$ with structure groupoid $(\xcL\xcW(\txG,\Pi),\om_{\xcL\xcW(\txG,\Pi)})$ admits {\em two} Godement quotients (or shadows),\ and thus gives rise to a {\bf double principal groupoid-bundle object in} $\Bun(\Si)/\Si$:
\qq\nn
\alxydim{@C=.5cm@R=.5cm}{ & \xcF\cong\xcP/\xcL\xcW(\txG,\Pi) \ar[ddl]_{\pi_\xcF} & & & & & & \xcL\xcW(\txG,\Pi) \ar@{=>}[dl] \ar@/_1.pc/[ddlllll]_{\varrho} \\ & & & & & & \txG & \\ \Si & & \xcP \ar[uul]_{\xcD} \ar[ddl]^{\xcD^*} \ar[ll]_{\pi_\xcP} \ar[urrrr]_{\mu} \ar[drrrr]^{\mu^*} & & &  \\ & & & & & & \txG^* & \\ & \xcF^*\cong\xcP/\xcL\xcW^*(\txG,\Pi) \ar[uul]^{\pi_{\xcF^*}} & & & & & & \xcL\xcW^*(\txG,\Pi) \ar@{=>}[ul] \ar@/^1.pc/[uulllll]^{\varrho^*} }\,.
\qqq
Given that the typical fibre $\xcL\xcW(\txG,\Pi)$ of $\xcP$ is locally diffeomorphic to the product manifold $\txG\x\txG^*$,\ and equipped with a non-degenerate 2-form $\om_{\xcL\xcW(\txG,\Pi)}$ (which admits---under obvious circumstances---a prequantisation to a Poincar\'e-type bundle),\ it is tempting to think of the above structure as an arena for various field-theoretic dualities---now amenable to gerbe-theoretic analysis,\ whenever prequantisable---formerly considered in the literature in the present Poisson--Lie setting,\ such as the duality between the Poisson $\si$-model with target $(\txG^*,\Pi^*)$ and the fully gauged WZNW $\si$-model with target $\txG$ \cite{Alekseev:1995py,Falceto:2001eh} (see:\ Ex.\,\ref{eg:AMM-5}),\ the open-closed duality between Poisson $\si$-models for the dual targets $(\txG,\Pi)$ and $(\txG^*,\Pi^*)$ \cite{Calvo:2003kv},\ or the intricate Poisson $T$-duality between the respective dynamical (PAG) $\si$-models \cite{Klimcik:1995ux,Alekseev:1995ym}.
\eit
Clearly,\ a systematic exploration of the physics of groupoidal gauge theory based on the notion of principaloid bundle and its Godement shadow,\  initiated in the present work,\ encompasses many conceptual and formal challenges.\ We hope to address them in our future work.

\appendix

\section{Generalities}

In this appendix,\ we gather some general conventions and results used in the main text.

\becon
Let $M_a$ and $N_a$,\ with $a\in\{1,2\}$,\ be smooth manifolds and let $f_a\colo M_a\to N_a$ be smooth maps.\ We write
\qq\nn
f_1\x f_2\colo M_1\x M_2\too N_1\x N_2,\ (m_1,m_2)\longmapsto\bigl(f_1(m_1),f_2(m_2)\bigr)\,.
\qqq
\econ

\becon
Let $M$ and $N_a$,\ with $a\in\{1,2\}$,\ be smooth manifolds and let $f_a\colo M\to N_a$ be smooth maps.\ We write 
\qq\nn
(f_1,f_2)\colo M\too N_1\x N_2,\ m\longmapsto\bigl(f_1(m),f_2(m)\bigr)\,.
\qqq
\econ

\becon\label{con:fibr-prod}
Let $N$ and $M_a$,\ with $a\in\{1,2\}$,\ be smooth manifolds and let $f_a\colo M_a\to N$ be smooth maps.\ We write
\qq\nn
M_1\x_N M_2\equiv M_1\,\fibx{f_1}{f_2}M_2=\bigl\{\ (m_1,m_2)\in M_1\x M_2 \quad\vert\quad f_1(m_1)=f_2(m_2)\ \bigr\}\,.
\qqq
\econ

\berop
Adopt the notation of Conv.\,\ref{con:fibr-prod}.\ Whenever one of the maps $f_a$ is a surjective submersion,\ $M_1\,\fibx{f_1}{f_2}M_2$ is a smooth submanifold of $M_1\x M_2$.
\eerop

\bethe[The Godement Criterion]\cite{Serre:1964,Fernandes:2024}\label{thm:Godement}
Let $M$ be a smooth manifold and let $\sim$ be an equivalence relation on M,\ with graph
\qq\nn
\alxydim{@C=.75cm@R=1.cm}{ \cR_\sim \ar@{^{(}->}[r] & M\x M \ar[dl]_{\pr_1} \ar[dr]^{\pr_2} \\ M & & M}\,.
\qqq
There exists a smooth structure on the quotient 
\qq\nn
M//\sim=\{\ [m]_\sim \quad\vert\quad m\in M \ \}\,, 
\qqq
compatible with the quotient topology,\ and such that $\pi\colo M\too M//\sim$ is a submersion,\ iff the graph $\cR_\sim$ is a proper submanifold of $M\x M$ and the restriction of the projection $\pr_1\colo M\x M\too M$ to $\cR_\sim$ is a submersion.
\ethe

\becon
Let $\cC$ be a category with source $s$ and target $t$.\ We shall write $A\in\cC$ if $A$ belongs to the object class of $\cC$,\ and $f\in\cC(A,B)$ if $f$ belongs to the morphism class of $\cC$,\ with $(s,t)(f)=(A,B)$.
\econ

\section{Essential facts about Lie groupoids and algebroids}

The present appendix contains elements of the theory of Lie groupoids and algebroids---definitions with examples,\ propositions and theorems---which constitute the conceptual core of the physically oriented considerations in the main text.

\bedef\label{def:grpd}
A \textbf{groupoid} is a small category with all morphisms invertible.\ Thus,\ it is a septuple $\Gr=(\obj\Gr\equiv M,\morf\Gr\equiv\xcG,s, t,\Id,\txm\equiv.,\Inv\equiv(\cdot)^{-1})$---to be abbreviated as $\grpd{\xcG}{M}$---which is composed of a pair of sets: 
\bit
\item the \textbf{object set} $M$; 
\item the \textbf{arrow set} $\xcG$,
\eit
and a quintuple of \textbf{structure maps}: 
\bit
\item the \textbf{source} ({\bf map}) $ s\colo \xcG\to M$;
\item the \textbf{target} ({\bf map}) $  t\colo \xcG\to M$; 
\item the \textbf{unit} ({\bf map}) $\Id\colo M\to\xcG,\ m\mapsto\Id_m$;
\item the \textbf{inverse} ({\bf map}) $\Inv\colo \xcG\to\xcG,\  g\mapsto \Inv(
g)\equiv g^{-1}$;
\item the \textbf{multiplication} ({\bf map}) $\txm\colo \xcG\fibx{s}{t}\xcG\to\xcG,\ ( g, h)\mapsto \txm( g, h)\equiv g.h$,
\eit 
subject to the conditions (in force whenever the expressions are well-defined):
\bit
\item[(i)] $ s( g. h)= s(
h),\   t( g. h)=  t(
g)$;
\item[(ii)] $( g. h). k= g.( h.k)$;
\item[(iii)] $\Id_{  t( g)}. g=
g= g.\Id_{ s( g)}$
;
\item[(iv)] $ s( g^{-1})=  t( g),\   t(
g^{-1})= s( g),\  g
. g^{-1}=\Id_{  t( g)},\
g^{-1}. g=\Id_{ s(g)}$.
\eit

A \textbf{morphism} between two groupoids $\Gr_A,\ A\in\{1,2\}$ is a functor $\Phi\colo \Gr_1\to\Gr_2$.

A \textbf{Lie groupoid} is a groupoid whose object and arrow sets are smooth manifolds,\ whose structure maps are smooth,\ and whose source and target maps are surjective submersions.\ A morphism between two Lie groupoids is a functor between them with smooth object and morphism components. 

A Lie groupoid is called 
\bit
\item {\bf transitive} if the {\bf anchor} $(t,s)\colo\xcG\to M\x M$ is surjective;
\item {\bf locally trivial} if $(t,s)\colo\xcG\to M\x M$ is a surjective submersion.
\eit
\exdef

\beg\label{eg:deloop} 
Every Lie group $(\txG,\txm_\txG,\Inv_\txG,e)$ gives rise to a Lie groupoid 
\qq\nn
\bbB\txG\equiv\bigl(\{\bullet\},\txG,\bullet,\bullet,e,\txm_\txG,\Inv_\txG\bigr)\,, 
\qqq
with the singleton $\{\bullet\}$ as the object manifold.\ It is termed the {\bf delooping} ({\bf groupoid}) of $\txG$.
\eeg

\beg\label{eg:pairgrpd}
Every manifold $M\in\Man$ gives rise to a Lie groupoid 
\qq\nn
{\rm Pair}(M)\equiv\bigl(M,M\x M,\pr_2,\pr_1,\D,\pr_{1,4},\t\bigr)\,,
\qqq 
with unit $\D\colo M\to M\x M,\ m\mapsto(m,m)$,\ multiplication $\pr_{1,4}\colo(M\x M)\ \fibx{\pr_2}{\pr_1}(M\x M)\to M\x M,\ ((m_4,m_2),(m_2,m_1))\mapsto(m_4,m_1)$,\ and inverse $\t\colo M\x M\to M\x M,\ (m_1,m_2)\mapsto(m_2,m_1)$.\ It is called the {\bf pair groupoid} of $M$.
\eeg

\beg\label{eg:subm-grpd}
Given $M,Y\in\Man$ and a surjective submersion $\pi\colo Y\twoheadrightarrow M$,\ there arises a Lie groupoid
\qq\nn
{\rm Pair}_\pi(Y)\equiv\bigl(Y,Y\fibx{\pi}{\pi}Y,\pr_2,\pr_1,\D,\pr_{1,4},\t\bigr)\subset{\rm Pair}(Y)\,,
\qqq 
defined in the notation of Ex.\,\ref{eg:pairgrpd}.\ It is called the {\bf submersion groupoid} of $\pi$.
\eeg

\beg\label{eg:actgrpd}
Given a Lie group $\txG$ and a manifold $M$ on which $\txG$ acts smoothly as $\la\colo\txG\to\Diff(M)$,\ there arises a Lie groupoid 
\qq\nn
\txG\,\lx_\la M\equiv\bigl(M,\txG\x M,\pr_2,\widetilde\la\equiv\ev\circ(\la\x\id_M),(e,\cdot),(\txm_\txG\circ\pr_{1,3},\pr_4),(\Inv_\txG\circ\pr_1,\widetilde\la)\bigr)\,,
\qqq
defined in terms of the evaluation map $\ev\colo\Diff(M)\x M\to M,\ (f,m)\mapsto f(m)$.\ It is called the {\bf action groupoid} of $\la$.\ By a mild abuse of the notation,\ we identify $\la$ and $\widetilde\la$ in this work.
\eeg

\beg\cite{Weinstein:1987,Coste:1987}\label{eg:symplgrpd}
A Lie groupoid $(M,\xcG,s,t,\Id,\txm,\Inv)$ is termed {\bf symplectic} if there exists a closed non-degenerate 2-form $\om\in\Om^2(\xcG)$,\ which makes $(\xcG,\om)$ a symplectic manifold,\ and such that the graph of $\txm$ is a lagrangean submanifold in $(\xcG^{\x 3},(\pr_1^*+\pr_2^*-\pr_3^*)\om)$.\ As shown in \cite[Thm.\,1.1,\ Chap.\,II]{Coste:1987},\ the object manifold $M$ then carries a canonical Poisson bivector $\Pi\in\G(\bigwedge^2\txT M)$,\ relative to which $s$ is Poisson,\ and $t$ is anti-Poisson.\ Furthermore,\ $\Id(M)$ is lagrangean.
\eeg

\bedef[\cite{Moerdijk:2003mm}]\label{def:bisec}
Adopt the notation of Def.\,\ref{def:grpd}.\ A ({\bf global}) {\bf bisection of} $\Gr$ is a section $\b\colo M\to\xcG$ of the surjective submersion $ s\colo \xcG\to M$ such that the induced map 
\qq\nn
t_*\b\equiv  t\circ\b\colo M\too M
\qqq
is a diffeomorphism.\ Equivalently,\ it is a submanifold $S\subset\xcG$ with the property that both restrictions:\ $ s\rstr_S$ and $t\rstr_S$ are diffeomorphisms.\ We shall denote the set of bisections as 
\qq\nn
\BisGr\equiv\bB(\xcG)\,.
\qqq

A {\bf local bisection of} $\Gr$ is a local section $\b\colo O\to\xcG$ of $ s$ over an open subset $O\subset M$,\ such that the induced map 
\qq\nn
t_*\b\equiv  t\circ\b\colo O\too  (t\circ\b)(O)
\qqq
is a diffeomorphism.\ We shall denote the set of local bisections as 
\qq\nn
{\rm Bisec}_{\rm loc}(\Gr)\equiv\bB_{\rm loc}(\xcG)\,.
\qqq
\exdef

\bedef
Adopt the notation of Def.\,\ref{def:bisec}.\ The {\bf group of bisections} of a Lie groupoid $\grpd{\xcG}{M}$ is the canonical structure of a group on $\bB(\xcG)$.\ Its binary operation is defined as
\qq\nn
\cdot\colo \bB(\xcG)\x\bB(\xcG)\too\bB(\xcG),\ (\b_2,\b_1)\longmapsto\b_2\bigl(t_*\b_1(\cdot)\bigr).\b_1(\cdot)\equiv\b_2\cdot\b_1\,.
\qqq
The neutral element is $\Id$,\ also termed the {\bf unit bisection},\ and the corresponding inverse is 
\qq\nn
\Inv\colo \bB(\xcG)\too\bB(\xcG),\ \b\longmapsto\Inv\circ\b\circ\bigl(  t_*\b\bigr)^{-1}\equiv\b^{-1}\,.
\qqq

Similarly,\ the {\bf group of local bisections} of $\grpd{\xcG}{M}$ is the canonical structure of a group on $\bB_{\rm loc}(\xcG)$.\ Its binary operation is defined as
\qq\nn
\cdot\colo \bB_{\rm loc}(\xcG)\x\bB_{\rm loc}(\xcG)\too\bB_{\rm loc}(\xcG),\ (\b_2,\b_1)\longmapsto\b_2\bigl(t_*\b_1(\cdot)\bigr).\b_1(\cdot)\rstr_{(t_*\b_1)^{-1}({\rm dom}(\b_2))\cap{\rm dom}(\b_1)}\equiv\b_2\cdot\b_1\,.
\qqq
The neutral element is $\Id$,\ and the corresponding inverse is 
\qq\nn
\Inv\colo \bB_{\rm loc}(\xcG)\too\bB_{\rm loc}(\xcG),\ \b\longmapsto\Inv\circ\b\circ\bigl( t_*\b\bigr)^{-1}\rstr_{t_*\b({\rm dom}(\b))}\equiv\b^{-1}\,.
\qqq
\exdef

\beg
Adopt the notation of Ex.\,\ref{eg:deloop}.\ There exists a canonical isomorphism ${\rm Bisec}(\bbB\txG)\cong\txG$.
\eeg

\beg
Adopt the notation of Ex.\,\ref{eg:pairgrpd}.\ There exists a canonical isomorphism ${\rm Bisec}({\rm Pair}(M))\cong\Diff(M)$.
\eeg

\beg\label{eg:flatbisec}
Adopt the notation of Ex.\,\ref{eg:actgrpd}.\ The group ${\rm Bisec}(\txG\,\lx_\la M)$ contains a subgroup 
\qq\nn
\txG\cong\iota(\txG)\subset{\rm Bisec}(\txG\,\lx_\la M)\,,
\qqq
given by the image of the monomorphism
\qq\nn
\iota\colo\txG\too{\rm Bisec}(\txG\,\lx_\la M),\ g\longmapsto(g,\cdot)\,.
\qqq
We shall refer to bisections in that image as {\bf flat}.
\eeg

\beg\cite{Coste:1987,Rybicki:2001}\label{eg:lagrbisec} 
For the symplectic groupoid from Ex.\,\ref{eg:symplgrpd},\ 
the group $\bB(\xcG)$ contains a subgroup 
\qq\nn
\bB(\xcG,\om)&:=&\bigl\{ \ \b\in\bB(\xcG) \quad\vert\quad \b(M)\subset\xcG \tx{ is lagrangean}\ \bigr\}\cr\cr 
&\equiv&\bigl\{ \ \b\in\bB(\xcG) \quad\vert\quad \b^*\om=0\ \bigr\}\subset\bB(\xcG)\,.
\qqq
These bisections are called {\bf lagrangean},\ and the subgroup $\bB(\xcG,\om)$ shall be referred to as the {\bf lagrangean group} of $\xcG$.
\eeg

\bedef\label{def:B-compl}
Adopt the notation of Def.\,\ref{def:bisec}.\ A Lie groupoid $\grpd{\xcG}{M}$ with the property:
\qq\nn
\forall\ g\in\xcG\ \exists\ \b\in\bB(\xcG)\colo\b\bigl(s(g)\bigr)=g
\qqq 
shall be termed {\bf $\bB(\xcG)$-complete}.
\exdef

\beg
All source- resp.\ target-connected Lie groupoids are $\bB(\xcG)$-complete (see:\ \cite[Thm.\,3.1]{Zhong:2009}),\ and so is also every action groupoid from Ex.\,\ref{eg:actgrpd}. 
\eeg

\begin{propanition}\label{prop:beta-split}
Adopt the notation of Def.\,\ref{def:bisec}.\ Every bisection $\b\in\bB_{\rm loc}(\xcG)$ determines,\ over each point $m\in{\rm dom}(\b)$,\ a direct-sum decomposition
\qq\nn
\txT_{\b(m)}\xcG\cong\txT_{\b(m)}\bigl(\b({\rm dom}(\b))\bigr)\oplus\bigl(\ker\,\txT s\bigr)_{\b(m)}\,.
\qqq
We shall call the corresponding pair $(\chi_\b,\upsilon_\b)$ of complementary projectors,
\qq\nn
\chi_\b(m)=\txT_{\b(m)}(\b\circ s)\,,\qquad\qquad\upsilon_\b(m)=\id_{\txT_{\b(m)}\xcG}-\chi_\b(m)\,,
\qqq
the {\bf splitting} of $\txT\xcG\rstr_{\b({\rm dom}(\b))}$ {\bf induced by} $\b$.
\end{propanition}

\bedef\cite[Sec.\,3.1]{Crainic:2015msp}\label{def:J1B}
Adopt the notation of Def.\,\ref{def:bisec}.\ The {\bf first-jet groupoid} of $\xcG$ is the groupoid
\qq\nn
\bigl(M,J^1\xcG,\pi_{J^1},\t,j^1\Id,j^1\txm,j^1\Inv\bigr)
\qqq
with the arrow manifold $J^1\xcG$ consisting of first jets $j^1_m\b$ of local bisections $\b\in\bB_{\rm loc}(\xcG)$ (${\rm dom}(\b)\ni m$) over $M$,\ and with source $\pi_{J^1}(j^1_m\b)=m$ and target $\t(j^1_m\b)=t_*\b(m)$,\ identity $j^1\Id(m)=j^1_m\Id$,\ multiplication $j^1\txm(j^1_{t_*\b_1(m)}\b_2,j^1_m\b_1)=j^1_m(\b_2\cdot\b_1)$,\ and inverse $j^1\Inv(j^1_m\b)=j^1_{t_*\b(m)}(\b^{-1})$. 
\exdef

\berop
Adopt the notation of Def.\,\ref{def:J1B}.\ There exists a group homomorphism 
\qq\label{eq:j1-hom}
j^1\colo\bB(\xcG)\too\bB(J^1\xcG),\ \b\longmapsto j^1_\cdot\b\,.
\qqq
\eerop

\bedef\label{def:tstar}
Adopt the notation of Def.\,\ref{def:bisec}.\ The {\bf shadow action of} $\bB(\xcG)$ {\bf on} $M$ is 
\qq\nn
t_*\colo \bB(\xcG)\x M\too M,\ (\b,m)\longmapsto t\bigl(\b(m)\bigr)\,.
\qqq
We refer by the same name to and use the same symbol for the group homomorphism 
\qq\label{eq:BisGr-act-obj}
t_*\colo \bB(\xcG)\too{\rm Diff}(M)\,.
\qqq 
\exdef

\bedef\label{def:bisec-act}
Adopt the notation of Def.\,\ref{def:bisec}.\ The {\bf left-multiplication of} $\xcG$ {\bf by} $\bB(\xcG)$ is the left action
\qq\nn
L\colo \bB(\xcG)\x\xcG\too\xcG,\ \bigl(\b, g\bigr)\longmapsto\b\bigl(  t\bigl( g\bigr)\bigr). g\equiv L_\b\bigl( g\bigr)\equiv \b\lact g\,.
\qqq
The {\bf right-multiplication of} $\xcG$ {\bf by} $\bB(\xcG)$ is the right action
\qq\nn
R\colo  \xcG\x\bB(\xcG)\too\xcG,\ \bigl( g,\b\bigr)\longmapsto g.\bigl(\b^{-1}\bigl(s\bigl( g\bigr)\bigr)\bigr)^{-1}\equiv R{}_\b\bigl( g\bigr)\equiv g\ract\b\,.
\qqq
The {\bf conjugation of} $\xcG$ {\bf by} $\bB(\xcG)$ is the \emph{left} action
\qq\nn
C\colo \bB(\xcG)\x\xcG\too\xcG,\ (\b, g)\longmapsto\b\bigl(  t( g)\bigr). g.\b\bigl( s( g)\bigr)^{-1}\equiv C_\b( g)\equiv \b\lact g\ract\b^{-1}\,.
\qqq
\exdef

\bedef\label{def:gr-mod}
Adopt the notation of Def.\,\ref{def:grpd}.\ A \textbf{right-$\xcG$-module space},\ or a \textbf{right-$\xcG$-module} for short,\ is a triple $(X,\mu,\varrho)$ composed of 
\bit
\item a smooth manifold $X$;
\item a smooth map $\mu\colo X\to M$,\ called the (\textbf{right}) \textbf{moment map};
\item a smooth map
\qq\nn
\varrho\colo  X\,\fibx{\mu}{t}\xcG\too
X,\ (x, g)\longmapsto\varrho(x, g)\equiv\varrho_{ g}(x)\equiv x\mact g\,,
\qqq
termed the (\textbf{right}) \textbf{action} ({\bf map}),
\eit 
subject to the relations (in force whenever the expressions are well-defined):
\bit
\item[(GrM1)] $\mu(x\mact g)= s( g)$;
\item[(GrM2)] $x\mact\Id_{\mu(x)}=x$;
\item[(GrM3)] $(x\mact g)\mact h=x\mact( g. h)$.
\eit

Similarly,\ a {\bf left-$\xcG$-module} ({\bf space}) is a triple $(X,\mu,\la)$ composed of 
\bit
\item a smooth manifold $X$;
\item a smooth map $\mu\colo X\to M$,\ called the (\textbf{left}) \textbf{moment map};
\item a smooth map
\qq\nn
\la\colo \xcG\,\fibx{s}{\mu}X\too
X,\ (g,x)\longmapsto\la(g,x)\equiv\la_g(x)\equiv g\mlact x\,, 
\qqq
termed the (\textbf{left}) \textbf{action} ({\bf map}),
\eit 
subject to the relations (in force whenever the expressions are well-defined):
\bit
\item[(GlM1)] $\mu(g\mlact x)= t(g)$;
\item[(GlM2)] $\Id_{\mu(x)}\mlact x=x$;
\item[(GlM3)] $h\mlact(g\mlact x)=(h.g)\mlact x$.
\eit

A right action $\varrho$ is termed \textbf{free} if the following implication obtains:
\qq\nn
x\mact g=x\qquad\Longrightarrow\qquad g=
\Id_{\mu(x)}\,,
\qqq
so that,\ in particular,\ the \textbf{isotropy group} $\xcG_m= s^{-1}(\{m\})\cap  t^{-1}(\{m\})$ of $m\in M$ acts freely (in the usual sense) on the fibre $\mu^{-1}(\{m\})$.\ A free left action is defined analogously.

A right action $\varrho$ is termed \textbf{transitive} if for any two points $x,x'\in X$ there exists an arrow $ g\in\xcG$ such that $x'=x\mact g$.\ A transitive left action is defined analogously.

The subset 
\qq\nn
\xcG\mlact x:=\bigl\{\ g\mlact x \quad\vert\quad g\in s^{-1}\bigl(\mu(x)\bigr) \ \bigr\}
\qqq
is called the {\bf left $\xcG$-orbit} of $x\in X$.\ {\bf Right $\xcG$-orbits} are defined analogously.

Let $\Gr_a\equiv\grpd{\xcG_a}{M_a},\ a\in\{1,2\}$ be a pair of Lie groupoids,\ and let $(X_a,\mu_a,\varrho_a)$ be the respective
right-$\xcG_a$-modules.\ A \textbf{morphism} between the latter is a pair $(\Theta,\Phi)$ consisting of a smooth manifold map $\Theta\colo  X_1\to X_2$ together with a functor $\Phi\colo \Gr_1\to\Gr_2$ for which the following diagrams commute
\qq
&\alxydim{@C=1.cm@R=1.cm}{ X_1 \ar[r]^{\Theta} \ar[d]_{\mu_1} &
	X_2 \ar[d]^{\mu_2}\cr M_1 \ar[r]_{\Phi} &
	M_2}\,,& \label{diag:mu-Th-mu}\\\cr\cr
&\alxydim{@C=1.5cm@R=1.cm}{ X_1\ \fibx{\mu_1}{t_1}\,\xcG_1
	\ar[r]^{\Theta\x\Phi} \ar[d]_{\varrho_1} &  X_2\
	\fibx{\mu_2}{t_2}\,\xcG_2 \ar[d]^{\varrho_2}\cr  X_1
	\ar[r]_{\Theta} &  X_2}\,.& \label{diag:Th-intertw}
\qqq
For every Lie groupoid $\Gr\equiv\grpd{\xcG}{M}$,\ right $\xcG$-modules together with the corresponding morphisms with $\Phi=\id_\Gr$ form a {\bf category of right $\xcG$-modules} $\Man_\xcG$.\ Its $\hom$-sets shall be denoted as
\qq\nn
C^\infty_\xcG(X_1,X_2)\equiv\Hom_{\Man_\xcG}\bigl(\bigl(X_1,\mu_1,\varrho_1\bigr),\bigl(X_2,\mu_2,\varrho_2\bigr)\bigr)\,,
\qqq
with elements referred to as {\bf smooth $\xcG$-equivariant maps} between $X_1$ and $X_2$.\ In the particular case of $(X_2,\varrho_2)=(X_1,\varrho_1)\equiv(X,\varrho)$ and smoothly invertible maps,\ we employ the notation
\qq\nn
\Diff_{\varrho(\xcG)}(X)=\{\ f\in C^\infty_\xcG(X,X) \quad\vert\quad \exists\ f^{-1}\in C^\infty_\xcG(X,X) \ \}
\qqq
\exdef

\beg\label{eg:st-act}
There exists canonical $\xcG$-module structures on the object manifold $M$:\ the left-$\xcG$-module structure
\qq\nn
\bigl(M,\id_M,t\circ\pr_1\bigr)\,,
\qqq
with the {\bf target action}
\qq\nn
(t\circ\pr_1)(g,m)=t(g)\,,
\qqq
and the right-$\xcG$-module structure
\qq\nn
\bigl(M,\id_M,s\circ\pr_2\bigr)\,,
\qqq
with the {\bf source action}
\qq\nn
(s\circ\pr_2)(m,g)=s(g)\,,
\qqq
\eeg

\beg\label{eg:fibred-act}
There exist two commuting canonical $\xcG$-module structures on the arrow manifold $\xcG$:\ the left-$\xcG$-module structure 
\qq\nn
(\xcG,  t,l\equiv.)\,,
\qqq
with the {\bf left-fibred action}
\qq\label{eq:Lgrpdact}
l\bigl( h, g\bigr)\equiv l_{ h}\bigl( g\bigr)= h. g\,,
\qqq
and the right-$\xcG$-module structure 
\qq\label{eq:can-R-Gr-mod}
(\xcG, s,r\equiv.)\,,
\qqq
with the {\bf right-fibred action}
\qq\label{eq:Rgrpdact}
r\bigl( g, h\bigr)\equiv r_{ h}\bigl( g\bigr)= g. h\,.
\qqq
\eeg

\berop\cite[Prop.\,2.8]{Strobl:2025}\label{prop:rGequiv-LB}
Adopt the notation of Defs.\,\ref{def:bisec} and \ref{def:gr-mod}.
\qq\nn
\Diff_{r(\xcG)}(\xcG)=L\bigl(\bB(\xcG)\bigr)\,.
\qqq
\eerop

\bethe\cite[Thm.\,1.5.12]{MacKenzie:2005}\label{thm:orb-sub}
Adopt the notation of Def.\,\ref{def:gr-mod}.\ The orbit 
\qq\nn
\xcG\mlact m:=t\bigl(s^{-1}\bigl(\{m\}\bigr)\bigr)
\qqq
of an arbitrary point $m\in M$ is a submanifold of $M$.\ The subset
\qq\nn
\xcG_{\xcG\mlact m}^{\xcG\mlact m}:=s^{-1}(\xcG\mlact m)\cap t^{-1}(\xcG\mlact m)
\qqq
is a smooth submanifold of $\xcG$,\ and a locally trivial Lie subgroupoid of $\grpd{\xcG}{M}$.
\ethe

\begin{propanition}\cite[Sec.\,2.3]{Arias:2011}\label{prop:grpd-act-grpd}
Adopt the notation of Def.\,\ref{def:gr-mod}.\ Every left $\xcG$-module $(X,\mu,\la)$ canonically determines a Lie groupoid 
\qq\nn
\xcG\,\lx_\la X\equiv\bigl(X,\xcG\,\fibx{s}{\mu}X,\pr_2,\la,(\Id\circ\mu,\id_X),(\Inv\circ\pr_1,\la),(\txm\circ\pr_{1,3},\pr_4)\bigr)
\qqq
with source $\widetilde s\equiv\pr_2$,\ target $\widetilde t\equiv\la$,\ unit
\qq\nn
\widetilde\Id\equiv(\Id\circ\mu,\id_X)\colo X\too\xcG\,\fibx{s}{\mu}X,\ x\longmapsto\bigl(\Id_{\mu(x)},x\bigr)\,,
\qqq
inverse 
\qq\nn
\widetilde\Inv\equiv(\Inv\circ\pr_1,\la)\colo\xcG\,\fibx{s}{\mu}X\too\xcG\,\fibx{s}{\mu}X,\ (g,x)\longmapsto\bigl(g^{-1},\la(g,x)\bigr)\,,
\qqq
and multiplication 
\qq\nn
\widetilde\txm\equiv(\txm\circ\pr_{1,3},\pr_4)\colo(\xcG\,\fibx{s}{\mu}X)\ \fibx{\pr_2}{\la}(\xcG\,\fibx{s}{\mu}X)\to\xcG\,\fibx{s}{\mu}X,\ \bigl(\bigl(h,\la(g,x)\bigr),(g,x)\bigr)\mapsto(h.g,x)\,.
\qqq
It is called the {\bf action groupoid} (of $(X,\mu,\la)$).\ In particular,\ for $\Gr=\bbB\txG$ from Ex.\,\ref{eg:deloop} and a $\txG$-manifold $(X,\la)$,\ we recover the action groupoid $\txG\,\lx_\la X$ from Ex.\,\ref{eg:actgrpd}.
\end{propanition}

\berop\label{prop:actgrpd-morph-grpd}
Adopt the notation of Prop./Def.\,\ref{prop:grpd-act-grpd}.\ The moment map $\mu$ determines a canonical Lie-groupoid morphism 
\qq\nn
\Phi[\mu]\colo\xcG\,\lx_\la X\too\Gr\,.
\qqq
For $(X,\la,\mu)=(M,t\circ\pr_1,\id_M)$,\ the morphism is an isomorphism.
\eerop

\berop\label{prop:G-in-B_act}
Adopt the notation of Defs.\,\ref{def:bisec} and \ref{def:gr-mod}.\ The action $\la\colo \xcG\,\fibx{s}{\mu}X\to X$ of $\xcG$ canonically induces a left action $B\la\colo\bB(\xcG)\x X\to X$ of $\bB(\xcG)$ determined by the commutative diagram
\qq\nn
\alxydim{@C=1.75cm@R=1.25cm}{ \bB(\xcG)\x M\,\fibx{\id_M}{\mu}X \ar[r]^{\qquad\ev\x\id_X} & \xcG\,\fibx{s}{\mu}X \ar[d]^{\la} \\ \bB(\xcG)\x X \ar[u]^{\id_{\bB(\xcG)}\x(\mu,\id_X)} \ar[r]_{\quad B\la} & X }\,,
\qqq
in which $\ev\colo \bB(\xcG)\x M\to\xcG,\ (\b,m)\mapsto\b(m)$ is the {\bf evaluation map}.\ An analogous statement holds true for right $\xcG$-modules.
\eerop

\bedef\cite[Sec.\,1.2]{Moerdijk:1991}\label{def:princ-gr-bun} 
Adopt the notation of Def.\,\ref{def:gr-mod}.\ A \textbf{right principal $\xcG$-bundle} is a quintuple $(\breP,\Si,\pi_{\breP},\mu,\varrho)$ composed of a pair of smooth manifolds: 
\bit
\item the \textbf{total space} $\breP$;
\item the \textbf{base} $\Si$,
\eit 
and a triple of smooth maps: 
\bit
\item a surjective submersion $\pi_{\breP}\colo \breP\to\Si$,\
termed the \textbf{bundle projection}; 
\item the \textbf{moment map} $\mu\colo \breP\to M$;
\item the \textbf{action} ({\bf map}) $\varrho\colo \breP\fibx{\mu}{t}\xcG\to\breP$,
\eit
with the following properties:
\bit
\item[(PGr1)] $(\breP,\mu,\varrho)$ is a right $\xcG$-module;
\item[(PGr2)] $\pi_{\breP}$ is $\xcG$-invariant in the sense made precise by the commutative diagram
\qq\nn
\alxydim{@C=1.cm@R=1.cm}{\breP\fibx{\mu}{t}
	\xcG \ar[r]^{\quad\varrho} \ar[d]_{\pr_1} & \breP
	\ar[d]^{\pi_{\breP}}\cr \breP \ar[r]_{\pi_{\breP}} & \Si}\,;
\qqq
\item[(PGr3)] the map
\qq\nn
(\pr_1,\varrho)\colo \breP\fibx{\mu}{t}\xcG\too\breP\fibx{\pi_{\breP}}{\pi_{\breP}}\breP,\ (p,g)\longmapsto(p,p\mact g)
\qqq
is a diffeomorphism,\ so that $\xcG$ acts freely and transitively on $\pi_{\breP}$-fibres.\ The smooth inverse of $(\pr_1,\varrho)$ takes the form
\qq\label{eq:div-map}
(\pr_1,\varrho)^{-1}=:(\pr_1,\phi_{\breP})\,,\qquad\qquad\phi_{\breP}\colo \breP\fibx{\pi_{\breP}}{\pi_{\breP}}\breP\too\xcG\,,
\qqq
and $\phi_{\breP}$ is called the \textbf{division map}.
\eit
\noindent The Lie groupoid $\xcG$ is termed the {\bf structure groupoid} of $\breP$ in this setting.

We represent a right principal $\xcG$-bundle by the simplified diagram 
\qq\label{diag:princGrdiag}\qquad\qquad
\alxydim{@C=.75cm@R=1cm}{ \breP \ar@{->>}[d]_{\pi_\breP} \ar[rd]^{\mu} & & \xcG \ar@{=>}[ld] \ar@/_1.pc/[ll]_{\varrho} \\ \Si & M & }\,,
\qqq
in which the remaining structure is implicit.

Let $\Gr_a\equiv\grpd{\xcG_a}{M_a},\ a\in\{1,2\}$ be a pair of Lie groupoids and let $(\breP{}_a,\Si_a,\pi_{\breP{}_a},\mu_a,\varrho_a)$ be the respective right principal $\xcG_a$-bundles.\ A \textbf{morphism} between the two bundles is a triple $(\Theta,f,\Phi)$ composed of a map $f\colo\Si_1\to\Si_2$ between the bases,\ and a morphism $(\Theta,\Phi)$ between the corresponding right $\xcG_a$-modules $(\breP{}_a,\mu_a,\varrho_a)$ which covers $f$ in the sense expressed by the commutative diagram 
\qq\nn
\alxydim{@C=1.cm@R=1.cm}{ \breP{}_1 \ar[r]^{\Theta} \ar[d]_{\pi_{\breP{}_1}} & \breP{}_2
	\ar[d]^{\pi_{\breP{}_2}}\cr \Si_1 \ar[r]_{f} & \Si_2 }\,.
\qqq

{\bf Left principal $\xcG$-bundles} $(\breP,\Si,\pi_\breP,\mu,\la)$ (and morphisms between them) are defined analogously.\ The corresponding diagrams take the self-explanatory form
\qq\nn
\alxydim{@C=.75cm@R=1cm}{ \xcG \ar@{=>}[rd] \ar@/^1.pc/[rr]^{\la} & & \breP \ar@{->>}[d]^{\pi_\breP} \ar[ld]_{\mu} \\ & M & \Si}\,,
\qqq
\exdef

\beg\cite[Remark 5.34(1)]{Moerdijk:2003mm}\label{eg:triv-grpd-bndle}
There exists a canonical structure of a right principal $\xcG$-bundle on $\grpd{\xcG}{M}$,\ given by 
\qq\nn
U_\xcG:=(\xcG,M,t,s,r)\,,
\qqq 
where $r$ is the right-fibred action \eqref{eq:Rgrpdact}.\ We call this structure the \textbf{unit bundle of} $\grpd{\xcG}{M}$.
\eeg

\bedef\cite{Pradines:1977,Hilsum:1983,Hilsum:1987,Haefliger:1984}\label{def:bibndl}
Adopt the notation of Def.\,\ref{def:gr-mod}.\ Let $\grpd{\xcG_a}{M_a},\ a\in\{1,2\}$ be Lie groupoids.\ A {\bf $(\xcG_1,\xcG_2)$-bibundle} is a quintuple $(\widehat P;\mu_1,\la_1;\mu_2,\varrho_2)$ composed of 
\bit
\item a smooth manifold $\widehat P$,\ termed the {\bf total space};
\item smooth maps $\mu_a\colo\widehat P\to M_a,\ a\in\{1,2\}$,\ termed the {\bf left} ($a=1$) and {\bf right} ($a=2$) {\bf moment map},\ respectively;
\item a smooth map $\la_1\colo\xcG_1\,\fibx{s_1}{\mu_1}\widehat P\to\widehat P$,\ termed the {\bf left action},\ and 
\item a smooth map $\varrho_2\colo\widehat P\fibx{\mu_2}{t_2}\xcG_2\to\widehat P$,\ termed the {\bf right action}\,,
\eit
such that
\bit 
\item $(\widehat P,\mu_1,\la_1\equiv\mlact)$ is a left $\xcG_1$-module;
\item $(\widehat P,\mu_2,\varrho_2\equiv\mact)$ is a right $\xcG_2$-module;
\item the two actions commute with one another,\ {\it i.e.},\ we have,\ for all $(g_1,p,g_2)\in\xcG_1\,\fibx{s_1}{\mu_1}\widehat P\fibx{\mu_2}{t_2}\xcG_2$,\ the identity $(g_1\mlact p)\mact g_2=g_1\mlact(p\mact g_2)$;
\item each moment map is invariant with respect to the other action,\ {\it i.e.},\ we have the identities $\mu_1(p\mact g_2)=\mu_1(p)$ and $\mu_2(g_1\mlact p)=\mu_2(p)$.
\eit

Whenever $(\widehat P,M_2,\mu_2,\mu_1,\la_1)$ is a (left) principal $\xcG_1$-bundle (with base $M_2\cong\widehat P/\xcG_1$),\ and $(\widehat P,M_1,\mu_1,\mu_2,\varrho_2)$ is a (right) principal $\xcG_2$-bundle (with base $M_1\cong\widehat P/\xcG_2$),\ in the sense of Def.\,\ref{def:princ-gr-bun},\ we call $(\widehat P;\mu_1,\la_1;\mu_2,\varrho_2)$ a ({\bf bi}){\bf principal $(\xcG_1,\xcG_2)$-bibundle},\ and depict it by the $W$-shaped diagram:
\qq\label{diag:W-diagram}
\alxydim{@C=.75cm@R=1.cm}{ \xcG_1 \ar@{=>}[rd] \ar@/^1.pc/[rr]^{\la_1} & & \widehat P \ar[dl]_{\mu_1} \ar[rd]^{\mu_2} & & \xcG_2 \ar@{=>}[ld] \ar@/_1.pc/[ll]_{\varrho_2} \\ & M_1 & & M_2 & }\,.
\qqq
\exdef

\beg\label{eg:can-biprinc}
Every Lie groupoid $\grpd{\xcG}{M}$ carries a canonical structure of a biprincipal $(\xcG,\xcG)$-bundle relative to the left- and right-fibred actions from Ex.\,\ref{eg:fibred-act},\ captured by the $W$-diagram
\qq\nn
\alxydim{@C=.75cm@R=1.cm}{ \xcG \ar@{=>}[rd] \ar@/^1.pc/[rr]^{l} & & \xcG \ar[dl]_{t} \ar[rd]^{s} & & \xcG \ar@{=>}[ld] \ar@/_1.pc/[ll]_{r} \\ & M & & M & }\,,
\qqq
\eeg

\bedef\cite{Strobl:2025}\label{def:Trident-oid}
Adopt the notation of Def.\,\ref{def:bibndl},\ and assume $(\widehat P;\mu_1,\la_1;\mu_2,\varrho_2)$ biprincipal.\ Let $\Si$ be a manifold,\ and let $\grpd{\xcG_3}{M_3}$ be a Lie groupoid.\ We call the sextuple $(\xcG_1,\widehat P,\xcG_2;\pi_{\widehat P},\Si;\xcG_3)$ a ({\bf left}) {\bf trident} with {\bf base} $\Si$ and fibre $\xcG_3$ if the following conditions are satisfied:
\bit
\item $\widehat P$ is the total space of a fibre bundle $\pi_{\widehat P}\colo\widehat P\to\Si$ with base $\Si$ and typical fibre $\xcG_3$;
\item the right $\xcG_2$-action $\varrho_2\equiv\mact$ preserves $\pi_{\widehat P}$-fibres,\ {\it i.e.},\ $\pi_{\widehat P}(p\mact g_2)=\pi_{\widehat P}(p)$ for all $(p,g_2)\in\widehat P\fibx{\mu_2}{t_2}\xcG_2$;
\item the Lie groupoid $\grpd{\xcG_1}{M_1}$ is the total space of a fibre-bundle object in the category of Lie groupoids,\ with base ${\rm Pair}(\Si)$ and typical fibre $\grpd{\xcG_3}{M_3}$;
\item the (left-)$\xcG_1$-module structure on $\widehat P$ covers the canonical left ${\rm Pair}(\Si)$-module structure on $\Si$,\ and is modelled on the canonical left $\xcG_3$-module structure on $\xcG_3$ (in a local trivialisation),\ as captured by the diagram:
\qq\nn
\alxydim{@C=.75cm@R=1.cm}{ \xcG_3 \ar@{=>}[rd] \ar@/^1.pc/[rr]^{l_3} \ar@{^{(}.>}[dd] & & \xcG_3 \ar[dl]^{t_3} \ar@{^{(}.>}[dd] \\ & M_3 \ar@{^{(}.>}[dd] & & \\ \xcG_1 \ar@{=>}[rd]\ar@/^1.pc/[rr]^(.46){\la_1} |!{[ur];[dr]}\hole \ar[dd]_{\pi_{\xcG_1}} & & \widehat P \ar[dl]^{\mu_1} \ar[dd]^{\pi_{\widehat P}} \\ & M_1 \ar[dd]^(.6){\pi_{M_1}} & & \\ \Si\x\Si \ar@{=>}[rd] \ar@/^1.pc/[rr]^(.4){\pr_1\circ\pr_1} |!{[ur];[dr]}\hole & & \Si \ar@{=}[dl] \\ & \Si & }\,.
\qqq
\eit
The trident is represented by the following diagram:
\qq\label{diag:trident}
\alxydim{@C=.75cm@R=1cm}{ & \xcG_3 \ar@{^{(}.>}[dr] & & & \\ \xcG_1 \ar@{=>}[rd] \ar@/^1.pc/[rr]^{\la_1} & & \widehat P \ar[dl]_{\mu_1} \ar[rd]^{\mu_2} \ar[dd]^{\pi_{\widehat P}} & & \xcG_2 \ar@{=>}[ld] \ar@/_1.pc/[ll]_{\varrho_2} \\ & M_1 & & M_2 & \\ & & \Si & &}\,.
\qqq 
\exdef

\bedef\label{def:Liealgbrd}
Let $X$ be a smooth manifold.\ A ({\bf real}) {\bf Lie algebroid over} $X$ ({\bf of rank} $N\in\bN^\x$) is a triple $((\cE,X,\bR^N,\pi_\cE),\a_\cE,[\cdot,\cdot]_\cE)$ composed of
\bit
\item a vector bundle $(\cE,X,\bR^N,\pi_\cE)$,\ with total space $\xcE$,\ base $X$,\ typical fibre $\bR^N$,\ and base projection $\pi_\cE\colo\cE\to X$;
\item a (vertical) vector-bundle morphism 
\qq\nn
\alxydim{@C=1.cm@R=1.cm}{ \cE \ar[r]^{\a_\cE\quad} \ar[d]_{\pi_\cE} & \txT X \ar[d]^{\pi_{\txT X}}\\ X \ar@{=}[r] & X}\,,
\qqq
termed the {\bf anchor} ({\bf map});
\item a binary operation $[\cdot,\cdot]_\cE\colo \G(\cE)\x\G(\cE)\to\G(\cE)$,
\eit
satisfying the following conditions:
\bit
\item $[\cdot,\cdot]_\cE$ is a Lie bracket;
\item $\forall \vep_1,\vep_2\in\G(\cE)\ \forall f\in C^\infty(X,\bR)\colo [\vep_1,f\lact\vep_2]_\cE=f\lact[\vep_1,\vep_2]_\cE+\a_\cE(\vep_1)(f)\lact\vep_2$ (the {\bf Leibniz rule}).
\eit
\exdef

\beg\label{eg:tanLiealgbrd}
Fix a Lie groupoid $\grpd{\xcG}{M}$.\ Let 
\qq\nn
\G(\txT\xcG)_{\rm R}=\bigl\{\  \xcV\in\G(\ker\,\txT s) \quad\vert\quad \forall g\in\xcG\ \forall h\in s^{-1}(\{t(g)\})\colo \txT_h r_{g}\bigl(\xcV(h)\bigr)=\xcV(h.g) \ \bigr\}
\qqq
be the set of {\bf right-invariant vector fields on} $\xcG$,\ defined in terms of the right-fibred action from Ex.\,\ref{eq:Rgrpdact}.\ The ({\bf right}) {\bf tangent algebroid of} $\xcG$ is the vector bundle 
\qq\nn
\pr_1\colo\Lie(\xcG)\equiv\xcE:=\Id^*\ker\,\txT s=M\,\fibx{\Id}{\pi_{\txT\xcG}}\ker\,\txT s\too M
\qqq 
with anchor 
\qq\nn
\a_\xcE=\txT t\circ\pr_2\,,
\qqq 
and Lie bracket 
\qq\nn
[\cdot,\cdot]_\xcE=\iota_{\rm R}\circ[\cdot,\cdot]_{\txT\xcG}\circ\bigl(\iota_{\rm R}^{-1}\x\iota_{\rm R}^{-1}\bigr)
\qqq
induced by the canonical isomorphism $\iota_{\rm R}\colo \G(\txT\xcG)_{\rm R}\to\G(\xcE),\ \xcV\mapsto(\cdot,\xcV(\Id(\cdot)))$.
\eeg

\beg
Adopt the notation of Ex.\,\ref{eg:deloop}.\ The Lie algebroid $\Lie(\bbB\txG)$  is the tangent Lie algebra $\ggt\equiv{\rm Lie}(\txG)$ of the Lie group $\txG$.
\eeg

\beg
Adopt the notation of Ex.\,\ref{eg:pairgrpd}.\ The Lie algebroid $\Lie({\rm Pair}(M))$ is the tangent bundle $\txT M$,\ with anchor $\id_{\txT M}$,\ and with the Lie bracket given by the commutator of vector fields on $M$.
\eeg

\beg\label{eg:actalgbrd}
Adopt the notation of Ex.\,\ref{eg:actgrpd}.\ The Lie algebroid $\ggt\,\lx_\la M\equiv\Lie(\txG\,\lx_\la M)$  is a vector bundle canonically isomorphic to the trivial bundle
\qq\nn
\alxydim{@C=1.5cm@R=1.5cm}{ \ggt \ar@{^{(}~>}[r] & \ggt\x M \ar[d]_{\pr_2} \\ & M }\,,
\qqq
with the anchor canonically induced by the fundamental vector field $\cK\equiv-\txT\la\circ(\cdot,\brd0_{\txT M})\colo\ggt\to\G(\txT M)$ of $\la$,\ and ---for a choice $\{t_a\}_{a\in\ovl{1,\dim\,\txG}}$ of basis in $\ggt$---evaluating on $(f^a(\cdot)\, t_a,\cdot)\in\G(\ggt\x M)$ as
\qq\nn
\a_{\ggt\,\lx_\la M}\bigl(f^a(\cdot)\, t_a,\cdot\bigr)=-f^a(\cdot)\,\cK_{t_a}(\cdot)\,,
\qqq
and with the Lie bracket canonically induced by that on $\ggt$---with basis presentation $[t_a,t_b]_\ggt=f_{ab}^{\ \ c}\,t_c$---and the commutator on $\G(\txT M)$ as
\qq\nn
[f_1^a\,t_a,f_2^b\,t_b]_{\ggt\,\lx_\la M}=\bigl(f_2^b\,\cK_{t_b}\bigl(f_1^a\bigr)-f_1^b\,\cK_{t_b}\bigl( f_2^a\bigr)-f_1^b\,f_2^c\,f_{bc}^{\ \ a}\bigr)\,t_a\,.
\qqq
\eeg

\beg\cite[Chap.\,II,\,Prop.\,2.1]{Coste:1987}\label{eg:cotang-Liealgd}
The tangent algebroid of the symplectic Lie groupoid from Ex.\,\ref{eg:symplgrpd} is canonically isomorphic to the {\bf cotangent Lie algebroid}
\qq\nn
\bigl(\bigl(\txT^*M,M,\bR^{\dim\,M},\pi_{\txT^*M}\bigr),\Pi^\#,\{\cdot,\cdot\}\bigr)\,,
\qqq
with anchor 
\qq\label{eq:anchor-Poisson}
\Pi^\#\colo\txT^*M\to\txT M\,,
\qqq
canonically induced by the Poisson bivector $\Pi$,\ and Lie bracket
\qq\nn
\{\cdot,\cdot\}\colo\G(\txT^*M)\x\G(\txT^*M)\too\G(\txT^*M),\ (\eta_1,\eta_2)\longmapsto\imath_{\Pi^\#(\eta_1)}\sfd\eta_2-\imath_{\Pi^\#(\eta_2)}\sfd\eta_1+\sfd\bigl(\eta_1\wedge\eta_2(\Pi)\bigr)\,,
\qqq 
aka the {\bf Koszul bracket} \cite{Koszul:1985bra}.\ The isomorphism identifies a section $\a\in\G(\xcE)$ with the differential 1-form $\eta[\a]\in\Om^1(M)$ such that \footnote{Note the transition from left-invariant vector fields on $\xcG$,\ considered in the original paper,\ to right-invariant vector fields employed in our reasoning.} \cite[II\ \S 2]{Coste:1987}
\qq\label{eq:smplalgbrd-secs-as-forms}
t^*\eta[\a]=\imath_\a\om\,.
\qqq
\eeg

\bethe\cite{Rybicki:2002ALG}\label{thm:LieBis}
Adopt the notation of Defs.\,\ref{def:bisec},\ and of Ex.\,\ref{eg:tanLiealgbrd}.\ The Lie algebra of the (Fr\'echet--)Lie group $\bB(\xcG)$ is isomorphic to the Lie algebra $\G_{\rm c}(\xcE)$ of smooth compactly supported sections of $\xcE\equiv\Lie(\xcG)$.
\ethe

\begin{propanition}\label{prop:algbd-act-algbd}
Adopt the notation of Def.\,\ref{prop:grpd-act-grpd}.\ The tangent algebroid $\xcE\,\lx_\la X\equiv\Lie(\xcG\,\lx_\la X)$ is a vector bundle canonically isomorphic to the pullback bundle
\qq\nn
\alxydim{@C=1.cm@R=1.cm}{ \mu^*\xcE\equiv X\,\fibx{\mu}{\pr_1}\Id^*\ker\,\txT s \ar[r]^{\qquad\pr_2} \ar[d]_{\pi_{\xcE\,\lx_\la X}\equiv\pr_1} & \Id^*\ker\,\txT s\equiv\xcE \ar[d]^{\pr_1\equiv\pi_\xcE} \\ X \ar[r]_{\mu} & M }
\qqq
with the anchor $\a_{\xcE\,\lx_{\la}X}$ given by the vector-bundle morphism
\qq\nn
\alxydim{@C=5cm@R=1.cm}{ \mu^*\xcE \ar[r]^{\a_{\xcE\,\lx_\la X}\equiv \txT\la\circ(\pr_2,\brd0_{\txT X}\circ\pr_1)} \ar[rd]_{\pi_{\xcE\,\lx X}} & \txT X \ar[d]^{\pi_{\txT X}} \\ & X }\,,
 \qqq
and a Lie bracket canonically induced,\ through the Leibniz rule,\ from that on $\G(\xcE)$.\ This Lie algebroid is called the {\bf action algebroid} (over $(X,\mu,\la)$).
\end{propanition}
\beroof
The statement of the theorem follows straightforwardly from the following observation:
\qq\nn
(V,v)\in\txT_{(g,x)}\bigl(\xcG\,\fibx{s}{\mu}X\bigr)\subset\txT_g\xcG\oplus\txT_x X\qquad\Longleftrightarrow\qquad\txT_g s(V)=\txT_x\mu(v)\,,
\qqq
put in conjunction with $(\ker\,\txT\widetilde s)_{(g,x)}\equiv\ker\,\txT_{(g,x)}\pr_2\equiv(\ker\,\txT s)_g\subset \txT_g\xcG\subset\txT_g\xcG\oplus\txT_x X$,\ 
and so also $(\widetilde\Id^*\ker\,\txT\widetilde s)_x\equiv(\ker\,\txT\widetilde s)_{(\Id_{\mu(x)},x)}\equiv(\ker\,\txT s)_{\Id_{\mu(x)}}\equiv\xcE_{\mu(x)}$,\ and subsequently combined with the identification $\txT\widetilde t\equiv\txT\la$.
\eroof

\brem\label{rem:algbd-act-algbd}
The assignment $\G(\mu^*\xcE)\ni\cV\to\a_{\xcE\,\lx X}\circ\cV\in\G(\txT M)$ shall be termed the {\bf fundamental vector field} for $\la$.
\erem

\berop\label{prop:Bis-on-E}
Adopt the notation of Defs.\,\ref{def:tstar} and \ref{def:bisec-act},\ and of Ex.\,\ref{eg:tanLiealgbrd}.\ There is a canonical action $\unl{\txT C}\colo\bB(\xcG)\to{\rm GL}(\xcE)$ 
of the group $\bB(\xcG)$ on the tangent algebroid $\xcE\equiv\Lie(\xcG)$ by vector-bundle isomorphisms.\ It lifts the restriction of the conjugation $C$ to the identity bisection $\Id(M)\subset\xcG$,\ and so covers the shadow action $t_*$ in the sense captured by the commutative diagram,\ written for $\b\in\bB(\xcG)$,
\qq\label{eq:Bis-on-E}
\alxydim{@C=1.cm@R=1.cm}{ \xcE \ar[r]^{\unl{\txT C}_\b} \ar[d]_{\pi_\xcE} & \xcE \ar[d]^{\pi_\xcE}\\ M \ar[r]_{t_*\b} & M}\,.
\qqq
\eerop
\beroof
Obvious (see,\ {\it e.g.}:\ \cite[Sec.\,3.1]{Crainic:2015msp}).
\eroof

\berop\cite[Sec.\,3.1]{Crainic:2015msp}\label{prop:J1GonMandE}
Adopt the notation of Ex.\,\ref{def:J1B}.\ There is a canonical structure of (left) $J^1\xcG$-module on $\txT M$,\ with moment map $\pi_{\txT M}\colo\txT M\to M$,\ and action map
\qq\nn
\widetilde t\colo J^1\xcG\fibx{\pi_{J^1}}{\pi_{\txT M}}\txT M\too\txT M,\ \bigl(j^1_m\b,v\bigr)\longmapsto\bigl(\txT_{\b(m)}t\circ \txT_m\b\bigr)(v)\equiv\widetilde t{}_{j^1_m\b}(v)\,,
\qqq
written in terms of an arbitrary representative $\b\in\bB_{\rm loc}(\xcG)$ of $j^1_m\b$.\ There is a similar structure on the tangent algebroid $\xcE\equiv\Lie(\xcG)$,\ with moment map $\pi_\xcE\colo\xcE\to M$,\ and action
\qq\nn
\widetilde C\colo J^1\xcG\fibx{\pi_{J^1}}{\pi_\xcE}\xcE\too\xcE,\ \bigl(j^1_m\b,\vep\bigr)\longmapsto\bigl(\txT_{\b(m)}r_{\b(m)^{-1}}\circ \txT_{(\b(m),\Id_m)}\txm\bigr)\bigl(\bigl(\txT_m\b\circ \txT_{\Id_m}t\bigr)(\vep),\vep\bigr)\equiv\widetilde C{}_{j^1_m\b}(\vep)\,.
\qqq

For every $j^1_m\unl\b\in J^1\xcG$ such that there exists $\b\in\bB(\xcG)$ with the property $j^1(\b)(m)=j^1_m\unl\b$,\ written in terms of the group homomorphism \eqref{eq:j1-hom},\ the following identities hold true
\qq\label{eq:tildetC-4-glob}
\widetilde t{}_{j^1_m\unl\b}=\txT_m(t_*\b)\,,\qquad\qquad\qquad\widetilde C{}_{j^1_m\unl\b}=\unl{\txT C}_\b\rstr_{\xcE_m}\,.
\qqq
\eerop
\beroof
The statement of the proposition is a faithful transcription of the explicit construction from the paper quoted,\ from which also the two identities follow straightforwardly.
\eroof

\beg[\cite{Fernandes:2014}]\label{eg:MC}
Adopt the notation of Def.\,\ref{eg:tanLiealgbrd}.\ The {\bf right-invariant Maurer--Cartan form on} $\xcG$ is the $\ker\,\txT  s$-foliated 1-form with values in $\xcE\equiv\Lie(\xcG)$ given in  
\qq\nn
\alxydim{@C=1.cm@R=1.cm}{ \ker\,\txT s \ar[r]^{\ \MCR} \ar[d]_{\pi_{\txT\xcG}\rstr_{\ker\,\txT s}} & \xcE \ar[d]^{\pi_\xcE\equiv\pr_1}\\ \xcG \ar[r]_{t\ } & M}\,,\qquad\quad\MCR(g)(w)=\bigl(t(g),\txT_g r_{g^{-1}}(w)\bigr)\,,\quad w\in(\ker\,\txT s)_g\,.
\qqq
Thus,\ $\MCR(g)\in(\ker\,\txT s)^*_g\ox_\bR(\ker\,\txT s)_{\Id_{t(g)}}$.
\eeg

\section{Simplicial tools,\ and the Bott--Schulman--Stasheff cohomology}

Below,\ we list the requisites of the simplicial model of equivariant cohomology of Lie groupoids and its sheaf-theoretic resolution employed in the discussion of multiplicative BSS extensions of de Rham cohomology classes and their geometrisations.

\becon
Let $\triangle$ be the standard simplex category,\ with objects $[n]=\{0,1,2,\ldots,n\}\subset\bN$ (see,\ {\it e.g.}:\ \cite{Kan:1957}).\ We denote the standard generators of its $\hom$-sets (of order-preserving maps) as
\qq\nn
\d^{(n+1)}_i\colo[n]\too[n+1],\ k\longmapsto\left\{ \barr{cl} k & {\rm if}\ k<i \\ k+1 & {\rm if}\ k\geq i \earr \right.\,,\qquad i\in\ovl{0,n}
\qqq
({\bf coface maps}),\ and
\qq\nn
\si^{(n)}_j\colo[n+1]\too[n],\ k\longmapsto\left\{ \barr{cl} k & {\rm if}\ k\leq j \\ k-1 & {\rm if}\ k>j \earr \right.\,,\qquad j\in\ovl{0,n}
\qqq
({\bf codegeneracy maps}).
\econ

\bedef
For $n\in\bN$,\ the {\bf topological $n$-simplex} is the topological space
\qq\nn
\triangle^n:=\bigl\{ \ (t_1,t_2,\ldots,t_{n+1})\in\bR_{\geq 0}^{\x n+1} \quad\vert\quad \sum_{k=1}^{n+1}\,t_k=1 \ \bigr\}\subset\bR^{\x n+1}\,.
\qqq
For $i\in\ovl{0,n+1}$,\ the {\bf $i^{\rm th}$ $n$-face inclusion} is the map
\qq\nn
\p_i^{(n+1)}\colo\triangle^n\too\triangle^{n+1},\ (t_1,t_2,\ldots,t_{n+1})\longmapsto(t_1,t_2,\ldots,t_i,0,t_{i+1},t_{i+2},\ldots,t_{n+1})\,.
\qqq
\exdef

\bedef\label{def:simpl-obj}
Let $\cC$ be a category.\ A {\bf simplicial object} in $\cC$ is a presheaf $C_\bullet\colo\triangle^{\rm op}\to\cC$.\ The functorial images of the generators of the $\hom$-sets $\triangle([k],[l]),\ k,l\in\bN$ are called {\bf face maps} $d^{(n+1)}_i=C_\bullet(\d^{(n+1)}_i),\ i\in\ovl{0,n}$ and {\bf degeneracy maps} $s^{(n)}_j=C_\bullet(\si^{(n)}_j),\ j\in\ovl{0,n}$ (defined for all $n\in\bN$),\ respectively.\ Given simplicial objects $C^a_\bullet,\ a\in\{1,2\}$,\ a {\bf simplicial morphism} between them is a natural transformation $\eta\colo C^1_\bullet\Rightarrow C^2_\bullet$. 

In particular,\ a simplicial object $M_\bullet\colo\triangle^{\rm op}\to\Man$ in the category $\Man$ of smooth manifolds is termed a {\bf simplicial manifold}.\ Its (smooth) face maps determine the {\bf Dupont operators}
\qq\nn
\D^{(\bullet)}_{(n)}:=\sum_{k=0}^{n+1}\,(-1)^k\,d_k^{(n+1)*}\colo\Om^\bullet(M_n)\too\Om^\bullet(M_{n+1})
\qqq
on differential forms on its component objects $M_n=M_\bullet([n])\in\Man$.
\exdef

\beg\cite{Segal:1968}\label{eg:face-nerve-LieGrpd}
Adopt the notation of Defs.\,\ref{def:grpd} and \ref{def:simpl-obj}.\ The {\bf nerve} of a Lie groupoid $\grpd{\xcG}{M}$ is a simplicial manifold $\xcG_\bullet\equiv N_\bullet(\grpd{\xcG}{M})$ with component objects $\xcG_0\equiv N_0(\grpd{\xcG}{M})=M$ and $\xcG_n\equiv N_n(\grpd{\xcG}{M})=\xcG\fibx{s}{t}\xcG\fibx{s}{t}\cdots\fibx{s}{t}\xcG\subset\xcG^{\x n}$ ($n>0$),\ face maps ($m>0$)
\qq\nn
&d^{(1)}_0=t\,,\qquad\qquad d^{(1)}_1=s\,,&\cr\cr
&d^{(m+1)}_i(g_{m+1},g_m,\ldots,g_1)=\left\{ \barr{cl} (g_{m+1},g_m,\ldots,g_2) & {\rm if}\ i=0 \\ (g_{m+1},g_m,\ldots,g_{i+2},g_{i+1}\circ g_i,g_{i-1},g_{i-2},\ldots,g_1) & {\rm if}\ i\in\ovl{1,m} \\ 
(g_m,g_{m-1},\ldots,g_1) & {\rm if}\ i=m+1 \earr \right.\,,&
\qqq
and degeneracy maps
\qq\nn
&s^{(0)}_0=\Id\,,&\cr\cr
&s^{(m)}_j(g_m,g_{m-1},\ldots,g_1)=\left\{ \barr{cl} (g_m,g_{m-1},\ldots,g_1,\Id_{s(g_1)}) & {\rm if}\ j=0 \\ (g_m,g_{m-1},\ldots,g_{j+1},\Id_{t(g_j)},g_j,g_{j-1},\ldots,g_1) & {\rm if}\ j\in\ovl{1,m} \earr \right.\,.&
\qqq
\eeg

\beg\label{eg:Cech-nerve}
Let $M\in\Man$,\ and let $\cO=\{O_i\}_{i\in I}$ be an open cover of $M$.\ The manifold $\check{Y}_\cO:=\bigsqcup_{i\in I}\,O_i$ (endowed with the standard disjoint-sum topology and smooth structure inherited from the submanifolds $O_i\subset M$) submerses $M$ as
\qq\nn
\check{\pi}\colo\check{Y}_\cO\twoheadrightarrow M\,,\ (x,i)\longmapsto x\,.
\qqq
The nerve $N_\bullet({\rm Pair}_{\check\pi}(\check Y{}_\cO))\equiv N_\bullet\cO$ of the submersion groupoid ${\rm Pair}_{\check\pi}(\check Y{}_\cO)$ (see:\ Ex.\,\ref{eg:subm-grpd}),\ with $N_n\cO\cong\bigsqcup_{i_0,i_1,\ldots,i_n\in I}\,O_{i_0 i_1\cdots i_n}$,\ where $O_{i_0 i_1\cdots i_n}\equiv O_{i_0}\cap O_{i_1}\cap\cdots\cap O_{i_n}$,\ is termed the {\bf \v Cech nerve} of $\cO$.\ The corresponding Dupont operators are known as the {\bf \v Cech coboundary operators}.
\eeg

\bedef\cite{Segal:1974}\label{def:fat-geom-real}
Adopt the notation of Def.\,\ref{def:simpl-obj}.\ Let $X_\bullet\colo\triangle^{\rm op}\to\Top$ be a simplicial object in the category $\Top$ of topological spaces,\ with objects $X_n\equiv X_\bullet([n])$ and face maps $\xi^{(n+1)}_i=X_\bullet(\d^{(n+1)}_i),\ i\in\ovl{0,n}$.\ A ({\bf fat}) {\bf geometric realisation} of $X_\bullet$ is the quotient topological space
\qq\nn
\Vert X_\bullet\Vert:=\bigsqcup_{n\in\bN}\,\bigl(X_n\x\triangle^n\bigr)/\sim
\qqq
obtained through identifications---written for $(x,t)\in X_{n+1}\x\triangle^n$---
\qq\nn
\bigl(\xi^{(n+1)}_i(x),t\bigr)\sim\bigl(x,\p_i^{(n+1)}(t)\bigr)\,.
\qqq
\exdef

\berop\label{prop:simpl-ext-princmom}
Adopt the notation of Def.\,\ref{def:princ-gr-bun} and Ex.\,\ref{eg:face-nerve-LieGrpd}.\ For every right principal $\xcG$-bundle $(\breP,\Si,\pi_{\breP},\mu,\varrho)$ with a submersively surjective moment map $\mu$,\ the division map $\phi_\breP$ canonically determines an extension of $\mu$ to a simplicial epimorphism 
\qq\label{eq:simpl-epi-ext}
\vep_\breP\colo N_\bullet\bigl({\rm Pair}_{\pi_\breP}(\breP)\bigr)\twoheadrightarrow N_\bullet(\grpd{\xcG}{M})\,.
\qqq
\eerop
\beroof
The principality of $\varrho$ can equivalently be understood as the isomorphy of the groupoids\footnote{The appearance of the opposite groupoid in the codomain is intuitively clear:\ It is ultimately encoded in the following identities $(s\circ\imath_1)(p_2,p_1)\equiv\pr_1(p_2,\phi_\breP(p_2,p_1))=p_2\equiv\widetilde t(p_2,p_1)$ and $(t\circ\imath_1)(p_2,p_1)\equiv\varrho(p_2,\phi_\breP(p_2,p_1))=p_1\equiv\widetilde s(p_2,p_1)$.}
\qq\nn
\imath\colo{\rm Pair}_{\pi_\breP}(\breP)\xrightarrow{\ \cong\ }\bigl(\breP\rx\xcG\bigr)^{\rm op}\,,
\qqq
with the object component of $\imath$ given by  $\imath_0=\id_{\breP}$,\ and the morphism component induced from $\phi_\breP$ as
\qq\nn
\imath_1=(\pr_1,\phi_\breP)\equiv(\pr_1,\varrho)^{-1}\,.
\qqq
The above implies the existence of a simplicial extension
\qq\nn
\widetilde\imath\colo N_\bullet\bigl({\rm Pair}_{\pi_\breP}(\breP)\bigr)\xrightarrow{\ \cong\ } N_\bullet\bigl(\bigl(\breP\rx\xcG\bigr)^{\rm op}\bigr)\,,\qquad\bigl(\widetilde\imath{}_0,\widetilde\imath{}_1\bigr)\equiv\imath\,.
\qqq
Explicitly,\ we find
\qq\nn
\widetilde\imath{}_n&=&\bigl(\imath_1\x\id_{\xcG^{\x n-1}}\bigr)\circ\bigl(\id_\breP\x\imath_1\x\id_{\xcG^{\x n-2}}\bigr)\circ\cdots\circ\bigl(\id_{\breP^{\x n-2}}\x\imath_1\x\id_\xcG\bigr)\circ\bigl(\id_{\breP^{\x n-1}}\x\imath_1\bigr)\cr\cr 
&\equiv&\bigl(\pr_1,\phi_\breP\circ\pr_{1,2},\phi_\breP\circ\pr_{2,3},\ldots,\phi_\breP\circ\pr_{n,n+1}\bigr)
\qqq
in the convenient presentation 
\qq\nn
N_n\bigl(\bigl(\breP\rx\xcG\bigr)^{\rm op}\bigr)=\breP\fibx{\mu}{t\circ\pr_1}\xcG_n\equiv\breP\fibx{\mu}{t\circ\pr_1} N_n\bigl(\grpd{\xcG}{M}\bigr)\,,\quad n\in\bN^\x\,,
\qqq
In the latter,\ we have the surjective submersions
\qq\nn
\mu_n:=\pr_{2,3,\ldots,n+1}\colo  N_n\bigl(\bigl(\breP\rx\xcG\bigr)^{\rm op}\bigr)\twoheadrightarrow N_n(\grpd{\xcG}{M})\,,
\qqq
which combine with the surjective submersion
\qq\nn
\mu_0:=\mu\colo N_0\bigl(\bigl(\breP\rx\xcG\bigr)^{\rm op}\bigr)\equiv\breP\twoheadrightarrow M\equiv N_0(\grpd{\xcG}{M})
\qqq
to yield a simplicial epimorphism
\qq\nn
\widetilde\mu\ :\  N_\bullet\bigl(\bigl(\breP\rx_\varrho\xcG\bigr)^{\rm op}\bigr)\twoheadrightarrow N_\bullet(\grpd{\xcG}{M})\,.
\qqq
Altogether,\ then,\ we arrive at the sought-after simplicial epimorphism
\qq\nn
\vep_\breP:=\widetilde\mu\circ\widetilde\imath\colo  N_\bullet\bigl({\rm Pair}_{\pi_\breP}(\breP)\bigr)\twoheadrightarrow N_\bullet(\grpd{\xcG}{M})\,.
\qqq
\eroof

\bedef\cite{Bott:1976bss,Dupont:1976,Dupont:1988}\label{def:BSS-complex}
Adopt the notation of Ex.\,\ref{eg:face-nerve-LieGrpd}.\ The {\bf Bott--Schulman--Stasheff} ({\bf BSS}) {\bf complex} of a Lie groupoid $\grpd{\xcG}{M}$ is the cochain bicomplex 
\qq\nn
\bigl(\Om^{\bullet_1}\bigl(\xcG_{\bullet_2}\bigr),\sfd^{(\bullet_1)}_{(\bullet_2)},\D_{(\bullet_2)}^{(\bullet_1)}\bigr)
\qqq
with the two commuting coboundary operators:\ the de Rham differentials $\sfd^{(\bullet_1)}_{(\bullet_2)}\colo\Om^{\bullet_1}(\xcG_{\bullet_2})\to\Om^{\bullet_1+1}(\xcG_{\bullet_2})$,\ and the Dupont operators $\D_{(\bullet_2)}^{(\bullet_1)}\colo\Om^{\bullet_1}(\xcG_{\bullet_2})\to\Om^{\bullet_1}(\xcG_{\bullet_2})$ (see:\ Def.\,\ref{def:simpl-obj}).\ Its total cohomology,\ {\it i.e.},\ the cohomology of the diagonal complex ${\rm Tot}^\bullet(\Om^{\bullet_1}(\xcG_{\bullet_2}),\sfd^{(\bullet_1)}_{(\bullet_2)},\D^{(\bullet_1)}_{(\bullet_2)})$ with components
\qq\nn
{\rm Tot}^r\bigl(\Om^{\bullet_1}(\xcG_{\bullet_2}),\sfd^{(\bullet_1)}_{(\bullet_2)},\D^{(\bullet_1)}_{(\bullet_2)}\bigr)=\bigoplus_{p+q=r}\,\Om^p(\xcG_q)\,,
\qqq
and the {\bf BSS coboundary operator}
\qq\nn
D_{(r)}\rstr_{\Om^p(\xcG_q)}=\D^{(p)}_{(q)}+(-1)^q\,\sfd^{(p)}_{(q)}\,,
\qqq
is called the {\bf BSS cohomology} of $\grpd{\xcG}{M}$.\ We denote its groups as
\qq\nn
H^r_{\rm BSS}(\grpd{\xcG}{M})=\tfrac{\ker\,D_{(r)}}{\im\,D_{(r-1)}}\,.
\qqq
\exdef

\bedef\label{def:mult-k-form}
Adopt the notation of Def.\,\ref{def:BSS-complex}.\ A $k$-form $\rho\in\Om^k(\cG_1)$ is called {\bf multiplicative} if $\rho\in\ker\,\D^{(k)}_{(1)}$.
\exdef

\bedef\label{def:simpl-cover}
Adopt the notation of Def.\,\ref{def:simpl-obj}.\ Let $M_\bullet\colo\triangle^{\rm op}\to\Man$ be a simplicial manifold with face maps $\{d^{(n+1)}_i\ \vert\ i\in\ovl{0,n}\}_{n\in\bN}$.\ A family $\{\cO_{M_n}\}_{n\in\bN}$ of open covers $\cO_{M_n}=\{O^{(n)}_{i_n}\}_{i_n\in I_n}$ of the respective objects $M_n\in\Man$ is called {\bf simplicial} if the index sets $I_n,\ n\in\bN$ compose a simplicial set $I_\bullet\colo\triangle^{\rm op}\to\Set$ with face maps $\iota^{(n+1)}_i\equiv I_\bullet(\d^{(n+1)}_i)\colo I_{n+1}\to I_n$ such that the relation
\qq\nn
d^{(n+1)}_j\bigl(O^{(n+1)}_{i_n}\bigr)\subset O^{(n)}_{\iota^{(n+1)}_j(i_n)}
\qqq
holds true for all $j\in\ovl{0,n+1}$ and $n\in\bN^\x$.
\exdef

\bethe\cite{Tu:2006}\label{thm:simpl-ref}
For every simplicial manifold,\ and every collection of open covers of its objects,\ there exists a refinement of the latter which is simplicial in the sense of Def.\,\ref{def:simpl-cover}.
\ethe

\berop\label{prop:tricompl}
Adopt the notation of Def.\,\ref{def:simpl-obj}.\ Let $M_\bullet\colo\triangle^{\rm op}\to\Man$ be a simplicial manifold with face maps $\{d^{(n+1)}_i\ \vert\ i\in\ovl{0,n}\}_{n\in\bN}$.\ Given a simplicial family $\{\cO_{M_n}\}_{n\in\bN}$ of open covers $\cO_{M_n}=\{O^{(n)}_{i_n}\}_{i_n\in I_n}$ of the respective objects $M_n\in\Man$,\ the Dupont operators $\D^{(\bullet_1)}_{(\bullet_2)}\colo\Om^{\bullet_1}(M_{\bullet_2})\to\Om^{\bullet_1}(M_{\bullet_2+1})$ canonically extend to \v{C}ech cochains with values in the sheaves $\unl{\Om^{\bullet_1}}(M_{\bullet_2})$ of locally smooth $\bullet_1$-forms on $M_{\bullet_2}$ as
\qq\nn
\D^{(p,q)}_{(r)}&\colo&\vC^p\bigl(\cO_{M_r},\unl{\Om^q}\bigr)\too\vC^p\bigl(\cO_{M_{r+1}},\unl{\Om^q}\bigr)\cr\cr 
&\colo&\om\equiv(\om_{j_0 j_1\ldots j_p})\longmapsto\left(\sum_{k=0}^{r+1}\,(-1)^k\,d_k^{(r+1)*}\om_{\iota^{(r+1)}_k(i_0) \iota^{(r+1)}_k(i_1)\ldots \iota^{(r+1)}_k(i_p)}\equiv\bigl(\D^{(p,q)}_{(r)}\om\bigr)_{i_0 i_1\ldots i_p}\right)\,,
\qqq 
and thus give rise to a cochain tricomplex  
\qq\nn
\bigl(\vC^{\bullet_1}\bigl(\cO_{M_{\bullet_3}},\unl{\Om^{\bullet_2}}\bigr),\vd^{(\bullet_1,\bullet_2)}_{(\bullet_3)},\sfd^{(\bullet_1,\bullet_2)}_{(\bullet_3)},\D^{(\bullet_1,\bullet_2)}_{(\bullet_3)}\bigr)\,.
\qqq
\eerop

\section{Gerbes,\ characters,\ equivariance and descent}

In this appendix,\ elements of the theory of gerbes are introduced,\ which are instrumental in the discussion of the gauging of rigid Lie-groupoidal symmetries of the PAG $\si$-model.

\bedef\label{def:Deligne}
Let $X\in\Man$.\ The {\bf degree-2 Deligne complex} of $X$ is the cochain complex
\qq\nn
\xcD^{\bullet}(2;X)\qquad\colo\qquad\brd0\xrightarrow{\hspace{45pt}}\unl{\uj}{}_X\xrightarrow{\ \, \sfd^{(0)}=-\sfi\sfd\log\ \, }\unl{\Om^1}(X)\xrightarrow{\quad\ \sfd^{(1)}=\sfd\quad\ }\unl{\Om^2}(X)
\qqq
of sheaves of abelian groups:\ $\unl{\uj}{}_X\equiv\unl{\uj}$ (locally smooth $\uj$-valued functions on $X$),\ $\unl{\Om^1}(X)\equiv\unl{\Om^1}$ (locally smooth real 1-forms on $X$),\ and $\unl{\Om^2}(X)\equiv\unl{\Om^2}$ (locally smooth real 2-forms on $X$).\ Given an open cover $\cO=\{O_i\}_{i\in I}$ of $X$,\ the corresponding {\bf \v{C}ech--Deligne bicomplex} is the semi-bounded bicomplex
{\small\qq\nn
\alxydim{@C=1.5cm@R=1.cm}{ 0 \ar@{-->}[r] & 0 \ar@{-->}[r] & 0 \ar@{-->}[r] & 0 \ar@{-->}[r] & \cdots \\ \vC^0\bigl(\cO,\unl{\Om^2}\bigr) \ar[r]_{\vd^{(0,2)}} \ar@{-->}[u] & \vC^1\bigl(\cO,\unl{\Om^2}\bigr) \ar[r]_{\vd^{(1,2)}} \ar@{-->}[u] & \vC^2\bigl(\cO,\unl{\Om^2}\bigr) \ar[r]_{\vd^{(2,2)}} \ar@{-->}[u] & \vC^3\bigl(\cO,\unl{\Om^2}\bigr) \ar[r]_{\qquad\vd^{(3,2)}} \ar@{-->}[u] & \cdots \\ \vC^0\bigl(\cO,\unl{\Om^1}\bigr) \ar[r]_{\vd^{(0,1)}} \ar[u]^{\sfd^{(0,1)}} & \vC^1\bigl(\cO,\unl{\Om^1}\bigr) \ar[r]_{\vd^{(1,1)}} \ar[u]^{\sfd^{(1,1)}} & \vC^2\bigl(\cO,\unl{\Om^1}\bigr) \ar[r]_{\vd^{(2,1)}} \ar[u]^{\sfd^{(2,1)}} & \vC^3\bigl(\cO,\unl{\Om^1}\bigr) \ar[r]_{\qquad\vd^{(3,1)}} \ar[u]^{\sfd^{(3,1)}} & \cdots \\ \vC^0\bigl(\cO,\unl{\uj}\bigr) \ar[r]_{\vd^{(0,0)}} \ar[u]^{\sfd^{(0,0)}} & \vC^1\bigl(\cO,\unl{\uj}\bigr) \ar[r]_{\vd^{(1,0)}} \ar[u]^{\sfd^{(1,0)}} & \vC^2\bigl(\cO,\unl{\uj}\bigr) \ar[r]_{\vd^{(2,0)}} \ar[u]^{\sfd^{(2,0)}} & \vC^3\bigl(\cO,\unl{\uj}\bigr) \ar[r]_{\qquad\vd^{(3,0)}} \ar[u]^{\sfd^{(3,0)}} & \cdots}
\qqq}
\hspace{-5pt}with the horizontal differentials $\vd^{(p,q)}=\vd^{(p)}_{\cS_q}$ given by the \v{C}ech coboundary operators (see:\ Ex.\,\ref{eg:Cech-nerve}) of the \v{C}ech complexes with values in the respective sheaves $(\cS_0,\cS_1,\cS_2)=(\unl{\uj},\unl{\Om^1},\unl{\Om^2})$,\ and with the vertical differentials $\sfd^{(p,q)}$ given by the obvious extensions of the differentials $\sfd^{(q)}$.

Its diagonal complex
\qq\nn
A^{\bullet}(\cO)\qquad\colo\qquad\alxydim{@C=1.25cm@R=1.25cm}{ \brd0 \ar[r] & A^0(\cO) \ar[r]^{D^{(0)}} & A^1(\cO) \ar[r]^{D^{(1)}} & A^2(\cO) \ar[r]^{D^{(2)}} & A^3(\cO) \ar[r]^{D^{(3)}} & \cdots}
\qqq
has groups of cochains 
\qq\nn
A^0(\cO)&=&\vC^0\bigl(\cO,\unl{\uj}\bigr)\cr\cr 
A^1(\cO)&=&\vC^0\bigl(\cO,\unl{\Om^1})\bigr)\oplus\vC^1\bigl(\cO,\unl{\uj}\bigr)\cr\cr 
A^2(\cO)&=&\vC^0\bigl(\cO,\unl{\Om^2}\bigr)\oplus\vC^1\bigl(\cO,\unl{\Om^1}\bigr)\oplus\vC^2\bigl(\cO,\unl{\uj}\bigr)\cr\cr 
A^3(\cO)&=&\vC^1\bigl(\cO,\unl{\Om^2}\bigr)\oplus\vC^2\bigl(\cO,\unl{\Om^1}\bigr)\oplus\vC^3\bigl(\cO,\unl{\uj}\bigr)\qquad\ldots
\qqq
and the {\bf Deligne differentials}
\qq\nn
D^{(0)}(f_i)&=&\bigl(-\sfi\,f_i^{-1}\,\sfd f_i,f_j^{-1}\cdot f_i\rstr_{O_{i j}}\bigr)\cr\cr 
D^{(1)}(\Pi_i,\chi_{ij})&=&\bigl(\sfd\Pi_i,-\sfi\,\chi_{ij}^{-1}\,\sfd\chi_{ij}+(\Pi_j-\Pi_i)\rstr_{O_{ij}},\chi_{jk}^{-1}\cdot\chi_{ik}\cdot\chi_{ij}^{-1}\rstr_{O_{ijk}}\bigr)\cr\cr 
D^{(2)}(B_i,A_{ij},g_{ijk})&=&\bigl(\sfd A_{ij}-(B_j-B_i)\rstr_{O_{ij}},-\sfi\,g_{ijk}^{-1}\,\sfd g_{ijk}+(A_{jk}-A_{ik}+A_{ij})\rstr_{O_{ijk}},g_{jkl}^{-1}\cdot g_{ikl}\cdot g_{ijl}^{-1}\cdot g_{ijk}\rstr_{O_{ijkl}}\bigr)\cr\cr
{\it etc.}
\qqq
Its cohomology 
\qq\nn
\bH^\bullet\bigl(\cO,\xcD^{\bullet}(2;X)\bigr)\equiv H^\bullet\bigl(A^\bullet(\cO),D^{(\bullet)}\bigr)
\qqq
is termed the ({\bf real}) {\bf degree-2 Beilinson--Deligne cohomology} of $X$ associated to $\cO$.\ Under refinement of the cover ({\it i.e.},\ for $\cO$ sufficiently fine---{\it e.g.},\ good),\ it yields the ({\bf real}) {\bf degree-2 Beilinson--Deligne cohomology} of $X$
\qq\nn
\bH^\bullet\bigl(X,\xcD^{\bullet}(2;X)\bigr)=\underset{\scriptsize{\cO}}{\varinjlim}\,\bH^\bullet\bigl(\cO,\xcD^{\bullet}(2;X)\bigr)
\qqq
\exdef

\berop\label{prop:BD-vs-Huj}
Adopt the notation of Def.\,\ref{def:Deligne}.\ There exist canonical isomorphisms
\qq\nn
\bH^0\bigl(X,\xcD^{\bullet}(2;X)\bigr)\cong H^0\bigl(X,\uj\bigr)\,,\qquad\qquad\bH^1\bigl(X,\xcD^{\bullet}(2;X)\bigr)\cong H^1\bigl(X,\uj\bigr)\,.
\qqq
\eerop

\bethe\label{thm:W3-tors}
Adopt the notation of Def.\,\ref{def:Deligne}.\ Every class $[(B_i,A_{ij},g_{ijk})]\in\bH^2(X,\xcD^{\bullet}(2;X))$ defines a 3-form $H\in Z^3_{\rm dR}(X)$ with restrictions $H\rstr_{O_i}=\sfd B_i$,\ called the {\bf curvature} of $[(B_i,A_{ij},g_{ijk})]$.\ The set $\cW^3(X;H)$ of classes of curvature $H$ is a torsor under an action of $\cW^3(X;0)$ given by
\qq\nn
&&\cW^3(X;0)\x\cW^3(X;H)\too\cW^3(X;H)\cr\cr  &&\bigl([(\b_i,\a_{ij},\g_{ijk})],[(B_i,A_{ij},g_{ijk})]\bigr)\longmapsto[(B_i+\b_i,A_{ij}+\a_{ij},g_{ijk}\cdot\g_{ijk})]\,.
\qqq
There is a canonical isomorphism
\qq\nn
\cW^3(X;0)\cong H^2\bigl(X,\uj\bigr)\,.
\qqq
\ethe

\bedef\label{def:Grb}
Let $X\in\Man$,\ and let $H\in\Om^3(X)$ be a de Rham 3-cocycle with periods in $2\pi\bZ$.\ A {\bf gerbe} of {\bf curvature} $H$ over $X$ is a sextuple $\cG\equiv(Y,B,L,\cA_L,\mu_L)$ composed of
\bit
\item a surjective submersion $Y\xrightarrow{\pi_Y}X$;
\item a 2-form $B\in\Om^2(Y)$,\ termed the {\bf curving} of $\cG$,\ satisfying the identity $\sfd B=\pi_Y^*H$;
\item a principal $\bC^\x$-bundle $L\xrightarrow{\pi_L}Y\x_X Y$  with principal connection 1-form $\cA_L\in\Om^1(L)$ satisfing the identity $\sfd\cA_L=\pi_L^*(\pr_2^*-\pr_1^*)B$;
\item a connection-preserving isomorphism $\mu_L\colo\pr_{1,2}^*L\ox\pr_{2,3}^*L\xrightarrow{\cong}\pr_{1,3}^*L$ of principal $\bC^\x$-bundles over $Y\x_X Y\x_X Y$ (the tensor product is Brylinski's contracted product from \cite[Sec.\,2.1]{Brylinski:1993ab}),\ satisfying the coherence condition 
\qq\nn
\pr_{1,3,4}^*\mu_L\circ\bigl(\pr_{1,2,3}^*\mu_L\ox\id\bigr)=\pr_{1,2,4}^*\mu_L\circ\bigl(\id\ox\pr_{2,3,4}^*\mu_L\bigr)
\qqq
over $Y\x_X Y\x_X Y\x_X Y$.
\eit

Given gerbes $\cG_a,\ a\in\{1,2\}$ as above,\ a {\bf 1-isomorphism} $\Phi\colo\cG_1\xrightarrow{\cong}\cG_2$ is a quadruple $\Phi\equiv(Z,E,\cA_E,\a_E)$ composed of
\bit
\item a surjective submersion $Z\xrightarrow{\pi_Z}Y_{1,2}\equiv Y_1\x_X Y_2$;
\item a principal $\bC^\x$-bundle $E\xrightarrow{\pi_E}Z$  with principal connection 1-form $\cA_E\in\Om^1(E)$ satisfing the identity $\sfd\cA_E=\pi_E^*\pi_Z^*(\pr_2^*B_2-\pr_1^*B_1)$;
\item a connection-preserving isomorphism $\a_E\colo \pi_Z^{\x 2\,*}\pr_{1,3}^*L_1\ox\pr_2^*E\xrightarrow{\cong}\pr_1^*E\ox\pi_Z^{\x 2\,*}\pr_{2,4}^*L_2$ of principal $\bC^\x$-bundles over $Z\x_X Z$,\ satisfying the coherence condition expressed by the commutative diagram
\qq\nn
\alxydim{@C=3.75cm@R=1.cm}{ \pi_Z^{\x 3\,*}\pr_{1,3}^*L_1\ox\pi_Z^{\x 3\,*}\pr_{3,5}^*L_1\ox\pr_3^*E \ar[r]^{\hspace{35pt}\pi_Z^{\x 3\,*}\pr_{1,3,5}^*\mu_{L_1}\ox\id} \ar[d]_{\id\ox\pr_{2,3}^*\a_E} & \pi_Z^{\x 3\,*}\pr_{1,5}^*L_1\ox\pr_3^*E \ar[dd]^{\pr_{1,3}^*\a_E} \\ \pi_Z^{\x 3\,*}\pr_{1,3}^*L_1\ox\pr_2^*E\ox\pi_Z^{\x 3\,*}\pr_{4,6}^*L_2 \ar[d]_{\pr_{1,2}^*\a_E\ox\id} & \\ \pr_1^*E\ox\pi_Z^{\x 3\,*}\pr_{2,4}^*L_2\ox\pi_Z^{\x 3\,*}\pr_{4,6}^*L_2 \ar[r]_{\hspace{35pt}\id\ox\pi_Z^{\x 3\,*}\pr_{2,4,6}^*\mu_{L_2}} & \pr_1^*E\ox\pi_Z^{\x 3\,*}\pr_{2,6}^*L_2 }
\qqq
over $Z\x_X Z\x_X Z$.
\eit

Given 1-isomorphisms $\Phi_b\colo\cG_1\xrightarrow{\cong}\cG_2,\ b\in\{1,2\}$ as above,\ a {\bf 2-isomorphism} $\psi\colo\Phi_1\overset{\cong}{\Longrightarrow}\Phi_2$ is an equivalence class of triples $(W,\b)$ composed of
\bit
\item a surjective submersion $W\xrightarrow{\pi_W}Z_1\x_{Y_{1,2}}Z_2$;
\item a connection-preserving isomorphism $\b\colo\pi_W^*\pr_1^*E_1\xrightarrow{\cong}\pi_W^*\pr_2^*E_2$ of principal $\bC^\x$-bundles over $W$,\ subject to the coherence condition expressed by the commutative diagram
\qq\nn
\alxydim{@C=2.5cm@R=1.cm}{ \pi_W^{\x 2}{}^*\pr_{1,3}^*\bigl(\pi_{Z_1}^{\x 2\,*}\pr_{1,3}^*L_1\ox\pr_2^*E_1\bigr) \ar[r]^{\pi_W^{\x 2\,*}\pr_{1,3}^*\a_{E_1}} \ar[d]_{\id\ox\pr_2^*\b} & \pi_W^{\x 2\,*}\pr_{1,3}^*\bigl(\pr_1^*E_1\ox\pi_{Z_1}^{\x 2\,*}\pr_{2,4}^*L_2\bigr) \ar[d]^{\pr_1^*\b\ox\id} \\ \pi_W^{\x 2\,*}\pr_{2,4}^*\bigl(\pi_{Z_2}^{\x 2\,*}\pr_{1,3}^*L_1\ox\pr_2^*E_2\bigr) \ar[r]_{\pi_W^{\x 2\,*}\pr_{2,4}^*\a_{E_2}} & \pi_W^{\x 2\,*}\pr_{2,4}^*\bigl(\pr_1^*E_2\ox\pi_{Z_2}^{\x 2\,*}\pr_{2,4}^*L_2\bigr) }
\qqq
over $W\x_X W$.
\eit
We consider pairs $(W_c,\b_c),\ c\in\{1,2\}$ as above equivalent if there exists a manifold $\widetilde W$ together with surjective submersions $\pi_c\colo\widetilde W\to W_c$ obeying $\pi_{W_2}\circ\pi_2=\pi_{W_1}\circ\pi_1$ and such that $\pi_2^*\b_2=\pi_1^*\b_1$.
\exdef

\bethe\cite{Stevenson:2000wj,Waldorf:2007mm}\label{thm:bgrb-bicat}
Adopt the notation of Def.\,\ref{def:Grb}.\ Gerbes over a given base $X\in\Man$,\ together with gerbe 1-isomorphisms between them,\ and gerbe 2-isomorphisms between the latter,\ form a monoidal bicategory $\bgrb_\nabla(X)$ (in the sense of \cite{Benabou:1994bi}) with duality.\ Moreover,\ every smooth map $f\in C^\infty(X_1,X_2)$ canonically induces a (strict monoidal) 2-functor $f^*\colo\bgrb_\nabla(X_2)\to\bgrb_\nabla(X_1)$.
\ethe

\berop\label{prop:gerbes-as-H2}
Adopt the notation of Defs.\,\ref{def:Deligne} and \ref{def:Grb},\ and of Thm.\,\ref{thm:W3-tors}.\ There is a one-to-one correspondence between 1-isomorphism classes of gerbes over $X$ and classes in $\bH^2(X,\xcD^{\bullet}(2;X))$.\ Under this correspondence,\ 1-isomorphism classes of gerbes of curvature $H$ are identified with elements in $\cW^3(X;H)$.\ In particular,\ such gerbes are classified by the cohomology group $H^2(X,\uj)$.
\eerop

\berop\label{prop:1iso-H1}
Adopt the notation of Def.\,\ref{def:Deligne}.\ Let the 2-cocycles $b_a\in\ker\,D^{(2)},\ a\in\{1,2\}$ represent the respective gerbes $\cG_a\in\bgrb_\nabla(X)$ under the correspondence from Prop.\,\ref{prop:gerbes-as-H2}.\ A 1-isomorphism $\Phi\colo\cG_1\xrightarrow{\cong}\cG_2$ is then represented by a 1-cochain $p\in A^1(\cO)$ such that 
\qq\nn
b_2=b_1+D^{(1)}p\,,
\qqq
and a 2-isomorphism $\varphi\colo\Phi_1\overset{\cong}{\Rightarrow}\Phi_2$ between two such 1-isomorphisms $\Phi_b\colo\cG_1\xrightarrow{\cong}\cG_2,\ b\in\{1,2\}$,\ given by the respective 1-cochains $p_b\in A^1(\cO)$,\ is represented by a 0-cochain $f\in A^0(\cO)$ such that 
\qq\nn
p_2=p_1+D^{(0)}f\,.
\qqq
Accordingly,\ 2-isomorphism classes of 1-isomorphisms between given gerbes $\cG_a,\ a\in\{1,2\}$ are classified by the cohomology group $H^1(X,\uj)$.
\eerop

\bedef\cite{Cheeger:1985}\label{def:CS-char}
Let $X\in\Man$,\ and let $C_k(X)$ and $Z_k(X)$,\ for $k\in\bN$,\ denote the abelian groups of (smooth) $k$-chains and $k$-cycles in $X$,\ respectively.\ For $n\in\bN$,\ a {\bf differential character} of {\bf degree} $n$ on $X$ is a pair $(\chi,\om)$ composed of an abelian-group homomorphism
\qq\nn
\chi\colo Z_{n-1}(X)\too\uj
\qqq
and a closed $n$-form $\om_\chi$,\ termed the {\bf curvature} of $\chi$,\ which satisfies the identity
\qq\nn
\chi\circ\p(\cdot)=\exp\left(\sfi\int_\cdot\,\om_\chi\right)\,.
\qqq
The group of differential characters of degree $n$ is denoted as
\qq\nn
\widehat H{}^n\bigl(X,\uj\bigr)\,.
\qqq 
\exdef

\bethe\cite[Prop.\,1.5.7]{Brylinski:1993ab}\label{thm:CH-BD}
Adopt the notation of Defs.\,\ref{def:Deligne} and \ref{def:CS-char}.\ There exists a canonical bijection
\qq\nn
\bH^2\bigl(X,\xcD^{\bullet}(2;X)\bigr)\cong\widehat H{}^2\bigl(X,\uj\bigr)\,.
\qqq
\ethe

\bedef\label{def:holG}
Adopt the notation of Def.\,\ref{def:Grb}.\ The image of the hypercohomology class representing a gerbe $\cG$ under the bijection from Thm.\,\ref{thm:CH-BD} is termed the {\bf holonomy} of $\cG$,\ and denoted as $\Hol_\cG$.
\exdef

\bedef\cite{Stevenson:2000wj,Gawedzki:2010rn}\label{def:desc-bicat}
Adopt the notation of Thm.\,\ref{thm:bgrb-bicat}.\ Let $Y,X\in\Man$,\ and let $\pi\colo Y\twoheadrightarrow X$ be a surjective submersion.\ Denote $Y^{[n]}:=\{ (y_1,y_2,\ldots,y_n)\in Y^{\x n}\,\vert\,\forall_{i,j\in\ovl{1,n}}\ \pi(y_i)=\pi(y_j)\}$,\ and---for every $k\in\ovl{1,n-1}$---$\pr_{i_1 i_2\cdots i_k}=(\pr_{i_1},\pr_{i_2},\ldots,\pr_{i_k})$,\ where $\pr_l\colo Y^{[n]}\to Y$ is the canonical projection onto the $l$-th cartesian factor.\ The {\bf descent bicategory} $\gt{Desc}(\pi)$ for $\pi$ is the bicategory in which 
\bit
\item 0-cells are triples $(\cG,\Phi,\varphi)$ composed of a gerbe $\cG\in\bgrb_\nabla(Y)$,\ a gerbe 1-isomorphism 
\qq\nn
\Phi\colo\pr_1^*\cG\xrightarrow{\ \cong\ }\pr_2^*\cG
\qqq 
over $Y^{[2]}$,\ and a gerbe 2-isomorphism 
\qq\nn
\varphi\colo\pr_{2,3}^*\Phi\circ\pr_{1,2}^*\Phi\xLongrightarrow{\cong}\pr_{1,3}^*\Phi
\qqq 
over $Y^{[3]}$,\ the latter satisfying the coherence identity
\qq\nn
\pr_{1,3,4}^*\varphi\bullet(\Id_{\pr_{3,4}^*\Phi}\circ\pr_{1,2,3}^*\varphi)=\pr_{1,2,4}^*\varphi\bullet(\pr_{2,3,4}^*\varphi\circ\Id_{\pr_{1,2}^*\Phi})
\qqq
over $Y^{[4]}$;
\item 1-cells are pairs $(\Psi,\psi)\colo(\cG_1,\Phi_1,\varphi_1)\xrightarrow{\cong}(\cG_2,\Phi_2,\varphi_2)$ composed of a gerbe 1-isomorphism
\qq\nn
\Psi\colo\cG_1\xrightarrow{\ \cong\ }\cG_2
\qqq
over $Y$,\ and a gerbe 2-isomorphism 
\qq\nn
\psi\colo\pr_2^*\Psi\circ\Phi_1\xLongrightarrow{\cong}\Phi_2\circ\pr_1^*\Psi
\qqq
over $Y^{[2]}$,\ the latter satisfying the coherence identity
\qq\nn
\pr_{1,3}^*\psi\bullet\bigl(\id_{\pr_3^*\Psi}\circ\varphi_1\bigr)=\bigl(\varphi_2\circ\id_{\pr_1^*\Psi}\bigr)\bullet\bigl(\id_{\pr_{2,3}^*\Phi_2}\circ\pr_{1,2}^*\psi\bigr)\bullet\bigl(\pr_{2,3}^*\psi\circ\id_{\pr_{1,2}^*\Phi_1}\bigr)
\qqq
over $Y^{[3]}$;
\item 2-cells $\xi\colo(\Psi_1,\psi_1)\overset{\cong}{\Rightarrow}(\Psi_2,\psi_2)$ in $\gt{Desc}(\pi)((\cG_1,\Phi_1,\varphi_1),(\cG_2,\Phi_2,\varphi_2))\ni(\Psi_1,\psi_1),(\Psi_2,\psi_2)$ are given by gerbe 2-isomorphisms 
\qq\nn
\xi\colo\Psi_1\xLongrightarrow{\ \cong\ }\Psi_2
\qqq
satisfying the coherence identity
\qq\nn
\bigl(\id_{\Phi_2}\circ\pr_1^*\xi\bigr)\bullet\psi_1=\psi_2\bullet\bigl(\pr_2^*\xi\circ\id_{\Phi_1}\bigr)\,.
\qqq
\eit
\exdef

\bedef\cite{Gawedzki:2010rn}\label{def:CDD}
Adopt the notation of Def.\,\ref{def:Deligne} and Ex.\,\ref{eg:face-nerve-LieGrpd}.\ Consider the family of \v{C}ech--Deligne complexes
$(A^{\bullet_1}(2;\xcG_{\bullet_2}),D^{(\bullet_1)}_{(\bullet_2)})$ associated with a {\em simplicial} family $\cO_{\xcG_\bullet}\equiv\{\cO_{\xcG_n}\}_{n\in\bN}$ of open covers $\cO_{\xcG_n}=\{O^{(n)}_{i_n}\}_{i_n\in I_n}$---in the sense of Def.\,\ref{def:simpl-cover}---of the components $\xcG_n\in\Man$ of the nerve $\xcG_\bullet\equiv N_\bullet(\grpd{\xcG}{M})$.\ We shall call the bicomplex 
\qq\nn
\left(A^{\bullet_1}\bigl(\cO_{\xcG_{\bullet_2}}\bigr),D^{(\bullet_1)}_{(\bullet_2)},\unl\D{}^{(\bullet_1)}_{(\bullet_2)}\right)
\qqq
with the vertical differentials
\qq\nn
&\unl\D{}^{(0)}_{(\bullet)}=\D^{(0,0)}_{(\bullet)}\,,\qquad\qquad\unl\D{}^{(1)}_{(\bullet)}=\D^{(0,1)}_{(\bullet)}\oplus\D^{(1,0)}_{(\bullet)}\,,&\cr\cr &\unl\D{}^{(2)}_{(\bullet)}=\D^{(0,2)}_{(\bullet)}\oplus\D^{(1,1)}_{(\bullet)}\oplus\D^{(2,0)}_{(\bullet)}\,,\qquad\qquad\unl\D{}^{(3)}_{(\bullet)}=\D^{(1,2)}_{(\bullet)}\oplus\D^{(2,1)}_{(\bullet)}\oplus\D^{(3,0)}_{(\bullet)}\qquad\qquad{\it etc.}&
\qqq
given in terms of the extensions of Dupont operators from\footnote{We implicitly pass from the additive notation to the multiplicative one for sections of $\unl{\uj}$.} Prop.\,\ref{prop:tricompl},\ the {\bf \v{C}ech--Deligne--Dupont} ({\bf \v{C}DD}) {\bf bicomplex} of $\grpd{\xcG}{M}$ associated to $\cO_{\xcG_\bullet}$.\ Its total cohomology,\ {\it i.e.},\ the cohomology of the diagonal complex 
\qq\nn
{\rm Tot}^\bullet\left(A^{\bullet_1}\bigl(\cO_{\xcG_{\bullet_2}}\bigr),D^{(\bullet_1)}_{(\bullet_2)},\unl\D{}^{(\bullet_1)}_{(\bullet_2)}\right)
\qqq 
with components
\qq\nn
{\rm Tot}^r\left(A^{\bullet_1}\bigl(\cO_{\xcG_{\bullet_2}}\bigr),D^{(\bullet_1)}_{(\bullet_2)},\unl\D{}^{(\bullet_1)}_{(\bullet_2)}\right)=\bigoplus_{p+q=r}\,A^p\bigl(\cO_{\xcG_q}\bigr)\,,
\qqq
and the {\bf \v{C}DD differentials}
\qq\nn
\nabla_{(r)}\rstr_{A^p(\cO_{\xcG_q})}=D^{(p)}_{(q)}+(-1)^p\,\unl\D{}^{(p)}_{(q)}\,,
\qqq
shall be referred to as the {\bf \v{C}DD cohomology} of $\grpd{\xcG}{M}$ associated to $\cO_{\xcG_\bullet}$.
\exdef

\berop\label{prop:ind-coc}
Adopt the notation of Def.\,\ref{def:CDD}.\ The \v{C}DD bicomplex canonically determines---for a sufficiently fine (simplicial) $\cO_{\xcG_\bullet}$---cochain complexes:
\qq\nn
\bigl(\G_\xcG^\bullet,\d_\xcG^{(\bullet)}\bigr)&:=&\bigl(H^0\bigl(\xcG_\bullet,\uj\bigr),[\unl{\D}{}^{(0)}_{(\bullet)}]\bigr)\,,\cr\cr 
\bigl(P_\xcG^\bullet,\widehat\d{}_\xcG^{(\bullet)}\bigr)&:=&\bigl(H^1\bigl(\xcG_\bullet,\uj\bigr),[\unl{\D}{}^{(1)}_{(\bullet)}]\bigr)\,,
\qqq
and
\qq\nn
\bigl(W_\xcG^\bullet,\widetilde\d{}_\xcG^{(\bullet)}\bigr):=\bigl(H^2\bigl(\xcG_\bullet,\uj\bigr),[\unl{\D}{}^{(2)}_{(\bullet)}]\rstr\bigr)\,,
\qqq
with the respective coboundary operators induced by the vertical differentials as
\qq\nn
[\unl{\D}{}^{(k)}_{(n)}]\colo H^k\bigl(\xcG_n,\uj\bigr)\cong\bH^k\bigl(\xcG_n,\xcD^{\bullet}(2;\xcG_n)\bigr)\ni[\xi]\longmapsto[\unl{\D}{}^{(k)}_{(n)}\xi]\in\bH^k\bigl(\xcG_{n+1},\xcD^{\bullet}(2;\xcG_{n+1})\bigr)\cong H^k\bigl(\xcG_{n+1},\uj\bigr)
\qqq
for $k\in\{0,1\}$,\ and
\qq\nn
[\unl{\D}{}^{(2)}_{(n)}]\colo H^2\bigl(\xcG_n,\uj\bigr)\cong\cW^3(\xcG_n;0)\ni[b]\longmapsto[\unl{\D}{}^{(2)}_{(n)}b]\in\cW^3(\xcG_{n+1};0)\cong H^2\bigl(\xcG_{n+1},\uj\bigr)\,,
\qqq
and with cohomology groups to be denoted as
\qq\nn
h^\bullet\bigl(\xcG,\uj\bigr)=H^\bullet\bigl(\G_\xcG^\bullet,\d_\xcG^{(\bullet)}\bigr)\,,\qquad\qquad\widehat h{}^\bullet\bigl(\xcG,\uj\bigr)=H^\bullet\bigl(P_\xcG^\bullet,\widehat\d{}_\xcG^{(\bullet)}\bigr)\,,
\qqq
and
\qq\nn
\widetilde h{}^\bullet\bigl(\xcG,\uj\bigr)=H^\bullet\bigl(W_\xcG^\bullet,\widetilde\d{}_\xcG^{(\bullet)}\bigr)\,.
\qqq
\eerop
\beroof
Obvious.\ (See:\ Prop.\,\ref{prop:BD-vs-Huj} and Thm.\,\ref{thm:W3-tors}.)
\eroof

\bedef\cite{Gawedzki:2010rn,Gawedzki:2012fu}\label{def:Gequiv-grb}
Adopt the notation of Def.\,\ref{def:mult-k-form} and Thm.\,\ref{thm:bgrb-bicat},\ and let $\rho\in\Om^2(\xcG)$ be multiplicative.\ A {\bf $\rho$-twisted $\xcG$-equivariant gerbe} is a triple $(\cG,\Upsilon,\g)$ composed of
\bit
\item a gerbe $\cG\in\bgrb_\nabla(M)$;
\item a gerbe 1-isomorphism $\Upsilon\colo d^{(1)*}_0\cG\xrightarrow{\cong}d^{(1)*}_1\cG\ox\cI_\rho$ over $\xcG$;
\item a gerbe 2-isomorphism $\g\colo(d^{(2)*}_2\Upsilon\ox\Id_{\cI_{d^{(2)*}_0\rho}})\circ d^{(2)*}_0\Upsilon\overset{\cong}{\Longrightarrow}d^{(2)*}_1\Upsilon$ over $\xcG_2$\,,
\eit
satisfying the coherence identity (with the self-explanatory subscripts on the identity 2-isomorphisms dropped for the sake of clarity)
\qq\label{eq:gamma-coh}
d^{(3)*}_2\g\bullet\bigl(\id\circ d^{(3)*}_0\g\bigr)=d^{(3)*}_1\g\bullet\bigl(\bigl(d^{(3)*}_3\g\ox\id\bigr)\circ\id\bigr)
\qqq
over $\xcG_3$.\ The pair $(\Upsilon,\g)$ is called a {\bf $\rho$-twisted $\xcG$-equivariant structure on} $\cG$.\ Whenever $\rho=0$,\ we call the corresponding $0$-twisted $\xcG$-equivariant gerbes and $0$-twisted $\xcG$-equivariant structures on them simply $\xcG${\bf -equivariant}.

A $\xcG${\bf -equivariant 1-isomorphism} between a $\rho_1$-twisted $\xcG$-equivariant gerbe $(\cG_1,\Upsilon_1,\g_1)$ and a $\rho_2$-twisted $\xcG$-equivariant gerbe $(\cG_2,\Upsilon_2,\g_2)$ is a pair $(\Phi,\varphi)$ composed of
\bit
\item a gerbe 1-isomorphism $\Phi\colo\cG_1\xrightarrow{\cong}\cG_2$ over $M$;
\item a gerbe 2-isomorphism $\varphi\colo(d^{(1)*}_1\Phi\ox\Id_{\cI_{\rho_1}})\circ\Upsilon_1\overset{\cong}{\Longrightarrow}\Upsilon_2\circ d^{(1)*}_0\Phi$ over $\xcG$,
\eit 
satisfying the coherence identity (written in the shorthand notation as above)
\qq\label{eq:Gequiv-iso-coh}
d^{(2)*}_1\varphi\bullet\bigl(\id\circ\g_1\bigr)=\bigl(\g_2\circ\id\bigr)\bullet\bigl(\id\circ d^{(2)*}_0\varphi\bigr)\bullet\bigl(\bigl(d^{(2)*}_2\varphi\ox\id\bigr)\circ\id\bigr)
\qqq
over $\xcG_2$.

A $\xcG${\bf -equivariant 2-isomorphism} between $\xcG$-equivariant 1-isomorphisms $(\Phi_c,\varphi_c),\ c\in\{1,2\}$ as above is a gerbe 2-isomorphism $\psi\colo\Phi_1\overset{\cong}{\Longrightarrow}\Phi_2$ over $M$,\ satisfying the coherence identity (written as the ones above)
\qq\nn
\varphi_2\bullet\bigl(\bigl(d^{(1)*}_1\psi\ox\id\bigr)\circ\id\bigr)=\bigl(\id\circ d^{(1)*}_0\psi\bigr)\bullet\varphi_1
\qqq
over $\xcG$.

Given a multiplicative 2-form $\rho\in\Om^2(\xcG)$,\ the corresponding $\rho$-twisted $\xcG$-equivariant gerbes together with $\xcG$-equivariant 1-isomorphisms between them,\ and $\xcG$-equivariant 2-isomorphisms between the latter compose the {\bf bicategory of $\rho$-twisted $\xcG$-equivariant gerbes} 
\qq\nn
\bgrb_\nabla(M)^{(\xcG,\rho)}\,.
\qqq 
Within the latter,\ there exists a full sub-bicategory 
\qq\nn
\bgrb_\nabla(M)^{(\xcG,0)}\equiv\bgrb_\nabla(M)^\xcG\,,
\qqq
termed the {\bf bicategory of $\xcG$-equivariant gerbes}.
\exdef

\section{A proof of Theorem \ref{thm:gauge-desc-aid}.}\label{app:proof-thm-aidesc}

Throughout this appendix,\ we assume the {\em reduced} form of the transition 1-cocycle:\ $\b_{ij}\colo O_{ij}\to\bB_\rho(\xcG)\subset\bB(\xcG)$,\ and consider smooth maps $\widetilde\b{}_{ij}\colo O_{ij}\x M\to\xcG$ from \eqref{eq:bij-tilde},\ alongside diffeomorphisms
\qq\nn
B_{ij}:=\xcF\t_i\circ\xcF\t_j^{-1}\rstr_{O_{ij}\x M}\equiv\bigl(\pr_1,t\circ\widetilde\b{}_{ij}\bigr)\colo O_{ij}\x M\too O_{ij}\x M,\ (\si,m)\longmapsto\bigl(\si,t_*\bigl(\b_{ij}(\si)\bigr)(m)\bigr)\,.
\qqq
The latter satisfy the obvious gluing identities
\qq\nn
B_{ij}\circ B_{jk}\rstr_{O_{ijk}\x M}=B_{ik}\rstr_{O_{ijk}\x M}\,,
\qqq
which reflect the 1-cocycle identities \eqref{eq:cocycle}.
\belem\label{lem:augs-glue}
The curvings $\kappa[A_i]$ from \Reqref{eq:kappAi} satisfy the identities
\qq\label{eq:augs-glue}
\kappa[A_j]\rstr_{O_{ij}\x M}-B_{ij}^*\kappa[A_i]\rstr_{O_{ij}\x M}=\widetilde\b{}_{ij}^*\rho\,.
\qqq
\elem
\begin{lemproof}
At a given point $(\si,m)\in O_{ij}\x M$,\ the pullback  
\qq\nn
\widetilde\b{}_{ij}^*\rho(\si,m)\equiv\rho\circ\wedge{}^2 \txT_{(\si,m)}\widetilde\b{}_{ij}
\qqq
localises the multiplicative 2-form $\rho$ on the bisection $\b_{ij}(\si)(M)\subset\xcG$,\ and so we may invoke Prop.\,\ref{prop:mult-on-bisec}.\ The relevant splitting (see:\ Prop.\,\ref{prop:beta-split}) is engendered by the decomposition 
\qq\nn
\txT_{(\si,m)}\widetilde\b{}_{ij}=\txT_m\bigl(\b_{ij}(\si)\bigr)+\txT_\si\widetilde\b{}_{ij}(\cdot,m)\,,
\qqq
with $\im\,\txT_\si\widetilde\b{}_{ij}(\cdot,m)\subset(\ker\,\txT s)_{\widetilde\b{}_{ij}(\si,m)}$.\ Here,\ it is to be understood that $\txT_\si\widetilde\b{}_{ij}(\cdot,m)$ (resp.\ $\txT_m(\b_{ij}(\si))$) acts on the first (resp.\ the second) summand in $\txT_\si O_{ij}\oplus\txT_m M\equiv\txT_{(\si,m)}(O_{ij}\x M)$.\ Hence,\ identity \eqref{eq:rho-on-bisec} takes the form
\qq\nn
\widetilde\b{}_{ij}^*\rho(\si,m)&=&\bigl(\kappa\circ\theta_{\rm R}\circ \txT_\si\widetilde\b{}_{ij}(\cdot,m)\circ\pr_1\bigr)\bigl(\txT_m\bigl(t_*\b_{ij}(\si)\bigr)\circ\pr_2\bigr)\cr\cr 
&&-\bigl(\kappa\circ\theta_{\rm R}\circ \txT_\si\widetilde\b{}_{ij}(\cdot,m)\circ\pr_2\bigr)\bigl(\txT_m\bigl(t_*\b_{ij}(\si)\bigr)\circ\pr_1\bigr)\cr\cr 
&&+\bigl(\kappa\circ\theta_{\rm R}\circ \txT_\si\widetilde\b{}_{ij}(\cdot,m)\circ\pr_1\bigr)\bigl(\txT_{\widetilde\b{}_{ij}(\si,m)}t\circ\txT_\si\widetilde\b{}_{ij}(\cdot,m)\circ\pr_2\bigr)\cr\cr 
&\equiv&\bigl(\kappa\circ\theta_{\rm R}\circ \txT_\si\widetilde\b{}_{ij}(\cdot,m)\circ\pr_1\bigr)\bigl(\txT_m\bigl(t_*\b_{ij}(\si)\bigr)\circ\pr_2\bigr)\cr\cr 
&&-\bigl(\kappa\circ\theta_{\rm R}\circ \txT_\si\widetilde\b{}_{ij}(\cdot,m)\circ\pr_2\bigr)\bigl(\txT_m\bigl(t_*\b_{ij}(\si)\bigr)\circ\pr_1\bigr)\cr\cr 
&&+\bigl(\kappa\circ\theta_{\rm R}\circ\txT_\si\widetilde\b{}_{ij}(\cdot,m)\circ\pr_1\bigr)\bigl(\a_\xcE\circ\theta_{\rm R}\circ\txT_\si\widetilde\b{}_{ij}(\cdot,m)\circ\pr_2\bigr)\,,
\qqq
with $\pr_1$ and $\pr_2$ projecting canonically onto the first and second vector argument of the 2-form,\ respectively.\ The $\bB_\rho(\xcG)$-equivariance of $\kappa$ and $\a_\xcE$ (see:\ Prop.\,\ref{prop:Spencer-on-holo}) enables us to rewrite the above as 
\qq\nn
\widetilde\b{}_{ij}^*\rho(\si,m)&=&\bigl(\kappa\circ\widetilde\b{}_{ij}^*\vartheta_{\rm L}(\si,m)\circ\pr_1\bigr)(\pr_2)-\bigl(\kappa\circ\widetilde\b{}_{ij}^*\vartheta_{\rm L}(\si,m)\circ\pr_2\bigr)(\pr_1)\cr\cr 
&&+\bigl(\kappa\circ\widetilde\b{}_{ij}^*\vartheta_{\rm L}(\si,m)\circ\pr_1\bigr)\bigl(\a_\xcE\circ\widetilde\b{}_{ij}^*\vartheta_{\rm L}(\si,m)\circ\pr_2\bigr)
\qqq
using the shorthand notation (see:\ Prop.\,\ref{prop:J1GonMandE})
\qq\nn
\widetilde\b{}_{ij}^*\vartheta_{\rm L}(\si,m):=\unl{\txT C}_{\b_{ij}(\si)^{-1}}\circ\theta_{\rm R}\circ \txT_\si\widetilde\b{}_{ij}(\cdot,m)=\txT_{\Id_{t_*(\b_{ij}(\si))(m)}}C_{\b_{ij}(\si)^{-1}}\circ\theta_{\rm R}\circ\txT_\si\widetilde\b{}_{ij}(\cdot,m)\,.
\qqq
Identity \eqref{eq:comom-asymm} leads to the final rewriting:
\qq\nn
\widetilde\b{}_{ij}^*\rho(\si,m)&=&\bigl(\kappa\circ\widetilde\b{}_{ij}^*\vartheta_{\rm L}(\si,m)\circ\pr_1\bigr)(\pr_2)-\bigl(\kappa\circ\widetilde\b{}_{ij}^*\vartheta_{\rm L}(\si,m)\circ\pr_2\bigr)(\pr_1)\cr\\
&&+\tfrac{1}{2}\,\bigl(\bigl(\kappa\circ\widetilde\b{}_{ij}^*\vartheta_{\rm L}(\si,m)\circ\pr_1\bigr)\bigl(\a_\xcE\circ\widetilde\b{}_{ij}^*\vartheta_{\rm L}(\si,m)\circ\pr_2\bigr) \label{eq:rho-pull-bij} \\ \cr 
&&-\bigl(\kappa\circ\widetilde\b{}_{ij}^*\vartheta_{\rm L}(\si,m)\circ\pr_2\bigr)\bigl(\a_\xcE\circ\widetilde\b{}_{ij}^*\vartheta_{\rm L}(\si,m)\circ\pr_1\bigr)\bigr)\,.\nn
 \qqq
	
On the other hand,\ we obtain (for $\pr_a\equiv\txT_{(\si,m)}\pr_a,\ a\in\{1,2\}\}$)
\qq\nn
B_{ij}^*\kappa[A_i](\si,m)&\equiv&\kappa[A_i]\circ \txT_{(\si,m)}\bigl(\pr_1,t\circ\widetilde\b{}_{ij}\bigr)^{\ox 2}\cr\cr 
&=&\kappa[A_i]\circ\bigl(\txT_\si\id_\Si,\txT_{\widetilde\b{}_{ij}(\si,m)}t\circ \txT_\si\widetilde\b{}_{ij}(\cdot,m)+\txT_m\bigl(t_*\b_{ij}(\si)\bigr)\bigr)^{\ox 2}\cr\cr 
&=&\kappa[A_i]\circ\bigl(\txT_\si\id_\Si,\a_\xcE\circ\theta_{\rm R}\circ \txT_\si\widetilde\b{}_{ij}(\cdot,m)+\txT_m\bigl(t_*\b_{ij}(\si)\bigr)\bigr)^{\ox 2}\cr\cr 
&=&\kappa[A_i]\circ\bigl(\txT_\si\id_\Si,\txT_m\bigl(t_*\b_{ij}(\si)\bigr)\circ\bigl(\txT_m\bigl(t_*\b_{ij}(\si)^{-1}\bigr)\circ\a_\xcE\circ\theta_{\rm R}\circ\txT_\si\widetilde\b{}_{ij}(\cdot,m)+\txT_m\id_M\bigr)\bigr)^{\ox 2}\cr\cr 
&=&\kappa[A_i]\circ\bigl(\txT_\si\id_\Si,\txT_m\bigl(t_*\b_{ij}(\si)\bigr)\circ\bigl(\txT_m\id_M+\a_\xcE\circ\widetilde\b{}_{ij}^*\vartheta_{\rm L}(\si,m)\bigr)\bigr)^{\ox 2}\cr\cr 
&=&\kappa[A_i]\circ\bigl(\txT_\si\id_\Si\x \txT_m\bigl(t_*\b_{ij}(\si)\bigr)\bigr)^{\ox 2}\circ\bigl(\txT_\si\id_\Si,\txT_m\id_M+\a_\xcE\circ\widetilde\b{}_{ij}^*\vartheta_{\rm L}(\si,m)\bigr)^{\ox 2}\,,
\qqq
where \eqref{eq:Bis-on-E-over-TM} has been used in the penultimate step.\ Taking into account the gluing law \eqref{eq:glue-Aij} and the $\bB(\xcG)$-equivariance of $\kappa$ and $\a_\xcE$ (see:\ Prop.\,\ref{prop:Spencer-on-holo}),\ we begin our analysis of the above expression from 
\qq\nn
&&\kappa[A_i]\circ\bigl(\id_{\txT_\si\Si}\x \txT_m\bigl(t_*\b_{ij}(\si)\bigr)\bigr)^{\ox 2}\cr\cr 
&\equiv&\bigl(t_*\b_{ij}(\si)\bigr)^*\kappa\overset{\wedge}{\circ} A_i\bigl(\si,\bigl(t_*\b_{ij}(\si)\bigr)(m)\bigr)+\tfrac{1}{2}\,\corr{\kappa\circ A_i\bigl(\si,\bigl(t_*\b_{ij}(\si)\bigr)(m)\bigr)\,\overset{\wedge}{,}\,\a_\xcE\circ A_i\bigl(\si,\bigl(t_*\b_{ij}(\si)\bigr)(m)\bigr)}\cr\cr 
&=&\bigl(t_*\b_{ij}(\si)\bigr)^*\kappa\overset{\wedge}{\circ}\txT_{\Id_m}C_{\b_{ij}(\si)}\circ\bigl(A_j+\widetilde\b{}_{ij}^*\vartheta_{\rm L}\bigr)(\si,m)\cr\cr 
&&+\tfrac{1}{2}\,\corr{\kappa\circ \txT_{\Id_m}C_{\b_{ij}(\si)}\circ\bigl(A_j+\widetilde\b{}_{ij}^*\vartheta_{\rm L}\bigr)(\si,m)\,\overset{\wedge}{,}\,\a_\xcE\circ \txT_{\Id_m}C_{\b_{ij}(\si)}\circ\bigl(A_j+\widetilde\b{}_{ij}^*\vartheta_{\rm L}\bigr)(\si,m)}\cr\cr 
&=&\kappa\overset{\wedge}{\circ}A_j(\si,m)+\kappa\overset{\wedge}{\circ}\widetilde\b{}_{ij}^*\vartheta_{\rm L}(\si,m)+\tfrac{1}{2}\,\corr{\kappa\circ\bigl(A_j+\widetilde\b{}_{ij}^*\vartheta_{\rm L}\bigr)(\si,m)\,\overset{\wedge}{,}\,\a_\xcE\circ\bigl(A_j+\widetilde\b{}_{ij}^*\vartheta_{\rm L}\bigr)(\si,m)}\cr\cr 
&\equiv&\kappa[A_j](\si,m)+\kappa\overset{\wedge}{\circ}\widetilde\b{}_{ij}^*\vartheta_{\rm L}(\si,m)+\tfrac{1}{2}\,\corr{\kappa\circ A_j\,\overset{\wedge}{,}\,\a_\xcE\circ\widetilde\b{}_{ij}^*\vartheta_{\rm L}}(\si,m)+\tfrac{1}{2}\,\corr{\kappa\circ\widetilde\b{}_{ij}^*\vartheta_{\rm L}\,\overset{\wedge}{,}\,\a_\xcE\circ A_j}(\si,m)\cr\cr 
&&+\tfrac{1}{2}\,\corr{\kappa\circ\widetilde\b{}_{ij}^*\vartheta_{\rm L}\,\overset{\wedge}{,}\,\a_\xcE\circ\widetilde\b{}_{ij}^*\vartheta_{\rm L}}(\si,m)\,.
\qqq
This we subsequently reduce to the form 
\qq\nn
&&\kappa[A_i]\circ\bigl(\id_{\txT_\si\Si}\x \txT_m\bigl(t_*\b_{ij}(\si)\bigr)\bigr)^{\ox 2}\cr\cr 
&=&\kappa[A_j](\si,m)+\kappa\overset{\wedge}{\circ}\widetilde\b{}_{ij}^*\vartheta_{\rm L}(\si,m)+\corr{\kappa\circ A_j\,\overset{\wedge}{,}\,\a_\xcE\circ\widetilde\b{}_{ij}^*\vartheta_{\rm L}}(\si,m)+\tfrac{1}{2}\,\corr{\kappa\circ\widetilde\b{}_{ij}^*\vartheta_{\rm L}\,\overset{\wedge}{,}\,\a_\xcE\circ\widetilde\b{}_{ij}^*\vartheta_{\rm L}}(\si,m)
\qqq
with the help of identity \eqref{eq:comom-asymm}.\ At this stage,\ it remains to precompose the above with the tensor square of the linear operator 
\qq\nn
\bigl(\txT_\si\id_\Si,\txT_m\id_M+\a_\xcE\circ\widetilde\b{}_{ij}^*\vartheta_{\rm L}(\si,m)\bigr)=\id_{\txT_{(\si,m)}(\Si\x M)}+\a_\xcE\circ\widetilde\b{}_{ij}^*\vartheta_{\rm L}(\si,m)\equiv\id_\txT+\a_\xcE\circ\widetilde\b{}_{ij}^*\vartheta_{\rm L}(\si,m)
\qqq
to obtain the final expression for $B_{ij}^*\kappa[A_i](\si,m)$.\ To this end,\ we calculate---invoking the foliated nature of the various forms involved and Rem.\,\ref{rem:kappAi-expl} along the way,\ and dropping $(\si,m)$ for the sake of transparency---
\qq\nn
&&\bigl(\kappa[A_j]+\corr{\kappa\circ A_j\,\overset{\wedge}{,}\,\a_\xcE\circ\widetilde\b{}_{ij}^*\vartheta_{\rm L}}\bigr)\circ\bigl(\id_\txT+\a_\xcE\circ\widetilde\b{}_{ij}^*\vartheta_{\rm L}\bigr)^{\ox 2}\cr\cr 
&=&\kappa[A_j]\circ\bigl(\id_\txT\ox\id_\txT+\id_\txT\ox\a_\xcE\circ\widetilde\b{}_{ij}^*\vartheta_{\rm L}+\a_\xcE\circ\widetilde\b{}_{ij}^*\vartheta_{\rm L}\ox\id_\txT\bigr)+\corr{\kappa\circ A_j\,\overset{\wedge}{,}\,\a_\xcE\circ\widetilde\b{}_{ij}^*\vartheta_{\rm L}}\circ\bigl(\id_\txT\ox\id_\txT\bigr)\cr\cr 
&=&\kappa[A_j]+\bigl(\kappa\overset{\wedge}{\circ}A_j\bigr)\circ\bigl(\id_\txT\ox\a_\xcE\circ\widetilde\b{}_{ij}^*\vartheta_{\rm L}+\a_\xcE\circ\widetilde\b{}_{ij}^*\vartheta_{\rm L}\ox\id_\txT\bigr)+\corr{\kappa\circ A_j\,\overset{\wedge}{,}\,\a_\xcE\circ\widetilde\b{}_{ij}^*\vartheta_{\rm L}}=\kappa[A_j]\,.
\qqq
Similarly,\ we obtain
\qq\nn
&&\bigl(\kappa\overset{\wedge}{\circ}\widetilde\b{}_{ij}^*\vartheta_{\rm L}+\tfrac{1}{2}\,\corr{\kappa\circ\widetilde\b{}_{ij}^*\vartheta_{\rm L}\,\overset{\wedge}{,}\,\a_\xcE\circ\widetilde\b{}_{ij}^*\vartheta_{\rm L}}\bigr)\circ\bigl(\id_\txT+\a_\xcE\circ\widetilde\b{}_{ij}^*\vartheta_{\rm L}\bigr)^{\ox 2}\cr\cr 
&=&\bigl(\kappa\overset{\wedge}{\circ}\widetilde\b{}_{ij}^*\vartheta_{\rm L}\bigr)\circ\bigl(\id_\txT\ox\id_\txT+\id_\txT\ox\a_\xcE\circ\widetilde\b{}_{ij}^*\vartheta_{\rm L}+\a_\xcE\circ\widetilde\b{}_{ij}^*\vartheta_{\rm L}\ox\id_\txT\bigr)\cr\cr 
&&+\tfrac{1}{2}\,\corr{\kappa\circ\widetilde\b{}_{ij}^*\vartheta_{\rm L}\,\overset{\wedge}{,}\,\a_\xcE\circ\widetilde\b{}_{ij}^*\vartheta_{\rm L}}\circ\bigl(\id_\txT\ox\id_\txT\bigr)=\kappa\overset{\wedge}{\circ}\widetilde\b{}_{ij}^*\vartheta_{\rm L}-\tfrac{1}{2}\,\corr{\kappa\circ\widetilde\b{}_{ij}^*\vartheta_{\rm L}\,\overset{\wedge}{,}\,\a_\xcE\circ\widetilde\b{}_{ij}^*\vartheta_{\rm L}}\,.
\qqq
Thus,\ altogether,\ we arrive at the formula
\qq\nn
B_{ij}^*\kappa[A_i]=\kappa[A_j]+\kappa\overset{\wedge}{\circ}\widetilde\b{}_{ij}^*\vartheta_{\rm L}-\tfrac{1}{2}\,\corr{\kappa\circ\widetilde\b{}_{ij}^*\vartheta_{\rm L}\,\overset{\wedge}{,}\,\a_\xcE\circ\widetilde\b{}_{ij}^*\vartheta_{\rm L}}\,,
\qqq
from which the desired equality \eqref{eq:augs-glue} ensues through a direct comparison with \eqref{eq:rho-pull-bij} .
\end{lemproof}
In the light of the above lemma,\ we obtain 1-isomorphisms
\qq\nn
&&\widetilde\Phi{}_{ij}:=\xcF\t_j^*\bigl(\widetilde\b{}_{ij}^*\Upsilon\ox\Id_{\cI_{B_{ij}^*\kappa[A_i]}}\bigr)\rstr_{\pi_\xcF^{-1}(O_{ij})}\colo\widetilde\cG[A_i]\rstr_{\pi_\xcF^{-1}(O_{ij})}=\xcF\t_j^*\bigl(B_{ij}^*\pr_2^*\cG\ox\cI_{B_{ij}^*\kappa[A_i]}\bigr)\rstr_{\pi_\xcF^{-1}(O_{ij})}\cr\cr 
&=&\xcF\t_j^*\bigl(\widetilde\b{}_{ij}^*d^{(1)*}_0\cG\ox\cI_{B_{ij}^*\kappa[A_i]}\bigr)\rstr_{\pi_\xcF^{-1}(O_{ij})}\xrightarrow{\ \cong\ }\xcF\t_j^*\bigl(\widetilde\b{}_{ij}^*d^{(1)*}_1\cG\ox\cI_{\widetilde\b{}_{ij}^*\rho+B_{ij}^*\kappa[A_i]}\bigr)\rstr_{\pi_\xcF^{-1}(O_{ij})}\cr\cr &=&\xcF\t_j^*\bigl(\pr_2^*\cG\ox\cI_{\widetilde\b{}_{ij}^*\rho+B_{ij}^*\kappa[A_i]}\bigr)\rstr_{\pi_\xcF^{-1}(O_{ij})}=\widetilde\cG[A_j]\rstr_{\pi_\xcF^{-1}(O_{ij})}\ox\xcF\t_j^*\bigl(\cI_{\widetilde\b{}_{ij}^*\rho+B_{ij}^*\kappa[A_i]-\kappa[A_j]}\rstr_{O_{ij}\x M}\bigr)\cr\cr 
&=&\widetilde\cG[A_j]\rstr_{\pi_\xcF^{-1}(O_{ij})}\,.
\qqq

Use the $\widetilde\b{}_{ij}$ to define smooth maps
\qq\nn
\widetilde\b{}_{ijk}:=\bigl(\widetilde\b{}_{ij}\circ B_{jk},\widetilde\b{}_{jk}\bigr)\rstr_{O_{ijk}\x M}\colo O_{ijk}\x M\too\xcG_2\,,
\qqq
satisfying the identities
\qq\label{eq:tilbijk-ids}
&d^{(2)}_0\circ\widetilde\b{}_{ijk}=\widetilde\b{}_{ij}\circ B_{jk}\rstr_{O_{ijk}\x M}\,,&\\ \cr
&d^{(2)}_1\circ\widetilde\b{}_{ijk}=\widetilde\b{}_{ik}\rstr_{O_{ijk}\x M}\,,\qquad\qquad d^{(2)}_2\circ\widetilde\b{}_{ijk}=\widetilde\b{}_{jk}\rstr_{O_{ijk}\x M}\,,&\nn
\qqq
the middle one reflecting the 1-cocycle condition \eqref{eq:cocycle}.\ We now find 2-isomorphisms
\qq\nn
&&\widetilde\varphi{}_{ijk}:=\xcF\t_k^*\bigl(\widetilde\b{}_{ijk}^*\g\ox\id_{\Id_{\cI_{B_{ik}^*\kappa[A_i]}}}\bigr)\rstr_{\pi_\xcF^{-1}(O_{ijk})}\colo\widetilde\Phi{}_{jk}\circ\widetilde\Phi{}_{ij}\rstr_{\pi_\xcF^{-1}(O_{ijk})}\cr\cr&\equiv&\bigl(\xcF\t_k^*\widetilde\b{}_{jk}^*\Upsilon\ox\Id_{\cI_{\xcF\t_j^*\kappa[A_j]}}\bigr)\circ\bigl(\xcF\t_j^*\widetilde\b{}_{ij}^*\Upsilon\ox\Id_{\cI_{\xcF\t_i^*\kappa[A_i]}}\bigr)\rstr_{\pi_\xcF^{-1}(O_{ijk})}\cr\cr
&=&\xcF\t_k^*\bigl(\bigl(\widetilde\b{}_{jk}^*\Upsilon\ox\Id_{\cI_{B_{jk}^*\kappa[A_j]}}\bigr)\circ\bigl(B_{jk}^*\widetilde\b{}_{ij}^*\Upsilon\ox\Id_{\cI_{B_{ik}^*\kappa[A_i]}}\bigr)\bigr)\rstr_{\pi_\xcF^{-1}(O_{ijk})}\cr\cr
&=&\xcF\t_k^*\bigl(\bigl(\widetilde\b{}_{ijk}^*d^{(2)*}_2\Upsilon\ox\Id_{\cI_{B_{jk}^*\kappa[A_j]}}\bigr)\circ\bigl(\widetilde\b{}_{ijk}^*d^{(2)*}_0\Upsilon\ox\Id_{\cI_{B_{ik}^*\kappa[A_i]}}\bigr)\bigr)\rstr_{\pi_\xcF^{-1}(O_{ijk})}\cr\cr
&=&\xcF\t_k^*\bigl(\bigl(\widetilde\b{}_{ijk}^*\bigl(\bigl(d^{(2)*}_2\Upsilon\ox\Id_{\cI_{d^{(2)*}_0\rho}}\bigr)\circ d^{(2)*}_0\Upsilon\bigr)\ox\Id_{\cI_{B_{ik}^*\kappa[A_i]}}\bigr)\bigr)\rstr_{\pi_\xcF^{-1}(O_{ijk})}\cr\cr 
&\xLongrightarrow{\ \cong\ }&\xcF\t_k^*\bigl(\widetilde\b{}_{ijk}^*d^{(2)*}_1\Upsilon\ox\Id_{\cI_{B_{ik}^*\kappa[A_i]}}\bigr)\rstr_{\pi_\xcF^{-1}(O_{ijk})}=\xcF\t_k^*\bigl(\widetilde\b{}_{ik}^*\Upsilon\ox\Id_{\cI_{B_{ik}^*\kappa[A_i]}}\bigr)\rstr_{\pi_\xcF^{-1}(O_{ijk})}\cr\cr 
&\equiv&\widetilde\Phi{}_{ik}\rstr_{\pi_\xcF^{-1}(O_{ijk})}\,.
\qqq

Finally,\ introduce smooth maps
\qq\nn
\widetilde\b{}_{ijkl}:=\bigl(\widetilde\b{}_{ij}\circ B_{jl},\widetilde\b{}_{jk}\circ B_{kl},\widetilde\b{}_{kl}\bigr)\rstr_{O_{ijkl}\x M}\colo O_{ijkl}\x M\too\xcG_3\,,
\qqq
satisfying the identities
\qq\label{eq:tilbijkl-ids}
&d^{(3)}_0\circ\widetilde\b{}_{ijkl}=\widetilde\b{}_{ijk}\circ B_{kl}\rstr_{O_{ijkl}\x M}\,,&\\ \cr
&d^{(3)}_1\circ\widetilde\b{}_{ijkl}=\widetilde\b{}_{ijl}\rstr_{O_{ijkl}\x M}\,,\qquad\qquad d^{(3)}_2\circ\widetilde\b{}_{ijkl}=\widetilde\b{}_{ikl}\rstr_{O_{ijkl}\x M}\,,\qquad\qquad d^{(3)}_3\circ\widetilde\b{}_{ijkl}=\widetilde\b{}_{jkl}\rstr_{O_{ijkl}\x M}\,,&\nn
\qqq
to verify\footnote{In order to keep track of the various identity morphisms appearing in the formula,\ consult \cite[App.\,H]{Gawedzki:2012fu},\ taking into account the `reverse' indexing convention for the face maps adopted in that work.}
\qq\nn
&&\widetilde\varphi{}_{ikl}\bullet\bigl(\id_{\widetilde\Phi{}_{kl}}\circ\widetilde\varphi{}_{ijk}\bigr)\rstr_{\pi_\xcF^{-1}(O_{ijkl})}\equiv\bigl(\xcF\t_l^*\widetilde\b{}_{ikl}^*\g\ox\id_{\Id_{\cI_{\xcF\t_i^*\kappa[A_i]}}}\bigr)\bullet\bigl(\id_{\widetilde\Phi{}_{kl}}\circ\bigl(\xcF\t_k^*\widetilde\b{}_{ijk}^*\g\ox\id_{\Id_{\cI_{\xcF\t_i^*\kappa[A_i]}}}\bigr)\bigr)\rstr_{\pi_\xcF^{-1}(O_{ijkl})}\cr\cr 
&=&\xcF\t_l^*\bigl(\bigl(\widetilde\b{}_{ikl}^*\g\ox\id_{\Id_{\cI_{B_{il}^*\kappa[A_i]}}}\bigr)\bullet\bigl(\id_{\widetilde\b{}_{kl}^*\Upsilon\ox\Id_{\cI_{B_{kl}^*\kappa[A_k]}}}\circ\bigl(B_{kl}^*\widetilde\b{}_{ijk}^*\g\ox\id_{\Id_{\cI_{B_{il}^*\kappa[A_i]}}}\bigr)\bigr)\rstr_{\pi_\xcF^{-1}(O_{ijkl})}\cr\cr 
&=&\xcF\t_l^*\bigl(\bigl(\widetilde\b{}_{ikl}^*\g\ox\id_{\Id_{\cI_{B_{il}^*\kappa[A_i]}}}\bigr)\bullet\bigl(\id_{\widetilde\b{}_{kl}^*\Upsilon\ox\Id_{\cI_{B_{kl}^*(B_{ik}^*\kappa[A_i]+\widetilde\b{}_{ik}^*\rho)}}}\circ\bigl(B_{kl}^*\widetilde\b{}_{ijk}^*\g\ox\id_{\Id_{\cI_{B_{il}^*\kappa[A_i]}}}\bigr)\bigr)\rstr_{\pi_\xcF^{-1}(O_{ijkl})}\cr\cr 
&=&\xcF\t_l^*\bigl(\bigl(\widetilde\b{}_{ikl}^*\g\ox\id_{\Id_{\cI_{B_{il}^*\kappa[A_i]}}}\bigr)\bullet\bigl(\id_{\widetilde\b{}_{kl}^*\Upsilon\ox\Id_{\cI_{\widetilde\b{}_{ikl}^*d^{(2)*}_0\rho}\ox\cI_{B_{il}^*\kappa[A_i]}}}\circ\bigl(B_{kl}^*\widetilde\b{}_{ijk}^*\g\ox\id_{\Id_{\cI_{B_{il}^*\kappa[A_i]}}}\bigr)\bigr)\rstr_{\pi_\xcF^{-1}(O_{ijkl})}\cr\cr 
&=&\xcF\t_l^*\bigl(\bigl(\widetilde\b{}_{ikl}^*\g\ox\id_{\Id_{\cI_{B_{il}^*\kappa[A_i]}}}\bigr)\bullet\bigl(\bigl(\id_{\widetilde\b{}_{kl}^*\Upsilon\ox\Id_{\cI_{\widetilde\b{}_{ikl}^*d^{(2)*}_0\rho}}}\circ B_{kl}^*\widetilde\b{}_{ijk}^*\g\bigr)\ox\id_{\Id_{\cI_{B_{il}^*\kappa[A_i]}}}\bigr)\rstr_{\pi_\xcF^{-1}(O_{ijkl})}\cr\cr 
&=&\xcF\t_l^*\bigl(\widetilde\b{}_{ikl}^*\g\bullet\bigl(\widetilde\b{}_{ikl}^*\id_{d^{(2)*}_2\Upsilon\ox\Id_{\cI_{d^{(2)*}_0\rho}}}\circ B_{kl}^*\widetilde\b{}_{ijk}^*\g\bigr)\ox\id_{\Id_{\cI_{B_{il}^*\kappa[A_i]}}}\bigr)\rstr_{\pi_\xcF^{-1}(O_{ijkl})}\cr\cr 
&=&\xcF\t_l^*\bigl(\widetilde\b{}_{ijkl}^*\bigl(d^{(3)*}_2\g\bullet\bigl(\id_{(d^{(2)}_2\circ d^{(3)}_2)^*\Upsilon\ox\Id_{\cI_{(d^{(2)}_0\circ d^{(3)}_2)^*\rho}}}\circ d^{(3)*}_0\g\bigr)\bigr)\ox\id_{\Id_{\cI_{B_{il}^*\kappa[A_i]}}}\bigr)\rstr_{\pi_\xcF^{-1}(O_{ijkl})}\cr\cr 
&=&\xcF\t_l^*\bigl(\widetilde\b{}_{ijkl}^*\bigl(d^{(3)*}_1\g\bullet\bigl(\bigl(d^{(3)*}_3\g\ox\id_{\Id_{\cI_{(d^{(2)}_0\circ d^{(3)}_0)^*\rho}}}\bigr)\circ\id_{(d^{(2)}_0\circ d^{(3)}_0)^*\Upsilon}\bigr)\bigr)\ox\id_{\Id_{\cI_{B_{il}^*\kappa[A_i]}}}\bigr)\rstr_{\pi_\xcF^{-1}(O_{ijkl})}\cr\cr 
&=&\xcF\t_l^*\bigl(\bigl(\widetilde\b{}_{ijl}^*\g\bullet\bigl(\bigl(\widetilde\b{}_{jkl}^*\g\ox\id_{\Id_{\cI_{B_{jl}^*\widetilde\b{}_{ij}^*\rho}}}\bigr)\circ\id_{B_{jl}^*\widetilde\b{}_{ij}^*\Upsilon}\bigr)\bigr)\ox\id_{\Id_{\cI_{B_{il}^*\kappa[A_i]}}}\bigr)\rstr_{\pi_\xcF^{-1}(O_{ijkl})}\cr\cr 
&=&\xcF\t_l^*\bigl(\bigl(\widetilde\b{}_{ijl}^*\g\ox\id_{\Id_{\cI_{B_{il}^*\kappa[A_i]}}}\bigr)\bullet\bigl(\bigl(\bigl(\widetilde\b{}_{jkl}^*\g\ox\id_{\Id_{\cI_{B_{jl}^*\widetilde\b{}_{ij}^*\rho}}}\bigr)\circ\id_{B_{jl}^*\widetilde\b{}_{ij}^*\Upsilon}\bigr)\ox\id_{\Id_{\cI_{B_{il}^*\kappa[A_i]}}}\bigr)\bigr)\rstr_{\pi_\xcF^{-1}(O_{ijkl})}\cr\cr 
&=&\xcF\t_l^*\bigl(\bigl(\widetilde\b{}_{ijl}^*\g\ox\id_{\Id_{\cI_{B_{il}^*\kappa[A_i]}}}\bigr)\bullet\bigl(\bigl(\widetilde\b{}_{jkl}^*\g\ox\id_{\Id_{\cI_{B_{jl}^*\widetilde\b{}_{ij}^*\rho+B_{il}^*\kappa[A_i]}}}\bigr)\circ\id_{B_{jl}^*\widetilde\b{}_{ij}^*\Upsilon\ox\Id_{\cI_{B_{il}^*\kappa[A_i]}}}\bigr)\bigr)\rstr_{\pi_\xcF^{-1}(O_{ijkl})}\cr\cr 
&=&\xcF\t_l^*\bigl(\bigl(\widetilde\b{}_{ijl}^*\g\ox\id_{\Id_{\cI_{B_{il}^*\kappa[A_i]}}}\bigr)\bullet\bigl(\bigl(\widetilde\b{}_{jkl}^*\g\ox\id_{\Id_{\cI_{B_{jl}^*(\widetilde\b{}_{ij}^*\rho+B_{ij}^*\kappa[A_i])}}}\bigr)\circ\id_{B_{jl}^*\widetilde\b{}_{ij}^*\Upsilon\ox\Id_{\cI_{B_{il}^*\kappa[A_i]}}}\bigr)\bigr)\rstr_{\pi_\xcF^{-1}(O_{ijkl})}\cr\cr 
&=&\xcF\t_l^*\bigl(\bigl(\widetilde\b{}_{ijl}^*\g\ox\id_{\Id_{\cI_{B_{il}^*\kappa[A_i]}}}\bigr)\bullet\bigl(\bigl(\widetilde\b{}_{jkl}^*\g\ox\id_{\Id_{\cI_{B_{jl}^*\kappa[A_j]}}}\bigr)\circ\id_{B_{jl}^*(\widetilde\b{}_{ij}^*\Upsilon\ox\Id_{\cI_{B_{ij}^*\kappa[A_i]}})}\bigr)\bigr)\rstr_{\pi_\xcF^{-1}(O_{ijkl})}\cr\cr 
&\equiv&\widetilde\varphi_{ijl}\bullet\bigl(\widetilde\varphi_{jkl}\circ\id_{\widetilde\Phi{}_{ij}}\bigr)\rstr_{\pi_\xcF^{-1}(O_{ijkl})}\,.
\qqq
Above,\ we have invoked identities \eqref{eq:augs-glue},\ \eqref{eq:tilbijk-ids} and \eqref{eq:tilbijkl-ids},\ in conjunction with the strict 2-functoriality of pullback and of the tensor product on the bicategory of gerbes (see:\ \cite{Waldorf:2007mm}).\ This completes the reconstruction of the descent datum.

In the light of Prop.\,\ref{prop:principoidle-connPhi-cech},\ the three families:
\qq\nn
&\widetilde\cG[A^\Phi_i]\equiv\widetilde\cG[\Theta^\Phi]\rstr_{\pi_\xcF^{-1}(O_i)}:=\xcF\t_i^*\bigl(\pr_2^*\cG\ox\cI_{\kappa[A^\Phi_i]}\bigr)\,,\quad i\in I\,,&\cr\cr 
&\widetilde\Phi{}^\Phi_{ij}:=\xcF\t_j^*\bigl(\widetilde\b{}_{ij}^*\Upsilon\ox\Id_{\cI_{B_{ij}^*\kappa[A^\Phi_i]}}\bigr)\rstr_{\pi_\xcF^{-1}(O_{ij})}\,,\quad i,j\in I&\cr\cr
&\widetilde\varphi{}^\Phi_{ijk}:=\xcF\t_k^*\bigl(\widetilde\b{}_{ijk}^*\g\ox\id_{\Id_{\cI_{B_{ik}^*\kappa[A^\Phi_i]}}}\bigr)\rstr_{\pi_\xcF^{-1}(O_{ijk})}\,,\quad i,j,k\in I&
\qqq
are in the same structural relations as their counterparts for $\Phi=\id_\xcP$ composing the descent datum \eqref{eq:desc-dat}.\ Hence,\ they,\ too,\ give rise to a descent datum
\qq\nn
\bigl(\{\widetilde\cG[A^\Phi_i]\}_{i\in I},\{\widetilde\Phi{}^\Phi_{ij}\}_{i,j\in I},\{\widetilde\varphi{}^\Phi_{ijk}\}_{i,j,k\in I}\bigr)\in\gt{Desc}(\check{\pi}_\xcF)\,,
\qqq
with the corresponding descended gerbe $\unl{\cG[\Theta^\Phi]}$.

Passing to the second part of the theorem,\ we start by noting that the diffeomorphism $\xcF_*(\Phi)\in\Diff(\xcF)$ restricts to each open set $\pi_\xcF^{-1}(O_{i_1 i_2\cdots i_n}),\ n\in\bN^\x$,\ and so it lifts canonically to the \Cv ech nerve $N_\bullet\pi_\xcF^{-1}\cO$ of the pullback cover $\pi_\xcF^{-1}\cO\equiv\{\pi_\xcF^{-1}(O_i)\}_{i\in I}$ of $\xcF$.\ The strict 2-functoriality of pullback then ensures that $\xcF_*(\Phi)^*\unl{\cG[\Theta^\Phi]}$ descends from the pullback \Cv ech descent datum
\qq\nn
\bigl(\{\xcF_*(\Phi)^*\widetilde\cG[A^\Phi_i]\}_{i\in I},\{\xcF_*(\Phi)^*\widetilde\Phi{}^\Phi_{ij}\}_{i,j\in I},\{\xcF_*(\Phi)^*\widetilde\varphi{}^\Phi_{ijk}\}_{i,j,k\in I}\bigr)\in\gt{Desc}(\check{\pi}_\xcF)\,.
\qqq
Thus,\ for the gauge invariance \eqref{eq:gauge-inv-descgrb},\ it is necessary and sufficient to establish the existence of a coherent 1-isomorphism
\qq\nn
\bigl(\{\widetilde\Psi{}_i\}_{i\in I},\{\widetilde\psi{}_{ij}\}_{i,j\in I}\bigr)\colo\cr\cr 
\bigl(\{\xcF_*(\Phi)^*\widetilde\cG[A^\Phi_i]\}_{i\in I},\{\xcF_*(\Phi)^*\widetilde\Phi{}^\Phi_{ij}\}_{i,j\in I},\{\xcF_*(\Phi)^*\widetilde\varphi{}^\Phi_{ijk}\}_{i,j,k\in I}\bigr)\xrightarrow{\ \cong\ }\bigl(\{\widetilde\cG{}_i\}_{i\in I},\{\widetilde\Phi{}_{ij}\}_{i,j\in I},\{\widetilde\varphi{}_{ijk}\}_{i,j,k\in I}\bigr)\,.
\qqq
Let $\Phi$ have {\em reduced} local data $\g_i\colo O_i\to\bB_\rho(\xcG)\subset\bB(\xcG)$ (see:\ Cor.\,\ref{cor:Requiv-principoidle-Auts}),\ and consider smooth maps
$\widetilde\g{}_i\colo O_i\x M\too\xcG$ from \eqref{eq:gammi-tilde},\ alongside diffeomorphisms
\qq\nn
\G_i:=\xcF\t_i\circ\xcF_*(\Phi)\circ\xcF\t_i^{-1}\equiv\bigl( \pr_1,t\circ\widetilde\g{}_i\bigr)\colo O_i\x M\too O_i\x M,\ (\si,m)\longmapsto\bigl(\si,t_*\bigl(\g_i(\si)\bigr)(m)\bigr)\,.
\qqq
These satisfy the gluing identities
\qq\label{eq:BGij-GBij}
B_{ij}\circ\G_j\rstr_{O_{ij}\x M}=\G_i\circ B_{ij}\rstr_{O_{ij}\x M}\,,
\qqq
which reflect the gauge 1-coboundary identities \eqref{eq:gau-coboundary}.
\belem\label{lem:augs-trafo}
The curvings $\kappa[A_i]$ of \Reqref{eq:kappAi} are related to those for the corresponding gauge-transforms $A^\Phi_i$ by the identities
\qq\nn
\kappa[A_i]\rstr_{O_i\x M}-\G_i^*\kappa[A_i^\Phi]\rstr_{O_i\x M}=\widetilde\g{}_i^*\rho\,.
\qqq
Moreover,\ the latter  satisfy the identities
\qq\nn
\kappa[A^\Phi_j]\rstr_{O_{ij}\x M}-B_{ij}^*\kappa[A^\Phi_i]\rstr_{O_{ij}\x M}=\widetilde\b{}_{ij}^*\rho\,.
\qqq
\elem
\begin{lemproof}
The first statement follows from a direct calculation analogous to that of Lemma \ref{lem:augs-glue}.\ The second one is implied by the same Lemma taken in conjunction with Prop.\,\ref{prop:principoidle-connPhi-cech}.
\end{lemproof}
Using the last result,\ we find 1-isomorphisms
\qq\nn
&&\widetilde\Psi{}_i:=\xcF\t_i^*\bigl(\widetilde\g{}_i^*\Upsilon\ox\Id_{\cI_{\G_i^*\kappa[A_i^\Phi]}}\bigr)\colo\xcF_*(\Phi)^*\widetilde\cG[A^\Phi_i]=\xcF\t_i^*\bigl(\G_i^*\pr_2^*\cG\ox\cI_{\G_i^*\kappa[A^\Phi_i]}\bigr)\cr\cr 
&=&\xcF\t_i^*\bigl(\widetilde\g{}_i^*d^{(1)*}_0\cG\ox\cI_{\G_i^*\kappa[A^\Phi_i]}\bigr)\xrightarrow{\ \cong\ }\xcF\t_i^*\bigl(\widetilde\g{}_i^*d^{(1)*}_1\cG\ox\cI_{\widetilde\g{}_i^*\rho+\G_i^*\kappa[A^\Phi_i]}\bigr)=\xcF\t_i^*\bigl(\pr_2^*\cG\ox\cI_{\kappa[A_i]}\bigr)=\widetilde\cG[A_i]\,.
\qqq
Define smooth maps
\qq\nn
\widetilde{\b\g}_{ij}:=\bigl(\widetilde\b{}_{ij}\circ\G_j,\widetilde\g{}_j\bigr)\rstr_{O_{ij}\x M}\colo O_{ij}\x M\too\xcG_2
\qqq
and 
\qq\nn
\widetilde{\g\b}_{ij}:=\bigl(\widetilde\g{}_i\circ B_{ij},\widetilde\b{}_{ij}\bigr)\rstr_{O_{ij}\x M}\colo O_{ij}\x M\too\xcG_2\,,
\qqq
which satisfy the identities
\qq
&d^{(2)}_0\circ\widetilde{\b\g}_{ij}=\widetilde\b{}_{ij}\circ\G_j\rstr_{O_{ij}\x M}\,,\qquad\qquad d^{(2)}_0\circ\widetilde{\g\b}_{ij}=\widetilde\g{}_i\circ B_{ij}\rstr_{O_{ij}\x M}\,,&\cr\cr
&d^{(2)}_1\circ\widetilde{\b\g}_{ij}=\widetilde\b{}_{ij}.\widetilde\g{}_j\rstr_{O_{ij}\x M}=\widetilde\g{}_i.\widetilde\b{}_{ij}\rstr_{O_{ij}\x M}=d^{(2)}_1\circ\widetilde{\g\b}_{ij}\,,\label{eq:tilbgij-ids}\\ \cr 
&d^{(2)}_2\circ\widetilde{\b\g}_{ij}=\widetilde\g{}_j\rstr_{O_{ij}\x M}\,,\qquad\qquad d^{(2)}_2\circ\widetilde{\g\b}_{ij}=\widetilde\b{}_{ij}&\nn
\qqq
We may now postulate---using the above together with \eqref{eq:BGij-GBij},\ and invoking Lemma \ref{lem:augs-trafo} along the way---the form of the anticipated 2-isomorphisms:
\qq\nn
&&\psi_{ij}:=\xcF\t_j^*\bigl(\widetilde{\g\b}_{ij}^*\g^{-1}\bullet\widetilde{\b\g}_{ij}^*\g\ox\id_{\Id_{\cI_{\G_j^*B_{ij}^*\kappa[A^\Phi_i]}}}\bigr)\colo\widetilde\Psi{}_j\circ\xcF_*(\Phi)^*\widetilde\Phi{}^\Phi_{ij}\rstr_{\pi_\xcF^{-1}(O_{ij})}\cr\cr &\equiv&\xcF\t_j^*\bigl(\bigl(\widetilde\g{}_j^*\Upsilon\ox\Id_{\cI_{\G_j^*\kappa[A_j^\Phi]}}\bigr)\circ\bigl(\G_j^*\widetilde\b{}_{ij}^*\Upsilon\ox\Id_{\cI_{\G_j^*B_{ij}^*\kappa[A^\Phi_i]}}\bigr)\bigr)\rstr_{\pi_\xcF^{-1}(O_{ij})}\cr\cr 
&=&\xcF\t_j^*\bigl(\bigl(\widetilde{\b\g}_{ij}^*d_2^{(2)\,*}\Upsilon\ox\Id_{\cI_{\G_j^*(B_{ij}^*\kappa[A^\Phi_i]+\widetilde\b{}_{ij}^*\rho)}}\bigr)\circ\bigl(\widetilde{\b\g}_{ij}^*d_0^{(2)\,*}\Upsilon\ox\Id_{\cI_{\G_j^*B_{ij}^*\kappa[A^\Phi_i]}}\bigr)\bigr)\rstr_{\pi_\xcF^{-1}(O_{ij})}\cr\cr 
&=&\xcF\t_j^*\bigl(\bigl(\widetilde{\b\g}_{ij}^*\bigl(\bigl(d_2^{(2)\,*}\Upsilon\ox\Id_{\cI_{d_0^{(2)\,*}\rho)}}\bigr)\circ d_0^{(2)\,*}\Upsilon\bigr)\ox\Id_{\cI_{\G_j^*B_{ij}^*\kappa[A^\Phi_i]}}\bigr)\bigr)\rstr_{\pi_\xcF^{-1}(O_{ij})}\xLongrightarrow{\ \cong\ }\cr\cr 
&&\xcF\t_j^*\bigl(\widetilde{\b\g}_{ij}^*d_1^{(2)\,*}\Upsilon\ox\Id_{\cI_{\G_j^*B_{ij}^*\kappa[A^\Phi_i]}}\bigr)\rstr_{\pi_\xcF^{-1}(O_{ij})}=\xcF\t_j^*\bigl(\widetilde{\g\b}_{ij}^*d_1^{(2)\,*}\Upsilon\ox\Id_{\cI_{\G_j^*B_{ij}^*\kappa[A^\Phi_i]}}\bigr)\rstr_{\pi_\xcF^{-1}(O_{ij})}\xLongrightarrow{\ \cong\ }\cr\cr 
&&\xcF\t_j^*\bigl(\bigl(\widetilde{\g\b}_{ij}^*\bigl(\bigl(d_2^{(2)\,*}\Upsilon\ox\Id_{\cI_{d_0^{(2)\,*}\rho}}\bigr)\circ d_0^{(2)\,*}\Upsilon\bigr)\ox\Id_{\cI_{\G_j^*B_{ij}^*\kappa[A^\Phi_i]}}\bigr)\bigr)\rstr_{\pi_\xcF^{-1}(O_{ij})}\cr\cr 
&=&\xcF\t_j^*\bigl(\bigl(\bigl(\widetilde\b{}_{ij}^*\Upsilon\ox\Id_{\cI_{B_{ij}^*\widetilde\g{}_i^*\rho}}\bigr)\circ B_{ij}^*\widetilde\g{}_i^*\Upsilon\bigr)\ox\Id_{\cI_{\G_j^*B_{ij}^*\kappa[A^\Phi_i]}}\bigr)\rstr_{\pi_\xcF^{-1}(O_{ij})}\cr\cr 
&=&\xcF\t_j^*\bigl(\bigl(\bigl(\widetilde\b{}_{ij}^*\Upsilon\ox\Id_{\cI_{B_{ij}^*\widetilde\g{}_i^*\rho+\G_j^*B_{ij}^*\kappa[A^\Phi_i]}}\bigr)\circ\bigl( B_{ij}^*\widetilde\g{}_i^*\Upsilon\ox\Id_{\cI_{\G_j^*B_{ij}^*\kappa[A^\Phi_i]}}\bigr)\bigr)\rstr_{\pi_\xcF^{-1}(O_{ij})}\cr\cr 
&=&\xcF\t_j^*\bigl(\bigl(\bigl(\widetilde\b{}_{ij}^*\Upsilon\ox\Id_{\cI_{B_{ij}^*\kappa[A_i]}}\bigr)\circ B_{ij}^*\bigl( \widetilde\g{}_i^*\Upsilon\ox\Id_{\cI_{\G_i^*\kappa[A^\Phi_i]}}\bigr)\bigr)\rstr_{\pi_\xcF^{-1}(O_{ij})}\equiv\widetilde\Phi{}_{ij}\circ\widetilde\Psi{}_i\rstr_{\pi_\xcF^{-1}(O_{ij})}\,.
\qqq

The very last thing to be verified is the coherence identity \eqref{eq:psi-phi-coh} for the 2-isomorphisms introduced above,\ which reads\footnote{We have dropped---for the sake of transparency---all easily deductible indices on the various identity 2-isomorphisms appearing in the formula,\ as well as the obvious restrictions to $O_{ijk}\x M$.}
\qq\nn
&&\xcF\t_k^*\bigl(\widetilde{\g\b}_{ik}^*\g^{-1}\bullet\widetilde{\b\g}_{ik}^*\g\ox\id\bigr)\bullet\bigl(\id\circ\xcF_*(\Phi)^*\xcF\t_k^*\bigl(\widetilde\b{}_{ijk}^*\g\ox\id\bigr)\bigr)\cr\cr 
&=&\bigl(\xcF\t_k^*\bigl(\widetilde\b{}_{ijk}^*\g\ox\id\bigr)\circ\id\bigr)\bullet\bigl(\id\circ\xcF\t_j^*\bigl(\widetilde{\g\b}_{ij}^*\g^{-1}\bullet\widetilde{\b\g}_{ij}^*\g\ox\id\bigr)\bigr)\bullet\bigl(\xcF\t_k^*\bigl(\widetilde{\g\b}_{jk}^*\g^{-1}\bullet\widetilde{\b\g}_{jk}^*\g\ox\id\bigr)\circ\id\bigr)\,.
\qqq 
Using the strict 2-functoriality of pullback to strip the above formula of $\xcF\t_k^*$,\ we end up with an equivalent one:
\qq\nn
&&\bigl(\widetilde{\g\b}_{ik}^*\g^{-1}\bullet\widetilde{\b\g}_{ik}^*\g\ox\id\bigr)\bullet\bigl(\id\circ\G_k^*\bigl(\widetilde\b{}_{ijk}^*\g\ox\id\bigr)\bigr)\cr\cr 
&=&\bigl(\bigl(\widetilde\b{}_{ijk}^*\g\ox\id\bigr)\circ\id\bigr)\bullet\bigl(\id\circ B_{jk}^*\bigl(\widetilde{\g\b}_{ij}^*\g^{-1}\bullet\widetilde{\b\g}_{ij}^*\g\ox\id\bigr)\bigr)\bullet\bigl(\bigl(\widetilde{\g\b}_{jk}^*\g^{-1}\bullet\widetilde{\b\g}_{jk}^*\g\ox\id\bigr)\circ\id\bigr)\,.
\qqq
In order to analyse the latter,\ we simplify the notation further by writing
\qq\nn
&\widetilde{\g\b}_{ij}\equiv\bigl(\g_i\circ t\circ\b_{ij},\b_{ij}\bigr)\,,\qquad\qquad\widetilde{\b\g}_{ij}\equiv\bigl(\b_{ij}\circ t\circ\g_j,\g_j\bigr)\,,\qquad\qquad\widetilde\b{}_{ijk}\equiv\bigl(\b_{ij}\circ t\circ\b_{jk},\b_{jk}\bigr)\,,&\cr\cr 
&\G_k\equiv t\circ\g_k\,,\qquad\qquad B_{jk}\equiv t\circ\b_{jk}\,,\qquad\qquad{\rm {\it etc.}}&
\qqq
We thus arrive at the fully fledged form
\qq\nn
&&\bigl(\bigl(\g_i\circ t\circ\b_{ik},\b_{ik}\bigr)^*\g^{-1}\bullet\bigl(\b_{ik}\circ t\circ\g_k,\g_k\bigr)^*\g\ox\id\bigr)\bullet\bigl(\id\circ\bigl(\bigl(\b_{ij}\circ t\circ\b_{jk}\circ t\circ\g_k,\b_{jk}\circ t\circ\g_k\bigr)^*\g\ox\id\bigr)\bigr)\cr\cr 
&=&\bigl(\bigl(\bigl(\b_{ij}\circ t\circ\b_{jk},\b_{jk}\bigr)^*\g\ox\id\bigr)\circ\id\bigr)\cr\cr 
&&\bullet\bigl(\id\circ \bigl(\bigl(\g_i\circ t\circ\b_{ij}\circ t\circ\b_{jk},\b_{ij}\circ t\circ\b_{jk}\bigr)^*\g^{-1}\bullet\bigl(\b_{ij}\circ t\circ\g_j\circ t\circ\b_{jk},\g_j\circ t\circ\b_{jk}\bigr)^*\g\ox\id\bigr)\bigr)\cr\cr 
&&\bullet\bigl(\bigl(\bigl(\g_j\circ t\circ\b_{jk},\b_{jk}\bigr)^*\g^{-1}\bullet\bigl(\b_{jk}\circ t\circ\g_k,\g_k\bigr)^*\g\ox\id\bigr)\circ\id\bigr)\,.
\qqq
On the right-hand side,\ we note the composition 
\qq\nn
&&\bigl(\bigl(\bigl(\b_{ij}\circ t\circ\b_{jk},\b_{jk}\bigr)^*\g\ox\id\bigr)\circ\id\bigr)\bullet\bigl(\id\circ \bigl(\bigl(\g_i\circ t\circ\b_{ij}\circ t\circ\b_{jk},\b_{ij}\circ t\circ\b_{jk}\bigr)^*\g^{-1}\ox\id\bigr)\bigr)\cr\cr 
&\equiv&\bigl(\g_i\circ t\circ\b_{ij}\circ t\circ\b_{jk},\b_{ij}\circ t\circ\b_{jk},\b_{jk}\bigr)^*\bigl[\bigl(\bigl(\pr_{2,3}^*\g\ox\id\bigr)\circ\id\bigr)\bullet\bigl(\id\circ \bigl(\pr_{1,2}^*\g^{-1}\ox\id\bigr)\bigr)\bigr]\cr\cr 
&=&\bigl(\g_i\circ t\circ\b_{ij}\circ t\circ\b_{jk},\b_{ij}\circ t\circ\b_{jk},\b_{jk}\bigr)^*\bigl((\id_\xcG\x\txm)^*\g^{-1}\bullet(\txm\x\id_\xcG)^*\g\ox\id\bigr)\cr\cr 
&=&\bigl(\g_i\circ t\circ\b_{ik},\b_{ik}\bigr)^*\g^{-1}\bullet\bigl(\bigl(\g_i.\b_{ij}\bigr)\circ t\circ\b_{jk},\b_{jk}\bigr)^*\g\ox\id\,,
\qqq
rewritten conveniently with the help of identity \eqref{eq:gamma-coh} in the equivalent form
\qq\nn
\bigl(\bigl(\pr_{2,3}^*\g\ox\id\bigr)\circ\id\bigr)\bullet\bigl(\id\circ \pr_{1,2}^*\g^{-1}\bigr)=(\id_\xcG\x\txm)^*\g^{-1}\bullet(\txm\x\id_\xcG)^*\g\,.
\qqq
The first term in the thus obtained expression cancels out against the first term on the left-hand side of the equality under verification,\ and so the latter reduces to 
\qq\nn
&&\bigl(\bigl(\b_{ik}\circ t\circ\g_k,\g_k\bigr)^*\g\ox\id\bigr)\bullet\bigl(\id\circ\bigl(\bigl(\b_{ij}\circ t\circ\b_{jk}\circ t\circ\g_k,\b_{jk}\circ t\circ\g_k\bigr)^*\g\ox\id\bigr)\bigr)\cr\cr 
&=&\bigl(\bigl(\bigl(\g_i.\b_{ij}\bigr)\circ t\circ\b_{jk},\b_{jk}\bigr)^*\g\ox\id\bigr)\bullet\bigl(\id\circ\bigl(\bigl(\b_{ij}\circ t\circ\g_j\circ t\circ\b_{jk},\g_j\circ t\circ\b_{jk}\bigr)^*\g\ox\id\bigr)\bigr)\cr\cr 
&&\bullet\bigl(\bigl(\bigl(\g_j\circ t\circ\b_{jk},\b_{jk}\bigr)^*\g^{-1}\bullet\bigl(\b_{jk}\circ t\circ\g_k,\g_k\bigr)^*\g\ox\id\bigr)\circ\id\bigr)\,.
\qqq
Similarly,\ we rewrite the composition of the second term and the third term on the right-hand side of the equality as
\qq\nn
&&\bigl(\id\circ\bigl(\bigl(\b_{ij}\circ t\circ\g_j\circ t\circ\b_{jk},\g_j\circ t\circ\b_{jk}\bigr)^*\g\ox\id\bigr)\bigr)\bullet\bigl(\bigl(\bigl(\g_j\circ t\circ\b_{jk},\b_{jk}\bigr)^*\g^{-1}\ox\id\bigr)\circ\id\bigr)\cr\cr 
&\equiv&\bigl(\b_{ij}\circ t\circ\g_j\circ t\circ\b_{jk},\g_j\circ t\circ\b_{jk},\b_{jk}\bigr)^*\bigl(\bigl(\id\circ\pr_{1,2}^*\g\ox\id\bigr)\bullet\bigl(\bigl(\pr_{2,3}^*\g^{-1}\ox\id\bigr)\circ\id\bigr)\bigr)\cr\cr 
&=&\bigl(\b_{ij}\circ t\circ\g_j\circ t\circ\b_{jk},\g_j\circ t\circ\b_{jk},\b_{jk}\bigr)^*\bigl((\txm\x\id_\xcG)^*\g^{-1}\bullet(\id_\xcG\x\txm)^*\g\ox\id\bigr)\cr\cr 
&=&\bigl(\bigl(\b_{ij}.\g_j\bigr)\circ t\circ\b_{jk},\b_{jk} \bigr)^*\g^{-1}\bullet\bigl(\b_{ij}\circ t\circ\g_j\circ t\circ\b_{jk},\g_j.\b_{jk}\bigr)^*\g\ox\id
\qqq
upon invoking another equivalent rephrasal of identity \eqref{eq:gamma-coh} given by
\qq\nn
\bigl(\id\circ \pr_{1,2}^*\g\bigr)\bullet\bigl(\bigl(\pr_{2,3}^*\g^{-1}\ox\id\bigr)\circ\id\bigr)=(\txm\x\id_\xcG)^*\g^{-1}\bullet(\id_\xcG\x\txm)^*\g\,.
\qqq
The first term in the ensuing expression cancels out against the first term on the right-hand side of the equality under verification (in the above partially reduced form) by \eqref{eq:gau-coboundary},\ and so we end up with
\qq\nn
&&\bigl(\bigl(\b_{ik}\circ t\circ\g_k,\g_k\bigr)^*\g\ox\id\bigr)\bullet\bigl(\id\circ\bigl(\bigl(\b_{ij}\circ t\circ\b_{jk}\circ t\circ\g_k,\b_{jk}\circ t\circ\g_k\bigr)^*\g\ox\id\bigr)\bigr)\cr\cr 
&=&\bigl(\bigl(\b_{ij}\circ t\circ\g_j\circ t\circ\b_{jk},\g_j.\b_{jk}\bigr)^*\g\ox\id\bigr)\bullet\bigl(\bigl(\bigl(\b_{jk}\circ t\circ\g_k,\g_k\bigr)^*\g\ox\id\bigr)\circ\id\bigr)\,.
\qqq
In the last step,\ we invoke identity \eqref{eq:cocycle} and,\ once more,\ identity \eqref{eq:gau-coboundary} to rewrite the above equality-to-be as
\qq\nn
&&\bigl(\b_{ij}\circ t\circ\g_j\circ t\circ\b_{jk},\b_{jk}\circ t\circ\g_k,\g_k\bigr)^*\bigl((\txm\x\id_\xcG)^*\g\ox\id\bigr)\bullet\bigl(\id\circ\bigl(\pr_{1,2}^*\g\ox\id\bigr)\bigr)\cr\cr 
&=&\bigl(\b_{ij}\circ t\circ\g_j\circ t\circ\b_{jk},\b_{jk}\circ t\circ\g_k,\g_k\bigr)^*\bigl((\id_\xcG\x\txm)^*\g\ox\id\bigr)\bullet\bigl(\pr_{2,3}^*\g\ox\id\bigr)\circ\id\bigr)\,,
\qqq
whereupon it becomes clear that it is a pullback of (a trivial tensor extension of) identity \eqref{eq:gamma-coh} in its original form
\qq\nn
(\txm\x\id_\xcG)^*\g\bullet\bigl(\id\circ \pr_{1,2}^*\g\bigr)=(\id_\xcG\x\txm)^*\g\bullet\bigl(\bigl(\pr_{2,3}^*\g\ox\id\bigr)\circ\id\bigr)\,.
\qqq
\qed

\end{document}